\documentclass[
 reprint,
 amsmath,amssymb,
 aps,
 floatfix
]{revtex4-2}

\usepackage{graphicx}
\usepackage{subcaption}
\usepackage{hyperref}
\usepackage[linesnumbered, ruled, vlined, titlenotnumbered, noend]{algorithm2e}
\usepackage{float}
\usepackage{amsmath, amsfonts, amssymb, amsthm, color}
\usepackage{tikz-cd}
\usepackage[capitalise]{cleveref}
\crefname{algocf}{alg.}{algs.}
\Crefname{algocf}{Algorithm}{Algorithms}
\usepackage{adjustbox}
\usepackage{tikz-cd}
\usepackage{siunitx}

\usepackage{tabularx}
\usepackage{multirow}
\usepackage{ragged2e}
\usepackage{makecell}
\usepackage{appendix}
\usepackage{xspace}
\usepackage{booktabs}

\usepackage[T1]{fontenc}

\newcolumntype{L}{>{\RaggedRight\arraybackslash}X}

\newtheorem{definition}{Definition}
\newtheorem{heuristic}{Heuristic}
\newtheorem{proposition}{Proposition}

\newtheorem{remark}{Remark}
\crefname{heuristic}{Heuristic}{Heuristics}
\Crefname{heuristic}{Heuristic}{Heuristics}

\newcommand{\ket}[1]{|#1\rangle}
\newcommand{\N}{{\mathbb{N}}}
\newcommand{\Z}{{\mathbb{Z}}}

\newcommand{\C}{{\mathbb{C}}}
\newcommand{\Pauli}{{\bar P}}

\newcommand{\PauliWeight}{\omega}
\newcommand{\PauliOpt}{\Pauli_{\text{opt}}}

\newcommand{\Stab}{{\cal S}}
\newcommand{\EDM}{\ensuremath{\mathrm{EDM}}\xspace}
\newcommand{\ECM}{\ensuremath{\mathrm{ECM}}\xspace}

\newcommand{\pEDM}{p_{\EDM r}}
\newcommand{\pECM}{p_{\ECM r}}
\newcommand{\pflip}{p_{\text{flip}}}
\newcommand{\pfitEDM}{p_{\EDM}^{\text{fit}}}
\newcommand{\pfitECM}{p_{\ECM}^{\text{fit}}}
\newcommand{\codeC}{\cal{C}}

\newcommand{\MEK}{\ensuremath{\mathrm{MEK}}\xspace}
\newcommand{\CHfactory}{\ensuremath{\mathrm{CH2}}\xspace}

\newcommand{\CX}{\ensuremath{\mathrm{CX}}\xspace}
\newcommand{\CY}{\ensuremath{\mathrm{CY}}\xspace}
\newcommand{\CZ}{\ensuremath{\mathrm{CZ}}\xspace}

\newcommand{\AccessibleLogicalOperators}{{\Omega}_{w}}

\newcommand{\logicalWidth}{w}
\newcommand{\physicalWidth}{\bar w}

\DeclareMathOperator{\rank}{rank}
\newcommand{\logicalBasis}{{\cal B}}
\newcommand{\PauliGroup}{{\cal P}}

\newcommand{\ploss}{p_\text{loss}}
\newcommand{\pleak}{p_\text{leak}}
\newcommand{\Prob}{\text{Prob}}
\newcommand{\code}[1]{\ensuremath{\mathrm{Q{#1}}}\xspace}
\newcommand{\conf}[4]{{#1}\times\code{#2}+{#3}\times\text{#4}}

\newcommand{\FLE}{FLE}

\begin{document}

\title{Fault-Tolerant Quantum Computing with Trapped Ions: The Walking Cat Architecture}
\author{
Felix Tripier,
Woo Chang Chung,
Jacob Young,
Safwan Alam,
Bryce Bjork,
Aharon Brodutch,
Finn Lasse Buessen,
Nolan J. Coble,
Thomas Dellaert,
Dmitri Maslov,
Martin Roetteler,
Edwin Tham,
Mark Webster,
Min Ye,
John Gamble,
Andrii Maksymov,
J. P. Marceaux,
Nicolas Delfosse
}
\affiliation{
    IonQ Inc.
}
\date{\today}

\begin{abstract}
We propose a fault-tolerant quantum computer architecture for trapped-ion devices, which we call the walking cat architecture.
Our blueprint includes a compiler, a detailed description of all the quantum error-correction protocols, a micro-architecture, a sufficiently fast decoder, and thorough simulations.
The backbone of the architecture is a cat factory, producing cat states distributed throughout the machine, which are consumed to perform logical operations.
The walking cat architecture is based entirely on a modern quantum error-correction approach called low-density parity-check (LDPC) codes.
We design two factories that produce magic states directly in a quantum LDPC code, based respectively on the Meier, Eastin, and Knill scheme and on cat-based Clifford measurements.

We identify promising instances of the walking cat architecture, such as 
(1) a simple architecture based on a single LDPC code used both as a memory and as a magic factory, 
(2) a fast architecture based on fast logical gates relying on a [[70, 6, 9]] code we introduce here, equipped with Clifford-frame tracking for any 6-qubit Clifford gate, and
(3) a dense architecture based on a new [[102, 22, 9]] code encoding 22 logical qubits per memory block.
Our dense architecture provides a design with 110 logical qubits executing about one million $T$ gates per day using only 2,514 physical qubits, which counts all qubits used for correction of errors, leakage and loss, magic factories, cat factories, reservoirs, and routing qubits.
Using our fast architecture with 10,000 physical qubits, we estimate that the quantum Hamiltonian simulation of a Heisenberg model on 100 sites can be executed within one month, including all shots required to achieve chemical accuracy, suggesting that such a device could enter the regime of classically intractable physics simulations.

Our design relies on hardware components that have been experimentally demonstrated on small devices.
We emphasize simplicity over hypothetical performance to facilitate the practical realization of this machine.
Based on this approach, we believe that a fault-tolerant quantum computer with hundreds of logical qubits capable of running millions of logical gates can be built in the near term, providing a platform to explore a broad range of applications.
\end{abstract}

\maketitle

\onecolumngrid 

\clearpage
\twocolumngrid
\tableofcontents
\clearpage

\clearpage
\newpage

\part{Introduction and overview}
\label{part:Introduction and overview}
\section{Introduction}
\label{sec:introduction}

\begin{figure}
    \centering
    \includegraphics[width=0.8\linewidth]{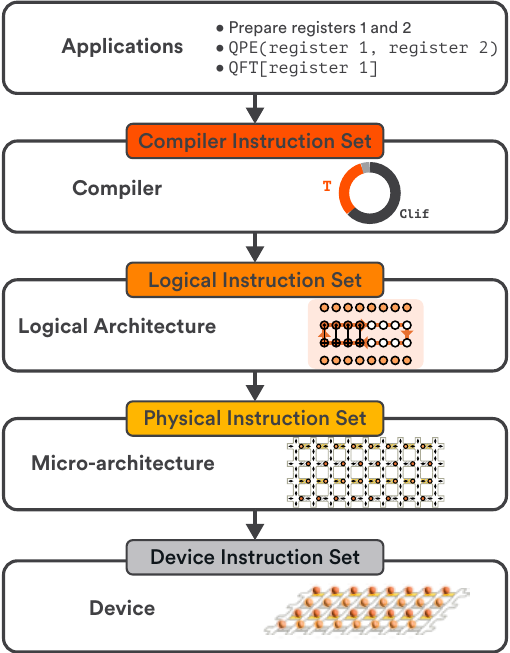}
    \caption{Layers of abstractions of an FTQC architecture.
    Depending on the context, one may consider a refined decomposition of some of the layers.
    The top layer provides a description of a quantum program. 
    It is mapped onto a sequence of logical instructions by the compiler.
    Following the convention of the quantum computing community, we use the term logical as a synonym for error-corrected and we refer to the operations used to design logical instructions as physical instructions.
    The physical instructions are mapped onto the device instruction by the micro-architecture.
    In the case of trapped ions, the device instructions represent the operations available in a QCCD quantum computer~\cite{kielpinski2002architecture, malinowski2023wire}.
    This paper focuses on the three inner layers of this stack.
    A more refined stack used in classical computer architecture can be found in~\cite{harris2021digital}.
    }
    \label{fig:intro_stack}
\end{figure}

Current quantum computers, broadly classified as Noisy Intermediate-Scale Quantum (NISQ) devices~\cite{preskill2018quantum}, can run thousands of gates on hundreds of physical qubits~\footnote{Following the convention of the quantum computing community, we refer to error-corrected qubits as logical qubits and the qubits used to encode logical qubits are called physical qubits. This term does not refer to a physical implementation of the qubits but to a model used to describe quantum error correction protocols.}, but rapidly accumulate errors during execution. While such devices are experimentally impressive, solving utility-scale problems will likely require a fault-tolerant quantum computer (FTQC) capable of reliably executing tens of millions of gates on thousands of qubits \cite{babbush2025grand}.
The transition from a NISQ device to a large-scale FTQC calls for more than just a substantial increase in gate and qubit count; it requires a paradigm shift in terms of architecture~\cite{shor1996fault}:
Whereas a NISQ application is compiled directly into the physical instructions available on the physical qubits, an FTQC application must be decomposed into logical instructions executed on logical qubits, which are themselves encoded into blocks of physical qubits that are corrected regularly to avoid the accumulation of errors.
This imposes stringent constraints on the design of an FTQC, requiring a fundamental rethinking of the architecture from compilation to micro-architecture, with an entirely new intermediate layer describing error correction and logical instructions as represented in \cref{fig:intro_stack}.

Quantum computation at the logical qubit level incurs significant qubit and gate overheads, but improved resource estimates suggest that practical applications of quantum computing may yet be closer than previously expected~\cite{babbush2025grand, reiher2017elucidating, beverland2022assessing, dalzell2023quantum, gidney2021factor}. They show, for instance, that an FTQC with about 100,000 physical qubits could be sufficient to break widely adopted cryptographic protocols--- uch as 2,048-bit RSA or 256-bit ECC---in a reasonable amount of time~\cite{gouzien2023performance, gidney2025factor, webster2026pinnacle, cain2026shor, babbush2026securing}.
Such estimations rely on the use of modern quantum error-correcting codes~\footnote{In classical information theory, the term ``modern'' typically refers to graph-based codes such as LDPC codes or turbo codes, in opposition to older ``algebraic'' constructions~\cite{richardson2008modern}. Similarly, by ``modern quantum error-correcting codes'' we mean high rate quantum LDPC codes, which have seen a growing adoption recently, in opposition to topological codes like surface codes.} such as quantum low-density parity-check (LDPC) codes~\cite{mackay2004sparse, breuckmann2021quantum}, more efficient operations~\cite{cohen2022low, gidney2024magic, cowtan2024css, swaroop2024universal, cross2024improved, cowtan2024ssip, baspin2025fast, xu2024fast, zheng2025high}, quantum algorithm improvements~\cite{babbush2026securing}, and significant expectations for hardware advances.

Despite great progress toward resource estimates for quantum computation~\cite{gottesman2013fault, strikis2023quantum, xu2024constant, yoder2025tour, webster2026pinnacle, cain2026shor}, an end-to-end blueprint for an FTQC architecture based on modern quantum error-correcting codes and designed with realistic engineering constraints in mind is still missing in the literature. 
In this work, we bridge the gap from NISQ devices to FTQCs by providing a detailed FTQC architecture for a trapped-ion quantum computer~\cite{bruzewicz2016scalable}, including compilation, quantum error correction, and micro-architecture, together with a sufficiently fast decoder to correct errors faster than they accumulate, based entirely on quantum LDPC codes. 
We rely extensively on architecture principles and techniques developed by the classical computer architecture community to leverage their extensive experience in computer design~\cite{patterson2016computer, harris2021digital}. 
In particular, we borrow the three design principles from~\cite{harris2021digital}: hierarchy, modularity, and regularity, which we complement with simplicity, crucial when building a new type of device with formidable complexity.
See~\cref{sec:Architecture design principles} for a discussion on the design principles we follow.

Our architecture for a trapped-ion FTQC is capable of running millions of gates on hundreds of logical qubits using only a few thousand physical qubits, and is based on hardware components already experimentally demonstrated in small devices.
We select the regime of hundreds of logical qubits and millions of gates because it is far beyond classical simulation capability, it surpasses any NISQ machine, and we expect that such a machine would empower the broad scientific community to explore new quantum computing applications.
We anticipate that a device utilizing our proposed architecture can be built in the near term, constituting a major advance in quantum computational power and being a stepping stone toward larger FTQCs.

The primary goal of this paper is to provide a blueprint to enable this breakthrough, including a complete description of the architecture components and sub-components, as well as thorough simulations of their performance.
The result is an FTQC design capable of executing a complex quantum program described in a high-level language, which is then mapped onto a sequence of low-level device instructions, which, in our case, refers to the instructions available in a trapped-ion quantum computer.
These low-level instructions are described by the Quantum Charge-Coupled Device (QCCD) architecture~\cite{kielpinski2002architecture, malinowski2023wire}, which allows us to move ions in a two-dimensional chip and perform single-qubit gates, two-qubit gates, and measurements in dedicated zones of the chip.
The QCCD architecture has been validated experimentally on devices with up to 98 qubits~\cite{pino2021demonstration, moses2023race, delaney2024scalable, dasu2026computing} and many trapped-ion experiments have been reported~\cite{monz2016realization, debnath2016demonstration, monz201114, nam2020ground, egan2021fault, egan2021scaling, postler2022demonstration, kranzl2022controlling, zhang2017observation, kamsap2017experimental, ryan2021realization, brown2023advances, chen2024benchmarking, ryan2024high, paetznick2024demonstration, wang2024fault, yamamoto2025quantum, daguerre2025experimental}. NISQ architectures for trapped ion quantum computers are considered in~\cite{kielpinski2002architecture, lekitsch2017blueprint, murali2020architecting, wu2021tilt, murali2022toward, schoenberger2024shuttling, malinowski2023wire} and certain aspects of FTQC architecture based on surface codes and color codes are discussed in~\cite{monroe2014large, brown2016co, leblond2023tiscc, schwerdt2024scalable, baek2025sdqc, jones2026architecting, lee2026ion}. We significantly expand this line of work by providing a unified picture from compilation to micro-architecture, using a modern approach based on quantum LDPC codes and achieving a more efficient encoding.

Our proposed FTQC architecture, which we call the {\em walking cat architecture}, is designed based on the three levels of abstraction corresponding to layers of the hierarchy in \cref{fig:intro_stack}: it includes a {\em compiler}, producing a sequence of logical instructions corresponding to an input quantum program, a {\em logical architecture}, providing a description of the logical operations available, and a {\em micro-architecture} mapping the logical instructions onto QCCD instructions.

The walking cat architecture relies on two essential features of trapped ions. 
First, we leverage the record fidelity of electronic qubit control, demonstrated experimentally with two-qubit gates fidelity above 99.99\%~\cite{hughes2025trapped} and single-qubit gates fidelity above 99.999\%~\cite{loschnauer2025scalable}.
These high-fidelity operations allow for the preparation of large resource states---which are post-selected with a small rejection probability to remove residual errors---in a regime where more noisy technologies would be facing rejection probabilities close to~1.
The resource states we rely on are physical cat states, after which we name our architecture.
Cat states appeared in fault-tolerant quantum computing as early as 1996, when Shor built the first fault-tolerant quantum error correction scheme~\cite{shor1996fault}.
However, they are typically reserved for small codes and rarely adopted in practical FTQC architectures for the reasons explained above.
Second, the ability to reliably move ions~\cite{kielpinski2002architecture, ransford2025helios} lets us transport qubits through the machine.
This unlocks the ability to implement non-local quantum error-correcting codes such as quantum LDPC codes, known to outperform surface codes~\cite{tremblay2022constant, bravyi2024high}, providing a simple alternative to the manufacture of long-range couplers required to implement these codes with superconducting qubits~\cite{bravyi2024high, webster2026pinnacle}.
We also leverage qubit transport to establish connections between physically separated logical qubits, allowing us to build fully connected logical qubits.

We discuss several instances of the walking cat architecture, achieving different tradeoffs between space, time and simplicity.
We showcase three carefully engineered quantum LDPC codes with parameters $[[102, 22, 9]]$, $[[70, 6, 9]]$ and $[[54, 2, 10]]$, which we refer to as $\code{102}$, $\code{70}$ and $\code{54}$, respectively.
\begin{enumerate}
\item Our simplest instance relies entirely on a single quantum error-correcting code, in contrast to other proposed architectures which use a variety of codes for memory, magic state cultivation, CCZ state production~\cite{yoder2025tour, webster2026pinnacle, cain2026shor}.
The code $\code{70}$ is carefully selected to allow for the implementation of the magic state distillation scheme of Meier, Eastin, and Knill, referred to as the MEK scheme~\cite{meier2013magic}.
This code is used to perform quantum error correction in the memory and to produce magic states, unifying the design of the memory block and the magic factory.
As a memory, it stores 6 logical qubits, and as a magic factory it produces 2 logical magic states which can be consumed to perform two $T$ gates.
The {\em single-code architecture} discussed in \cref{subsec:The single-code architecture} allows to adjust the allocation of resources to memory or magic state production during the computation to accommodate the compiler needs.
\item To obtain faster gates, we design a more efficient magic factory utilizing cat-based Clifford measurements in $\code{54}$, which we call \CHfactory.
Moreover, our architecture provides a quasi-instantaneous implementation of any Pauli operation and any 6-qubit logical Clifford gate by software-based frame-tracking inside any memory block based on $\code{70}$. For comparison, the surface code generally only allows for the frame-tracking of Pauli operations~\cite{litinski2019game}.
\item We propose a high-rate memory based on $\code{102}$, which encodes 22 logical qubits into 102 physical qubits, providing 2.75x more logical qubits than biplanar LDPC codes of~\cite{bravyi2024high} with comparable length and minimum distance ({\em e.g.}, $[[90, 8, 10]]$, or $[[108, 8, 10]]$). Moreover, we prove in \cref{prop:biplanar_weight_eight_obstruction} that such quantum LDPC codes cannot be implemented within the biplanar architecture, illustrating the greater flexibility of trapped ions.
For comparison, if we were to use surface codes, encoding 22 logical qubits with the same minimum distance would require 1782 physical qubits instead of 102 using $\code{102}$.
\end{enumerate}

\cref{tab:intro_example_table} illustrates the performance of different instances of the walking cat architecture with about 100 to 200 logical qubits, capable of implementing up to 1 to 10 million $T$ gates per day (including magic state production and $T$ gate implementation).
The $\conf{5}{102}{1}{CH2}$ configuration provides a memory of 110 logical qubits and executes up to a million $T$ gates per day using only around 2,500 qubits.
With 7,600 qubits, one can reach 220 logical qubits and more than 10 million $T$ gates per day.

\cref{tab:intro_example_table} corresponds to a target logical error rate of $10^{-10}$. A larger number of logical qubits can be obtained when targeting a higher logical rate or by varying the degree of parallelism of logical operations.

\begin{table}[ht]
\centering
\begin{tabular}{|l|r|r|r|}
\hline
{\bf Configuration} & \makecell{\bf Logical\\\bf qubits} & \bf T gates/day & \makecell{\bf Physical\\\bf qubits} \\
\hline
$\conf{17}{70}{3}{MEK}$   & 102 & 1.3M &  5,722 \\
$\conf{17}{70}{1}{CH2}$   & 102 & 1.1M &  5,280 \\
$\conf{5}{102}{1}{CH2}$   & 110 & 1.0M &  2,514 \\
\hline
$\conf{17}{70}{24}{MEK}$  & 102 & 10.4M & 11,847 \\
$\conf{17}{70}{9}{CH2}$   & 102 & 10.3M &  8,114 \\
$\conf{5}{102}{10}{CH2}$  & 110 & 10.5M &  5,475 \\
\hline\hline
$\conf{34}{70}{3}{MEK}$   & 204 & 1.3M & 10,625 \\
$\conf{34}{70}{1}{CH2}$   & 204 & 1.1M & 10,239 \\
$\conf{10}{102}{1}{CH2}$  & 220 & 1.0M &  4,540 \\
\hline
$\conf{34}{70}{24}{MEK}$  & 204 & 10.4M & 17,000 \\
$\conf{34}{70}{9}{CH2}$   & 204 & 10.3M & 13,264 \\
$\conf{10}{102}{10}{CH2}$ & 220 & 10.5M &  7,559 \\
\hline
\end{tabular}
\caption{Examples of instances of the walking cat architecture and their computational power. A expanded version of this table including the detailed qubit allocation can be found in \cref{tab:qubit_allocation_details}.}
\label{tab:intro_example_table}
\end{table}

We estimate that Shor's algorithm can be executed in an instance of a walking cat architecture with the configuration $\conf{34}{70}{9}{CH2}$, which consumes about 13,000 physical qubits to factor 30-bit numbers within less than a day.
This allows us to factor a number such as $1{,}071{,}514{,}531 = 32{,}749 \times 32{,}719$. For comparison, the record-size implementation of Shor's algorithm in a quantum device remains to be the factorization of $15=3 \times 5$~\cite{vandersypen2001experimental}. See~\cite{gidney2025_factor21} for a discussion on the difficulty of quantum factorization.

Rather than optimizing a specialized architecture for a specific application, such as breaking cryptosystems, the aim of this work is to lay the foundation for a general-purpose FTQC with greater flexibility.
We prioritize practical simplicity over performance in order to facilitate the fabrication of the actual machine, recognizing the integration complexity of a system at this scale.
In the future, we consider improving efficiency for specific applications by integrating specific components such as an optimized adder or the $8T$-to-CCZ distillation factory of~\cite{gidney2025factor}.
With hundreds of logical qubits and millions of $T$ gates available, we expect this machine to play a key role in exploring new applications in a broad range of scientific domains.

To illustrate the potential of our FTQC architecture, we compile the quantum Hamiltonian simulation of a Heisenberg model on a random regular graph with degree-seven over 100 sites on the walking cat architecture, which utilizes 162 logical qubits and about 10,000 physical qubits.
We estimate that the execution of this quantum algorithm, including all the shots needed to reach sufficient accuracy, would take about one month.
We expect this type of material science problem on high-degree regular graphs (\emph{e.g.}, 3D glasses) to be classically intractable and out of reach for NISQ machines, as it requires millions of quantum gates.

The remainder of this paper is organized as follows.
\cref{part:The architecture model} introduces the moving-qubit model, which describes the physical instruction set used to the build the logical architecture, and its noise model.
An overview of the logical architecture, its components and their sub-components, and a description of the logical instruction set is provided in \cref{part:The logical Architecture}.
Then, \cref{part:Implementation of the components} provides an in-depth description of the implementation of each component of the logical architecture, illustrated with simulations.
The micro-architecture is described in \cref{part:The Micro-architecture}.
Finally, \cref{part:Compilation and Applications} describes the compiler and provides resource estimates for quantum applications executed on instances of the walking cat architecture.


\section{Architecture design principles}
\label{sec:Architecture design principles}

This section discusses our design strategy, inspired by classical computer architecture.
We use the abstraction levels of \cref{fig:intro_stack} to design an architecture allowing for the decomposition of a quantum program into low-level device instructions so that it can be executed.

To enable the execution of quantum programs, we design a {\em logical architecture set}, providing a set of instructions that can be fault-tolerantly implemented on the logical qubits.
The logical instructions and their cost (execution time, qubit consumed, \emph{etc}.), accounting for the resources consumed by all error-correction subroutines, is given to the compiler.
The compiler is designed to be agnostic to the details of the error correction subroutines: a change in one of these subroutines requires only an update to the resource cost, not to the compiler, directly.

The logical architecture is built around a simplified model of for the QCCD architecture that we call the {\em moving-qubit model}, and detail fully in \cref{sec:The moving qubit model}.
The goal of this model is to capture the crucial features of the QCCD architecture, while removing technical details that are not expected to significantly affect the performance of our error correction protocols.
For example, in the QCCD architecture, implementing a two-qubit gate requires four steps: transporting the two target qubits to a gating zone, merging them into a two-qubit chain, applying the gate, and finally splitting the temporary chain.
This level of detail is unnecessary when designing a quantum error correction scheme.
Instead, the moving-qubit model assumes that qubits are placed on the sites of a square grid and that two-qubit gates can be implemented on nearest neighbors.

The moving-qubit model is designed to capture the ability to move qubits and the locality of two-qubit gates, and includes noise in all qubit operations, idling, and transport. We also factor in leakage and loss, which is not considered in many resource estimations.
We refer to the instructions available in the moving-qubit model as the \emph{physical instruction set}.

Once the logical architecture is designed, our final task in detailing all layers of abstraction in \cref{fig:intro_stack} is to map to the device instructions. While the most straightforward way to do this would be decomposing each instruction of the physical instruction set into instructions available within the constraints of the QCCD chip, doing so could alter the performance of the error-correction protocols and logical instructions built within the logical architecture layer.
To avoid this, the approach we take is to optimize the mapping of an entire component of the logical architecture and the logical instructions they provide directly onto the device instructions.
We refer to the low-level description of a logical architecture component as the {\em micro-architecture of a component}.
In this work, we propose a micro-architecture for the three most critical components of the logical architecture: the memory block, the magic factory and the cat factory. By imposing this hierarchy, we can explore the algorithmic impact of component performance targets, which then inform a co-design process between the micro-architecture and chip design.

Our architecture is designed based on the following HMRS (pronounce `hammers') principles, inspired by classical computer architecture.
\begin{itemize}
    \item {\bf Hierarchy}: The chip is tiled with components that execute sub-components. For the components, we use memory blocks to store logical qubits, magic factories, cat factories and Bell factories producing their respective states, and a qubit factory used to replace lost qubits due to ion loss.
    \item {\bf Modularity}: The components execute their functions independently, in parallel with other components, and they interface exclusively through resource states produced in a component and transported to another component where they are consumed.
    \item {\bf Regularity}: The chip is tiled with many copies of the same components and the memory block and the magic factory use identical sub-components for error correction, leakage detection, loss detection, and decoding.
    Moreover, our architecture is built around cyclic shifts of the qubits which move qubits along a ring. This primitive is easy to implement using trapped ions and extensively used in memory blocks, magic factories and cat factories.
    Finally, the walking cat architecture is entirely based on quantum LDPC codes, unlike previous work, which typically uses a combination of topological codes and LDPC codes.
    We introduce a unified error-correction framework to encompass all the code layouts used in our architecture including generalized bicycle (GB) codes~\cite{kovalev2013quantum}, bivariate bicycle (BB) codes~\cite{bravyi2024high} and cyclic hypergraph product (HGP) codes~\cite{aydin2025cyclic}. As a result many quantum LDPC codes can be immediately integrated in our architecture (see~\cref{app:sec_code_database_table}). 
    \item {\bf Simplicity}: To reduce the risks coming with the design of an entirely new type of computer, we prioritize simplicity.
    We design new magic factories, which do not rely on the widely adopted magic state cultivation~\cite{gidney2024magic, yoder2025tour, webster2026pinnacle, cain2026shor}, because cultivation is quite technical, involving color codes, surface codes and a merge color-surface codes.
    Moreover, we rely exclusively on cat states to perform logical measurements, whereas most recent works use complex resource states~\cite{cohen2022low, cowtan2024css, swaroop2024universal, cross2024improved, cowtan2024ssip, baspin2025fast}.
    The simplicity of our approach is well illustrated by the walking cat architectures using the single LDPC code in all the memory blocks and the magic factories.
\end{itemize}
The first three principles are adopted from classical computer architecture community~\cite{harris2021digital}.
Simplicity is added for our purpose to reduce the risks in the hardware design and manufacture of the machine.

Our architectural blueprint is substantially more detailed than typical resource estimations in several ways. 
Rather than giving a rough gate-count estimate for a specific algorithm, we fully specify a compiler, evaluate a complete logical architecture, and propose a concrete micro-architecture---critical details often omitted from high-level resource estimates.
We account for not only circuit-level noise but also leakage and qubit loss, both of which can significantly impact resource estimates.
We provide a noise sensitivity analysis to investigate the impact of varying different sources of noise (two-qubit gate noise, measurement noise, idle noise, transport noise, loss or leakage).
We estimate reloading requirements, and also describe all the routing paths required to reload qubits in each component of the architecture.
We include local reservoirs in the components where fast reloading is required to avoid stalling the whole computation, or seriously harming the code performance by letting the impact of a qubit loss spread.
Moreover, we include a global reservoir that stores freshly loaded qubits until they are distributed to refill the local reservoirs.
The tradeoff between the ion reloading speed and the global reservoir size is optimized using a discrete-time Markov chain, see \cref{sec:The qubit factory}.
Finally, we provide a streaming decoder fast enough for online decoding and we analyze its runtime distribution over millions of rounds of error correction. Our results show that our decoder is sufficiently fast and accurate for real-time decoding and its reaction time is small enough to avoid stalling logical operations in the walking cat architecture.

Throughout this paper, our estimate for the time to solution of quantum algorithms is based on resource counts within the moving-qubit model. 
Namely, we assume that preparation, gates and measurement all take 200 micro-seconds while a transport step is executed in 10 micro-seconds (see \cref{subsec:Noise strength and POC time} for details).
While directly simulating our micro-architecture at the device instruction level would give a more accurate estimate for the time to solution, we expect that progress in hardware design will lead to variations in the micro-architecture runtime that will make our result inaccurate rapidly after the publication of this paper.
To keep this paper relevant over the long term, we use the simplified time estimate based on the moving-qubit model which we expect to be more stable over time.
Moreover, the main message of this paper---that we expect to be able to build a breakthrough machine capable of running millions of gates on hundreds of logical qubits in the near-term---is not affected by small variations of the operation time.
The main goal of the micro-architecture presented in this work is to provide a detailed proof-point of the logical architecture to facilitate the fabrication of the actual machine.

\clearpage

\part{The moving-qubit model}
\label{part:The architecture model}
\section{Review of the fully connected model}
\label{subsec:Review of the fully connected model}

In this section, we review the fully connected model which is one of the most popular model for the simulation of quantum error correction codes and fault-tolerant gadgets.
This model assumes no restriction on qubit connectivity or operation parallelism, and errors are modeled as depolarizing noise inserted after each operation at the physical circuit level.
The main features of the fully connected model are summarized in \cref{tab:fully connected model operations}.

\begin{table}[h]
\centering
\begin{adjustbox}{max width=\columnwidth}
\begin{tabular}{|c|c|c|c|}
\hline
\textbf{Operation} & \textbf{Location} & \makecell{\textbf{Error rate} \\ \textbf{(uniform)}} & \makecell{\textbf{Error rate} \\ \textbf{(ionic)}} \\
\hline
Preparation  & Any qubit      & $p$      & $p/10$   \\ \hline
One-qubit gate & Any qubit      & $p$      & $p/10$   \\ \hline
Two-qubit gate    & Any pair & $p$      & $p$      \\ \hline
Measurement               & Any qubit      & $p$      & $p/10$   \\ \hline
Idle                      & Any qubit      & $p$      & $p/100$  \\
\hline
\end{tabular}
\end{adjustbox}
\caption{Operations in the fully connected model.}
\label{tab:fully connected model operations}
\end{table}

We consider a register of qubits equipped with the following operations: Preparation of a single-qubit state, single-qubit unitary gates, two-qubit unitary gates such as \CX (CNOT), \CY, \CZ, and measurement of a qubit in any basis.
We assume that measurements are non-destructive. The state of a qubit after measurement is either $\ket 0$ or $\ket 1$, depending on the outcome measured. We can reset a qubit left in $\ket 1$ to $\ket 0$ by applying an $X$ gate.
We assume that this reset if free as it can be done by relabeling the two qubit states in the classical control. This technique is sometimes referred to as physical Pauli frame tracking~\cite{raussendorf2001one, knill2005quantum, riesebos2017pauli, litinski2019game, on2023multilayered, paler2014software}.

We assume that the qubits are {\em fully connected} in the sense that two-qubit gates acting on any pair of qubits are available. 

For simplicity, we suppose that any operation is implemented in depth one, \textit{i.e.}, all operations take the same amount of time.
Moreover, we assume {\em full parallelism}, which means that any disjoint set of operations can be implemented concurrently.

With {\em uniform circuit-level noise}, single-qubit preparations, idle qubits, and unitary gates (single-qubit or two-qubit) are followed by depolarizing noise on their support. The noise rate is $p$ for preparations, idle qubits, and unitary operations. Measurement outcomes are flipped with probability $p$.

The {\em ionic circuit-level noise} is defined similarly but with noise rate $p$ for two-qubit gates, $p/10$ for single-qubit operations, and $p/100$ for idle steps. 
In the rest of this paper, we adopt this variant of the circuit-level noise because it is a reasonable model for trapped ions.

The main advantage of the fully connected model is that it is widely adopted in the literature, which facilitates comparison with previous work.
However, it does not include locality constraints of quantum operations and qubit transport. Moreover, important sources of noise are missing, such as qubit losses or leakages, which we consider next in \cref{sec:The moving qubit model}.

\section{The moving-qubit model}
\label{sec:The moving qubit model}

The {\em moving-qubit model} is a refined version of the fully connected model, capturing the main features of a two-dimensional grid of qubits, where qubits can be physically moved. 
In this model, they are susceptible to not only circuit-level noise, but also leakages and qubit losses.
The main features of the moving-qubit model are summarized in \cref{tab:moving qubit model operations}.

\begin{table}[h]
\centering
\begin{adjustbox}{max width=\columnwidth}
\begin{tabular}{|c|c|c|c|c|c|}
\hline
\textbf{Operation} & \textbf{Location} & \makecell{\textbf{Error} \\ \textbf{rate}} & \makecell{\textbf{Loss} \\ \textbf{rate}} & \makecell{\textbf{Leakage} \\ \textbf{rate}} & \makecell{\textbf{Time} \\ \textbf{(in POC)}} \\
\hline
Preparation                      & Any qubit         & $p/10$   & $\ploss$      & $\pleak$      & $1$       \\ \hline
One-qubit gate                          & Any qubit         & $p/10$   & $\ploss$      & $\pleak$      & $1$       \\ \hline
Two-qubit gate                          & Nearest neighbors & $p$      & $\ploss$      & $\pleak$      & $1$       \\ \hline
Measurement                      & Any qubit         & $p/10$   & $\ploss$      & $\pleak$      & $1$       \\ \hline
Leakage reset               & Any qubit         & $p/10$   & $\ploss$      & $\pleak$      & $1$       \\ \hline
Idle                             & Any qubit         & $p/100$  & $\ploss$      & $\pleak$      & $1$       \\ \hline
Transport step                   & Any qubit         & $p/2000$  & $\ploss/20$   & $\pleak/20$   & $1/20$    \\
\hline
\end{tabular}
\end{adjustbox}
\caption{Operations in the moving-qubit model. The operation time is measured in physical operation cycle or POC (see \cref{subsec:Operation time}).
Other variants of the moving qubit model can be considered by adjusting the relative strength of noise sources.
}
\label{tab:moving qubit model operations}
\end{table}

Our main motivation is to explore the design of fault-tolerant quantum computing architectures based on trapped ions~\cite{kielpinski2002architecture}.
The moving-qubit model may also provide insights for neutral atoms~\cite{bluvstein2024logical}, spin qubits~\cite{loss1998quantum}, and electrons floating on helium~\cite{castoria2025selective}.
We use this model to design and optimize a fault-tolerant quantum computing architecture.
In \cref{sec:micro-architecture}, we discuss a refined design for trapped ions based on a micro-architecture model that captures more detailed features of the chip.

\subsection{Operations available}

The {\em chip} is represented by an $L \times L$ square grid of sites.
Each {\em site} is either empty or it holds a {\em physical qubit}. A site cannot hold more than one physical qubit. When it is clear from the context, and no confusion is possible with the logical qubits, we refer to the physical qubits as qubits.

Qubits are moved through sequences of {\em transport steps}.
During a transport step, each qubit either remains in its current site or is moved to one of the four neighboring sites, with the constraint that after these moves any site still contains at most one qubit. We allow a transport step to swap neighboring qubits.

Our architecture is designed around specific transport paths that can be optimized in hardware.
The {\em cyclic shift} is one such transport operation during which a set of qubits move all together along a loop.
Several components of our architecture are built on top of cyclic shifts of the qubits.

In addition to transport, we consider the same operations as in the fully connected model, with the exception that two-qubit gates are only available between nearest-neighbor sites.
We call these {\em computational operations}.
We assume that transport and computational operations cannot be performed simultaneously.

\subsection{Operation time}
\label{subsec:Operation time}

We refer to the concurrent implementation of a set of disjoint computational operations as a {\em physical operation cycle} or {\em POC}, and the time it takes as the {\em POC time}.
The POC time of each operations of the moving-qubit model is reported in \cref{tab:moving qubit model operations}.

To keep the model simple, we assume that all computational operations take the same amount of time, equal to 1 POC.
The measurement time is set to 1 POC, including the leakage and loss measurement introduced in \cref{subsec:Loss and leakage measurement}.
We could assume that measurements without leakage and loss outcome are faster, but we do not distinguish these cases for simplicity.
The leakage reset also takes 1 POC and is treated as a computational operation.
Any set of computational operations acting non-trivially on disjoint sets of qubits can be implemented simultaneously in 1 POC.

We assume that transport is faster than computational operations, and a transport step is implemented in $1/20$ POC.
For long-distance transport, we use a highway capable of faster transport.
For simplicity, we assume that the highway transport is instantaneous. This assumption is discussed in \cref{sec:micro-architecture}.

\subsection{Noise model}
\label{subsec:noise-model}

We consider the same noise rates as in the ionic circuit-level noise for computational operations, which we complement with qubit loss and leakage to make the model more realistic.

During a transport step, we assume that all qubits, transported or idle, suffer from independent depolarizing noise with rate $p/2000$. This corresponds roughly to 1/20th of the idle noise per POC.
This model is motivated by the fact that transport noise is expected to be similar to idle noise for trapped ions~\cite{ransford2025helios} and the factor $1/20$ comes from the fact that a transport step is assumed to be 20 times faster than a computational operation (see \cref{subsec:Operation time}).

Qubit loss originates from the physical loss of the carrier of the quantum state (\textit{e.g.}, an atom or electron), which requires replacement. We define a {\em qubit loss} rate $\ploss$ per POC: after each POC, each qubit is lost independently with probability $\ploss$. During transport, each qubit, transported or not, incurs an independent loss probability $\ploss/20$ per transport step. Any non-measurement operation involving a lost qubit is replaced with the identity operation, and any two-qubit gate involving a lost qubit propagates the loss to the other qubit involved in the gate. 
One could consider a model where a loss does not always propagate through a two-qubit gate. By designing our architecture for the most harmful model where losses propagate the fastest, we guarantee that it is also viable for other loss models. 
Loss measurements are discussed in \cref{subsec:Loss and leakage measurement}.
The replacement of lost qubits requires reloading which is allowed during transport steps by inserting fresh qubits into empty sites on the boundary of the $L \times L$ grid, provided no site is doubly occupied at the end of the step.

We define a {\em qubit leakage} rate $\pleak$ per POC. Physically, leakage processes result from encoding qubits in objects which themselves possess higher-dimensional Hilbert spaces. A leaked qubit remains in the same site, but its state is mapped onto the orthogonal complement of the computation space, which we call the {\em leakage space}. After each POC, each qubit leaks independently with probability $\pleak$. During transport, we assume a leakage rate $\pleak/20$ per qubit per transport step. Any two-qubit gate involving a leaked qubit is replaced by the identity (\textit{i.e.}, the two-qubit gate is removed), and the noise associated with the gate remains. Any single-qubit gate involving a leaked qubit is similarly replaced by the identity with its associated noise applied. 
We assume that an {\em leakage reset} operation is available to reset a leaked qubit to the maximally mixed state $I/2$~\cite{hayes2020eliminating, miao2023overcoming}. If a leaked qubit experiences loss, it becomes a lost qubit. When a leaked qubit interacts with a lost qubit, the loss propagates to the leaked qubit.

\subsection{Loss and leakage measurements}
\label{subsec:Loss and leakage measurement}

In addition to measurement in the computational basis, which projects a qubit onto the state $\ket 0$ or $\ket 1$ and returns the corresponding outcome $0$ or $1$, we assume that our system is equipped with measurements informing us about qubit loss and leakage~\cite{sotirova2024high}.

The {\em loss measurement} of the qubit returns the outcome `lost' if this qubit is lost and the {\em leakage measurement} of the qubit returns the outcome `leaked' if the qubit is leaked.
Loss and leakage measurements are only available immediately after a qubit is measured in the computational basis, and, for simplicity, we assume that they are applied after every measurement.
Therefore, one can think of these operations as four-outcome measurements returning one of the four values 0, 1, `lost' or `leaked'.
We assume that the outcomes of the loss and leakage measurements are noiseless. More details on their implementation can be found in \cref{sec:micro-architecture}.

\subsection{Noise strength and physical operation cycle time}
\label{subsec:Noise strength and POC time}

In this work, the logical architecture and all its components are built for the moving-qubit model.
We focus on the noise regime with $p=10^{-4}$, $\pleak=10^{-5}$ and $\ploss = 10^{-7}$. 
We assume a POC time of $200 \mu s$.

Experiments on small trapped-ion devices with electronic qubit control show two-qubit gates with noise rate of the order of $10^{-4}$~\cite{hughes2025trapped} and single-qubit gates with noise rate around $10^{-5}$~\cite{loschnauer2025scalable}, we motivates us to prove the regime $p=10^{-4}$.

We motivate $p_{\mathrm{leak}} = 10^{-5}$ and $p_{\mathrm{idle}} = 10^{-6}$ with the assumption that the qubit levels are encoded inside the $D_{5/2}$ metastable states of ${}^{137}\mathrm{Ba}^{+}$, both of which decay at a $1/e$ lifetime of approximately 30 seconds~\cite{allcock2021omg}. The qubit levels can be chosen such that they are first-order insensitive to magnetic field noise and with low second-order field sensitivity, such that the encoded qubits can have long coherence times greatly exceeding their natural lifetime ($136\pm42~\mathrm{s}$ demonstrated in Ref.~\cite{shi2025long}) if erasure errors from decay to ground states are detected. Imperfections in gate mechanism can add additional leakage error rate on top of the $6.7\times 10^{-6}$ contribution from the 30~s lifetime, and we assume that less than 10~\% of the total two-qubit gate error budget comes from exciting unwanted spectator transitions based on preliminary atomic physics simulation. We assume $p_{\mathrm{loss}}=10^{-7}$ and provide a motivation based on experimentally feasible vacuum pressure and resulting collision rate between a trapped ion and a background gas particle in \cref{appendix:ion_loss_rate}. 

A detailed discussion on the operation time in trapped-ion devices is provided in \cref{sec:micro-architecture}, justifying our simplified model with $200~\mathrm{\mu s}$ per POC and $10~\mathrm{\mu s}$ per transport step.

\clearpage

\part{Overview of the logical architecture}
\label{part:The logical Architecture}
\section{The logical architecture components}
\label{sec:Component and sub-components of the walking cat architecture}

\begin{figure}
    \centering
    \includegraphics[width=0.98\linewidth]{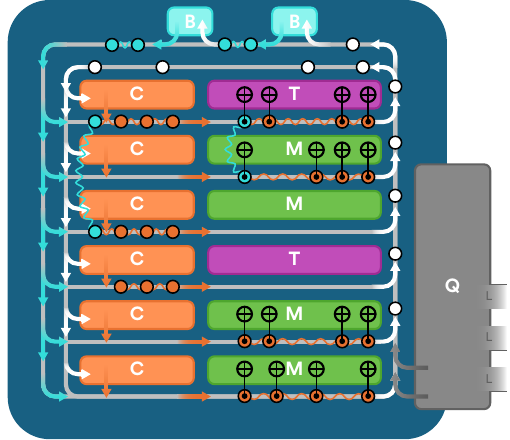}
    \caption{Simplified representation of a walking cat architecture. The logical qubits are stored in memory blocks (M), logical magic states are generated in magic factories (T), physical cat states are generated in cat factories (C) and physical Bell states are generated in Bell factories (B).
    The cat states are transported next to a memory block or a magic factory where they are used to perform a logical measurement.
    The Bell states are used to stitch two cat states together, which are then used to perform a joint logical measurement supported on two different blocks, allowing us to entangle logical qubits in different blocks and to distribute magic states.
    Each logical measurement consumes multiple cat states. 
    After being measured, the cat state qubits are sent back to Bell factories and cat factories to be reused.
    To compensate for qubit loss, we include a qubit factory (Q), consisting of a global reservoir and loading zones (L), which loads and stores extra qubits that can be delivered to all the components when a loss is detected.
    }
    \label{fig:architecture_overview_chip}
\end{figure}

A high-level view of a walking cat architecture is represented in \cref{fig:architecture_overview_chip}.
Quantum information is stored in memory blocks and logical circuits are implemented through two types of operations: logical magic state preparation and logical Pauli measurements.
This model of computation, called the {\em Pauli-based computation} model, introduced in~\cite{bravyi2016trading}, is widely adopted in fault-tolerant quantum computing architectures based on surface codes~\cite{litinski2019game, fowler2012surface, gidney2025factor} or quantum LDPC codes~\cite{yoder2025tour, cain2026shor, webster2026pinnacle}.
However, our design of the components of the architecture and our implementation of the logical instructions is fundamentally different from previous work.

All the quantum codes used in this paper are stabilizer codes~\cite{gottesman1997stabilizer}. For a review of the stabilizer formalism, see \cref{app:Background on stabilizer codes}.

Our memory is based on quantum LDPC codes implemented by moving qubits along cyclic shifts whereas superconducting qubit implementations rely on long-range couplers~\cite{yoder2025tour} and neutral atom implementations typically rely on long-distance transport of blocks of qubits~\cite{xu2024constant} or long-range Rydberg gates~\cite{poole2024architecture, pecorari2025high}.
Moreover, our memory is equipped with leakage and loss correction, ignored in these previous works.

Logical magic states are produced in magic factories and consumed to implement logical $T$ gates on memory qubits.
We design new classes of magic factories based on custom LDPC codes.
Previous architectures typically rely on magic state distillation in surface codes~\cite{bravyi2005universal, litinski2019magic} or magic state cultivation~\cite{gidney2024magic}.

Logical Pauli measurements are implemented using physical cat states produced in cat factories and transported next to the qubit to measure.
The idea of using cat states to perform fault-tolerant measurements originates from Shor's first paper on fault-tolerant quantum computing~\cite{shor1996fault} but is rarely used in fault-tolerant quantum computing architecture because building large cat states generally requires many attempts.
Recent experimental breakthroughs proved that trapped ions can achieve fidelity above $99.99\%$~\cite{hughes2025trapped}, unlocking the ability to prepare large cat states in few attempts.
Moreover, because one can move qubits, we can design dedicated cat factories in specific regions of the chip and transport the produced cat states wherever they are needed.
For comparison, previous work relies on lattice surgery consuming large ancilla patches of surface codes~\cite{litinski2019game} or complex ancilla qubit systems~\cite{cohen2022low, cowtan2024css, swaroop2024universal, cross2024improved, cowtan2024ssip, baspin2025fast, yuan2026parsimonious}.

To perform a logical measurement supported on two different blocks, that may be far from each other, we stitch two cat states together and send them each to one of the blocks, where they are consumed to perform the join measurement.
The stitching is performed using Bell states, produced in Bell factories and distributed to the cat factories.

\begin{figure}
    \centering
    \includegraphics[width=1\linewidth]{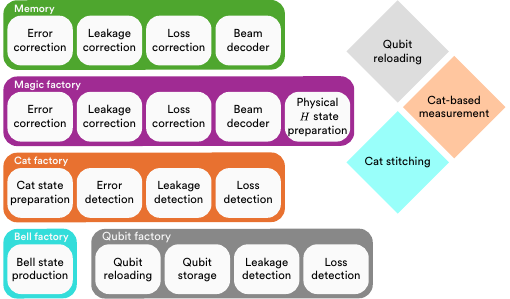}
    \caption{Representation of the main components in the walking cat architecture and their sub-components. The three types of interfaces between these components are represented with diamond shapes.}
    \label{fig:architecture_subcomponents}
\end{figure}

This section provides a qualitative overview of the main components of the architecture and their sub-components listed in \cref{fig:architecture_subcomponents}.
Concrete examples including numerical estimates of the performance of each components are provided in \cref{sec:Examples of walking cat architectures}.

\subsection{The memory component}

The {\em memory component} is designed to store logical information.
Its four sub-components are responsible for the execution of quantum error correction protocols.
The {\em error-correction sub-component} corrects errors that preserve the computational space, the {\em leakage correction sub-component} corrects leakages outside of the computational space and the {\em loss correction sub-component} corrects the loss of a qubit.
The {\em beam decoder sub-component} uses the measurement outcomes observed to determine the effect of errors, leakages and losses so that they can be reversed without affecting the result of the logical computation.

\begin{figure}
    \centering


    \includegraphics[width=1\linewidth]{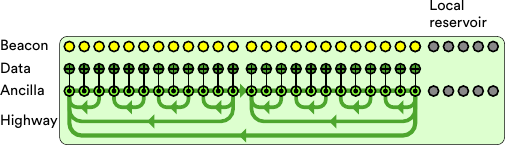}

    
    \caption{
    A memory block consists of four rows: a row of data qubits (dark green) to store the logical information, a row of ancilla qubits (light green) to perform error correction, a row of beacon qubits (yellow) used for the detection of qubit loss, and an initially empty row, called the ``highway'', that facilitates the transport of ancilla qubits during syndrome extraction. Syndrome extraction consists of alternating rounds of two-qubit gates and moves of the ancilla qubits along three types of cyclic shifts (\cref{subsec:three-ring-framework}). The highway is shown here with three ``rings'' to demonstrate that there are three fundamental shift operations used; in practice, these rings all utilize the shared highway row. Gray qubits represent a local reservoir used to rapidly replace lost qubits.
    }
    \label{fig:architecture_overview_memory}
\end{figure}

The logical information is stored in {\em memory blocks}, represented in \cref{fig:architecture_overview_memory}.
Each block stores $k$ logical qubits using an $[[n, k, d]]$ stabilizer code, and consumes a total of $3n+10$ physical qubits split into four functions as follows.
We use $n$ data qubits to encode the $k$ logical qubits, $n$ ancilla qubits to perform error and leakage correction, $n$ beacon qubits to correct losses~\cite{coble2025correction}, and an extra 10 qubits are kept in local reservoirs inside the memory block to refill lost qubits.
We select the stabilizer code among three extensive classes of quantum LDPC codes: generalized bicycle (GB) codes~\cite{kovalev2013quantum}, bivariate bicycle (BB) codes~\cite{bravyi2024high}, and cyclic hypergraph product (HGP) codes~\cite{aydin2025cyclic} (see also~\cite{tillich2013quantum} for the original construction for HGP codes), giving us a broad catalog of codes to choose from, including codes that outperform the surface codes for storage density~\cite{bravyi2024high, ye2025quantum, aydin2025cyclic}.

We use a syndrome extraction circuit, and a layout based on cyclic shifts of the ancilla qubits unifying the BB code layout of~\cite{tham2025distributed} and the cyclic HGP layout of~\cite{aydin2025cyclic} and the GB code layout of~\cite{siegel2024towards}.
We refer to our unifying picture as the three-ring framework.

In what follows, we refer to a round of syndrome extraction as a {\em syndrome extraction cycle} or SEC. One can think of the SEC time as the logical clock cycle time of a fault-tolerant quantum computer.

We introduce a local reservoir inside each memory block instead of using the global reservoir in order to reload the lost qubits as fast as possible without incurring any delay due to the transport from the global reservoir to the memory block.
The global reservoir is then used to refill the local memory reservoirs with less stringent time requirements.

All the codes considered in this work are decoded using the beam decoder, which can be used for any quantum LDPC code and achieves state-of-the-art performance for BB codes and cyclic HGP codes~\cite{ye2025beam}.
To make it practical, we introduce a streaming version of the beam decoder, capable of decoding on-the-fly by sliding a decoding window~\cite{bombin2023modular, skoric2023parallel, tan2023scalable}.
To confirm that our decoder is sufficiently fast to fulfill the requirements for our architecture, we perform a simulation of one million SEC for two different codes used throughout this paper.
We estimate the average runtime and the reaction time for the extraction of all logical measurement outcomes during a destructive measurement.
As in previous work, we observe that the beam decoder achieves a better logical error rate than BP-OSD~\cite{panteleev2021degenerate, roffe_decoding_2020, Roffe_LDPC_Python_tools_2022}.
These simulations confirm that the streaming variant of our decoder satisfies all the requirements of our architecture.

A complete description of the memory blocks, their syndrome extraction circuit and all the codes we use is provided in \cref{sec:Memory block}.
The loss correction and leakage correction gadgets are described in \cref{sec:Correction of losses and leakages}.
The streaming beam decoder is described in \cref{sec:decoder}.

\subsection{The magic factory component}
\label{subsec:magic-factory-component}

The purpose of the {\em magic factory component} is to produce logical resource states called magic states, which are consumed for the implementation of logical $T$ gates.
Our magic factory is also a memory so that it can store the produced logical magic state until it is consumed.
The magic factory includes five sub-components, the first four being identical to the memory sub-components.
What differentiates the magic factory from a memory block is the {\em physical $H$ state preparation sub-component}, which is used to inject a magic state into the code.
Like in a memory block, logical magic states are protected from errors, leakages and losses, and decoded using a beam decoder.
One may use different codes in magic factories and memory blocks, but these codes are selected from the same families: GB codes, BB codes, cyclic HGP codes.

\begin{figure}
    \centering
    \includegraphics[width=1\linewidth]{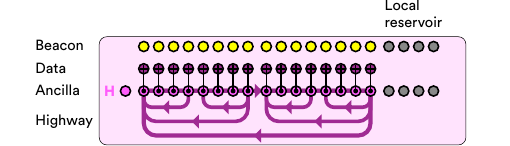}
    \caption{The magic factory is built on top of a memory block and therefore its high-level structure is identical even though the code used may be different from the memory block code. 
    The only difference is the H box on the left of the ancilla row, which represents an extra ancilla qubit used prepare a physical H state.
    The magic factory produces magic states and stores them until they are consumed for the implementation of $T$ gates.}
    \label{fig:architecture_magic_factory}
\end{figure}

The magic state distillation procedure produces a few high-fidelity logical magic states from multiple noisy copies of a logical magic state~\cite{knill2004fault, bravyi2005universal, bravyi2012magic, litinski2019magic}.
Magic state distillation has been extensively optimized for surface codes. However, because it is implemented at the logical level, distillation consumes a large number of qubits.
For example, the 15-to-1 distillation scheme~\cite{bravyi2005universal}, which distills one logical $T$ state from $15$ noisier logical $T$ states, consumes 15 patches of surface codes and additional patches to execute the distillation circuit.

Alternatives based on physical operations and verification have been considered to produce magic states in surface codes or color codes~\cite{goto2014step, chamberland2020very, itogawa2025efficient}, culminating with the cultivation scheme, which is far more efficient than surface code distillation~\cite{gidney2024magic}.
However, implementing this protocol is very technical as it involves color codes, surface codes, merged color-surface codes and a decoder capable of handling all these codes. 
Other variants of the cultivation scheme remove the need for a color code, but they require non-trivial qubit topology or connectivity~\cite{vaknin2025efficient, chen2025efficient, sahay2025fold}.
Moreover, to implement a $T$ gate using these magic states in our architecture, we would have to connect the output code of the cultivation scheme to another family of codes---the LDPC codes hosting our memory blocks---necessitating additional code connector gadgets and increasing architectural complexity.

To reduce the qubit overhead induced by the surface code, it is natural to consider switching from the surface code to higher rate codes.
Recent progress led to the design of constant-overhead magic distillation schemes in the asymptotic regime~\cite{krishna2018towards, golowich2025asymptotically, wills2025constant, nguyen2025good, golowich2025near}.
Unfortunately, these results cannot be used in the present work as they rely on very large codes and it is unclear if these codes perform well with circuit-level noise as they are not LDPC and making them fault tolerant might be resource intensive.
The tricycle codes of~\cite{menon2025magic} could be more practical but these three-fold product codes are not immediately compatible with our architecture focused on two-fold products.

To avoid the large overhead of topological codes and the complexity of the cultivation scheme, we design two magic factories producing magic states directly in quantum LDPC codes.
These two factories produce $H$-type magic states, referred to throughout as $H$ \emph{states}:
\begin{align}
    \ket H = \cos\left(\pi/8\right)\ket 0 + \sin\left(\pi/8\right)\ket 1,
\end{align}
which is the +1-eigenstate of the $H$ gate.
They provide the same computational power as $T$ states because these states are Clifford-equivalent: $\ket H = SH\ket T$.
We can also see the $H$ state as the $Y$ version of the $T$ state in the sense that $\ket H$ = $e^{-i \pi/8 Y}\ket 0$ whereas $\ket T = e^{-i \pi/8 Z}\ket+$.

Our first magic factory is based on the Meier, Eastin, and Knill (\MEK) distillation scheme~\cite{meier2013magic}, which produces 2 logical $H$ states from 10 physical $H$ states.
The main novelty in our approach is that the distillation is implemented  directly in a custom quantum LDPC codes compatible with our architecture.
This LDPC code is designed to have 6 logical qubits, which is enough to run the \MEK distillation circuit, and has the property that all $6$-qubit Clifford unitaries can be implemented by frame tracking, making the \MEK circuit faster to execute.

Our second magic factory, which we call \CHfactory, is implemented in a custom $[[n, 2, d]]$ LDPC code with a strongly transversal $H$ gate ($H^{\otimes n} = \bar{H}^{\otimes 2}$).
This property allows us to prepare two logical $H$ states and to verify them by direct measurement of the logical operator $\bar{H}^{\otimes 2}$.
The measurement of $\bar{H}^{\otimes 2}$ is performed using $n$-qubit cat states, leveraging the strongly transversal $H$ gate.
We want to keep $n$ small to minimize the size of the cat state consumed, while having a sufficiently large minimum distance $d$ to achieve a low logical error rate.
The code $\code{54}$ is selected for this task.

In both cases, the quantum LDPC code hosting the magic factory is selected from the three families of codes compatible with our memory design so that the same design can be shared between the memory and the magic factory.
This leads to a magic factory design with the same sub-components as the memory, with the exception of a physical $H$ state preparation which is used to inject an $H$ state in a logical qubit.
The magic factory consumes a total of $3n+11$ physical qubits, with a structure similar to the memory, although the underlying code might be different and may have a different code length $n$.

The \MEK factory and the \CHfactory factory are described in \cref{sec:magic_state_factory}.

\subsection{The cat factory component}

The {\em cat factory component} is dedicated to the production of cat states which are consumed in order to perform logical measurements.
The {\em cat state production sub-component} produces cat states, and these are then verified for the presence of errors, leakage or loss by the {\em error detection sub-component}, the {\em leakage detection sub-component} and the {\em loss detection sub-component}.

\begin{figure}
    \centering
    \includegraphics[width=1\linewidth]{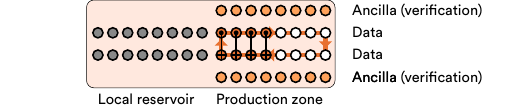}
    \caption{Like the memory block and magic factory, the cat factory consists of four rows of qubits.
    A cat state is prepared using a sequence of \CX gates and cyclic shifts in the production zone.
    Then, it is verified to detect errors, leakages and qubit losses using stabilizer measurements and cyclic shifts.
    Finally, it is sent to a memory block or a magic factory where it is consumed to measure a Pauli operator. After the measurement, its qubits are sent back to a cat factory and stored in the local reservoir.
    }
    \label{fig:architecture_overview_cat_factory}
\end{figure}

Many cat factories have been proposed to produce and verify cat states that are fault-tolerant with a level of protection that corresponds to the code distance~~\cite{shor1996fault, preskill1998fault, divincenzo2007effective, stephens2014efficient, yoder2017surface, prabhu2021fault, rodatz2025fault, khesin2026spidercat, peham2026optimizing}.
In this work, we propose a different approach which results in more efficient cat factories. 
We fix a verification threshold $\varepsilon$ and we verify that the cat state does not contain correlated errors which could degrade the code performance with probability larger than $\varepsilon$.

The cat factory is parametrized by the verification threshold $\varepsilon$ and the maximum size $w$ of the cat states it can produce.
The cat factory is built around a cyclic shift as shown in \cref{fig:architecture_overview_cat_factory} and it consumes $2w$ qubits in total.
The parameter $w$ is carefully optimized to keep the qubit cost of the cat factory small while allowing sufficiently large cat states to provide a broad set of logical measurements.

A cat state is prepared using a sequence of \CX gates and cyclic shifts of the cat qubits.
Then, cat state stabilizers are measured using to detect the presence of errors, leakages or losses on the cat state qubits.
The verification consumes one ancilla per cat qubit and is implemented using \CX gates and cyclic shifts.

Any cat state that does not pass verification is discarded to avoid injecting noise into the computation.
Its qubits are reused for another cat state preparation attempt.
Leaked qubits are reset and lost qubits are replaced by sending fresh qubits from the global reservoir to the cat factories.
The verification does not completely remove the possibility of a correlated error, a leaked qubit or a lost qubit in the cat state, but it makes the probability of such an event sufficiently small to guarantee that it does not significantly affect the logical error rate of the memory block.
The remaining noise on the cat qubits is corrected by the error-correction mechanisms of the memory and the magic factory.

To speed up the cat state production and to achieve at least one cat state per SEC, we use standard computer architecture techniques such as instruction pipelining~\cite{harris2021digital}.
The basic idea is to perform the cat state transport while the next cat state is produced.
This requires enough qubits to run these two stages in parallel.

The cat factory is described in \cref{sec:Cat factories}.
The transport overhead due to pipelining is discussed in \cref{subsec:Transport overhead}.

\subsection{The Bell factory component}
\label{subsec:The Bell factory component}

The role of the {\em Bell factory component} is to produce physical Bell states which are used to stitch together two cat states.
It is the simplest component as it only performs a single \CX after preparing two qubits in the state $\ket + \otimes \ket 0$.
This is the task accomplished by the {\em Bell production sub-component}.

Even though it is a trivial component, the Bell factory plays a key role in the walking cat architecture as it allows for the implementation of entangling logical operations between two distinct memory blocks and for the application of a logical $T$ gate in a memory consuming a magic state from a magic factory.

The Bell factory is described in \cref{sec:Bell_factories}.

\subsection{The qubit factory component}
\label{subsec:The qubit factory component}

The {\em qubit factory} produces new qubits to replace the lost ones.
It includes a {\em qubit loading sub-component} which is used to reload new qubits at regular time intervals and a {\em qubit storage sub-component} which stores newly loaded qubits in a reservoir, which we refer to as the {\em global reservoir} because its qubits are distributed to all the components of the chip.
We also include a {\em loss detection sub-component} and a {\em leakage detection sub-component} executed at regular interval to avoid dispatching qubits from the global reservoir when they are already lost or leaked.

New qubits are loaded at regular intervals in loading zones placed at the edge of the chip.
The number of loading zones must be adjusted to compensate for the qubit losses in the whole architecture during a computation.
Trapped-ion reloading was demonstrated experimentally, reaching hundreds of strontium-88 ions per second in~\cite{bruzewicz2016scalable} and tens of barium-138 ions per second in~\cite{johansen2022fast}.
Experimental demonstrations with neutral atoms reached hundreds of ytterbium atoms per second in~\cite{li2025fast} and were used to maintain a 3,000 qubit array of rubidium atoms with continuous reloading over two hours in~\cite{chiu2025continuous}.

The global reservoir holds reloaded qubits until a loss is detected in a memory block, a magic factory or a cat factory.
Even though losses occur in other places (during transport, in the Bell factory), they are only detected in these three components.
Once a loss is detected, the lost qubits are replaced by sending qubits from the global reservoir.

The global reservoir has a maximum capacity which cannot be exceeded.
Its size must be carefully optimized, together with the number of reloading zones, to avoid consuming too many resources (loading zones and reservoir qubits) while limiting the risk of system failure, which occurs if the global reservoir is empty and a qubit is requested by another component.
We consider this event as a system failure because qubit losses spread: a two-qubit gate targeting a lost qubit triggers the loss of the other qubit supporting the gate. 
As a result, a lost qubit that cannot be rapidly replaced might trigger an avalanche loss destroying logical information.
Alternatively, one could wait until the lost qubits are replaced but a long wait time without error correction would affect the logical error rate because idle noise, leakage and loss accumulate.
The global reservoir is designed to make such an event sufficiently rare so that it can be ignored.

The qubit factory is described in \cref{sec:The qubit factory}.

\subsection{Component interfaces}
\label{subsec:Interfaces}

The components of the walking cat architecture are connected through three types of interfaces enabled by qubit transport which we can see as a quantum version of a data bus.

The qubit factory must be connected to all the other components in order to support the reloading of their local reservoir.
The interface between the qubit factory and another component is triggered when a loss is detected.
It handles the routing of the qubits from the global reservoir to the local reservoir that needs to be refilled.

The Bell factory interacts with the cat factory to perform the stitching. When two cat states need to be merged in order to perform a logical measurement supported on a pair of blocks, the stitching interface is used.
It drives the production of Bell states and route their qubits to the two cat factories where they are consumed to merge cat states.

Logical measurements are orchestrated using the cat-based measurement interface which is responsible for transporting cat states (potentially stitched) from the cat factories to the memory blocks or the magic factories where measurements are requested. The cat state qubits are moved under the data qubits supporting the operator to measure and the measurement is executed by a round of two-qubit gates followed by the measurement of the cat state qubit.
Multiple round of cat states measurements are consumed for each logical measurements.

\section{Accessible logical operations}
\label{sec:Accessible logical gates}

This section introduces accessible logical Pauli operators, describing the available logical measurements and accessible logical Clifford gates which are the logical gates that can be implemented fully in software by frame-tracking.

In what follows, a {\em block} refers to either a memory block or a magic factory.
Each block comes with an associated cat factory as shown in \cref{fig:architecture_overview_chip}.
In this section, we consider a fixed block with $n$ data qubits, $k$ logical qubits using a stabilizer code.

\subsection{Logical operators and their physical representatives}
\label{subsec:Logical operators and their physical representatives}

To introduce a Clifford frame, we first need to review the notion of symplectic basis which is used to select a basis for the code space.

A {\em symplectic basis} of the block is defined to be a set of $n$-qubit Pauli operators denoted $\bar X_1, \bar Z_1, \dots, \bar X_k, \bar Z_k$, commuting with all the stabilizers of the code and satisfying the same relations (commutations and products) as the standard Pauli operators $X_1, Z_1, \dots, X_k, Z_k$.
For CSS codes, one can build a symplectic basis with the additional property that the operators $\bar X_i$ belong to $\{I, X\}^{\otimes n}$ and the operators $\bar Z_i$ belong to $\{I, Z\}^{\otimes n}$.
We refer to such a basis as a {\em CSS symplectic basis}.

A symplectic basis or a CSS symplectic basis can be computed by a modified Gaussian elimination in $O(n^3)$ bit-operations~(see \cite{nielsen2010quantum}, Section 10.5.7).
It can be stored as a $2k \times 2n$ binary matrix, whose rows represent the basis elements, together with a phase (power of $i$) for each row, consuming $O(n^2)$ bits in memory.

Given a symplectic basis $\logicalBasis = \{\bar X_1, \bar Z_1, \dots, \bar X_k, \bar Z_k\}$, one can define $\ket{\bar 0^k}$ to be the state of the code space fixed by the operators $\bar Z_1, \dots, \bar Z_k$. The commutation relations between the symplectic basis and the stabilizers guarantee that this state exists and is unique.
Then, from $\ket{\bar 0^k}$, define $\ket{\bar u} := \bar X_1^{u_1} \dots \bar X_k^{u_k} \ket{\bar 0^k}$ for all $u \in \Z_2^k$.
The states $\ket{\bar u}$ belong to the code space for all $u$ thanks to the commutation between the symplectic basis and the stabilizers.
One can prove that the $2^k$ states $\ket{\bar u}$ form an orthonormal basis of the code space.
This leads to an encoding map $U_\logicalBasis$ mapping the $k$-qubit state 
$\ket\psi := \sum_{u\in\Z_2^k} \alpha_u \ket{u}$, with $\alpha_u \in \C$, onto the code state 
\begin{align}
\overline{\ket{\psi}} := \sum_{u\in\Z_2^k} \alpha_u \ket{\bar u} \cdot
\end{align}

From the same symplectic basis $\logicalBasis$, we also build the transformation mapping any $k$-qubit Pauli operator $P = \prod_{i=1}^k X_i^{a_i} Z_i^{b_i}$ with $a_i, b_i \in \{0,1\}$ onto the $n$-qubit Pauli operator 
\begin{align}
\bar P := \prod_{i=1}^k \bar X_i^{a_i} \bar Z_i^{b_i}\,.
\end{align}
By construction, we have 
$
\bar P \ket{\bar{\psi}} = \overline{P \ket \psi}
$
which can be written as the following commutative diagram:
\begin{center}
\begin{tikzcd}
\lvert\psi\rangle 
    \arrow[r, "U_{\logicalBasis}"] \arrow[d, "P"']
  & \overline{\lvert\psi\rangle} 
    \arrow[d, "\bar P"] \\
P\lvert\psi\rangle \arrow[r, "U_{\logicalBasis}"']
  & \overline{P\lvert\psi\rangle}
\end{tikzcd}
\end{center}
This proves that applying $\bar P$ on the data qubits is equivalent to the application of $P$ on the logical qubits.
Moreover, for any stabilizer $S$, the operator $\bar PS$ has the same effect as $\bar P$ on the code states.
For this reason, we refer to $P \in \PauliGroup_k$ as a {\em logical operator} and any operator of the form $\bar PS \in \PauliGroup_n$, where $S \in \Stab$ is a stabilizer, is called a {\em physical representative} of $P$.

\subsection{Accessible logical Pauli operators}\label{subsec:accessible_logical_operators}

This subsection defines the set of logical Pauli measurements which are allowed in the logical instruction set.
We first define the set of accessible logical Pauli operators and then we explain how the cat factory size is tuned to make this set measurable using cat states. 

For each block, we introduce a parameter $\logicalWidth$, that we call the {\em logical width} of the block.
A logical operator $P\in \PauliGroup_k$, is said to be an {\em accessible logical Pauli operator} if its logical weight (\emph{i.e.}, its weight in $\PauliGroup_k$) is less than or equal to the logical width.
The set of accessible logical Pauli operators is denoted $\AccessibleLogicalOperators$.

To measure a logical Pauli operator $P\in \PauliGroup_k$, we must specify a physical representative $\bar{P}S \in \PauliGroup_n$ of the operator.
To maximize the number of logical Paulis which can be measured within our architecture, we choose a low-weight representative of the logical Pauli, $\bar{P}_\text{opt}$ (calculated using the weight-reduction algorithm of \cref{sec:Weight-Reduced Logical Pauli Operators}).
We define a map
\begin{align}
\Phi_\logicalBasis: 
& \AccessibleLogicalOperators \longrightarrow \PauliGroup_n ,\\
& P \longmapsto \bar{P}_\text{opt}.
\end{align}
In software, $\Phi_\logicalBasis$ can be stored as a lookup table containing $|\AccessibleLogicalOperators|$ $n$-qubit Pauli operators. This lookup table size is bounded by $4^k$ but remains small in practice. For the codes used later in this work, we have 
$|\AccessibleLogicalOperators|=15$ for $\code{54}$ (logical width $w=k=2$), 
$|\AccessibleLogicalOperators|=4,095$ for $\code{70}$ ($w=k=6$), and 
$|\AccessibleLogicalOperators|=43,725$ for $\code{102}$ ($w=3, k=22$).

Moreover, in our case, the measurement procedure relies on cat states, thus we must make sure that the cat factory of the block is capable of producing sufficiently large cat states.
The {\em block width}, denoted $\physicalWidth$, is defined to guarantee this property.
It is defined as the maximum weight of any representative $\bar{P}_\text{opt} = \Phi_\logicalBasis(P)$ of an accessible logical Pauli operator $P$.
To ensure that the accessible logical Pauli operators can be measured using the companion cat factory of the block, this cat factory is set to produce cat states with weight up to $\physicalWidth$.

Overall, to design a block and its companion factory in the walking cat architecture, we first select a symplectic basis for the code and we pick a logical width. We search for minimum-weight physical representatives for all accessible logical operators. Then, we compute the block width which is used to set the cat factory size.

The exact implementation of the measurement of a logical representative is discussed in \cref{sec:Logical measurements}.
The accessible logical Pauli operators of a block depend on the code used, the symplectic basis and the logical width.
We provide concrete examples in \cref{sec:Weight-Reduced Logical Pauli Operators}.
The section also sets out a weight-reduction algorithm generating low-weight logical representatives to optimize the block width required to reach a given logical width.

\subsection{Clifford frame-tracking}
\label{subsec:Clifford frame-tracking}

In a measurement-based setting, some logical operations can be implemented entirely in software, without any action on the qubits~\cite{raussendorf2001one, knill2005quantum}.
This strategy, called frame-tracking, is commonly used to implement Pauli operations in surface codes~\cite{riesebos2017pauli, litinski2019game, on2023multilayered, paler2014software} or to correct Pauli or even Clifford errors without any physical action on the qubits~\cite{chamberland2018fault}.
Here, we discuss a frame-tracking protocol for the implementation of a subset of Clifford gates that depends on the sets of accessible logical operators. 
This provides a broad set of fast logical Clifford gates.

Consider a unitary $k$-qubit Clifford gate $U$ followed by a sequence of measurements of Pauli operators $P_1, P_2, \dots, P_s$.
The unitary gate $U$ can be pushed to the end of the circuit and removed by replacing each Pauli operator $P_i$ by $U^\dagger P_i U$ without changing the quantum circuit outcome distribution.
The Clifford frame is introduced to keep track of this conjugation.

For each block, we store a pair $(\logicalBasis, U)$ where $\logicalBasis$ is a symplectic basis and $U$ is a $k$-qubit Clifford gate which we refer to as the {\em Clifford frame}.
The Clifford frame is initialized with $U = I$.

In the presence of a Clifford frame $(\logicalBasis, U)$, the measurement of a logical Pauli operator $P$ is performed in two steps.
The software first conjugates the target logical operator $P$ by the inverse of the current Clifford frame, computing $U^\dagger PU$. It then queries the lookup table $\Phi_\logicalBasis$ to obtain the physical representative
$
\bar P = \Phi_\logicalBasis(U^\dagger PU) \cdot
$
The classical control system then sends this string $\bar P$ to the hardware to be physically measured.
Measuring this physical representative $\bar P$ has the effect of measuring the logical operator $P$ preceded by the application of the logical Clifford gate $U$, without ever needing to physical apply an operation to the qubit to implement the gate $U$. This gate is applied simply by storing $U$ in the Clifford frame.
This leads to the Clifford gates by frame tracking.

The {\em frame-tracking implementation} of a $k$-qubit Clifford gate $V$ is defined to be the update of the Clifford frame $U \leftarrow VU$.

To execute a Clifford gate by frame tracking, we store two components:
\begin{enumerate}
    \item The lookup table representing $\Phi_\logicalBasis$.
    \item The Clifford frame $U$, stored as a $2k \times 2k$ binary symplectic matrix with $2k$ phases. This tracks the dynamic software state and consumes $O(k^2)$ bits. 
\end{enumerate}
In this setting, implementing a frame update requires only multiplying the parameter $U$ by $V$ ($U \leftarrow VU$), taking $O(k^3)$ classical bit-operations and reading a value in a lookup table.

\subsection{Accessible logical Clifford gates}
\label{subsec:Accessible logical Clifford gates}

To check whether a measurement sequence can be implemented within our architecture, it suffices to check that each logical operator of the sequence belongs to the accessible set $\AccessibleLogicalOperators$.
The frame-tracking makes the situation more complicated because a frame update may change the set of accessible logical operators.
Then, the accessible set at a given time step depends on the entire history of Clifford updates preceding this time step.
To avoid this issue, we focus on a subset of Clifford gates that map accessible logical Pauli operators to accessible logical Pauli operators.

We define an {\em accessible logical Clifford gate} to be a unitary Clifford operation $V$ which preserves the accessible logical Paulis of \cref{subsec:accessible_logical_operators} under conjugation, that is
$
V \AccessibleLogicalOperators V^{\dagger} = \AccessibleLogicalOperators
$.
This is the subset of logical Clifford gates that we implement by frame-tracking in our architecture.

For some stabilizer codes, we can set the logical width to $k$ while preserving a relatively small block width.
Then, all logical Paulis are accessible and the whole $k$-qubit Clifford group is accessible.
This is the case for example with $\code{54}$ and $\code{70}$ for which we get a physical width of 16 and 18 respectively.
In the case of the code $\code{102}$, we select a logical width of $3$, which induces a block width of 30.
Then, any logical Pauli operator with weight up to 3 is accessible.
All logical SWAPs and single-qubit Clifford operators are guaranteed to be accessible.
This is because SWAP and single-qubit Clifford operations preserve the weight of Pauli operators.

Any Clifford operations which are not accessible are implemented using the techniques set out in \cref{subsec:Derived logical operations and enrichments}.

\section{Logical instruction set}
\label{sec:Logical instruction set architecture}

This section describes the logical instruction set, that is the abstracted set of logical operations provided by a walking cat architecture.

Our logical instruction set consists of logical state preparations (zero, plus, magic state), logical Pauli measurements, and a subset of logical Clifford gates implemented entirely in software by frame updates.

Throughout this section, we consider an instance of the walking cat architecture. Each block comes with corresponding sets of accessible logical Pauli operators and accessible logical Clifford operators as defined in \cref{sec:Accessible logical gates}.

\subsection{The logical instruction set}
\label{subsec:The logical instruction set}

The logical instruction set is made of the following {\em logical instructions}.
\begin{itemize}
    \item (LZ) {\bf Logical zero preparation}: Prepare or reset all the logical qubits of a block in the state $\ket{0}$.
    \item (LP) {\bf Logical plus preparation}: Prepare or reset all the logical qubits of a block in the state $\ket{+}$.
    \item (LT) {\bf Logical magic state preparation}: Prepare a pair of logical qubits in a magic factory in the state $\ket{H}$.
    \item (LM1) {\bf In-block logical measurement}: Measure an accessible logical operator in a block.
    \item (LM2) {\bf Inter-block logical measurement}:
    Measure the product of two accessible logical operators in two different blocks.
    \item (CLIF) {\bf In-block logical Clifford gate}: Apply an accessible Clifford operation by frame-tracking in a block.
\end{itemize}
A block must be prepared with (LZ), (LP) or (LT) before any of the other logical instructions are implemented.

The logical instruction set is defined relative to the set of accessible operations.
Logical measurements are selected in the set of accessible logical operators to guarantee that sufficiently large cat states are available to measure their representatives.

The logical Clifford operations (CLIF) implemented by frame-tracking are required to be accessible so that if we provide an implementation of each of the instructions of the logical instruction set, then any sequence of operations from the logical instruction set can be implemented, independently of the previous instructions.

A magic factory is a memory block with the extra ability to prepare magic states. Therefore, all the logical instructions are available for magic factories but the logical instruction (LT) is not available in memory blocks.
A magic factory can also be used as a memory block if needed.

At the beginning of the computation, all magic factories are available.
Any available magic factory can be allocated to prepare magic states using the instruction (LT).
The logical magic state preparation is not guaranteed to succeed. If it succeeds, we obtain two magic states. 
If it fails and one still needs a magic state, we can reallocate the magic factory for another attempt.
A magic factory remains allocated until the end of the preparation attempt.
If a magic factory containing magic states is allocated for a preparation attempt, its magic states are lost.

Both (LM1) and (LM2) consume cat states.
The instruction (LM2) allows for joint logical measurements between two memory blocks, or a memory block and a magic factory, or two magic factories.
We refer to (LM2) as a {\em two-block instruction}.
All the other instructions are {\em single-block instructions}.

In this work, we design walking cat architectures with accessible sets providing all the logical instructions required for the implementation of \CX gates between any pair of qubits with the circuit of \cref{fig:architecture_overview_Clifford_gates_by_meas}(c).
Together with the instruction (LT), this makes the logical instruction set universal for quantum computation.

Our logical instruction set is similar to the Pauli-based computation model introduced in~\cite{bravyi2016trading}. However, instead of including arbitrary Pauli measurements, we only allow for some specific Pauli measurements corresponding to accessible logical operators.

Another related logical gate set is the logical instruction set of the surface code~\cite{dennis2002topological, raussendorf2007fault, fowler2012surface} or Floquet codes~\cite{hastings2021dynamically, gidney2021fault, paetznick2023performance}.
Our architecture offers a broader set of logical measurements like a $Y$ measurement without consuming any extra logical ancilla.
The ability to measure a logical $Y$ unlocks the full single-qubit Clifford group by frame-tracking.

In some cases, we even have access to all logical measurements inside a block including multiple logical qubits, like with the code $\code{70}$ used later in this work, for which we can measure arbitrary 6-qubit logical Pauli operators in a block.
The ability to measure all 6-qubit logical Pauli operators allows for a frame-tracking implementation of arbitrary 6-qubit Clifford gates acting on a block.

Finally, we also have more flexibility than typical logical instruction sets because inter-block logical measurements are available between any pair of blocks whereas, for example, they are limited to nearest-neighbor blocks in the surface code.
The connectivity induced on the logical qubits, discussed in \cref{subsec:Logical connectivity and parallelism} is not limited by locality.

\begin{figure*}
    \centering
    \includegraphics[width=1\linewidth]{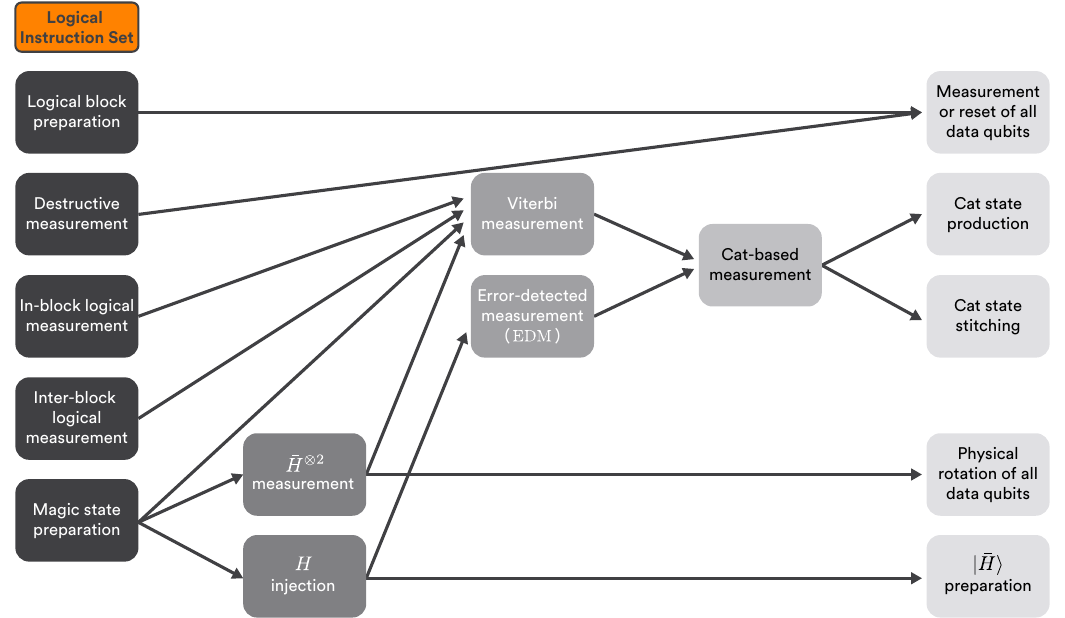}
    \caption{Decomposition of the logical operations from the logical instruction set.}
    \label{fig:logical_isa_decomposition}
\end{figure*}

\subsection{Destructive measurements}
\label{subsec:Destructive measurements}

This subsection introduces two additional logical instructions.
These logical instructions are not necessary for universality as the previous set is already sufficient, but they provide alternative logical measurements that are sometimes faster than (LM1) and (LM2).

This work focuses on CSS codes and therefore $X$-type and $Z$-type operators play a special role.
For each block, denote by $L_X$ and $L_Z$ the sets of logical operators admitting at least one physical representative in the set of $\{I, X\}^{\otimes n}$, and $\{I, Z\}^{\otimes n}$, respectively.
If the Clifford frame of the block is trivial, all the operators of $L_X$ can be measured simultaneously through a direct measurement of the data qubits in the $X$ basis in a single POC.
The same holds for $L_Z$ by measuring the data qubits in the $Z$ basis in one POC.
If the Clifford frame is non-trivial, we can still measure all the data qubits in the $X$ basis (or the $Z$ basis), but the set $L_X$ (or $L_Z$) of logical operators whose outcomes can be extracted from these data qubit measurements must be adjusted.
This is done by updating the sets $L_X$ and $L_Z$ and replacing them by $U^\dagger L_X U$, and $U^\dagger L_Z U$, respectively.

This provides two new logical instructions.
\begin{itemize}
    \item (DMX) {\bf Destructive $X$ measurement}: Measure all the logical operators of the set $U^\dagger L_X U$ in a block with Clifford frame $U$.
    \item (DMZ) {\bf Destructive $Z$ measurement}: Measure all the logical operators of the set $U^\dagger L_Z U$ in a block with Clifford frame $U$.
\end{itemize}
When using these operations, one must be careful with the exact logical measurements implemented because the sets of measured logical operators depend on the logical frame of the block and these sets may be modified during the computation.

A destructive measurement returns the outcome of a complete set of commuting measurements at once but the block is destroyed and must be re-prepared using (LZ), (LP) or (LT) before being used again.
The timing of all these operations is discussed in \cref{sec:Examples of walking cat architectures}.

The main advantage of destructive measurements is their speed: they can extract many logical measurement outcomes at once with a single POC, immediately followed by running the decoder.

\subsection{Implementation of the logical instruction set}
\label{subsec:Implementation of the logical instruction set}

A complete description of the implementation of each instruction of the logical instruction set is provided in this paper.
\cref{fig:logical_isa_decomposition} illustrates the primitive operations used to implement these logical instructions.

We use a standard implementation of the logical block preparation (LZ) and (LP). To initialize a block in the logical zero state, it suffices to prepare all the data qubits of a block in the state $\ket 0$ and to perform a round of syndrome extraction.
The preparation of logical plus states is similar, starting with the data qubits in the $\ket +$ state.

We also use a standard approach for the destructive measurements (DMX) and (DMZ). In (DMX), the data qubits are measured in the $X$ basis and the logical outcomes are extracted by the decoder. The logical instruction (DMZ) is implemented similarly.
We simulate a destructive logical measurement with the streaming beam decoder in \cref{sec:decoder}, demonstrating that the logical outcomes can be extracted by the decoder before the beginning of the next SEC.

Logical measurements are performed using cat states produced by the companion cat factory of the block.
Cat states are post-selected in the cat factory to guarantee that the noise introduced on the data qubits can be corrected by the following SECs without significantly affecting the logical error rate~\cite{shor1996fault}.
The outcome of a cat-based measurement cannot be corrected by the code itself. 
We design three protocols for this task:
An {\em error-detected measurement} or \EDM, which detects errors on the measurement outcome using a fixed number of cat-based measurements,
an {\em error-corrected measurement} or \ECM, which corrects errors on the measurement outcome using a fixed number of cat-based measurements,
and 
a {\em Viterbi measurement} which is an optimal adaptive protocol relying on the Viterbi algorithm~\cite{viterbi2003error} to interrupt the cat-based measurement sequence as soon as the logical outcome has reached a target likelihood.
\cref{sec:Logical measurements} provides detailed descriptions and simulations of \EDM, \ECM and Viterbi measurements.

We select Viterbi measurements for the implementation of in-block and inter-block logical measurements because their average runtime is more favorable than \ECM.
{\EDM}s are used in a subroutine of the magic factories where restarts upon error detection are possible without disrupting the entire computation.

The logical magic state preparation (LT) is performed inside a magic factory using Viterbi measurements, an $H$ injection circuit which injects a physical $H$ state on a logical qubit and an $\bar{H}^{\otimes 2}$ measurement circuit which performs the measurement of a logical $\bar{H}^{\otimes 2}$ gate using a cat state.
The $H$ injection circuit and the $\bar{H}^{\otimes 2}$ measurement circuit are described in \cref{sec:magic_state_factory}.

The implementation of Clifford gates by frame-tracking was covered in \cref{subsec:Clifford frame-tracking}.

\subsection{The \texorpdfstring{$T$}{T} gate and the double \texorpdfstring{$T$}{T} gates}
\label{subsec:The $T$ gate and double $T$ gate}

\begin{figure}
    \centering
    \subfloat[]{

    \includegraphics[width=.75\linewidth]{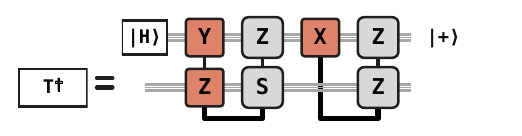}
}

    \subfloat[]{
    \includegraphics[width=.75\linewidth]{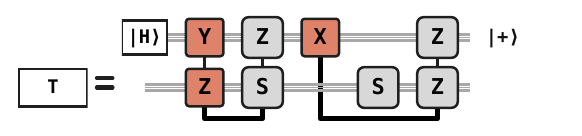}
}

    \subfloat[]{
    \includegraphics[width=1\linewidth]{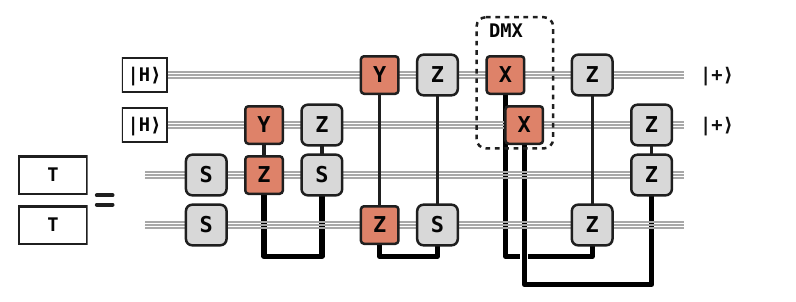}
}

    \subfloat[]{
    \includegraphics[width=1\linewidth]{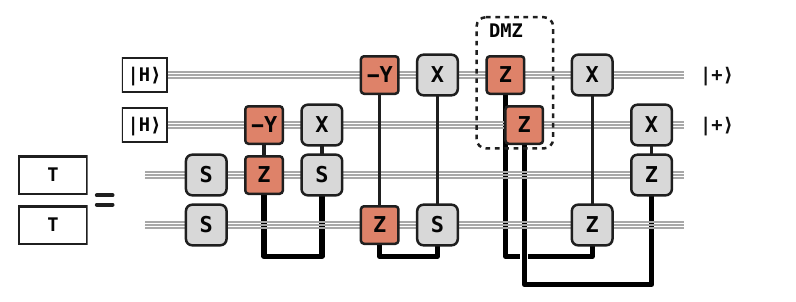}
}
    \caption{
    Orange square boxes represent Pauli measurements and rounded corner gray boxes represent Clifford gates. 
    A Clifford gate connected to a Pauli measurement by a thick line represents a Clifford gate conditioned on the measurement outcome being non-trivial.
    (a) $T^\dagger$ gate implemented using an $H$ state, Pauli measurements and Clifford gates.
    (b) $T$ gate implemented using an $H$ state, Pauli measurements and Clifford gates.
    (c) and (d) A pair of $T$ gates implemented using two $H$ states, Pauli measurements and Clifford gates, where the final measurement of the two $H$ states is performed simultaneously using a destructive measurement (DMX) in (c) and (DMZ) in~(d).
    }
    \label{fig:architecture_overview_T_gates_by_meas}
\end{figure}

This section describes implementations of $T$ and $T^\dagger$ based on instructions from the logical instruction set consuming a magic state and pipelined versions of these operations.

Our starting point is the implementation of the gate $T^\dagger$ based on an $H$ state described in \cref{fig:architecture_overview_T_gates_by_meas}.
For the sake of completeness, we provide a proof that this sequence of Pauli measurements and Clifford gates implements $T^\dagger$.

\begin{proposition}  [Measurement-based $T^\dagger$ gate]
\label{prop:Tdagger_gate_by_measurements}
The circuit of \cref{fig:architecture_overview_T_gates_by_meas}(a) implements the gate $T^\dagger$.
\end{proposition}

The proof of this proposition is straightforward, and closely follows Fig.~7 of~\cite{litinski2019game}. It is deferred to \cref{app:Proofs of circuit identities}.

\begin{proposition} [Measurement-based $T$ gate]
\label{prop:T_gate_by_measurements}
The circuit of \cref{fig:architecture_overview_T_gates_by_meas}(b) implements the gate $T$.
\end{proposition}

\begin{proof}
The $T$ gate implementation in \cref{fig:architecture_overview_T_gates_by_meas}(b) is derived from the $T^\dagger$ implementation in \cref{prop:Tdagger_gate_by_measurements} using $T = S T^\dagger$.
\end{proof}

In this work, we consider magic factories producing pairs of $H$ states. 
The implementation of the two corresponding $T$ gates can be pipelined to reduce their runtime using destructive measurements in the circuits of \cref{fig:architecture_overview_T_gates_by_meas}(c) and (d).

\begin{proposition} [Double $T$ gate]
\label{prop:double_T_gate_by_measurements}
The circuits of \cref{fig:architecture_overview_T_gates_by_meas}(c) and (d) implement the gate $T \otimes T$.
\end{proposition}

One can design a double $T^\dagger$ gate in a similar way starting from the circuit of \cref{prop:Tdagger_gate_by_measurements}.

\begin{proof}
The circuit of \cref{fig:architecture_overview_T_gates_by_meas}(c) is constructed by applying the circuit of \cref{prop:T_gate_by_measurements} to two magic states and two target qubits and grouping together the $X$ measurement of the two input $H$ states.

The circuit of \cref{fig:architecture_overview_T_gates_by_meas}(d) is derived from the circuit (c) by inserting two $H$ gates right after the preparation of each of the two $H$ states and pushing these $H$ gates through the whole circuit.
The insertion of these $H$ gates does not change the operation performed by the circuit because $H \ket H = \ket H$.
\end{proof}

If the Clifford frame $U$ is such that $U^\dagger L_X U = L_X$ or $U^\dagger L_Z U = L_X$ then, one can implement the two logical $X$ measurements in \cref{fig:architecture_overview_T_gates_by_meas}(c) in a single SEC using a destructive measurement (DMX) or (DMZ) respectively.
Similar conditions holds for the double $T$ gate based on (DMZ).
In this work, we use the double $T$ gate for magic states generated with the \CHfactory factory introduced later because the Clifford gates it includes satisfy the condition state above.
However, the magic states generated using the \MEK factory are consumed using two sequential $T$ gate using the circuit of \cref{fig:architecture_overview_T_gates_by_meas}(b).
The execution time for these variants of $T$ gates is analyzed in \cref{sec:Examples of walking cat architectures}.

\subsection{Logical connectivity and parallelism}
\label{subsec:Logical connectivity and parallelism}

The walking cat architecture is designed to ensure maximum connectivity and parallelism at the block level.

To keep the discussion simple, we assume throughout this subsection that the SEC time of all the blocks of the architecture is identical.

The quantum computation is executed as a sequence of instructions from the logical instruction set.
Initially, all the blocks are available.
At the beginning of each SEC, a set of logical instructions is assigned to the blocks available.
These blocks then become unavailable for the duration of the instruction.
The execution time of each logical instruction is a random variable depending on the implementation of the instruction and the block details.

An available block can be assigned any single-block instruction, except (LT) which can only be assigned to a magic factory.
Moreover, we say that the walking cat architecture has {\em full block connectivity} because any two-block operation (LM2) can be assigned to any pair of available blocks, independently of the distance between the blocks.

The walking cat architecture is said to have {\em full block parallelism} because at a given time step, we can execute simultaneously any combination of logical instructions acting on different blocks.
Therefore, with $N$ blocks, we can execute simultaneously up to $N$ single-block instructions or $\lfloor N/2 \rfloor$ two-block instructions.
We can also execute simultaneously a combination of $s$ single-block instructions and $t$ two-block instructions with $s+2t \leq N$ as long as these instructions are supported on different blocks.

To ensure the full block connectivity and parallelism, we pipeline the cat state factories, the cat qubit routing and the Bell qubit routing. 
This guarantees that cat states are produced fast enough to support the full block parallelism and to keep up with the SEC time,
and Bell states can be distributed fast enough to stitch arbitrary pairs of cat states providing the full block connectivity.
The cat factory pipelining and the qubit routing are analyzed in \cref{sec:Examples of walking cat architectures}.
In this work, when we report the total number of physical qubits in a walking cat architecture, we include all the physical qubits consumed for the cat factory pipelining and the qubit routing.

\subsection{Derived logical operations and enrichments}
\label{subsec:Derived logical operations and enrichments}

\begin{figure}[bh!]
    \centering
    \subfloat[]{
    \includegraphics[width=\linewidth]{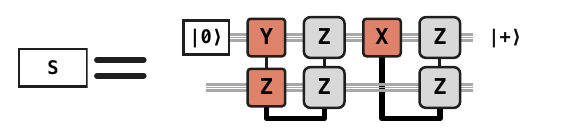}
}

    \subfloat[]{
    \includegraphics[width=\linewidth]{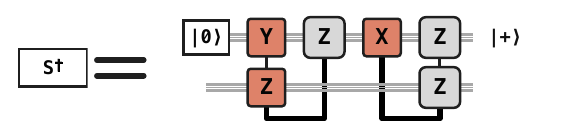}
}

    \subfloat[]{
    \includegraphics[width=\linewidth]{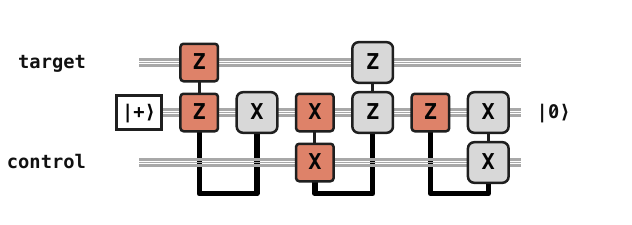}
}
    
    \caption{
    We use the notation of \cref{fig:architecture_overview_T_gates_by_meas}.
    (a) $S^\dagger$ gate implemented Pauli measurements and Pauli gates.
    (b) $S$ gate implemented Pauli measurements and Pauli gates.
    Both $S$ and $S^{\dagger}$ gadgets shown here closely follow Fig.~11(b) of~\cite{litinski2019game}.
    (c) \CX gate implemented using Pauli measurements and Pauli gates (circuit from \cite{chao2020optimization}).
    }
    \label{fig:architecture_overview_Clifford_gates_by_meas}
\end{figure}

Additional logical operations can be derived from the logical instruction set.
For example, all logical Pauli operations can be implemented by frame-tracking as part of the logical instruction (CLIF). In other words, the set of accessible logical Clifford gates always contains the full Pauli group.

In \cref{subsec:The $T$ gate and double $T$ gate}, we discussed the implementation of the $T$ gate, the $T^\dagger$ gate and their double version using Pauli measurements and conditional Clifford operations.
Unitary Clifford gates can be implemented in a similar way using Pauli measurements and conditional Pauli operations from the logical instruction set.
\cref{fig:architecture_overview_Clifford_gates_by_meas} shows the implementation of the $S$ gate, the $S^\dagger$ gate and the \CX gate.
The implementation of the \CY and \CZ gates as a sequence of logical instructions can be derived from the \CX implementation by conjugating the target qubit by a single-qubit rotation.

Additional logical operations can be added to our architecture, providing a broader logical instruction set at the price of making the architecture more complex.
Even though in this paper we favor the simplest design based on the previously described logical instructions, we discuss enrichments of the walking cat architecture based on transversal gates and permutation-based gates in \cref{sec:Enriched walking cat architecture}.
\section{Instances of the walking cat architecture}
\label{sec:Examples of walking cat architectures}

In this section, we discuss specific instantiations of the walking cat architecture.
Based on the analysis and simulations of all the components of the logical architecture discussed in \cref{part:Implementation of the components}, we estimate the compute capacity and the resource cost of these examples.

All the estimates reported in this section are based on the moving-qubit model with noise rate $p=10^{-4}$, $\pleak=10^{-5}$ and $\ploss=10^{-7}$ and POC time of $200~\mu$s and transport step of $10~\mu$s described in~\cref{subsec:Noise strength and POC time}.

\subsection{Walking cat architecture configurations}
\label{subsec:Walking cat architecture configurations}

A walking cat {\em architecture configuration} is defined by selecting the number of memory blocks and their type and the number of magic factories and their type.

Here, we focus on the three specific codes introduced previously $\code{54}$, $\code{70}$ and $\code{102}$.
These three codes all achieve a logical error rate per SEC of the order of $10^{-10}$ with a SEC time around 30 POC, that is about 6 ms (see~\cref{tab:loss_leakage_summary}).
In principle, we can select any code from the three-ring framework which covers GB codes, BB codes and cyclic HGP codes.
More examples of codes compatible with our architecture are provided in \cref{app:sec_code_database_table}.

We consider two options for the memory code: $\code{102}$ or $\code{70}$.
For the magic factory, we use either the \MEK factory implemented inside $\code{70}$ or the \CHfactory factory implemented in $\code{54}$.
The performance of these two magic factories is analyzed in \cref{sec:magic_state_factory} and summarized in \cref{tab:Magic factory parameters}.

We use the notation 
\begin{align}
\conf{N_M}{102}{N_T}{CH2}
\end{align}
to represent the walking cat architecture configuration based on $N_M$ memory blocks built over $\code{102}$ and $N_T$ \CHfactory factories. Similar notations are used for other configurations.

For simplicity, we assume that all the memory blocks are based on the same code and all the magic factories are identical. This constraint can be relaxed to build other instances of the walking cat architecture.

\subsection{Component allocation}
\label{subsec:Component allocation}

As shown in \cref{fig:architecture_overview_chip}, we use one cat factory for each memory block and one cat factory for each magic factory to maximize the logical measurement parallelism.
The cat factories associated with the memory and magic factory may have different sizes because the corresponding block could be different.

For the memory block, the logical width is set to $6$ for the $\code{70}$ and $3$ for $\code{102}$, corresponding block width of $18$ and $30$ respectively.

The \MEK magic factory based on $\code{70}$ uses the same block width of 18.
The \CHfactory magic factory is a special case. We say that it has block width 54 because it requires a companion factory producing 54-qubit cat states to implement the \CHfactory scheme.

The cat state verification threshold is set to $\varepsilon = 10^{-10}$ which is sufficiently low to have a negligible effect on the logical error rate of the code.

\begin{table}[ht]
\centering
\begin{tabular}{|l|r|r|r|r|}
\hline
\makecell{\bf Block} & \makecell{\bf SEC\\\bf (POC)} & \makecell{\bf Block\\ \bf width} & \makecell{\bf Prod.\ time\\\bf (POC)} & \makecell{\bf Distrib.\ time\\\bf (POC)} \\
\hline
\code{102}          & 33.70 & 30 & 14 & 5.10 \\
\hline
\code{70}           & 27.70 & 18 & 14 & 3.50 \\
\hline
\code{54}           & 28.15 & 16 & 13 & 2.70 \\
\hline
\MEK           & 27.70 & 18 & 14 & 3.50 \\
\hline
\CHfactory           & 28.15 & 54 & 15 & 2.70 \\
\hline
\end{tabular}
\caption{Comparison between the SEC time and the cat state production and distribution time. We assume the each cat qubit is transported over at most $n$ steps during the distribution where $n$ is the length of the code used in the target block.
}
\label{tab:cat_factory_requirements}
\end{table}

The cat state factory must be fast enough to produce a cat state and deliver it to the associated block within 1 SEC. \cref{tab:cat_factory_requirements} shows that our cat factories are fast enough in all cases. These numbers are based on the results of \cref{sec:Cat factories}.

Following the analysis of \cref{sec:Bell_factories}, we allocate $\lceil N/3 \rceil$ Bell factories where $N$ is the total number of blocks (memories or magic factories).

We set the reservoir size to 200 physical qubits which is large enough for tens of blocks based on \cref{sec:The qubit factory}.

\subsection{Transport overhead}
\label{subsec:Transport overhead}

We previously argued that a cat factory can produce and distribute a cat state  fast enough to send a cat state to its companion block within less than one SEC.
However, once the cat state is consumed, the cat qubits must be sent back to the cat factory after going through the chip outer loop as shown in \cref{fig:architecture_overview_chip}.
To make sure that the cat factories have available qubits to start the production of a new cat state at the beginning of each SEC, we include additional cat qubits so that a cat can be prepared while others are transported.

To compute the length of the chip outer loop, we assume that each memory block and each magic factory occupy a $4 \times n$ rectangle where $n$ is the length of the underlying code.
Moreover, a cat factory for a block with block width $\physicalWidth$ occupies a $4 \times 2\physicalWidth$ rectangle.
We insert one empty horizontal row between any two block for cat state transport.
The outer loop is the smallest rectangle that contains all the architecture components.
Its length can be calculated for any walking cat architecture configuration.

Overall, the number of additional transport cat qubits required is obtained as
$\lceil t / f \rceil$
where $t$ is the time taken to go through the outer loop in SEC and $f$ is the cat factory flow, that is the number of qubits going through the cat factory per SEC. The cat factory flow is reported in \cref{tab:cat_heuristic}.
The number of transport Bell qubits is estimated similarly from the Bell factory flow that can be found in \cref{tab:summary_bell_factory}.

This approach provides an upper bound on the number of qubits required for cat qubit and Bell qubit transport.
Shorter transport paths are possible. One could for instance directly send these qubits back to their factory. 
This could reduce the total qubit count but transport only accounts for a small fraction of the total number of qubits, therefore we choose the simple solution provided by the outer loop transport.

\subsection{Qubit allocation}
\label{subsec:Qubit allocation}

\begin{figure}
    \centering
    \includegraphics[width=1\linewidth]{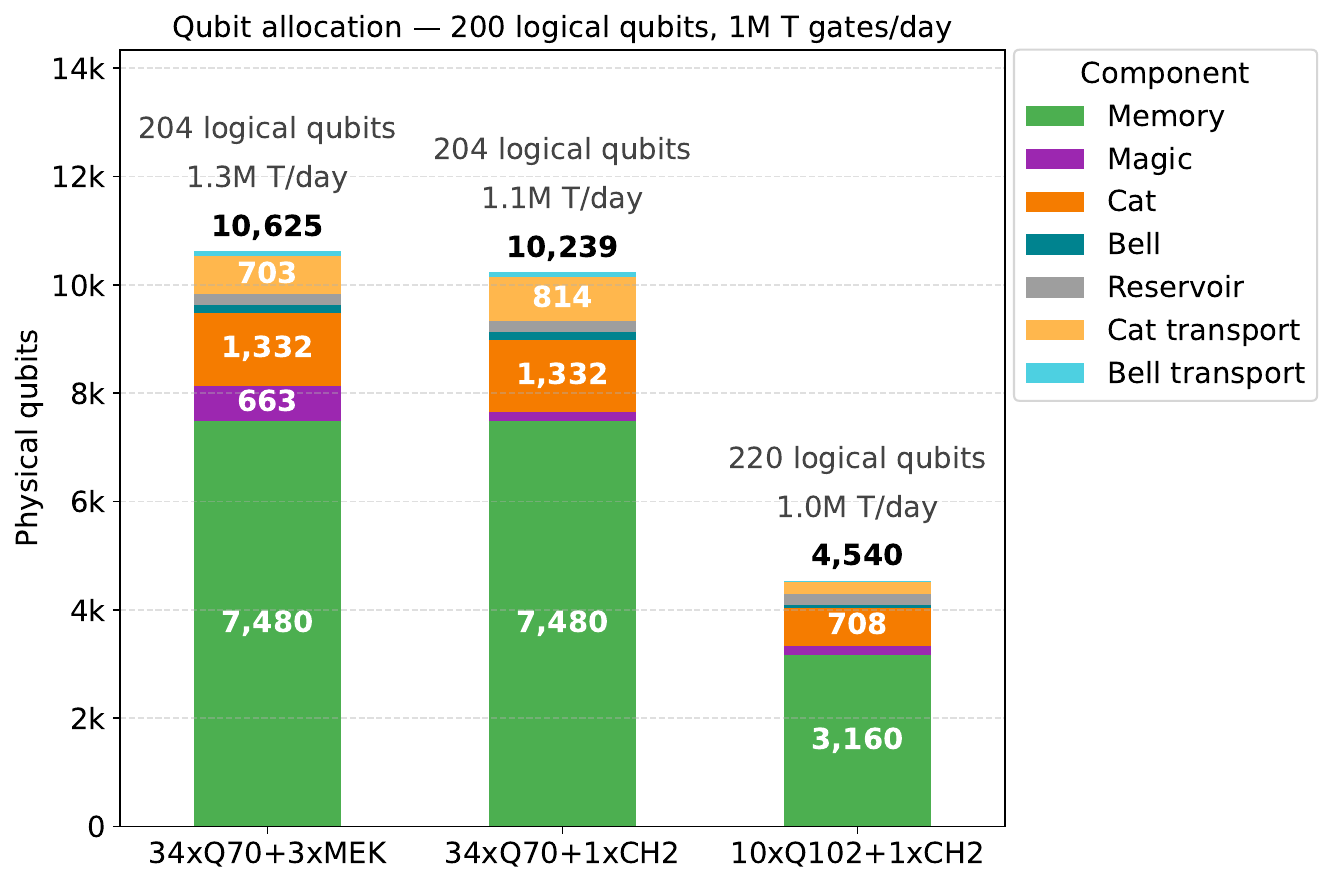}

    (a)

    \includegraphics[width=1\linewidth]{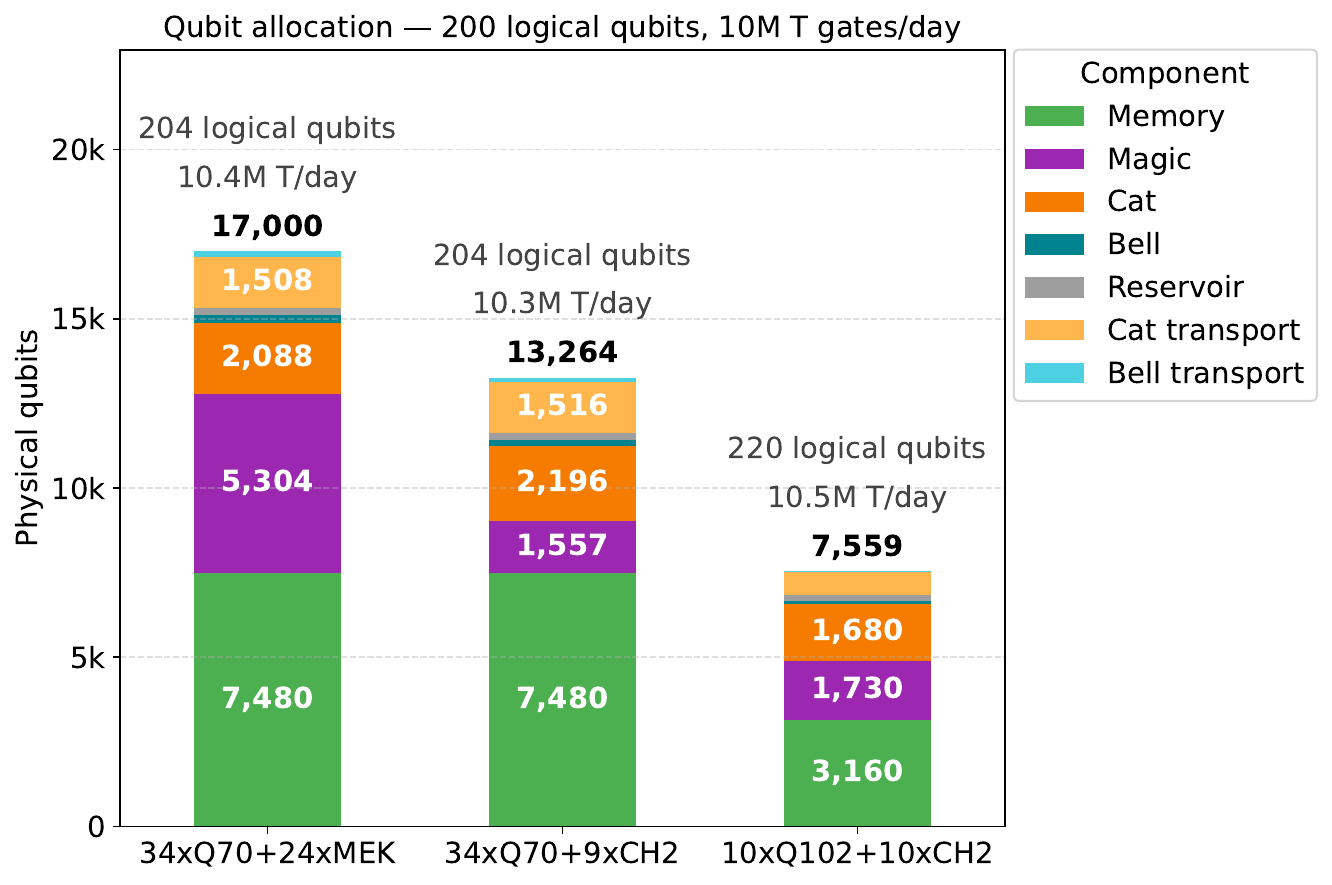}

    (b)
    \caption{Qubit allocations in instances of walking cat architectures with at least 200 logical qubits capable of executing up to 1M $T$ gates per day (a) and 10M $T$ gates per day (b). Here the $T$ gate time includes magic state preparation and implementation of a $T$ gate on a logical qubit in an arbitrary memory block. This $T$ gate time assumes the double $T$ gate circuit of \cref{fig:architecture_overview_T_gates_by_meas}.} 
    \label{fig:qubit_allocation_K200}
\end{figure}

\begin{table*}[ht]
\centering
\label{tab:qubit_summary_full}
\begin{tabular}{|l|r|r|r|r|r|r|r|r|r|r|}
\hline
\textbf{Configuration} & \makecell{ \bf Log.\\ \bf   qubits} & \makecell{ \bf  T gates/\\  \bf  day} &  \textbf{Memory} &  \textbf{Magic} &  \textbf{Cat} &  \textbf{Bell} &  \textbf{Reserv} & \makecell{  \bf Cat\\   \bf transp} & \makecell{  \bf Bell\\  \bf  transp} & \makecell{ \bf  Phys.\\  \bf  qubits} \\
\hline
$\conf{17}{70}{3}{MEK}$   & 102 & 1.3M & 3,740 (65\%) &   663 (12\%) &   720 (13\%) &  84 (1\%) & 200 (3\%) &   280 (5\%) &  35 (1\%) &  5,722 \\
$\conf{17}{70}{1}{CH2}$   & 102 & 1.1M & 3,740 (71\%) &   173 (3\%)  &   720 (14\%) &  72 (1\%) & 200 (4\%) &   339 (6\%) &  36 (1\%) &  5,280 \\
$\conf{5}{102}{1}{CH2}$   & 110 & 1.0M & 1,580 (63\%) &   173 (7\%)  &   408 (16\%) &  24 (1\%) & 200 (8\%) &   121 (5\%) &   8 (0\%) &  2,514 \\
\hline
$\conf{17}{70}{24}{MEK}$  & 102 & 10.4M & 3,740 (32\%) & 5,304 (45\%) & 1,476 (12\%) & 168 (1\%) & 200 (2\%) &   861 (7\%) &  98 (1\%) & 11,847 \\
$\conf{17}{70}{9}{CH2}$   & 102 & 10.3M & 3,740 (46\%) & 1,557 (19\%) & 1,584 (20\%) & 108 (1\%) & 200 (2\%) &   862 (11\%) &  63 (1\%) &  8,114 \\
$\conf{5}{102}{10}{CH2}$  & 110 & 10.5M & 1,580 (29\%) & 1,730 (32\%) & 1,380 (25\%) &  60 (1\%) & 200 (4\%) &   500 (9\%)  &  25 (0\%) &  5,475 \\
\hline\hline
$\conf{34}{70}{3}{MEK}$   & 204 & 1.3M & 7,480 (70\%) &   663 (6\%)  & 1,332 (13\%) & 156 (1\%) & 200 (2\%) &   703 (7\%) &  91 (1\%) & 10,625 \\
$\conf{34}{70}{1}{CH2}$   & 204 & 1.1M & 7,480 (73\%) &   173 (2\%)  & 1,332 (13\%) & 144 (1\%) & 200 (2\%) &   814 (8\%) &  96 (1\%) & 10,239 \\
$\conf{10}{102}{1}{CH2}$  & 220 & 1.0M & 3,160 (70\%) &   173 (4\%)  &   708 (16\%) &  48 (1\%) & 200 (4\%) &   235 (5\%) &  16 (0\%) &  4,540 \\
\hline
$\conf{34}{70}{24}{MEK}$  & 204 & 10.4M & 7,480 (44\%) & 5,304 (31\%) & 2,088 (12\%) & 240 (1\%) & 200 (1\%) & 1,508 (9\%) & 180 (1\%) & 17,000 \\
$\conf{34}{70}{9}{CH2}$   & 204 & 10.3M & 7,480 (56\%) & 1,557 (12\%) & 2,196 (17\%) & 180 (1\%) & 200 (2\%) & 1,516 (11\%) & 135 (1\%) & 13,264 \\
$\conf{10}{102}{10}{CH2}$ & 220 & 10.5M & 3,160 (42\%) & 1,730 (23\%) & 1,680 (22\%) &  84 (1\%) & 200 (3\%) &   670 (9\%) &  35 (0\%) &  7,559 \\
\hline\hline
$\conf{50}{70}{3}{MEK}$   & 300 & 1.3M & 11,000 (71\%) &   663 (4\%)  & 1,908 (12\%) & 216 (1\%) & 200 (1\%) & 1,325 (9\%)  & 162 (1\%) & 15,474 \\
$\conf{50}{70}{1}{CH2}$   & 300 & 1.1M & 11,000 (73\%) &   173 (1\%)  & 1,908 (13\%) & 204 (1\%) & 200 (1\%) & 1,482 (10\%) & 170 (1\%) & 15,137 \\
$\conf{14}{102}{1}{CH2}$  & 308 & 1.0M &  4,424 (72\%) &   173 (3\%)  &   948 (15\%) &  60 (1\%) & 200 (3\%) &   347 (6\%)  &  25 (0\%) &  6,177 \\
\hline
$\conf{50}{70}{24}{MEK}$  & 300 & 10.4M & 11,000 (50\%) & 5,304 (24\%) & 2,664 (12\%) & 300 (1\%) & 200 (1\%) & 2,294 (10\%) & 275 (1\%) & 22,037 \\
$\conf{50}{70}{9}{CH2}$   & 300 & 10.3M & 11,000 (60\%) & 1,557 (9\%)  & 2,772 (15\%) & 240 (1\%) & 200 (1\%) & 2,310 (13\%) & 200 (1\%) & 18,279 \\
$\conf{14}{102}{10}{CH2}$ & 308 & 10.5M &  4,424 (48\%) & 1,730 (19\%) & 1,920 (21\%) &  96 (1\%) & 200 (2\%) &   824 (9\%)  &  48 (1\%) &  9,242 \\
\hline
\end{tabular}
\label{tab:qubit_allocation_details}
\caption{Qubit allocation for instances of the walking cat architecture with 100, 200 and 300 logical qubits and 1 to 10 million $T$ gates per day.}
\end{table*}

Qubit allocation in six instances of the walking cat architecture with about 200 logical qubits running up to 1M or 10M $T$ gates per day are shown in \cref{fig:qubit_allocation_K200}.
See \cref{tab:intro_example_table} for the exact number of logical qubits and $T$ gates per day achievable by these configurations.
These numbers account for all the physical qubits used, including transport and local and global reservoirs.
As expected, increasing the $T$ gate requirements leads to a greater qubit allocation toward magic factories.
Moreover, we see that the \CHfactory magic factory consumes fewer qubits than the \MEK magic factory. The high density of logical qubits offered by $\code{102}$ is also clear.

The detailed qubit allocation for all the instances of the walking cat architecture presented in \cref{tab:intro_example_table} and additional instances with 300 logical qubits are provided in \cref{tab:qubit_allocation_details}.

\subsection{Logical instruction sets}
\label{subsec:Logical instruction sets}

The logical instruction set depends on the code and the logical width chosen.
\cref{tab:example_accessible_ops} shows the set of accessible logical Pauli operators and the set of accessible logical Clifford gates that is available for the logical instructions (LM1), (LM2) and (CLIF).
The notation $\PauliGroup_{22}^{\leq 3}$ refers to the set of all $22$-qubit Pauli operators with weight up to~$3$.
We use the notation $\mathrm{Clif}_m$ for the $m$-qubit Clifford group and $\mathrm{Clif}_1^{\otimes 22}$ refers to the products of single-qubit Clifford gates.

\begin{table}[ht]
\centering
\begin{tabular}{|l|c|c|c|}
\hline
 & $\code{54}$ & $\code{70}$ & $\code{102}$ \\
\hline
Logical width & 2 & 6 & 3 \\
\hline
LM1  & $\PauliGroup_2$ & $\PauliGroup_6$ & $\PauliGroup_{22}^{\leq 3}$ \\
\hline
LM2  & $\PauliGroup_2 \times \PauliGroup_2$ & $\PauliGroup_6 \times \PauliGroup_6$ & $\PauliGroup_{22}^{\leq 3} \times \PauliGroup_{22}^{\leq 3}$ \\
\hline
CLIF & $\mathrm{Clif}_2$ & $\mathrm{Clif}_6$ & $\mathrm{Clif}_1^{\otimes 22} + \mathrm{SWAPs}$ \\
\hline
\end{tabular}
\caption{Logical measurements and logical Clifford gates available by frame-tracking for three examples of codes with a specific choice of the logical width.
}
\label{tab:example_accessible_ops}
\end{table}

The time for the logical instructions available and derived logical operations is reported in \cref{tab:example_logical_op_times} for our two options for logical memories.

\begin{table}[ht]
\centering
\begin{tabular}{|l|r|r|}
\hline
\textbf{Operation} & $\code{70}$ & $\code{102}$ \\
\hline
LZ / LP        & 0.0055 & 0.0067 \\
\hline
LT (CH2)       & 0.0757 & 0.0757 \\
LT (MEK)       & 0.2643 & 0.2643 \\
\hline
LM1            & 0.0337 & 0.0414 \\
LM2            & 0.0342 & 0.0495 \\
\hline
CLIF           & 0      & 0      \\
SWAP           & 0      & 0      \\
\hline
DMX / DMZ      & 0.0055 & 0.0067 \\
\hline
T gate x2 (CH2) & 0.1507 & 0.1652 \\
T gate x2 (MEK) & 0.4000 & 0.4297 \\
\hline
\end{tabular}
\caption{Average logical operation times in seconds.
The magic factory prepares a pair of $T$ together.
For LT, we report the time for the preparation of both magic states.
The $T$ gate time given is also for two $T$ gates.
We assume that these $T$ gates are performed with the double T gate for the \CHfactory factory and with two consecutive $T$ gates for the \MEK factory.
We report the time in second and not in SEC because the architecture may involves codes with different SEC time in the magic factory and the memory block.
}
\label{tab:example_logical_op_times}
\end{table}

The time for logical measurements is estimated using the results of \cref{sec:Logical measurements}.
We assume that (LM1) and (LM2) are implemented using Viterbi measurements. We estimate the average number of SEC required for this measurement as a function of the weight of the physical representative of the logical operator measured. We use the block width as an upper bound for this weight (leading to a pessimistic estimate of the runtime).
An additional SEC is added for the decoding reaction time.
The logical measurement time in SEC is converted to seconds using the estimates for the SEC time provided in \cref{tab:loss_leakage_summary}.

When an inter-block logical measurement (LM2) is performed between two blocks hosting two different codes, we use the slowest of the two SEC time to estimate the runtime in SEC.
Extra idle steps must be inserted into the SEC of the slowest of the two codes to keep them synchronized during such an inter-block measurement.
We expect that this does not significantly impact the code performance because the SEC times are relative close to each other and the idle noise rate is sufficiently low.

\begin{table}[ht]
\centering
\begin{tabular}{|l|r|r|r|}
\hline
 & $\code{102}$ & $\code{70}$ & $\code{54}$ \\
\hline
$\code{102}$ & 0.0495 & 0.0490 & 0.0419 \\
$\code{70}$  & 0.0490 & 0.0342 & 0.0347 \\
$\code{54}$  & 0.0419 & 0.0347 & 0.0346 \\
\hline
\end{tabular}
\caption{Duration of an inter-block logical measurement (LM2) in seconds for all pairs of blocks.}
\label{tab:lm2_times_3x3}
\end{table}

The duration of inter-block measurements between all pairs of codes is reported in \cref{tab:lm2_times_3x3}.

The time for magic state preparation and $T$ gate depends on the magic factory analyzed in \cref{sec:magic_state_factory}.
The average time for magic state generation with (LT) is derived from the expected time per attempt and the failure rate per attempt reported in \cref{tab:Magic factory parameters}.
To obtain the double $T$ gate time, we add the time for two inter-block logical measurements between the magic factory code and the memory code and the time for a destructive measurement.
The time for two $T$ gates is estimated similarly.

\subsection{The single-code architecture}
\label{subsec:The single-code architecture}

\begin{figure}
    \centering
    \includegraphics[width=1\linewidth]{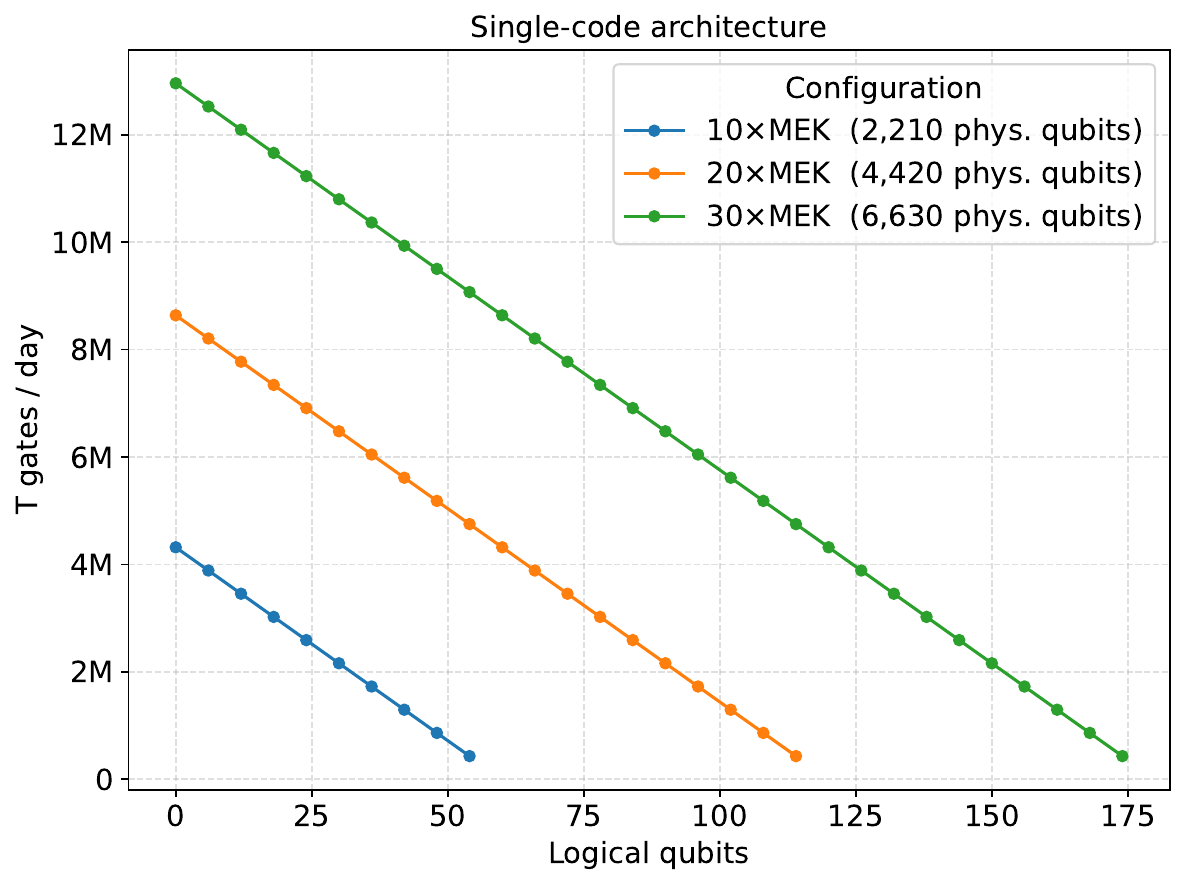}
    \caption{Tradeoff between the number of logical qubits and the magic state production achievable using the single-code architecture $N_T \times \MEK$ for $N_T = 10, 20, 30$. We can use $m$ of the $N_T$ \MEK factories to play the role of memory blocks, forming $6m$ logical qubits, and keep the remaining $N_T-m$ \MEK factories to produce magic states.}
    \label{fig:single_code_architecture}
\end{figure}

Here, we describe a particularly simple instance of the walking cat architecture that relies entirely on a single quantum error-correcting code obtained from the configurations of the form 
\begin{align}
N_T \times \MEK,
\end{align}
made with $N_T$ copies of the \MEK factory.

Each \MEK factory can be used either as a magic factory producing pairs of magic states or as a memory block storing six logical qubits in the code $\code{70}$.
A block may assume different roles at distinct stages of the computation to accommodate the varying needs for magic states and memory of the compiler.
The configuration $N_T \times \MEK$ can be used to achieve any of the configurations
\begin{align}
\conf{m}{70}{(N_T-m)}{MEK},
\end{align}
where $m$ may vary between 0 and $N_T$ during the computation.
\cref{fig:single_code_architecture} shows the tradeoffs achievable by varying $m$ for 10, 20 and 30 blocks.

To guarantee that no information is lost when allocating a memory block as a magic factory, one must make sure that this block does not contain any logical information.
To avoid losing magic states when a magic factory is made a memory block, it is better to consume its magic states before initializing it with a logical zero preparation (LZ) or a logical plus preparation (LP).

\clearpage

\part{Implementation of the logical instructions}
\label{part:Implementation of the components}
\section{The memory block}
\label{sec:Memory block}

In this section we introduce the \emph{three-ring} framework, which provides a unified architectural description of the three LDPC code families considered in this work: GB codes~\cite{kovalev2013quantum}, BB codes~\cite{bravyi2024high} and Cyclic HGP codes~\cite{aydin2025cyclic}. The key idea is to index qubits by elements of a finite abelian group with three cyclic factors and to realize syndrome extraction through cyclic transport along three corresponding ring families.

This framework is the basis of the memory component used throughout the walking cat architecture. It gives a common description of the code families compatible with our transport model.

The architecture examples presented in this work focus on three concrete memory blocks, denoted \code{102}, \code{70}, and \code{54}. We introduce the shorthand GB$w$ and BB$w$ for generalized bicycle and bivariate bicycle codes of check weight $w$. The \code{102}, \code{70}, and \code{54} codes are built from $[[102, 22, 9]]$ GB8, $[[70, 6, 9]]$ BB7, and $[[54, 2, 10]]$ GB8 codes, respectively. 

Their essential properties--alongside those of other three-ring codes--are summarized in \cref{tab:three_ring_code_examples}, while the schedule permutations and circuit-level distances of the syndrome-extraction circuits used in this work are collected in \cref{tab:memory_block_syndrome_schedules}.
Additional properties for the codes introduced in this work, including the lattice dimensions $\ell$ and $m$ and the generating polynomials, are given in \cref{app:sec_code_database_table}, which also contains many other examples of codes compatible with the walking cat architecture.

The rest of this section is organized as follows. \cref{subsec:three-ring-framework} introduces the common algebraic and transport picture. \cref{subsec:three-ring-syndrome-extraction} describes the syndrome-extraction circuit enabled by the three-ring layout.

\begin{table}[t]
\centering
\scriptsize
\setlength{\tabcolsep}{3pt}
\renewcommand{\arraystretch}{1.15}
\begin{adjustbox}{max width=\columnwidth}
\begin{tabular}{|l|c|c|c|c|c|}
\hline
\rule[-1.1ex]{0pt}{3.6ex}\textbf{Family} & \textbf{[[n,k,d]]} & \textbf{$r_{enc}$} & \textbf{$d_{\mathrm{circ}}$} & \textbf{$w_{\mathrm{check}}$} & \textbf{Reference} \\
\hline
GB & [[102, 22, 9]] & $1/5$ & $9$ & 8 & This work (\code{102}) \\
\hline
BB & [[70, 6, 9]] & $1/12$ & $9$ & 7 & This work (\code{70}) \\
\hline
GB & [[54, 2, 10]] & $1/27$ & $9$ & 8 & This work (\code{54}) \\
\hline
BB & [[30, 4, 5]] & $1/8$ & $\leq 5$ & 5 & \cite{ye2025quantum} \\
\hline
BB & [[48, 4, 7]] & $1/12$ & $\leq 7$ & 5 & \cite{ye2025quantum} \\
\hline
Cyclic HGP & [[450, 32, 8]] & $1/14$ & $\leq 8$ & 6 & \cite{aydin2025cyclic} \\
\hline
Cyclic HGP & [[882, 98, 8]] & $1/9$ & $\leq 8$ & 8 & \cite{aydin2025cyclic} \\
\hline
Cyclic HGP & [[882, 50, 10]] & $1/18$ & $\leq 10$ & 6 & \cite{aydin2025cyclic} \\
\hline
BB & [[72, 12, 6]] & $1/6$ & $\leq 6$ & 6 & \cite{bravyi2024high} \\
\hline
BB & [[90, 8, 10]] & $1/11$ & $\leq 8$ & 6 & \cite{bravyi2024high} \\
\hline
BB & [[144, 12, 12]] & $1/12$ & $\leq 10$ & 6 & \cite{bravyi2024high} \\
\hline
\end{tabular}
\end{adjustbox}
\caption{Representative three-ring codes compatible with the walking cat architecture. The first three rows correspond to the codes used in the walking cat architecture memory and magic factory code blocks. The parameters $[[n, k, d]]$ give the number of data qubits, logical qubits, and code distance; the data and ancilla registers together use $2n$ qubits before the beacon and local reservoir qubits of the full memory block are added. The encoding rate is $r_{enc} = k/n$. The circuit-level distance $d_{\mathrm{circ}}$ is the minimum number of faulty operations in the syndrome-extraction circuit required to generate an undetectable logical error. The check weight $w_{\mathrm{check}}$ gives the number of data neighbors visited by an ancilla during one SEC.}
\label{tab:three_ring_code_examples}
\end{table}

\subsection{The three-ring framework}
\label{subsec:three-ring-framework}

We formally define a three-ring code and its associated transport structure, which together constitute the three-ring framework.

\paragraph{Algebraic structure.}
Let $a,b,c \geq 1$ and define
\[
G = \Z_a \times \Z_b \times \Z_c,
\qquad n = abc.
\]
Each element $g = (u,v,z) \in G$ labels both a data qubit and an ancilla qubit. We denote these by $\delta(g)$ and $\alpha(g)$, respectively.
Accordingly, we write
\[
\mathcal{D} := \{\delta(g) : g \in G\},
\qquad
\mathcal{A} := \{\alpha(g) : g \in G\},
\]
for the full data and ancilla registers.

\paragraph{Physical layout and transport.}
Data and ancilla qubits are arranged in two parallel rows as depicted in \cref{fig:architecture_overview_memory}, with $\delta(u,v,z)$ aligned vertically above $\alpha(u,v,z)$. We refer to this arrangement as the \emph{one-dimensional layout}: one row contains the data qubits and the other row contains the ancilla qubits, and in each row the qubit labeled by $(u,v,z)$ is placed at position
\[
q = ubc + vc + z.
\]
Syndrome extraction alternates between nearest-neighbor two-qubit gates acting on aligned pairs and cyclic transport of ancilla qubits. In the walking cat architecture, the one-dimensional layout is augmented by two additional rows.
The first additional row corresponds to the beacon qubits which facilitate loss detection (described further in \cref{sec:Correction of losses and leakages}; see in particular \cref{subsec:atom-loss-gadget}).
The second additional row is an otherwise empty transport “highway” that accelerates ancilla motion during syndrome extraction.

Transport is implemented using three families of cyclic rings (throughout this paper, the term ``ring'' always refers to the shape, not the algebraic structure), represented in \cref{fig:architecture_overview_memory}(b), corresponding to the three factors of $G$.

In the one-dimensional layout, the labels {\em short}, {\em medium}, and {\em long} refer to the spatial period with which the corresponding transport pattern repeats along a row:

\begin{itemize}
    \item short rings, corresponding to shifts in the $z$ axis, have period $c$ in the one-dimensional layout, with a shift of $t$ corresponding to a displacement of $t$ sites,
    \item medium rings, corresponding to shifts in the $v$ axis, have period $bc$ in the one-dimensional layout, with a shift of $s$ corresponding to a displacement of $sc$ sites,
    \item long rings, corresponding to shifts in the $u$ axis, have period $abc$ in the one-dimensional layout, with a shift of $r$ corresponding to a displacement of $rbc$ sites.
\end{itemize}

\begin{figure}[t!]
    \centering
    \includegraphics[width=1\linewidth]{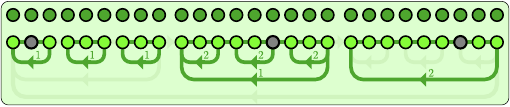}
    \caption{Examples of different ring shifts for $G=\mathbb{Z}_3\times\mathbb{Z}_3\times \mathbb{Z}_3$. There are three medium-ring blocks: $\mathcal{M}_0$ (the left 9 ancilla qubits), $\mathcal{M}_1$ (the middle 9 ancilla qubits), and $\mathcal{M}_2$ (the right 9 ancilla qubits). The shift operations in this figure are (1) $\mathrm{Shift}(\mathcal{M}_{0};\,0,0,1)$, where qubits in $\mathcal{M}_0$ shift by 1 on the short rings, (2) $\mathrm{Shift}(\mathcal{M}_{1};\,0,1,2)$, where qubits in $\mathcal{M}_1$ shift by 2 on the short rings and by $1\cdot 3$ on the medium ring, and (3) $\mathrm{Shift}(\mathcal{M}_{2};\,0,2,0)$, where qubits in $\mathcal{M}_2$ shift by $2\cdot 3$ on the medium ring. The gray qubit in each medium-ring block marks the qubit that was originally the left-most qubit in that block. Since $r=0$ for each shift, these operations can all be performed in parallel.}
    \label{fig: ring shift}
\end{figure}

It is sometimes useful to apply different cyclic shifts to disjoint subsets of ancillas associated with the three-ring structure. The long-ring case is trivial, since a long ring contains all $n$ ancillas and therefore does not single out a proper subset. By contrast, the ancillas in a given medium ring are often shifted independently, for example in the decomposed transport operations used in the syndrome-extraction circuit of \cref{subsec:three-ring-syndrome-extraction}. For each $u_0 \in \Z_a$, define the corresponding \emph{medium-ring block}
\[
\mathcal{M}_{u_0} := \{\alpha(u_0,v,z) : (v,z) \in \Z_b \times \Z_c\} \subseteq \mathcal{A}.
\]
In the one-dimensional layout, $\mathcal{M}_{u_0}$ occupies a contiguous block of $bc$ ancilla sites, corresponding to a single medium ring.
A \emph{three-ring transport} is specified by a choice of medium-ring block $\mathcal{M}_{u_0}$ together with a shift vector $(r,s,t) \in G$.
We denote it by $\mathrm{Shift}(\mathcal{M}_{u_0};\,r,s,t)$ and define its action on ancilla labels by
\[
\alpha(u,v,z) \mapsto
\begin{cases}
\alpha(u \oplus r,\, v \oplus s,\, z \oplus t), & u = u_0, \\
\alpha(u \oplus r,\, v,\, z), & u \neq u_0,
\end{cases}
\]
where $\oplus$ denotes modular addition in each coordinate. The long-ring component shifts every medium-ring block by $r$, while the medium- and short-ring components act only on the selected block $\mathcal{M}_{u_0}$. In particular, if $r=0$, then every ancilla outside $\mathcal{M}_{u_0}$ is fixed.
An example of three different ring shifts on medium-ring blocks is given in \cref{fig: ring shift}.

A transport on $\mathcal{M}_{u_0}$ consists of a global long-ring cycle together with medium- and short-ring shifts restricted to that block.

The three-ring framework includes the following code families:

\begin{enumerate}
    \item {\bf BB codes.}
    Let $S_m$ denote the $m \times m$ circulant matrix whose first row is $(0,1,0,\dots,0)$. For integers $\ell,m$, define
    \[
    x := S_\ell \otimes I_m,
    \qquad
    y := I_\ell \otimes S_m.
    \]
    BB codes are CSS codes with
    \[
    H_X = [A \mid B], 
    \qquad 
    H_Z = [B^T \mid A^T],
    \]
    where $A$ and $B$ are sums of monomials $x^i y^j$ with 
    $i \in \Z_\ell$ and $j \in \Z_m$.
    Equivalently, BB codes are three-ring codes over 
    $\Z_2 \times \Z_\ell \times \Z_m$~\cite{tham2025distributed}. A BB code with check weight $w$ is denoted BB$w$.

    \item {\bf GB codes.}
    GB codes~\cite{kovalev2013quantum} arise as the special case $m = 1$ of BB codes. In this case $x = S_\ell$ and $y = I_m$, so $A$ and $B$ are circulant matrices. In the three-ring picture the group is 
    $\Z_2 \times \Z_\ell \times \Z_1$, meaning one ring family is trivial. A GB code with check weight $w$ is denoted GB$w$.

    \item {\bf Cyclic HGP codes.}
    Cyclic HGP codes~\cite{aydin2025cyclic} may also be viewed as a special case of BB codes, in which $A$ is a sum only of powers of $x$ and $B$ is a sum only of powers of $y$.
\end{enumerate}

\subsection{Syndrome extraction circuit}
\label{subsec:three-ring-syndrome-extraction}
The syndrome extraction circuit described in this subsection applies only to three-ring codes with $a=2$, \emph{i.e.}, exactly the BB codes (and therefore also GB and cyclic HGP codes). Accordingly, we specialize to
\[
    G = \Z_2 \times \Z_b \times \Z_c.
\]
For $a>2$, the first cyclic factor would require a more general routing rule than the binary $A/B$ alternation used below.
In this $a=2$ setting, we also write $\mathcal{A}_X := \mathcal{M}_0$ and $\mathcal{A}_Z := \mathcal{M}_1$ for the two medium-ring blocks occupied by the X- and Z-check ancillas, respectively.

The syndrome extraction circuit described here mildly generalizes earlier cyclic-shift-based syndrome extraction circuits for BB memories \cite{tham2025distributed}, shuttling-based GB memories \cite{siegel2024towards}, and cyclic HGP codes \cite{aydin2025cyclic} by allowing the schedule-permutation freedom emphasized for BB codes in \cite{bravyi2024high}. This flexibility matters because different schedule permutations induce different hook-error propagation patterns. We therefore search over them for circuits that preserve the leading-order circuit-level suppression exponent of the underlying code. In the examples considered here, the chosen schedules satisfy $\lceil d_{\mathrm{circ}}/2 \rceil = \lceil d/2 \rceil$. Writing
\[
    A = A_1 + \cdots + A_r,
    \qquad
    B = B_1 + \cdots + B_s,
\]
as sums of monomials $x^i y^j$ (so each $A_\mu$ and $B_\nu$ is a single term of that form), a \emph{schedule permutation} is simply an ordering of the monomial terms that specifies the order in which an ancilla visits the corresponding data neighbors during one SEC. For example, when
\[
    A = A_1 + A_2,
    \qquad
    B = B_1 + B_2,
\]
a full weight-$4$ maximally parallel schedule can be written in the paired tuple form
\[
    \Sigma = \big((B_1,B_2^T),(A_1,A_1^T),(A_2,A_2^T),(B_2,B_1^T)\big).
\]
Each pair $(T_X^{(\tau)},T_Z^{(\tau)})$ specifies the monomial visited in round $\tau$ by the X-check ancillas and Z-check ancillas, respectively: the X ancillas follow the first entry of each pair, while the Z ancillas follow the second. In the schedule above, the X ancillas visit $B_1,A_1,A_2,B_2$ and the Z ancillas simultaneously visit $B_2^T,A_1^T,A_2^T,B_1^T$. In the polynomial representation, transposition acts on a monomial by inversion:
\[
    (x^i y^j)^T = (x^i y^j)^{-1} = x^{-i} y^{-j},
\]
with exponents taken modulo the ring sizes. This is because each monomial $x^i y^j$ is a permutation matrix, and the transpose of a permutation matrix is its inverse.

\begin{table*}[t]
    \centering
    {\setlength{\tabcolsep}{5pt}
    \renewcommand{\arraystretch}{1.1}
    \footnotesize
    \begin{adjustbox}{max width=\textwidth}
    \begin{tabular}{|c|c|c|c|}
        \hline
        Code & \textbf{Q102} & \textbf{Q70} & \textbf{Q54} \\
        \hline
        \makecell{Circuit-level \\ distance $d_{\mathrm{circ}}$} & $9$ & $9$ & $9$ \\
        \hline
        \makecell{Schedule \\ permutation} &
        \makecell[l]{$((B_1,B_4^T),(A_1,A_2^T),(A_3,A_4^T),(B_2,B_3^T),$ \\
        $(B_3,B_2^T),(A_4,A_3^T),(A_2,A_1^T),(B_4,B_1^T))$} &
        \makecell[l]{$((B_1,B_3^T),(A_1,A_3^T),(A_4,A_2^T),(B_2,B_2^T),$ \\
        $(A_3,A_1^T),(A_2,A_4^T),(B_3,B_1^T))$} &
        \makecell[l]{$((B_2,B_1^T),(A_1,A_4^T),(B_4,B_3^T),(A_3,A_2^T),$ \\
        $(A_2,A_3^T),(B_3,B_4^T),(A_4,A_1^T),(B_1,B_2^T))$} \\
        \hline
    \end{tabular}
    \end{adjustbox}
    }
    \caption{Summary of the syndrome-extraction schedules used for the memory blocks Q102, \code{70}, and \code{54}.
    The schedule-permutation row lists one known maximally parallel schedule for the circuit described in this subsection, written as a sequence of paired X- and Z-check terms.
    The circuit-level-distance row reports the corresponding $d_{\mathrm{circ}}$, the minimum number of faulty operations in the syndrome-extraction circuit required to produce an undetectable logical error.
    Unless noted otherwise, whenever we refer to the circuit for one of these codes, we mean the circuit defined by the listed schedule permutation.}
    \label{tab:memory_block_syndrome_schedules}
\end{table*}

A high-level description of the syndrome extraction circuit is given in \cref{alg:base_syndrome_extraction}.

\begin{figure*}[t]
    \centering
    \includegraphics[width=\textwidth]{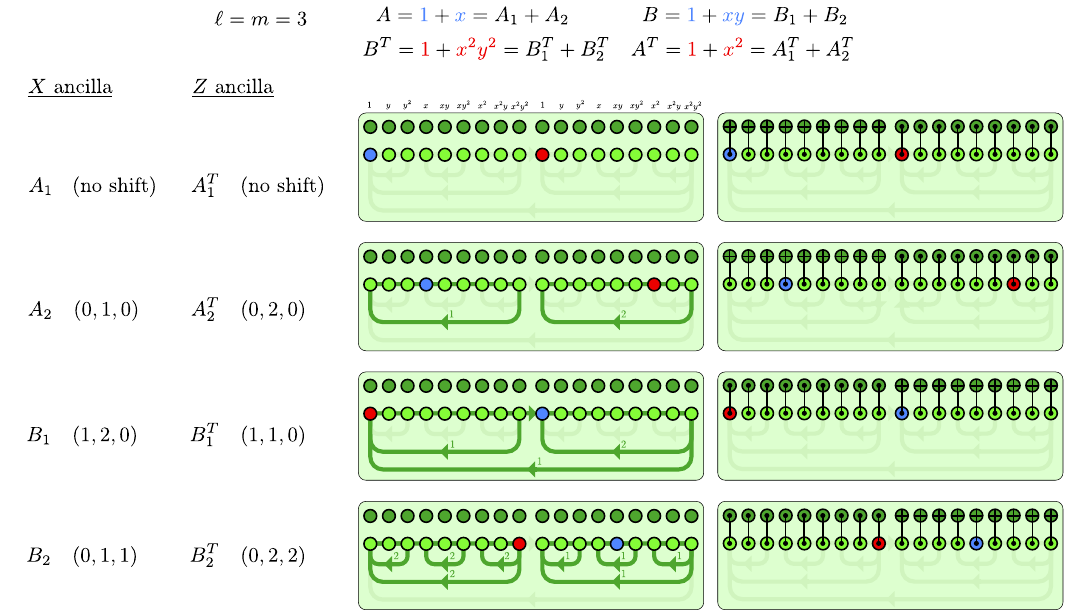}
    \caption{Example of a syndrome extraction circuit for a BB4 code with $\ell = m = 3$, defined by $A = 1 + x$ and $B = 1 + xy$. To simplify slightly, we removed the beacon qubits from the diagram. Syndrome extraction proceeds by iteratively aligning ancilla qubits with the data qubits they must entangle. One $X$ ancilla and one $Z$ ancilla are shown in blue and red respectively. They are tracked as their position changes.
    In round 1, the $(0,0,0)$ $X$ ancilla is already aligned with the $A_1 = 1$ term, so a sequence of $\CX$ gates is applied from the X ancilla qubits to the data qubits. In round 2, we align the $(0,0,0)$ ancilla with the $A_2 = x$ term by applying the $(0,1,0)$ medium-ring shift to the left medium-ring block. By symmetry, this simultaneously aligns all $X$ ancilla qubits with their corresponding data qubits. After a $(1,0,0)$ long-ring shift aligns the $X$ ancilla qubits with the right medium-ring block, rounds 3 and 4 proceed analogously. The $Z$ ancillas undergo the analogous transport sequence using the inverse terms of the generator polynomials and apply $\CZ$ gates to their aligned data qubits.}
    \label{fig:bb_l3_m3_qubit_rows}
\end{figure*}

A concrete example of a full SEC for a BB4 code with $\ell = 3$ and $m = 3$ is shown in \cref{fig:bb_l3_m3_qubit_rows}.

Supplementary Sec.~1 of Bravyi \emph{et al.}~\cite{bravyi2024high} proves that a valid maximally parallel schedule must be a sequence
\[
    \Sigma = \big((T_X^{(1)},T_Z^{(1)}), \dots, (T_X^{(w_{\mathrm{check}})},T_Z^{(w_{\mathrm{check}})})\big)
\]
of length $w_{\mathrm{check}}$, where $w_{\mathrm{check}}$ is the check weight and each pair is either of the form $(A_i,A_j^T)$ or of the form $(B_i,B_j^T)$. Fix a nominal starting point in which ancilla qubits are paired one-to-one with data qubits. Then each schedule pair $(T_X^{(\tau)},T_Z^{(\tau)})$ specifies two target ancilla alignments for round $\tau$: one for the X-check ancillas and one for the Z-check ancillas. The first round can be executed by relabeling the ancilla in software so that the initial alignment already matches $(T_X^{(1)},T_Z^{(1)})$, after which the scheduled parallel two-qubit gates are applied without any physical transport. For each subsequent round $\tau \in \{2,\dots,w_{\mathrm{check}}\}$, the circuit computes the common long-ring shift $r$ together with the medium- and short-ring components $(s_X,t_X)$ for $\mathcal{A}_X$ and $(s_Z,t_Z)$ for $\mathcal{A}_Z$, applies the resulting transport step, and then performs the scheduled parallel two-qubit gates for round $\tau$. Because each schedule pair is jointly $A$-type or jointly $B$-type, the long-ring component is shared between X and Z ancillas; by contrast, the $s$ and $t$ components generally differ between them.

The initial alignment of data and ancilla goes as follows. In the one-dimensional layout, $\alpha(0,0,0)$ is aligned with data qubit $0$, while $\alpha(1,0,0)$ is aligned with data qubit $n/2$. Equivalently, the ancillas $\alpha(0,v,z)$ with $(v,z)\in \Z_b \times \Z_c$ form the medium-ring block $\mathcal{A}_X = \mathcal{M}_0$ and occupy the first half of the ancilla row, while the ancillas $\alpha(1,v,z)$ form the medium-ring block $\mathcal{A}_Z = \mathcal{M}_1$ and occupy the second half. With this convention, each $A_i$ term implicitly refers to an alignment in which the X ancillas are matched, up to a permutation within that half, with the first $n/2$ data qubits, and each transposed term $A_i^T$ refers to the analogous alignment of the Z ancillas with the second $n/2$ data qubits. Passing to the support of the $B$ polynomial applies the long shift $(1,0,0)$, which swaps these two halves: the Z ancillas align with the first $n/2$ data qubits and the X ancillas align with the second $n/2$.

These shifts are computed from two consecutive schedule pairs. Given $(T_X^{(\tau-1)},T_Z^{(\tau-1)})$ and $(T_X^{(\tau)},T_Z^{(\tau)})$, we first compute the common long-ring shift $r$ from the family change between the two rounds: we set $r=1$ if the schedule pair changes family ($A$ versus $B$), and $r=0$ otherwise. We then compute the medium- and short-ring components separately for the X and Z ancillas. If, for ancilla type $q \in \{X,Z\}$, the terms in rounds $\tau-1$ and $\tau$ are $x^{u_-}y^{v_-}$ and $x^{u_+}y^{v_+}$, respectively, then $s_q = u_+ - u_-$ and $t_q = v_+ - v_-$, with the differences taken modulo the relevant ring sizes. $s_q$ and $t_q$ record the changes in the exponents of $x$ and $y$, respectively.

Thus, the circuit algorithm is essentially the same as the memory-block syndrome extraction circuit of Bravyi \emph{et al.}~\cite{bravyi2024high} with the only difference being the addition of explicit ancilla cyclic shift transport operations: between two rounds of two-qubit gates, we specify the three-ring cyclic shift transport operations needed to bring each ancilla next to the data qubit it must visit in the next round.

\begin{algorithm}[tbp]
\caption{Base SEC for a three-ring memory block with $a=2$}
\label{alg:base_syndrome_extraction}
\KwData{Data qubits, ancilla qubits, and a valid maximally parallel schedule $\Sigma = \big((T_X^{(1)},T_Z^{(1)}), \dots, (T_X^{(w_{\mathrm{check}})},T_Z^{(w_{\mathrm{check}})})\big)$ for a three-ring code with $a=2$}
Prepare all ancilla qubits in $\ket{+}$\;
Relabel the ancilla in software so that the initial alignment matches the first schedule pair in $\Sigma$\;
Apply the scheduled parallel gates for $(T_X^{(1)},T_Z^{(1)})$: $\CX$ from each aligned X-check ancilla to its data qubit, and $\CZ$ between each aligned Z-check ancilla and its data qubit\;
\For{$\tau \in \{2,\dots,w_{\mathrm{check}}\}$}{
    Compute the common long-ring shift $r$ from the family change between rounds $\tau-1$ and $\tau$, together with $(s_X,t_X)$ from $T_X^{(\tau-1)}$ and $T_X^{(\tau)}$ and $(s_Z,t_Z)$ from $T_Z^{(\tau-1)}$ and $T_Z^{(\tau)}$\;
    Apply the transport step consisting of a global long-ring cycle by $r$ on all ancilla qubits together with the medium/short shifts $(s_X,t_X)$ on the medium-ring block $\mathcal{A}_X$ and $(s_Z,t_Z)$ on the medium-ring block $\mathcal{A}_Z$\;
    Apply the scheduled parallel gates for $(T_X^{(\tau)},T_Z^{(\tau)})$: $\CX$ from each aligned X-check ancilla to its data qubit, and $\CZ$ between each aligned Z-check ancilla and its data qubit\;
}
Measure all ancilla qubits in the $X$ basis\;
\end{algorithm}

\newcommand{\algaddmark}{\mbox{\raisebox{0.15ex}{\footnotesize\texttt{[*]}}}\nobreakspace}
\newcommand{\algadd}[1]{\algaddmark #1}

\section{Correction of losses and leakages}
\label{sec:Correction of losses and leakages}

This section explains how the walking cat architecture deals with leakage and loss at the memory-block level. 

In \cref{subsec:noise-model,subsec:Loss and leakage measurement}, we introduced the moving-qubit loss and leakage model: the rates $\ploss$ and $\pleak$, the way these faults propagate through operations, and the measurement primitives that reveal them. These events are qualitatively different from the Pauli faults for which standard stabilizer-code decoding is designed. Loss removes the physical carrier of the qubit, whereas leakage transfers population outside the computational two-level subspace and can seed downstream correlated faults~\cite{aliferis2007localleakage}. To keep the memory blocks fault tolerant, the architecture must therefore detect these events quickly, and reset or replace them before the usual decoder handles the remaining Pauli noise~\cite{kang2023metastable}.

Qubit loss is handled with a dedicated loss-detection protocol that checks whether each data qubit is affected by a loss after each scheduled layer of two-qubit gates within the SEC and locally reloads the affected qubits when a loss is flagged. Leakage on ancilla qubits is detected with the leakage measurement introduced in \cref{subsec:Loss and leakage measurement}, while leakage on data qubits is detected with a teleportation-based \emph{leakage detection unit}~(LDU)~\cite{knill2005quantum,chow2024circuit,perrin2025quantum}. Any qubit flagged as leaked is then reset to the computational subspace. We then quantify the resulting memory-block performance, identify the dominant physical noise sources, and study the space-time trade-offs of the loss-detection protocol.

The main properties of the loss- and leakage-corrected memory blocks are summarized in \cref{tab:loss_leakage_summary}.

The rest of this section is organized as follows.
\Cref{subsec:simulation_loss_n_leakage} describes the simulation workflow used to model loss and leakage.
\Cref{subsec:atom-loss-gadget} introduces the loss-detection protocol: the beacon protocol.
\Cref{subsec:reload_sync_overhead} estimates the synchronization overhead caused by rare local reloads.
\Cref{subsec:leakage-correction-gadget} presents the leakage-detection and reset protocol for ancilla and data qubits.
\Cref{subsec:MemoryBlockPerf} reports the logical performance of the candidate memory blocks.
\Cref{subsec:Noise Sensitivity Analysis} identifies the dominant physical noise sources.
\Cref{subsec:Optimizing Beacon Protocol Space and Time Overhead} studies space-time optimizations of the loss-detection protocol.

\begin{table*}[t]
    \centering
    {\setlength{\tabcolsep}{5pt}
    \renewcommand{\arraystretch}{1.1}
    \footnotesize
    \begin{adjustbox}{max width=\textwidth}
    \begin{tabular}{|c|c|c|c|}
        \hline
        Code & \textbf{Q102} & \textbf{Q70} & \textbf{Q54} \\
        \hline
        \makecell{Physical \\ qubits} & 316 & 220 & 172 \\
        \hline
        \makecell{Logical error \\ rate / SEC} & $10^{-11}$ & $10^{-10}$ & $3\times 10^{-10}$ \\
        \hline
        \makecell{SEC time budget\\in POC} &
        {\scriptsize
        \begin{tabular}{@{}l r@{}}
        Ancilla reset & 1 \\
        2q gate layer & 8 \\
        Cyclic shift & 15.65 \\
        Loss and leakage checks & 8.05 \\
        Measurement & 1 \\
        \hline
        Total & 33.70
        \end{tabular}} &
        {\scriptsize
        \begin{tabular}{@{}l r@{}}
        Ancilla reset & 1 \\
        2q gate layer & 7 \\
        Cyclic shift & 11.65 \\
        Loss and leakage checks & 7.05 \\
        Measurement & 1 \\
        \hline
        Total & 27.70
        \end{tabular}} &
        {\scriptsize
        \begin{tabular}{@{}l r@{}}
        Ancilla reset & 1 \\
        2q gate layer & 8 \\
        Cyclic shift & 10.10 \\
        Loss and leakage checks & 8.05 \\
        Measurement & 1 \\
        \hline
        Total & 28.15
        \end{tabular}} \\
        \hline
        \makecell{Loss distribution \\ per SEC} &
        \begin{tabular}{@{}c|c@{}}
        \makecell{Number of \\ qubits lost} & Probability \\
        \hline
        $0$ & $0.999148$ \\
        $1$ & $3.20 \cdot 10^{-5}$ \\
        $2^{*}$ & $3.20 \cdot 10^{-10}$ \\
        $3$ & $8.18 \cdot 10^{-4}$ \\
        $4^{*}$ & $2.09 \cdot 10^{-8}$ \\
        $5^{*}$ & $2.64 \cdot 10^{-13}$ \\
        $6$ & $2.00 \cdot 10^{-6}$ \\
        $>6^{*}$ & $1.03 \cdot 10^{-10}$
        \end{tabular} &
        \begin{tabular}{@{}c|c@{}}
        \makecell{Number of \\ qubits lost} & Probability \\
        \hline
        $0$ & $0.999580$ \\
        $1$ & $1.30 \cdot 10^{-5}$ \\
        $2^{*}$ & $1.15 \cdot 10^{-10}$ \\
        $3$ & $4.07 \cdot 10^{-4}$ \\
        $4^{*}$ & $6.14 \cdot 10^{-9}$ \\
        $5^{*}$ & $4.65 \cdot 10^{-14}$ \\
        $6^{*}$ & $8.19 \cdot 10^{-8}$ \\
        $>6^{*}$ & $1.23 \cdot 10^{-11}$
        \end{tabular} &
        \begin{tabular}{@{}c|c@{}}
        \makecell{Number of \\ qubits lost} & Probability \\
        \hline
        $0$ & $0.999689$ \\
        $1$ & $1.50 \cdot 10^{-5}$ \\
        $2^{*}$ & $6.10 \cdot 10^{-11}$ \\
        $3$ & $2.96 \cdot 10^{-4}$ \\
        $4^{*}$ & $3.31 \cdot 10^{-9}$ \\
        $5^{*}$ & $1.83 \cdot 10^{-14}$ \\
        $6^{*}$ & $4.50 \cdot 10^{-8}$ \\
        $>6^{*}$ & $5.00 \cdot 10^{-12}$
        \end{tabular} \\
        \hline
    \end{tabular}
    \end{adjustbox}
    }
    \caption{Summary of the memory blocks studied in this section.
    The physical-qubit count includes data qubits, ancilla qubits used for syndrome extraction, beacon qubits used for loss correction, and a reservoir used for reloading.
    The logical error rates correspond to the operating points used throughout this section: $p=10^{-4}$, $\pleak = 10^{-5}$, and $\ploss = 10^{-7}$.
    The SEC time budget in POC decomposes the time spent during a single SEC.
    The ``Ancilla reset'' entry is conservatively counted as 1 POC, assuming the worst-case scenario that a qubit needs to be reset due to leakage (see \Cref{subsec:Review of the fully connected model,subsec:Operation time}).
    The loss distributions were obtained from Monte Carlo simulation with $10^6$ shots. Entries marked with an asterisk were not observed directly in the Monte Carlo sample and are instead estimated using the compound Poisson rare-event model described in \Cref{subsec:simulation_loss_n_leakage}.
    }
    \label{tab:loss_leakage_summary}
\end{table*}

\subsection{Simulation with loss and leakage}
\label{subsec:simulation_loss_n_leakage}

We simulate standard \texttt{stim} circuits~\cite{gidney2021stim} annotated with detectors and logical observables, and use these circuits as the common input to all of our simulations.
For noise models containing only Pauli errors, we use the standard \texttt{stim} Monte Carlo shot sampler.
To incorporate qubit loss and leakage, we pass the same annotated circuits to a bespoke stabilizer simulator extending this workflow.

The simulator executes the circuit gate by gate using a Pauli tableau representation while tracking, for every qubit, whether it is in the computational subspace, lost, or leaked.
Operations involving lost or leaked qubits follow the rules of the moving-qubit noise model described in \cref{subsec:noise-model}.

At the start of each simulated time step, the simulator samples independent loss and leakage events for every qubit at the prescribed rates scaled by the duration of that step in POCs.
All qubits appearing in the circuit are susceptible to these events, including qubits used inside loss- and leakage-detection gadgets.
When such gadgets are added to a nominal syndrome extraction circuit, we also include the additional gate and idle noise accompanying the physical operations required to execute the gadget.
Apart from these extensions, the simulator operates like a standard Monte Carlo shot sampler for Clifford circuits with Pauli noise.

The loss distributions reported in \Cref{tab:loss_leakage_summary} are obtained from the simulation workflow described above. For outcomes that were sampled in the $10^6$-shot Monte Carlo run, we report the empirical frequencies directly. For rarer outcomes that were not observed in that Monte Carlo sample, we estimate the corresponding probabilities with a compound Poisson model. We approximate the number of initial loss events in one SEC by a Poisson random variable $X$ with mean $\ploss \sum_t N_t \Delta_t$, where $N_t$ is the number of qubits exposed during simulated time step $t$ and $\Delta_t$ is its duration in POCs. Let $d_{\mathrm{POC}} = \sum_t \Delta_t$ denote the total SEC duration in POCs for the memory block under consideration. In this rare-event model, we approximate the time of an initial loss event as uniformly distributed over that SEC duration. For each initial loss event, we introduce a loss multiplicity random variable $Y$: with probability $1/d_{\mathrm{POC}}$, the event occurs in the final measurement layer of duration 1 POC and contributes $Y=1$ lost qubit; with the remaining probability, the event occurs during or before the loss detection protocol and contributes $Y=3$, corresponding to the loss of one beacon qubit, one data qubit, and the ancilla that most recently interacted with that data qubit. The total number of lost qubits in the model is then
\[
    L = \sum_{j=1}^{X} Y_j,
\]
where the $Y_j$ are independent copies of $Y$.

\subsection{Qubit loss detection protocol}
\label{subsec:atom-loss-gadget}
In this subsection, we introduce the walking cat architecture's loss-detection protocol.
Although the walking cat architecture applies to general moving-qubit families, the protocol analyzed here is tailored to trapped ions in the QCCD architecture.
In the trapped ion QCCD architecture, ions are transported in small wells that must be merged together to perform two-qubit gates, and then split apart to allow subsequent interactions with other ions.
In that setting, attempting an ion merge/split operation after one ion has been lost can dramatically heat the remaining computational ion~\cite{fallek2024rapid},
thereby likely completely losing the ion. Therefore, in our model, we always treat any merge/split operation involving a lost ion as a propagated loss. 
We can exploit this fact by pairing each data qubit with an aligned ancillary 
qubit that can be measured repeatedly for loss without disturbing the data qubit's quantum information.
We refer to these ancillary qubits as \emph{beacon qubits}, and the loss detection protocol that involves them as the \textit{beacon protocol}.
This approach was introduced for long chains in \cite{coble2025correction}, 
we extend that approach here to the moving-qubit model.

After each scheduled layer of two-qubit gates within the SEC, we merge and split every data qubit with its
aligned beacon qubit and then perform a loss-detection measurement on the beacon.
If the data qubit was already lost, the loss propagates to the beacon, so a `lost' outcome on the beacon serves as a proxy for data-qubit loss.
We check for loss immediately after each scheduled layer of two-qubit gates within the SEC,
before the loss can spread through subsequent interactions. Because this check does not require an additional physical two-qubit gate,
it reduces both time and noise overhead relative to earlier loss-detection schemes.

The syndrome extraction circuit extended to include the beacon protocol and leakage detection (which is detailed later in \cref{subsec:leakage-correction-gadget}) is summarized in \cref{alg:beacon_loss_round}, and the loss-triggered reloading and start-of-SEC data-leakage routines it invokes are collected in \cref{alg:beacon_loss_subroutines}.

Even a per-qubit loss probability as low as $\ploss = 10^{-7}$ per POC of computational operations is damaging at scale. For a device with $N \approx 10^4$ physical qubits, the expected rate of initial loss events per POC is $N\ploss \approx 10^{-3}$, so one initial loss event occurs every $\sim 1/(N\ploss) \approx 10^3$ POCs. If a loss is not immediately corrected, subsequent entangling operations spread the loss; in the worst case, the affected region can double after each scheduled layer of two-qubit gates within the SEC. Thus a single initial loss event can rapidly grow to a size where an unrecoverable logical error is inevitable, effectively limiting reliable execution to $\mathcal{O}(10^3)$ POCs. This is orders of magnitude below what is required to realize a computation with $10^6$ logical $T$ gates, motivating explicit loss detection and correction.

\Cref{fig:ion_loss_rate} shows that a high loss rate is a major issue. We plot the logical error rate as a function of the loss rate parameter using the moving-qubit noise model on all operations with fixed noise parameter $p=10^{-3}$, varying $\ploss$ between $10^{-8}$ and $10^{-3}$ for our \code{102} memory block, all with loss correction via the beacon protocol enabled. When $\ploss$ approaches $p$, ion loss strongly degrades logical performance, which may be an issue for technologies where loss is a dominant source of noise. By contrast, trapped ions can be confined more reliably due to their charge. Indeed, the low-loss operating point $\ploss = p/1000$ considered here is a conservative estimate relative to cryogenic trapped-ion experiments that report holding chains of over 100 ions for hours in a low-collision environment~\cite{pagano2019cryogenic}.

\begin{algorithm}[tbp]
\caption{Beacon-assisted syndrome extraction with loss and leakage handling. Lines or phrases marked with \texttt{[*]} are additions or modifications relative to \cref{alg:base_syndrome_extraction}.}
\label{alg:beacon_loss_round}
\algadd{\KwData{A data register $D$, an ancilla register $A$, a beacon register $B$, a local reservoir register $R$ containing fresh replacement qubits, and a valid maximally parallel schedule $\Sigma = \big((T_X^{(1)},T_Z^{(1)}), \dots, (T_X^{(w)},T_Z^{(w)})\big)$ for a three-ring code with $a=2$}}
\algadd{\textsc{RunDataLeakageDetection}$(D,A,R)$}\;
Prepare all ancilla qubits in $\ket{+}$\;
Relabel the ancilla in software so that the initial alignment matches the first schedule pair in $\Sigma$\;
Apply the scheduled parallel gates for $(T_X^{(1)},T_Z^{(1)})$: $\CX$ from each aligned X-check ancilla to its data qubit, and $\CZ$ between each aligned Z-check ancilla and its data qubit\;
\algadd{Merge and split each data qubit with its aligned beacon qubit, then perform a loss-detection measurement on the beacon qubit}\;
\algadd{Let $\mathcal{L}$ be the set of beacon qubits whose loss-detection measurement returns `lost'}\;
\algadd{\If{$\mathcal{L} \neq \emptyset$}{
  \algadd{\textsc{ReloadAfterLoss}$(\mathcal{L},D,A,B,R)$}\;
}
}
\For{$\tau \in \{2,\dots,w\}$}{
  Compute the common long-ring shift $r$ from the family change between rounds $\tau-1$ and $\tau$, together with $(s_X,t_X)$ from $T_X^{(\tau-1)}$ and $T_X^{(\tau)}$ and $(s_Z,t_Z)$ from $T_Z^{(\tau-1)}$ and $T_Z^{(\tau)}$\;
  Apply the transport step consisting of a long-ring cycle by $r$ on all long-ring blocks together with the medium/short shifts $(s_X,t_X)$ on $\mathcal{A}_X$ and $(s_Z,t_Z)$ on $\mathcal{A}_Z$\;
  Apply the scheduled parallel gates for $(T_X^{(\tau)},T_Z^{(\tau)})$: $\CX$ from each aligned X-check ancilla to its data qubit, and $\CZ$ between each aligned Z-check ancilla and its data qubit\;
  \algadd{Merge and split each data qubit with its aligned beacon qubit, then perform a loss-detection measurement on the beacon qubit}\;
  \algadd{Let $\mathcal{L}$ be the set of beacon qubits whose loss-detection measurement returns `lost'}\;
  \algadd{\If{$\mathcal{L} \neq \emptyset$}{
    \algadd{\textsc{ReloadAfterLoss}$(\mathcal{L},D,A,B,R)$}\;
  }
  }
}
Measure all ancilla qubits in the $X$ basis\;
\algadd{Replace any ancilla qubit flagged lost during readout by a fresh qubit prepared in $I/2$ from $R$}\;
\end{algorithm}

\begin{algorithm}[tbp]
\caption{Subroutines for loss-triggered reloading and data leakage detection}
\label{alg:beacon_loss_subroutines}
\KwData{A data register $D$, an ancilla register $A$, a beacon register $B$, and a local reservoir register $R$ containing fresh replacement qubits}
\SetKwProg{Fn}{Function}{}{}
\Fn{ReloadAfterLoss$(\mathcal{L},D,A,B,R)$}{
  Let $\mathcal{J}$ be the ancilla qubits that interacted in the current layer with the data qubits indexed by $\mathcal{L}$\;
  Eject the corresponding data qubits from the system if they are still present\;
  Replace lost data qubits and ancilla qubits by fresh qubits from $R$ prepared in $I/2$\;
  Replace the lost beacon qubits indexed by $\mathcal{L}$ by fresh qubits from $R$ prepared for the next loss check\;
}
\Fn{RunDataLeakageDetection$(D,A,R)$}{
  Prepare the ancilla qubits in $\ket{0}$\;
  Apply the teleportation-based leakage detection unit of \cref{fig:ancilla_leakage_ldu} in parallel between each data qubit and its aligned ancilla qubit\;
  Measure the old data qubits in the $X$ basis using the augmented loss/leakage readout\;
  If the computational outcome of that $X$-basis measurement is $1$, apply $Z$ to the aligned ancilla qubit via virtual frame tracking\;
  Physically exchange the data and ancilla rows through transport, equivalently swapping their roles, so that the teleported state now occupies the data row and the measured qubits become ancillas for the next SEC\;
  Let $\mathcal{K}$ be the set of indices whose old data-qubit measurement reports `leaked'\;
  Reset the output data qubits indexed by $\mathcal{K}$ to the maximally mixed state $I/2$\;
}
\end{algorithm}

\begin{figure}[tbp]
    \centering
    \includegraphics[width=0.9\linewidth]{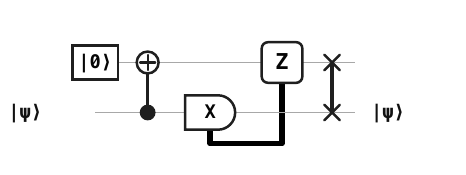}
    \caption{Teleportation-based data-qubit leakage detection unit shown as a physical state-transfer diagram~\cite{knill2005quantum,chow2024circuit}. The ancilla qubit is prepared in $\ket{0}$ and targeted by a \CX from the data qubit. The old data qubit is then measured in the $X$ basis using the augmented loss/leakage readout; if the computational outcome is $1$, a $Z$ correction is applied to the teleported state before a final exchange restores the data/ancilla ordering. In the moving-qubit model, this exchange can be implemented with physical transport operations rather than a SWAP gate.}
    \label{fig:ancilla_leakage_ldu}
\end{figure}

\begin{figure}[ht]
    \centering
    \includegraphics[width=1\linewidth]{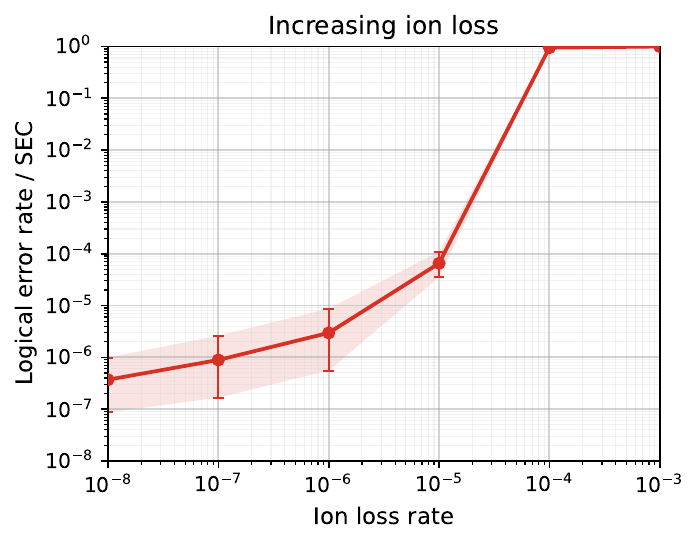}
    \caption{Logical error rate as a function of the loss rate parameter $\ploss$ for the \code{102} memory block with loss correction, using the moving-qubit noise model with fixed $p = 10^{-3}$.}
    \label{fig:ion_loss_rate}
\end{figure}

\subsection{Synchronization overhead from local reloading}
\label{subsec:reload_sync_overhead}

To limit the impact of reloading on the time of an SEC, we assume that the reservoir is placed so that each replacement requires only a few transport steps. For example, if one data qubit and one beacon qubit are missing, a short chain of fresh qubits can be shifted from the reservoir toward the left and then downward so that the two vacancies are filled locally. In the memory-block layouts considered here, we use a preliminary design assumption that this bounded reloading overhead is three transport steps, \emph{i.e.}, \ $3/20$ POC in the moving-qubit timing model.

The SEC times reported in \cref{tab:loss_leakage_summary} assume that no loss-triggered reloading occurs during that cycle. When a loss is detected, the affected block must execute the local reservoir refill before continuing, so its SEC becomes longer by at most
\[
    \Delta T_{\mathrm{reload}} = \frac{3}{20} \ \text{POC}
\]
under this preliminary three-transport-step local-refill assumption.

Let $q_{\mathrm{reload}}$ denote the probability that a given block requires at least one local reload during an SEC. At the operating points reported in \cref{tab:loss_leakage_summary}, we estimate this from the probability that at least one qubit is lost during the SEC:
\[
    \begin{aligned}
        q_{\mathrm{reload}}
        &= 1 - \Pr[\text{0 qubits lost}], \\
        &=
        \begin{cases}
            8.52 \times 10^{-4} & \text{for \code{102}}, \\
            4.20 \times 10^{-4} & \text{for \code{70}}, \\
            3.11 \times 10^{-4} & \text{for \code{54}}.
        \end{cases}
    \end{aligned}
\]

\begin{proposition}[Synchronization overhead from local reloading is bounded and small]
    Let $M_i$ denote the number of loss-triggered reload rounds executed by block $i$ during one SEC, let $\overline{m}_{\mathrm{reload}} = \mathbb{E}[M_i]$, and let $T_{\mathrm{SEC}}^{(0)}$ be the nominal SEC time of one memory block in the absence of loss-triggered reloading. Assume that each local reload adds at most $\Delta T_{\mathrm{reload}} = 3/20~\mathrm{POC}$. If $N_{\mathrm{blk}}$ memory blocks must remain synchronized, then their average SEC time satisfies
    \begin{equation}
        \overline{T}_{\mathrm{SEC}}^{(N_{\mathrm{blk}})}
        \leq
        T_{\mathrm{SEC}}^{(0)} + \frac{3}{20} N_{\mathrm{blk}} \overline{m}_{\mathrm{reload}}.
        \label{eq:reload_sync_bound}
    \end{equation}
    Equivalently, the relative SEC overhead obeys
    \[
        \frac{\overline{T}_{\mathrm{SEC}}^{(N_{\mathrm{blk}})} - T_{\mathrm{SEC}}^{(0)}}{T_{\mathrm{SEC}}^{(0)}}
        \leq
        \frac{3 N_{\mathrm{blk}} \overline{m}_{\mathrm{reload}}}{20 \, T_{\mathrm{SEC}}^{(0)}}.
    \]
    \label{prop:reload_sync_overhead}
\end{proposition}

\begin{proof}
    If block $i$ executes $M_i$ local reload rounds during the SEC, then its completion time is at most $T_{\mathrm{SEC}}^{(0)} + M_i \Delta T_{\mathrm{reload}}$. Define
    \[
        M_{\max} = \max_{1 \leq i \leq N_{\mathrm{blk}}} M_i.
    \]
    Synchronization forces every block to wait for the slowest block, so
    \[
        \overline{T}_{\mathrm{SEC}}^{(N_{\mathrm{blk}})}
        \leq
        T_{\mathrm{SEC}}^{(0)} + \Delta T_{\mathrm{reload}} \, \mathbb{E}[M_{\max}].
    \]
    Since
    \[
        M_{\max} \leq \sum_{i=1}^{N_{\mathrm{blk}}} M_i,
    \]
    taking expectations gives
    \[
        \mathbb{E}[M_{\max}]
        \leq
        \sum_{i=1}^{N_{\mathrm{blk}}} \mathbb{E}[M_i]
        =
        N_{\mathrm{blk}} \overline{m}_{\mathrm{reload}}.
    \]
    Substituting this into the previous inequality and using $\Delta T_{\mathrm{reload}} = 3/20~\mathrm{POC}$ yields \cref{eq:reload_sync_bound}.
\end{proof}

At the noise rates in \cref{tab:loss_leakage_summary}, reload rounds are extremely rare, so $\overline{m}_{\mathrm{reload}}$ differs from $q_{\mathrm{reload}} = \Pr[M_i \geq 1]$ only through events with two or more distinct reload rounds in one SEC, which occur at rate $\approx p_{\mathrm{loss}}^2$, which is effectively negligible. Substituting $\overline{m}_{\mathrm{reload}} \approx q_{\mathrm{reload}}$ into the relative-overhead bound from \cref{prop:reload_sync_overhead} and using
    \[
        T_{\mathrm{SEC}}^{(0)} \in \{33.70,\,27.70,\,28.15\}~\mathrm{POC}
    \]
and
    \[
        q_{\mathrm{reload}} \in \{8.52\times10^{-4},\,4.20\times10^{-4},\,3.11\times10^{-4}\}
    \]
for \code{102}, \code{70}, and \code{54} gives isolated-block overhead estimates of only $0.00038\%$, $0.00023\%$, and $0.00017\%$ for \code{102}, \code{70}, and \code{54}, respectively, under the preliminary three-transport-step local-refill assumption. Even for $N_{\mathrm{blk}}=100$ synchronized blocks, the estimated overhead remains below $0.04\%$: $0.038\%$ for \code{102}, $0.023\%$ for \code{70}, and $0.017\%$ for \code{54}.

This proposition clarifies the role of the local memory-block reservoir described in \cref{sec:Memory block}: it confines rare reload events to a short bounded delay. The shared reservoir, whose global replenishment model is developed later in \cref{sec:The qubit factory,subsec:reservoir DTMC}, can then refill the local one asynchronously. The resulting extra idle time is negligible both for runtime and for logical performance.

\subsection{Leakage correction gadget}
\label{subsec:leakage-correction-gadget}

\begin{figure}[ht]
        \centering
        \includegraphics[width=1\linewidth]{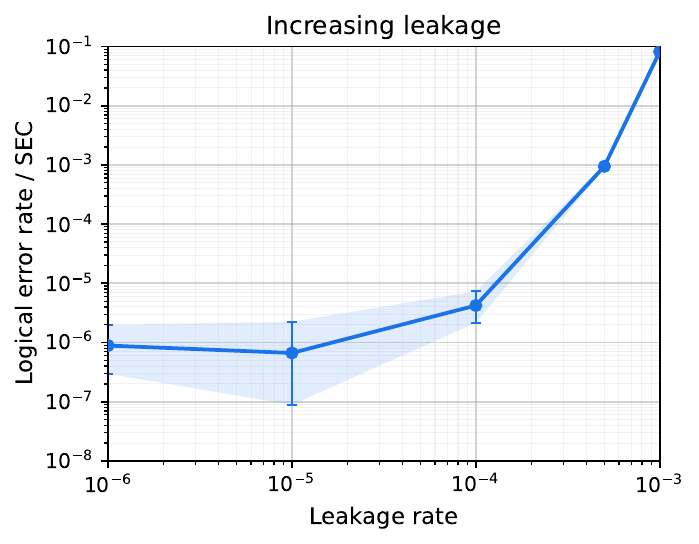}
        \caption{Logical error rate as a function of the leakage rate $\pleak$ for three codes with leakage correction, using the moving-qubit noise model with fixed $p = 10^{-3}$ and $\ploss = 0$.}
        \label{fig:increasing_leakage_rate}
    \end{figure}

\Cref{fig:increasing_leakage_rate} shows that a high leakage probability is a major issue. We plot the logical error rate as a function of the leakage rate using the moving-qubit noise model on all operations with fixed noise parameter $p=10^{-3}$ and $\ploss = 0$, varying the leakage rate $\pleak$ between $10^{-6}$ and $10^{-3}$ with leakage correction enabled.

When $\pleak$ approaches $p$, leakage strongly degrades logical performance, which may be an issue for technologies where leakage is a dominant source of noise. However, leakage affects code performance less than loss. A single leakage error can affect at most one qubit, whereas qubit loss, when detected only after each scheduled layer of two-qubit gates within the SEC, affects at least two qubits, such as a data qubit and the last ancilla with which it interacted.

Leakage detection is handled differently between ancilla and data qubits. For ancilla qubits, we use the leakage measurement introduced in \cref{subsec:Loss and leakage measurement}, obtained when the ancillas are measured at the end of each SEC. For data qubits, we use the teleportation-based {\em leakage detection unit} (LDU) shown in \cref{fig:ancilla_leakage_ldu}, run at the start of the SEC with ancillas reset to $\ket{0}$: a one-qubit teleportation gadget adapted from \cite{knill2005quantum,chow2024circuit}.

Whenever leakage is flagged, we apply a leakage reset that resets the leaked qubit to the computational manifold before the next SEC. In our simulations, this reset is modeled as replacement by a fresh maximally mixed qubit. This operation executes in one POC in the moving-qubit model; see~\cref{subsec:noise-model,subsec:Operation time}.

Because every qubit is checked for leakage at least once per SEC (ancillas when they are measured at the end of the SEC, and data at the start of the SEC through the teleportation-based LDU), the timelike support of one leakage event is bounded to at most two SECs: the SEC where leakage first occurs and the following SEC after reset.

\subsection{Memory block performance}
\label{subsec:MemoryBlockPerf}
\begin{figure}[ht]
    \centering
    \includegraphics[width=1\linewidth]{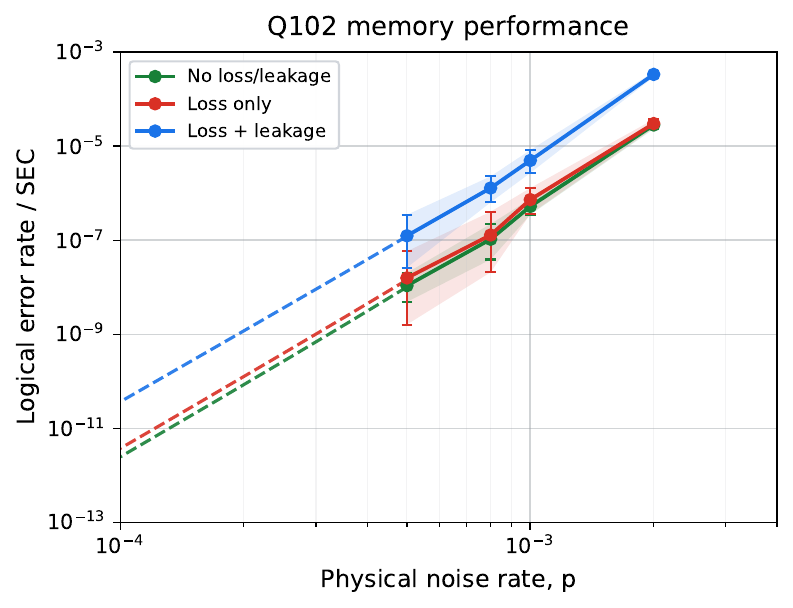}
    \caption{Logical error rate per SEC as a function of the physical error rate $p$ for the \code{102} memory block under the moving-qubit noise model, comparing the cases $\ploss = \pleak = 0$, $\ploss = p/1000$ and $\pleak = 0$, and $\ploss = p/1000$ and $\pleak = p/10$. Dashed segments indicate fit extrapolations obtained from the three-parameter ansatz $p^{\lceil d_{\mathrm{circ}}/2 \rceil}\exp(\alpha p^2 + \beta p + \zeta)$.}
    \label{fig:combined_loss_and_leakage_chungus}
\end{figure}

We summarize in \Cref{fig:combined_loss_and_leakage_chungus} three \code{102} curves comparing the no-loss/no-leakage reference configuration with two configurations that activate the loss- and leakage-handling protocol. All curves are simulated with circuit-level noise using the moving-qubit noise model.

The reference curve, with $\ploss = 0$ and $\pleak = 0$, is obtained from a standard Monte Carlo simulation using \texttt{stim}~\cite{gidney2021stim}. The remaining curves are generated with the loss- and leakage-aware simulator described in \cref{subsec:simulation_loss_n_leakage}, augmented to enforce the corresponding loss/leakage-handling protocol. In these protocol-enabled simulations, beacon measurements are explicitly included whenever $\ploss > 0$, together with the associated memory error channels during beacon measurement (idle dephasing, leakage, and loss), and teleportation-based LDU checks are explicitly included whenever $\pleak > 0$. By contrast, the reference curve does not include these additional protocol steps or their accompanying noise. Consequently, the effective multiplier reported in the table below should be interpreted as the combined cost of enabling the active loss/leakage-handling protocol at the stated $(\ploss,\pleak)$, rather than as an isolated penalty from loss and leakage alone.

\begin{table}[bp]
    \centering
    {\setlength{\tabcolsep}{3pt}
    \renewcommand{\arraystretch}{1.1}
    \footnotesize
    \begin{adjustbox}{max width=\columnwidth}
    \begin{tabular}{|c|c|c|c|c|c|c|}
        \hline
        \textbf{Code} &
        \makecell{\textbf{Rates} \\ \textbf{$(\ploss,\pleak)$}} &
        \makecell{\textbf{Logical error} \\ \textbf{at $p=10^{-3}$}} &
        \makecell{\textbf{Logical error} \\ \textbf{at $p=10^{-4}$}} &
        \textbf{$p_{\mathrm{eff}}$} &
        \makecell{\textbf{Ansatz} \\ \textbf{$(\alpha,\beta,\zeta)$}} &
        \textbf{$d_{\mathrm{circ}}$} \\
        \hline
        \multirow{3}{*}{\code{102}}
        & $(0,0)$ & $5.28\mathrm{E}{-}07$ & $2.50\mathrm{E}{-}12$ & $1.00$ & $(6810,656,19.3)$ & \multirow{3}{*}{$9$} \\
        \cline{2-6}
        & $(p/1000,0)$ & $7.34\mathrm{E}{-}07$ & $3.78\mathrm{E}{-}12$ & $1.07$ & $(-4210,501,19.7)$ & \\
        \cline{2-6}
        & $(p/1000,p/10)$ & $5.00\mathrm{E}{-}06$ & $2.81\mathrm{E}{-}11$ & $1.57$ & $({1.03\mathrm{E}{+}05},481,21.7)$ & \\
        \hline
        \multirow{2}{*}{\code{70}}
        & $(0,0)$ & $9.72\mathrm{E}{-}07$ & $7.27\mathrm{E}{-}11$ & $1.00$ & $({1.07\mathrm{E}{+}06},-3410,23.0)$ & \multirow{2}{*}{$9$} \\
        \cline{2-6}
        & $(p/1000,p/10)$ & $3.69\mathrm{E}{-}06$ & $9.71\mathrm{E}{-}11$ & $1.31$ & $({8.46\mathrm{E}{+}05},-1910,23.2)$ & \\
        \hline
        \multirow{2}{*}{\code{54}}
        & $(0,0)$ & $5.37\mathrm{E}{-}06$ & $1.58\mathrm{E}{-}10$ & $1.00$ & $({5.29\mathrm{E}{+}05},-1810,23.7)$ & \multirow{2}{*}{$9$} \\
        \cline{2-6}
        & $(p/1000,p/10)$ & $3.64\mathrm{E}{-}05$ & $3.03\mathrm{E}{-}10$ & $1.39$ & $({-3.76\mathrm{E}{+}04},235,24.1)$ & \\
        \hline
    \end{tabular}
    \end{adjustbox}
    }
    \caption{Logical error rates for the reference and protocol-enabled memory-block configurations.
    The factor $p_{\mathrm{eff}}$ is defined relative to the no-loss/no-leakage reference curve so that the full protocol-enabled configuration at the stated loss/leakage scaling is approximated by replacing the physical gate error rate $p$ by $p_{\mathrm{eff}} p$.
    $p_{\mathrm{eff}}$ absorbs both the effect of loss/leakage and the protocol overhead required to detect them.
    The fit-parameter column lists $(\alpha,\beta,\zeta)$, where available, for the ansatz
    $p^{\lceil d_{\mathrm{circ}}/2 \rceil}\exp(\alpha p^2 + \beta p + \zeta)$ used to extrapolate
    the $p=10^{-4}$ logical error.}
    \label{tab:loss_leakage_perf}
\end{table}

\subsection{Noise sensitivity analysis}
\label{subsec:Noise Sensitivity Analysis}

To identify the physical sources of error that dominate the logical error rate, we perform a one-at-time sensitivity analysis on \code{102} with $p=10^{-3}$, $\ploss = 10^{-6}$, and $\pleak = 10^{-4}$. We independently double each noise parameter while keeping all others fixed, and we report the resulting logical error rate normalized by the baseline value.

\begin{figure}[tb]
    \centering
    \includegraphics[width=0.85\linewidth]{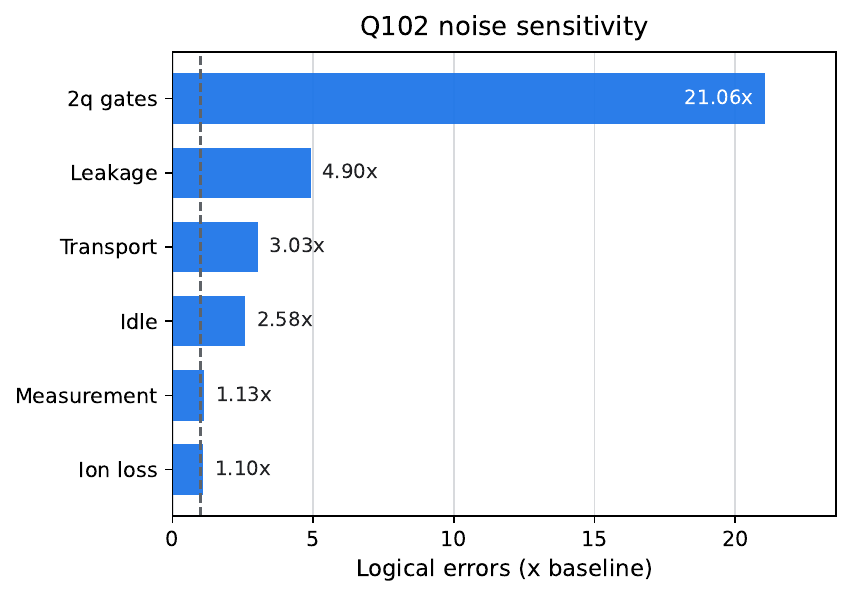}
    \caption{Sensitivity analysis showing the impact of doubling each noise parameter on the logical error rate.}
    \label{fig:sensitivity_analysis}
\end{figure}

The results are summarized in \Cref{fig:sensitivity_analysis}. The dominant contribution comes from two-qubit gate errors, whose doubling increases the logical error rate by $21.06\times$. The next largest effects come from leakage ($4.90\times$), transport ($3.03\times$), and idle errors ($2.58\times$). Doubling the measurement error rate produces a comparatively modest $1.13\times$ increase, and doubling the initial qubit loss rate $\ploss$ produces only a $1.10\times$ increase in this operating regime. This is because the beacon protocol localizes a detected loss to a rare flagged Clifford fault on at most two qubits: the affected data qubit and the ancilla that most recently interacted with it. Its leading-order impact is therefore comparable to a two-qubit flagged fault event at a rate of only $\ploss$. Scaling transport length corresponds to scaling the noise rates of all sources of noise occurring during transport: depolarizing noise, loss and leakage on all qubits (transported or idle). This aggregate effect still produces a substantial $3.03\times$ increase because roughly half of the execution time of a syndrome extraction round is spent in transport.

\subsection{Optimizing beacon protocol space and time overhead}
\label{subsec:Optimizing Beacon Protocol Space and Time Overhead}

\begin{figure}[tb]
    \centering
    \includegraphics[width=1\linewidth]{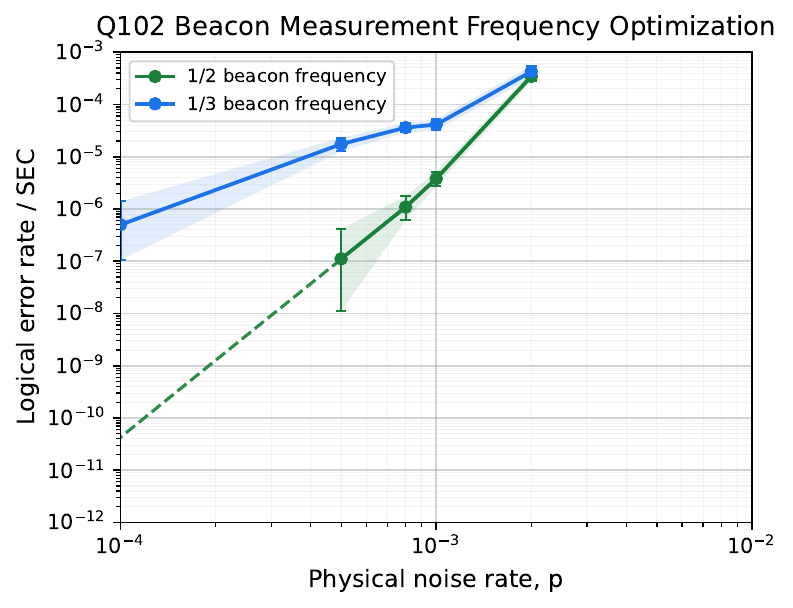}
    \caption{Logical error rate per SEC versus beacon measurement frequency for \code{102} under the moving-qubit noise model with $\ploss = p/1000$ and $\pleak = p/10$. Dashed segments indicate extrapolations obtained from the three-parameter ansatz $p^{\lceil d_{\mathrm{circ}}/2 \rceil}\exp(\alpha p^2 + \beta p + \zeta)$.}
    \label{fig:frequency_optimization}
\end{figure}

\begin{figure}[ht]
    \centering
    \includegraphics[width=1\linewidth]{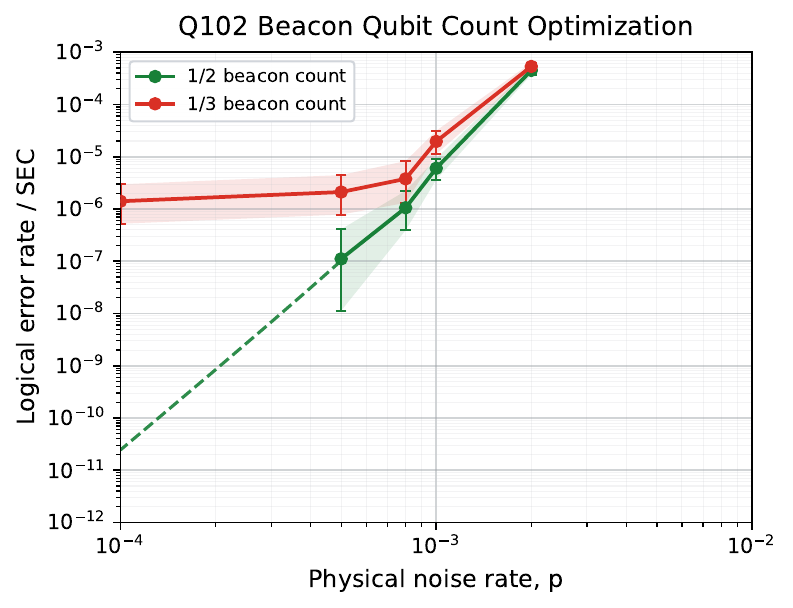}
    \caption{Logical error rate per SEC versus beacon qubit count for \code{102} under the moving-qubit noise model with $\ploss = p/1000$ and $\pleak = p/10$. Dashed segments indicate extrapolations obtained from the three-parameter ansatz $p^{\lceil d_{\mathrm{circ}}/2 \rceil}\exp(\alpha p^2 + \beta p + \zeta)$.}
    \label{fig:beacon_count_optimization}
\end{figure}

We next study two beacon-protocol trade-offs for \code{102} using the moving-qubit noise model with $\ploss = p/1000$ and $\pleak = p/10$: reducing the beacon-measurement frequency (\Cref{fig:frequency_optimization}) and reducing the number of beacon qubits measured per scheduled layer through beacon reuse (\Cref{fig:beacon_count_optimization}). In this regime, we can preserve a logical error rate per SEC near $10^{-10}$ either by measuring beacon qubits after every other scheduled layer of two-qubit gates (half frequency) or by measuring beacon qubits after every scheduled layer of two-qubit gates while checking only half of the data qubits at a time (half beacon count).

By contrast, reducing to one-third frequency or one-third beacon count is too aggressive. The reason is that reducing temporal or spatial beacon coverage increases the effective weight of each loss event by a constant factor that grows exponentially with the reduction factor. For example, at one-third frequency, a single undetected loss can spread to an effective weight-8 event before it is flagged and corrected. At that point, loss events are essentially always logical errors, and the logical error rate becomes lower bounded by the rate of loss.

\section{The cat factory}
\label{sec:Cat factories}

In this section, we design a cat factory for the walking cat architecture, tailored to the constraints of the moving-qubit model. 

Cat states are ubiquitous in quantum computing and quantum communications~\cite{shor1996fault, greenberger1989going}.
They are used in various fault-tolerant constructions as a means of measuring Pauli observables without propagating errors between the measured qubits~\cite{shor1996fault}.
For cat states to be useful, they should not have too many correlated errors. 
Previous works have proposed factories which produce cat states fault-tolerantly~\cite{shor1996fault, preskill1998fault, divincenzo2007effective, stephens2014efficient, yoder2017surface, prabhu2021fault, rodatz2025fault, khesin2026spidercat, peham2026optimizing}.

Fault-tolerance can be relaxed in practice and the cat states we use in our architecture are not fault-tolerant in the sense of these previous works.
They are built in such a way that, with the exception of a small set of errors with total probability $\varepsilon = 10^{-10}$, the produced cat states suffer from any given Pauli error $e$ with probability at most $p^{|e|}$, where $|e|$ is the weight of $e$.
This property guarantees that the cat state noise is quasi-independent, which means that it can be consumed safely to perform logical measurements without degrading the code minimum distance, and without significantly affecting its logical error rate.
In \cref{sec:Logical measurements}, we use these cat states for performing logical measurements while keeping the logical error rate in the regime of our desired value of $10^{-10}$.

The properties of the cat factories and the cat states they generate are summarized in \cref{tab:cat_heuristic}, so that the reader only interested in applying these results can refer to this table and skip the rest of this section.

The rest of this section is organized as follows.
\cref{subsec:Quasi independent cat states} introduces a notion of quasi-independent cat states.
The cat factories are designed in \cref{subsec:Design of cat factories}.
Heuristic estimates for their performance are proposed in \cref{subsec:CatHeuristics} and verified by comparison with numerical simulations in \cref{subsec:cat_numerical_results}.

\begin{table*}
\centering
\begin{tabular}{|c|c|}
\hline
\textbf{Parameter} & \textbf{Estimate} \\
\hline
\rule{0pt}{12pt}Required number of verification rounds $m$ & $\left\lceil \frac{\log(\varepsilon)}{2\log(2p)} \right\rceil$ \\[4pt]
\hline
$X$ error rate per cat state qubit & $p/2$ \\
\hline
$Z$ error rate per cat state qubit & $4(m+1)p/15$ \\
\hline
Rejection rate due to error detection & $(2m+1)wp$ \\
\hline
Rejection rate due to leakage detection & $\pleak w\left(\lceil\log_2(w)\rceil + w/40 + 6m + 4\right)$ \\
\hline
Rejection rate due to loss detection & $\ploss w\left(\lceil\log_2(w)\rceil + w/40 + 6m + 4\right)$ 
\\\hline
Qubit flow per cat factory & $w$ qubits per SEC
\\ \hline
 Production time&  \begin{tabular}{l c}
  POCs & $\lceil\log_2(w)\rceil+3m+3$    \\
     Transport steps &  $w/2+m-1$  
      
 \end{tabular} 
\\\hline
Loss distribution per attempt &
\begin{tabular}{@{}c|c@{}}
Number of qubits lost & Probability \\[2pt]
\hline
$1$  & $2w\ploss$ \\[2pt]
$4m$ & $8wm\ploss$ \\[2pt]
$8m$ & $\ploss\left(\lceil\log_2(w)\rceil + 1\right)/2$ \\[2pt]
$2w$ & $\ploss\left(\lceil\log_2(w)\rceil + 1\right)/2$
\end{tabular} \\
\hline
\end{tabular}
\caption{Heuristic estimates for the noise and rejection rate of the cat factory and the produced cat states as a function of the cat state weight $w$ and the target precision $\varepsilon$. The number of verification rounds, $m$ is determined by the desired precision, $\varepsilon$ and the physical error rate $p$. 
In this work, we use $\varepsilon=10^{-10}$ and $p=10^{-4}$, so we require $m=2$. The model is not expected to be accurate for error rates below $\varepsilon$.
The qubit flow is the number of qubits that go through the component per unit time.
The numerical simulations of \cref{subsec:cat_numerical_results} show that these heuristics are pessimistic and overestimate the impact of noise.
}
\label{tab:cat_heuristic}
\end{table*}

\subsection{Quasi-independent cat states}
\label{subsec:Quasi independent cat states}

We design cat factories within the moving-qubit model, producing $w$-qubit cat states 
$$
\frac{\ket{0}^{\otimes w} + \ket{1}^{\otimes w}}{\sqrt{2}} \cdot
$$
The produced cat state can be modeled as a perfect cat state up to some Pauli errors $e\in E$ occurring with probability $\Prob(e)$.
Here $E$ is the set of all possible Pauli errors on the final cat state. The noisy cat state is fully characterized by the probability distribution $\Prob(e)$, which is a property of the production procedure. 
For simplicity, we ignore leakage and loss which are removed by post-selection.

Any error $e$ can lead to one of the following distinct effects: a  $Z$ error leading to the state \[\frac{\ket{0}^{\otimes w}-\ket{1}^{\otimes w}}{\sqrt{2}},\] one of $w\choose k $ weight-$k$ $X$ errors, and a combination of a $Z$ error and one of the weight-$k$ $X$ errors. 

Note that: (1) An even number of $Z$ terms in $e$ is equivalent to no $Z$ terms. (2) A weight-$k$ $X$ error is equivalent to its complementary weight-$(w-k)$ $X$ error; (3) Any $Y$ term in $e$ will have the same effect as a $Z$ term when accounting for the $Z$ error and an $X$ term when accounting for the weight of the $X$ error.  

Cat states are used to measure Pauli observables on data blocks. The circuits used to implement these measurements (\emph{e.g.}, \cref{fig:CatBasedMeasurement} in \cref{sec:Logical measurements}) propagate weight-$k$ $X$ errors on the cat state to weight-$k$ $X$ errors in the data block. A $Z$ error on the cat state causes an error in the measurement outcome. We say that the cat state is fault-tolerant if it meets the following definition: 

\begin{definition}[Fault-tolerant cat state]
A noisy cat state is said to be fault-tolerant with noise rate $p$ if it is a perfect cat state up to Pauli errors $e\in E$, where the probability for an error $e$ is upper bounded by $p^{|e|}$. \label{def:FTcat-states}
\end{definition}

It is possible to add a notion of distance to the above definition, requiring that it only holds up to $|e|\leq\frac{d-1}{2}$ (see \cite{prabhu2021fault}). However, in the context of a specific fault-tolerant architecture, it is more accurate to assess performance based on error rates. We therefore choose to set a target precision $\varepsilon$, below which errors are too rare to impact performance. We construct our cat states so that they are fault tolerant above this cutoff rate. This is done by breaking the set of errors into two subsets $E_0$ and $E_1$ such that the total probability of errors in $E_0$ is below the target precision, and 
errors in $E_1$ meet the fault tolerant condition \cref{def:FTcat-states}.

\begin{definition}[$(\varepsilon,p)$-independent cat states]
A weight-$w$ noisy cat state is said to be $(\varepsilon,p)$-independent if it is a perfect cat state up to Pauli errors from two  sets, $E_0$ and $E_1$ such that:
\begin{itemize}
    \item $\Prob(E_0) \le \varepsilon$.
    \item For all $e\in E_1$, $\Prob(e)\le p^{|e|}$.
\end{itemize}
We call $\varepsilon$ the precision of the cat state. 
\label{def:QuasiFT-Cats}
\end{definition}

Below, we show that constructing $(\varepsilon, p)$-independent cat states can be significantly cheaper than constructing fault-tolerant cat states. This is of particular importance in our architecture, which consumes cat states that are typically larger than the code distance. For our specific architecture, we use $\varepsilon = 10^{-10}$. 

\subsection{Design of the cat factories}
\label{subsec:Design of cat factories}

\begin{algorithm}
    \caption{Even-weight cat state preparation}
 \label{algorithm:cat_prep}
\DontPrintSemicolon
\SetAlgoLined
\KwIn{
An even integer $w$.

A register of $w$ qubits in two rows. The positions of the top row are labeled $\{1,\dots, w/2\}$ and the bottom are $\{w/2+1,\dots,w\}$, both going from left to right.
}
\KwOut{
A weight-$w$ cat state.
}
    $c = \lfloor\log_2 (w)\rfloor-1$
    
    Initialize qubit in position $1$ in $\ket{+}$.

    Initialize qubits in positions $\{2,\dots,w\}$ in $\ket{0}$.

    \For {$i\in \{0,\dots,c\}$}{
        \For{$j\in\{1,\dots,2^i\}$}{
            \label{algoline:cnots_for_cat}
            Perform $\CX$ from the qubit in position $j$ to the qubit in position $j+(w/2)$.
        }
        \If{$i<c$}
        {Do a clockwise shift of $2^i$ steps.}
        }
    \If{$2^{c+1} < w$}{
    Do a clockwise shift of $w/2-2^{c}$ steps.
    
    \For{$j\in\{2^{c+1}-w/2+1,\dots,w/2\}$}{
    Perform $\CX$ from the qubit in position $j$ to the qubit in position $j+(w/2)$.}
    }
    \Return{all qubits} 
    
\end{algorithm}
\begin{figure}[h!]
    \centering
    \includegraphics[width=\linewidth]{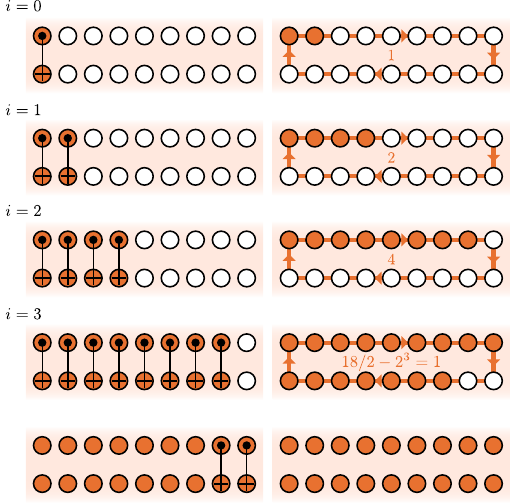}
    \caption{Visual depiction of the cat state preparation step  (\cref{algorithm:cat_prep}) for $w=18$ and $c=3$. The initial cat qubit (initialized in  $\ket{+}$) begins in the upper left corner. White circles represent qubits in the state  $\ket{0}$. During each iteration of the \textbf{for} loop, $\CX$ gates are applied from current cat qubits in the top row to their neighbors in the bottom row, and then the qubits undergo a clockwise shift to align all cat qubits with unentangled qubits. During the final iteration, the qubits only need to shift far enough so that the remaining unentangled qubits are aligned with current cat qubits; only these qubits are targeted by $\CX$ gates to complete the state preparation.}
    \label{fig: cat prep}
\end{figure}

\begin{algorithm}
    \caption{Even-weight cat state verification}
\label{algorithm:cat_verfication}
\DontPrintSemicolon
    \KwIn{ 
         Cat state weight, $w$, an even integer. 
         
         Number of verification rounds, $m$.
         
         A register of $2w$ qubits in four rows as follows:

                {Top row: $w/2$ ancilla qubits. 
                
                 Middle two rows: weight-$w$ cat state.
                 
                 Bottom row: $w/2$ ancilla qubits.}
        }
    \KwOut{Verified weight-$w$ cat state.}

    \For{$\_$ in 1 to $m$}{
    Initialize the ancilla qubits to $\ket{+}$.
    
    Perform \CZ between each qubit in top and bottom rows (ancilla) and the adjacent qubit in the middle rows (cat qubit). 

    Shift the cat state qubits one step clockwise.
    
    Perform \CZ between each qubit in top and bottom rows (ancilla) and the adjacent qubit in the middle rows (cat qubit). 

    Measure the ancilla qubits in the $X$ basis.

    If any ancilla measurement has a non-trivial outcome (1, `leaked' or `lost'), abort and return ``Cat state rejected''. 
    }
    
    Prepare all ancilla in $\ket{0}$. 

    Perform \CX between all cat qubits (control) and adjacent ancilla (target). 

    Measure the cat qubits in the $X$ basis.   
    
    If any cat qubit measurement has the outcome `leaked' or `lost', abort and return ``Cat state rejected''. 

    Change labels between ancilla and cat qubits. The middle two rows are now called ancilla and the top and bottom rows are now cat qubits. 

    \Return{Qubits in top and bottom rows.} 
\end{algorithm}

Our cat factories use a two-step procedure to produce even-weight cat states that are $(\varepsilon,q)$-independent, where $q$ is a multiple of the physical error rate $p$. 
\begin{enumerate}
    \item {\bf Cat state preparation:}  Prepare a cat state using an iterative process that doubles the size of the cat state with each POC (see \cref{algorithm:cat_prep} and \cref{fig: cat prep}). This step takes  $\lceil\log(w)\rceil+1$ POCs. The cat state at the output can have high-weight $X$ errors, a $Z$ error, leaked qubits and lost qubits. 
    \item {\bf Cat state verification:} Verify the cat state using $w$ ancilla qubits and $m$ verification rounds. Each verification round consists of checking for $X$ errors using $ZZ$ stabilizer checks. The stabilizer checks can also detect loss.  Each verification round takes three POCs  and one transport step. The final verification round is followed by a leakage detection round which takes 2 POCs (\cref{algorithm:cat_verfication}).
    
\end{enumerate}

The parameter $m$ determines the number of verification rounds and the values  $(\varepsilon, q)$. For our purposes we choose $m=2$, which is sufficient for $q=p=10^{-4}$ and $\varepsilon=10^{-10}$.  A heuristic model for choosing $m$ and for the expected error, rejection and loss rates is given in \cref{tab:cat_heuristic}. 

\begin{table}[h]
    \centering
    \begin{tabular}{|c|c|c|}\hline
        {\bf Step} &{\bf POCs} &{\bf Transport steps }\\\hline
         Preparation&  $\lceil\log_2(w)\rceil+1$& $(w/2-1)$ \\\hline
         $m=2$ verification rounds & $3m+2=8$& 2\\\hline
    \end{tabular}
    \caption{Time required to prepare a $(\varepsilon = 10^{-10}, p=10^{-4})$-independent, weight-$w$ cat state. }
    \label{tab:cat_state_cost}
\end{table}

Cat state preparation is done in a register consisting of two rows, each with $w/2$ qubits. The first (leftmost) qubit on the top row is labeled $Q_1$ and initialized in the state $\ket{+}$. All other qubits are initialized in $\ket{0}$. $Q_1$ is now a $w=1$ cat state. 
We grow this cat state by iteratively performing \CX operations between the qubits in the cat state and the qubits in the row below, shifting clockwise each time so that all cat qubits are on the left side of the top row (see \cref{algorithm:cat_prep} and \cref{fig: cat prep}). 
The total preparation time is $\lceil\log_2(w)\rceil+1 $ POCs and $w/2-1$ transport steps. Note that state initialization is executed simultaneously on all qubits. Similarly, all \CX gates in the {\it for loop} (\cref{algoline:cnots_for_cat})  are executed simultaneously.  We use the term {\em cat qubit} to denote qubits that are part of the cat state. 

It is easy to see that a single error can propagate into a high weight error during the preparation procedure. For example, in \cref{fig: cat prep}, an $X$ error on qubit 2 at $i=1$ would propagate to qubit 13 at $i=2$, and eventually to three more qubits, making it a weight-five $X$ error. 
Verification (\cref{algorithm:cat_verfication}) is used to detect errors and reject faulty cat states. Specifically, we only check for $X$-type errors using multiple verification rounds. At each round, we measure all  $ZZ$ stabilizers on neighboring qubits (in a cyclic manner). This can be done by introducing ancilla qubits above and below the two-row cat state, performing \CX between the ancilla and the cat qubit either above or below it, shifting clockwise, performing another round of \CX between ancilla and cat qubits above or below them, measuring the ancilla and checking the parity of the result.  One round of verification takes three POCs and one transport step. 
This verification needs to be repeated a few times to reduce the probability of an undetected high weight $X$ error. Based on the heuristic and simulation below, $m=2$ rounds are sufficient to achieve $\varepsilon = 10^{-10}$ when $p=10^{-4}$. 

Loss is detected during the ancilla measurements. Since all cat qubits interact with an ancilla during every verification round, any lost cat qubit will propagate to a lost ancilla that will be measured. Leakage detection on the ancilla is also done as part of the measurement. Leakage detection on the cat qubits cannot be performed directly. Instead, we use the LDU introduced in \cref{fig:ancilla_leakage_ldu}. As a result the roles are now swapped: the cat qubits are now the ancilla qubits and vice versa. This operation takes 2 POCs. 

The total depth for preparation and verification is $\lceil\log_2(w)\rceil+3m+3$ POCs and  $w/2+m-1$ transport steps. 
 
The verification procedure above reduces the probability of $X$ errors on the cat state. In principle, it is possible to include a second verification procedure to detect $Z$ errors. One possibility is to prepare two cat states and use one to measure the $X^{\otimes w}$ stabilizer on the other. This can be done in the factory or just before the cat state is consumed. The latter is particularly effective for catching transport-related errors. Within the context of this work, the impact of $Z$ errors in the cat state is small enough that we do not need this type of verification (see \cref{subsec:CatHeuristics} and \cref{sec:Logical measurements}). 

\subsection{Heuristic error estimates}
\label{subsec:CatHeuristics}
We now derive the expressions that lead to the heuristic model of \cref{tab:cat_heuristic}. 
We begin with models for $X$ and $Z$ errors, including the rejection rate due to these errors.  We then derive estimates for loss and leakage.  
The model is expected to be pessimistic and can be compared with the simulation results in \cref{subsec:cat_numerical_results} below.

We assume that leakage and loss errors are always detected and lead to the cat state being rejected.  
 Under this assumption, we want to show that a produced cat states can be modeled as a perfect cat state that goes through two error channels. The first induces a single qubit $X$-error on each qubit, with independent probability $p_x=p/2$; the second induces a single qubit $Z$-error on each qubit,  with independent probability $p_z=(4w+4mw)p/(15w)$. Under this model, the cat state is $(\varepsilon,q)$-independent for any $\varepsilon \le (2p)^{2m}$ and $q=\max(p_x,p_z)$.  At $m=2$, $p_z<p$ so the cat state is $(\varepsilon,p)$-independent for any $\varepsilon \le (2p)^{4}$. We show below that this model is expected to overestimate the error probabilities, up to a precision of $\varepsilon = p^{2m}$.  

In the derivation of the $X$ and $Z$ error models we assume that the only source of errors is two-qubit gates. This simplifies the derivation without having a significant impact on the result. We also ignore some of the transport steps as explained below.  

{\bf Rejection rate due to error detection - } There are $M_g = (w-1)+2mw$ two-qubit gates in the circuit, excluding leakage detection.  The rate of rejection due to an error detected in verification is therefore upper bounded by $M_gp<(2m+1)wp$.

As can be seen in the simulation results (\cref{fig:weight_30_cat_c}), this bound is relatively tight when $p$ is small, but becomes loose for large $p$. This is expected since we are using a sum of probabilities for independent events.

Rejection due to loss is negligible due to the magnitude of $\ploss$ (\cref{eq:cat_loss}), but leakage (\cref{eq:cat_leakage}) can be significant. Both are treated below. 

{\bf $X$-type errors -} A single $X$ error during preparation can propagate to a high-weight  $X$  error on the prepared cat state. An $X$ error of any weight anti-commutes with at least two $ZZ$ stabilizers during one round of verification. Each stabilizer check involves two two-qubit gates, so the probability of an undetected preparation error after $m$ rounds of verification is at most $(2p)^{2m}$. We can therefore ensure that $X$ errors during preparation will be detected with probability at least $1-(2p)^{2m}$. We can set any $\varepsilon > (2p)^{2m}$ as a cutoff above which we expect $X$ errors to only be the result of errors during verification.  For $p=10^{-4}$ and $\varepsilon = 10^{-10}$, it is sufficient to choose $m=2$. 

Any $X$ errors during verification will not propagate to more than one cat qubit. Moreover, $X$ errors during one verification round will be detected at the next round, unless two stabilizer measurements have an error. At sufficiently small $p$, the main source of $X$ errors is therefore the last two-qubit gate acting on each cat qubit.  For each of these, there are a total of four distinct error types that lead to an undetectable $X$ error in the cat state, each of these with probability  $p/15$. The probability for an $X$ error on each individual cat-qubit is therefore smaller than our heuristic value $p_x=p/2$. 

The total probability for an $X$ error of weight $k$ is 

\begin{equation}
    \Prob(\text{weight}~k~X~\text{error})={w \choose k}p_x^{k}.
    \label{eq:cat_px}
\end{equation}

This value can be compared with the simulation results in 
\cref{fig:weight_30_cat_a} and \cref{fig:weight_50_cat}. 

{\bf $Z$-type errors -} We bound the probability of a single undetected $Z$ error. If the error is during preparation, then it might either be the result of a $Z$ or $Y$ error, but the $Y$ error will be detected later (as an $X$ error), so the undetectable errors are only $ZI$ and $IZ$. The probability of an undetectable $Z$ fault during preparation is therefore $2(w-1)p/15<2wp/15$. 

Similarly, during the first $m-1$ rounds of verification the errors that will not be detected  are only of the form $ZZ$ and $ZI$ (where the second qubit is the ancilla) so again, the probability of an undetected $Z$ error is $4wp/15$. For the final round,~$w$ of the $YI$ and $YZ$ errors do not get detected because they are on a qubit that goes unchecked later. So we have $6wp/15$. The heuristic expression 
\begin{equation}
\label{eq:cat_z_error}
   (4w+4mw)p/15
\end{equation}
is the sum of all these contributions. 
 For $m=2$ this gives $12wp/15<wp$. 

 {\bf Leakage -}  Within the moving-qubit model, we assume that the leakage measurement outcome is noiseless. Our leakage detection steps will detect all leakage that can lead to errors. As a result, leakage does not contribute to the error probability, only to the rejection rate. During preparation there are $w$ qubits and the depth is $\lceil \log_2(w)\rceil +1 +(w/2-1)/20$ time steps. The depth of each verification round is $3+1/20$ and there are $2w$ qubits. Additionally, $2w$ qubits can leak during leakage detection which takes 2 POCs. 
 The total leakage probability is therefore at most 
 \begin{equation}
      \label{eq:cat_leakage}
 \pleak w\left(\lceil \log_2(w)\rceil+w/40+4+6m\right).
 \end{equation}
For simplicity, we ignored the $1/20$ terms which have a negligible contribution. 

A comparison of this heuristic with simulation for a weight-30 cat state is given in \cref{tab:cat_w30_loss_and_leakage}.

 {\bf Loss -}  Loss contributes to both the rejection rate (through loss detection) and to the requirements on the global reservoir. We assume that the final loss and leakage measurement outcomes are noiseless. Any loss before the final round of \CX gates would propagate to an ancilla and lead to rejection. 
 
 To calculate the rejection rate due to loss, we note that loss mechanisms are similar to leakage mechanisms.  The probability of rejection due to loss is therefore upper bounded by 

\begin{equation}
 \label{eq:cat_loss}
 \ploss w\left(\lceil \log_2(w)\rceil+w/40+4+8m\right).
 \end{equation}
This is two orders of magnitude lower than the leakage rate, so it does not contribute significantly to the rejection rate. However, since loss events can cascade to multiple qubits, the expected number of lost qubits per attempt can be significant.

Most loss events will cascade and cause a number of qubits to be lost. We can upper bound the number of qubits lost based on when the \emph{first loss event} (\FLE) occurs. During preparation, the probability of \FLE~at any POC is $w\ploss$ (where we use the fact that $\ploss$ is very small). During verification the probability of \FLE~at any POC is $2w\ploss$. A single qubit lost at the end of the preparation stage will propagate to $4m$ lost qubits at the end of verification. In a worst-case scenario $w/(2m)$ qubits lost during verification will lead to $2w$ lost qubits. \FLE~during verification will lead to at most $4m$ lost qubits. We therefore expect to have a probability of $8mw\ploss$ for \FLE~during verification leading to between one and $4m$ lost qubits.  Roughly half of the loss events during preparation will result in two lost qubits at the end of preparation, and up to $8m$ lost qubits at the end of verification. This happens with a probability of about  $\ploss[\lceil \log_2(w)\rceil+1]/2$. The other loss events can lead to at most $2w$ lost qubits with the same probability. Additionally, there is the probability of a single qubit lost during the final measurement. This happens with probability $2w\ploss$.
We therefore use the four peak heuristic in \cref{tab:cat_heuristic}. The expected loss rate in this model is 
\begin{equation}
    \ploss w \left(2+32m^2+(8m+2w)\left (\lceil \log_2(w)\rceil+1\right)\right).
\end{equation}
While this model is expected to significantly overestimate the loss rate (compare with \cref{fig:cat_loss}), it is sufficient for our purposes. 

Note that in both this heuristic and the simulations, we did not account for the possibility that verification stops because the state is rejected in an early verification round. As a result, they overestimate the number of lost qubits.

Our numerical results below support the heuristic models. 

\subsection{Numerical results} \label{subsec:cat_numerical_results} 
\begin{figure}
    \centering
    \subfloat[]{
    \includegraphics[width=0.95\linewidth]{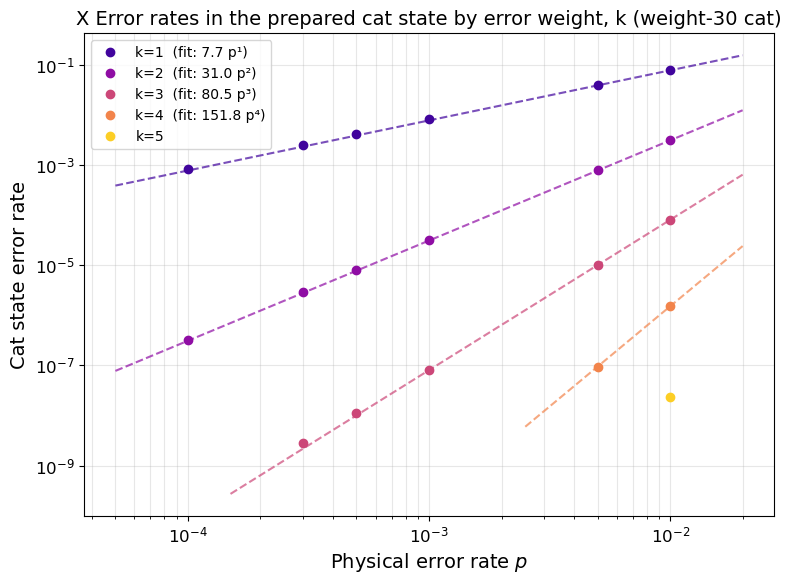}
    \label{fig:weight_30_cat_a}
}

   \subfloat[]{
\includegraphics[width=0.95\linewidth]{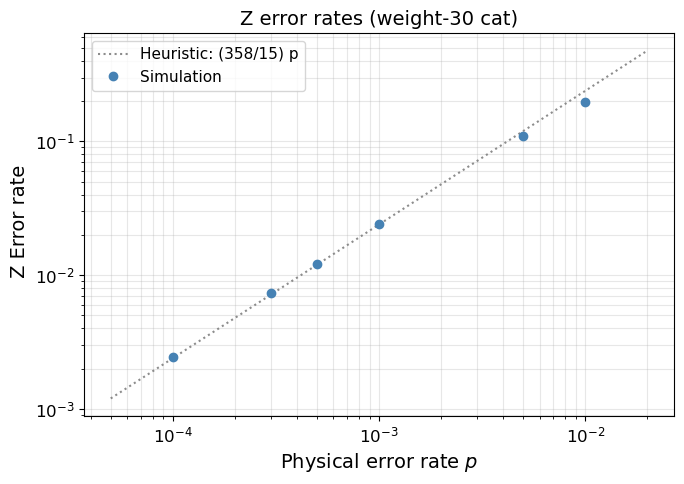}
\label{fig:weight_30_cat_b}
}

    \subfloat[]{
\includegraphics[width=0.95\linewidth]{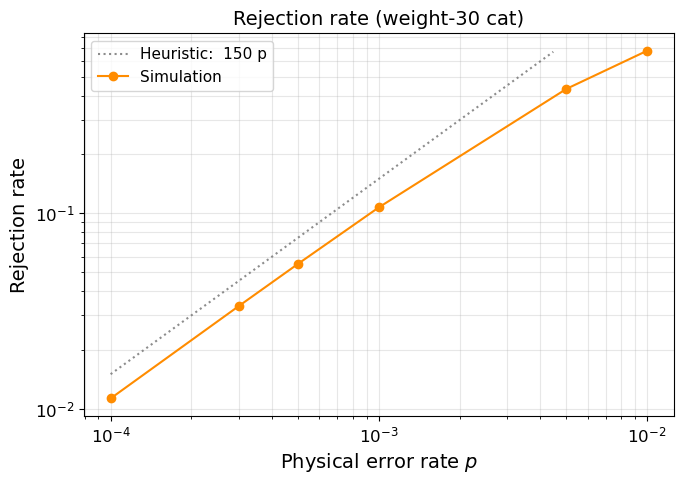}
\label{fig:weight_30_cat_c}
}

    \caption{Errors and rejection during verification for a weight-30 cat state with $m=2$.  Results from Monte Carlo simulations with $10^{10}$ shots taken at each  value of $p$. (a) $X$ error rates by weight, $k$, with a one parameter ($C_k$) fit to $C_kP^k$.  (b) $Z$ error rate compared to the heuristic upper bound.  (c) Rejection due to non-trivial measurement results during verification, compared with the heuristic upper bound.  }
    \label{fig:weight_30_cat}
\end{figure}

\begin{figure}
    \centering
    \includegraphics[width=0.9\linewidth]{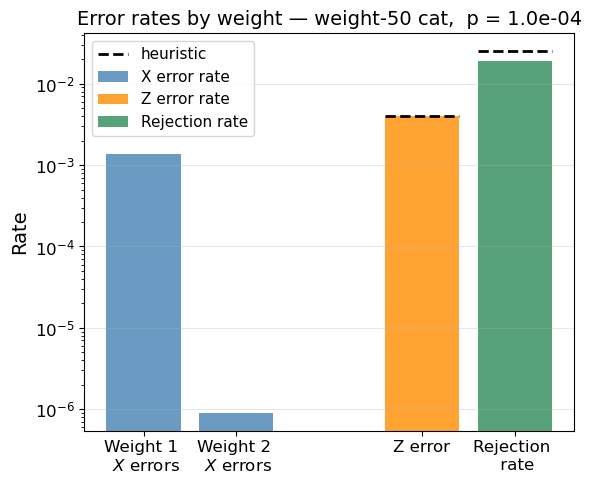}
    \caption{Simulation results (using $10^{10}$ shots) for a weight-50 cat state at $p=10^{-4}$ and $m=2$. Rejection rate here represents only the rejection due to non-trivial measurement results in verification.}
    \label{fig:weight_50_cat}
\end{figure}

\begin{figure}[tb!]
    \centering
    \includegraphics[width=1\linewidth]{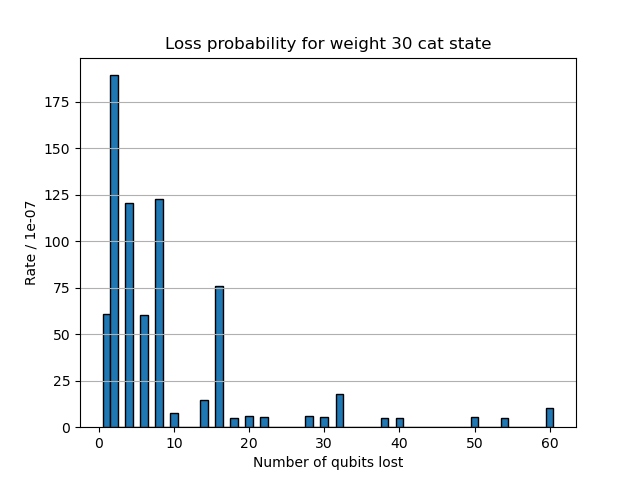}
    \caption{Qubit loss probability for a 30 qubit cat state. Note the four main areas for the distribution are $1$, $2-8$, $9-16$, and $17-60$ as predicted by the heuristic in \cref{tab:cat_heuristic}. The probability for a loss event is $729 \times 10^{-7}$ and the expected loss rate is $6640 \times 10^{-7}$.  }
    \label{fig:cat_loss}
\end{figure}

Numerical estimates for the performance of the cat factory were generated with Monte Carlo simulations using Stim \cite{gidney2021stim}. The simulations include all gates and idles between gates, but do not account for transport steps which are not expected to contribute significantly even for large cat states. Simulations for Pauli errors (\cref{fig:weight_30_cat},  and \cref{fig:weight_50_cat}) were done with $m=2$ and $10^{10}$ shots. Simulations for loss and leakage (\cref{fig:cat_loss} and \cref{tab:cat_w30_loss_and_leakage}) were done with $m=2$ and $10^8$ shots. 

Results for a weight-30 cat state are shown in \cref{fig:weight_30_cat}. The probability for $X$ errors by weight-$k$ (\cref{fig:weight_30_cat_a} follows $C_kp^k$ as expected, where the constant $C_k$ is fit to the data. Notably, $C_k<{w \choose k}$, so the heuristic model is an upper bound. 
The Z error rate on the final cat state  (\cref{fig:weight_30_cat_b} follows the heuristic estimate. The rejection rate (\cref{fig:weight_30_cat_c}, is lower than the heuristic bound, especially at relatively high error rates. This is expected since the heuristic is only expected to hold when $(1-p)\approx 1$. It should be noted that the choice of $m$ depends on $p$. Rejection rates at high error rates would therefore be significantly higher if we wanted to reach the same $\varepsilon$ for the approximate $(\varepsilon,p)$-independent cat states. 

In \cref{fig:weight_50_cat}, we show results for a weight-50 cat state prepared with a physical error rate of $p=10^{-4}$. The  $X$ errors decay exponentially with weight (as expected) and the rejection rates and $Z$ error rates are slightly below the heuristic estimate.

\begin{table}
    \centering
    \begin{tabular}{|c|c|c|}\hline
         &  {\bf Heuristic}& {\bf Simulation}\\\hline
         Rejection rate &  $7.5\cdot 10^{-3}$& $7.3\cdot10^{-3}$\\
 due to leakage detection& &\\\hline
         Rejection rate &  $7.5\cdot 10^{-5}$& $7.2\cdot 10^{-5}$\\
 due to loss detection& &\\\hline
 Expected number of qubits  & $1.8\cdot 10^{-3}$&$6.5\cdot10^{-4}$\\ 
 lost per attempt& &\\ \hline
    \end{tabular}
    \caption{Impact of leakage and loss for a weight-30 cat state.  Leakage rate and loss probability represent the rate at which at least one leakage and/or loss event occurs. The loss rate is the mean number of qubits lost per cat state production attempt. Note that the loss distribution heuristic (\cref{tab:cat_heuristic})  is expected to be very pessimistic.}
    \label{tab:cat_w30_loss_and_leakage}
\end{table}

\section{The Bell factories}
\label{sec:Bell_factories}

In this section, we describe our Bell factories and the cat state stitching protocol. The stitching protocol uses two cat states located in different cat factories, and a set of Bell states with a qubit in each of these cat factories to create a new cat state that is distributed across the two factories. This distributed cat state can then be used to measure logical operators that have support on the two data blocks adjacent to the cat factories. Important parameters and performance estimates of the Bell factories and stitching protocol are summarized in \cref{tab:summary_bell_factory}. The estimates in the table are expressed as functions of two parameters: the target precision of the cat state, $\varepsilon$, and the physical error rate, $p$. The requirements for the stitching protocol and the Bell factory are independent of the weight of the cat states. 

\begin{table*}
    \centering
    \begin{tabular}{|c|c|}
    \hline
    {\bf Parameter} & \bf{Estimate} \\
    \hline
        Production time for one  Bell pair  & 2POCs   \\ \hline
        Bell state rejection rate & 0 \\ \hline
        Bell states required for stitching & $m=\left\lceil \frac{\log(\varepsilon)}{2\log(2p)} \right\rceil$  \\\hline
        Stitching time   &  1POC\\ \hline
        Stitching rejection rate & \ $4mp$ \\ \hline
        Bell factories & $\lceil N/3 \rceil$  \\ 
        for $N$ cat factories &  \\ \hline
        Qubit flow per Bell factory & $2 \lceil 3m/2 \rceil$ qubits per cat state production round \\ \hline
    \end{tabular}
    \caption{Estimates of the performance of and requirements for the stitching protocol and the Bell factories. The stitching protocol uses two cat states of weights $w_1$ and $w_2$ and precision~$\varepsilon$, to create a cat state of weight $w=w_1+w_2$ with the same precision. Recall that the qubit flow measures the number of qubits going through the component per unit time.}
    \label{tab:summary_bell_factory}
\end{table*}

\subsection{Bell state production}

Preparing a Bell state, $\frac{1}{\sqrt{2}}(\ket{00}+\ket{11})$, is done using two qubits, one prepared in $\ket{+}$ and the second in $\ket{0}$, and then performing a $\CX$ gate. This takes two POCs and has an error probability of at most $p+2p/10$. Since the Bell pairs will be measured directly and used in a stitching protocol that rejects when detecting an error, leakage, or loss, they do not need to be verified. Error propagation from the Bell pairs to the cat state is discussed below.

\subsection{Stitching protocol}
\begin{algorithm}[t]
    \caption{Stitching two cat states of weights $w_1$ and $w_2$ into a single cat state of weight $w=w_1+w_2$.}
\label{algorithm:cat_stitch}
\DontPrintSemicolon

    \KwIn{
    
    Two cat states $C_1$, $C_2$ of weight $w_1$ and $w_2$.  Cat qubits in $C_1$ are labeled $Q_{1,\mu}, \mu \in \{ 0, \dots, w_1-1\}$; Cat qubits in $C_2$ are labeled $Q_{2,\nu}$, $\nu \in\{ 0,\dots, w_2-1\} $.
    
    A positive integer $m$, specifying the number of joint $ZZ$ measurements. 
    
    $m$ Bell pairs. Each Bell pair consists of  two qubits $B_{j,\ell}$, $j\in \{1,2\}$, $\ell \in \{0,\dots, m-1\}$. Each of these is located next to cat qubit $Q_{j,\ell}$.

        }

    \KwOut{Cat state with weight $w=w_1+w_2$.}

\For{each qubit $B_{j,\ell}$}
{Perform a \CZ between  $Q_{j,\ell}$ and $B_{j,\ell}$.}

Measure all $B_{j,\ell}$ in the $X$ basis and record the parity of the results from each Bell pair in $P_\ell, \ell \in \{0,\dots, m-1\}$.

\If{any measurement results are either `lost' or `leaked'}
{Reject cat state and abort. 

 Return {"Cat state rejected".}}

\If{all $P_\ell$ have the same value}{
$Parity = P_0$}
\Else
{Reject cat state and abort. 

 Return {"Cat state rejected".}}

    \Return{Cat qubits and $Parity$ } 
\end{algorithm}

The stitching protocol (see \cref{algorithm:cat_stitch}) uses Bell states to stitch two cat states $C_1$  and $C_2$ at two different locations into a single cat state. The new cat state is distributed across the same locations as $C_1$ and $C_2$, and uses the same qubits. If $C_1$ has weight $w_1$ and $C_2$ has weight $w_2$, then the new cat state will have weight $w=w_1+w_2$.
The protocol consists of a number ($m$) of $ZZ$ stabilizer measurements across the cat state, \textit{i.e.}, each $ZZ$ measurement is between one qubit in $C_1$ and one qubit in $C_2$. The cat state is accepted if all $ZZ$ measurements have the same parity, otherwise it is rejected. If the parity is even, the new stitched cat state is the standard cat state. If the parity is odd, the stitched cat state differs from the standard cat state only by a known $X$ error on all qubits of $C_1$ (or $C_2$), which can be tracked. The value of $m$ is chosen to ensure that the cat state is approximately $(\varepsilon,p)$-independent (see \cref{tab:summary_bell_factory}). 

The protocol has a depth of two POCs. The probability of an error on one of the $m$ parity checks is upper bounded by $p'=2p+2p/10+p_B$, where $p_B$ is the error rate for the Bell state (including transport errors). In \cref{tab:summary_bell_factory} we use the pessimistic estimate $p' = 4p$, which accounts for over a hundred transport steps.  The probability that the stitching protocol ends in rejecting the new cat state is at most  $mp'$.  The result of an error during stitching could be one of three types:  1. The parity is not measured correctly, leading to an $X$ error of weight $w_1$ on the stitched cat state; 2. An error propagates into one cat qubit on one cat state, leading to the same error on the stitched cat state; 3. An error propagates into one cat qubit at each location, leading to a correlated error on the stitched cat state.  
Error 1 can lead to a logical error. However, the cat is rejected unless all $m$ measurements have the same outcome. The probability that all $m$ measurements give the same incorrect outcome is at most $(p')^m$. We therefore choose $m$ so that $(p')^m < \varepsilon$. For our purposes, choosing $m=4$ provides sufficient margin at $\varepsilon=10^{-10}$.  We note that due to error type 3 above, the probability of a weight-two $X$ error can be of order $p$, which means the cat state is not $(\varepsilon,p)$-independent. However, the two parts of the cat states will be used on different memory blocks, and correlated errors on different memory blocks do not reduce fault-tolerant capability. 

\subsection{Required number of Bell factories}
\label{subsec:Required number of Bell factories}

For a walking cat architecture with $N$ blocks and $N$ cat factories, we have to stitch at most $N/2$ pairs of cat states during each round of cat production. Each stitching consumes $m$ Bell states, so we need to produce at most $Nm/2$ Bell states per round of cat state production.
Each Bell factory produces at least $\lceil 3m/2 \rceil$ Bell states per cat state production round because the cat state production time is lower bounded by $3m+3$ POCs (see \cref{tab:cat_heuristic}).
Therefore,
$
\lceil N/3 \rceil
$ Bell factories are sufficient for all required cat state stitching.

Finally, the qubit flow is upper bounded by $2 \lceil 3m/2 \rceil$ qubits per cat state production round.

\section{Logical measurements}
\label{sec:Logical measurements}
\begin{table}[b]
    \begin{centering}
    \begin{tabular}{|c|c|c|c|}
    \hline 
    \multirow{2}{*}{ \bf Weight ($\physicalWidth$)} & \multicolumn{2}{c|}{ \bf \# SEC (\EDM)} & \textbf{\# SEC (Viterbi)}\tabularnewline
    \cline{2-4} \cline{3-4} \cline{4-4} 
    & $\varepsilon=10^{-5}$ & $\varepsilon=10^{-10}$ & $\varepsilon=10^{-10}$\tabularnewline
    \hline 
    10 & 3 & 5 & 4.04\tabularnewline
    \hline 
    20 & 3 & 6 & 5.10\tabularnewline
    \hline 
    30 & 3 & 6 & 5.14\tabularnewline
    \hline 
    54 & 4 & 8 & 6.31\tabularnewline
    \hline 
    \end{tabular}
    \par\end{centering}
    \caption{Duration (in number of syndrome extraction cycles) for error-detected (\EDM) and adaptive error-corrected (Viterbi) logical measurements.
    In each logical measurement, physical cat-based measurements are repeated until a target logical measurement error rate ($\varepsilon$) is reached.
    Logical measurement durations are shown for logical representatives of different weights ($\physicalWidth$).
    Here, the physical gate noise rate is $p=10^{-4}$ and cat states are missing with probability $5\physicalWidth p$.
    We assume one extra SEC following the last cat-based measurement, to allow for classical reaction time, for instance from a decoder.}
    \label{tab:LMtime}
\end{table}
\Cref{subsec:The logical instruction set} defined logical measurements (LM1,~LM2) as a key element of the logical instruction set, through which logical operations are realized (see~\cref{subsec:Implementation of the logical instruction set}).
This section discusses implementations of these logical measurements using cat states (\cref{subsec:FTLM}), constructs simple heuristics with which to estimate their performance (\cref{subsec:Heuristics}), and validates those heuristics with numerical simulations (\cref{subsec:Numerics}).
\Cref{tab:LMtime} summarizes a key figure-of-merit, the expected duration of each variant of cat-based logical measurement.

\begin{figure}
    \centering
    \includegraphics[width=0.5\linewidth]{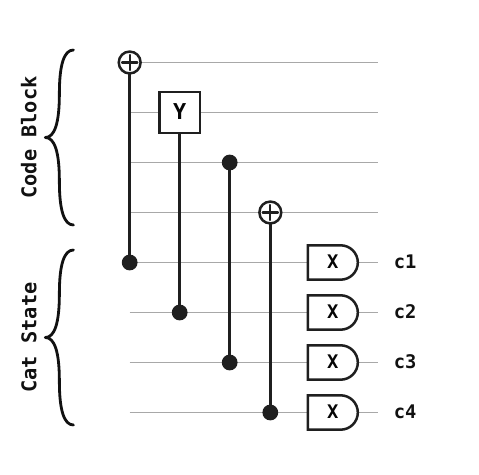}
    \caption{Using a cat state to measure $X_1 Y_2 Z_3 X_4$ on four data qubits. The outcome is $c_1\oplus c_2 \oplus c_3\oplus c_4$.}
    \label{fig:CatBasedMeasurement}
\end{figure}

Our implementation of logical measurements employs a physical cat state of size $\physicalWidth$ to measure one logical Pauli operator $P$ with representative $\Pauli$ of weight $\physicalWidth$ (\textit{i.e.}, a {\em cat-based measurement}).
\Cref{fig:CatBasedMeasurement} shows an example circuit gadget implementing a cat-based measurement of a $\physicalWidth=4$, $X$ Pauli operator.
Cat-based measurements of an arbitrary Pauli operator follow an almost identical measurement gadget, with \CX gates replaced by \CZ or \CY where appropriate.
In a single cat-based measurement, the outcome $b$ of measuring $\Pauli$ is given by the parity of the bits obtained from measuring individual qubits of the cat state: $b = c_1\oplus c_2\oplus c_3 ...$ (here $\oplus$ denotes addition modulo 2).

Our use of cat states for logical measurements follows the strategy of ``Shor-style'' syndrome measurements~\cite{shor1996fault}, that eschews use of a single bare ancillae in order to avoid introducing high-weight Pauli errors into the code block being measured.
Since a high-weight $X$-type error in the cat-state (as used in the circuit of~\cref{fig:CatBasedMeasurement}) similarly propagates into the target code block, we use $(\epsilon,p)$-independent cat states (\cref{def:QuasiFT-Cats}) verified with the procedure of~\cref{sec:Cat factories}, wherein correlated errors occur with vanishingly small probability (see~\cref{tab:cat_heuristic}).

\subsection{Fault-tolerant logical measurements}
\label{subsec:FTLM}
From any individual cat-based measurement, the value obtained is incorrect with probability $O(p)$, with $p$ the noise rate of physical \CX, \CY or \CZ gates used to implement the measurement gadget.
To ensure a logical measurement is reliable---and therefore any logical operation(s) derived from it---we repeat the cat-based measurement several times and aggregate their results $b_1,...,b_r$ to obtain a more robust estimate, $\hat b$.
We now define two fault-tolerant logical measurement procedures, comprised of repeated cat-based measurements.

\begin{definition}
    An $r$-round {\em error-detected measurement} (\EDM-$r$) employs $r$ cat-based measurements of a given Pauli operator $\Pauli$ to obtain bits $b_1$, $b_2$, ..., $b_r$.
    It returns an estimate of the outcome of $\Pauli$ as $\hat b = b_1 = b_2 = ... = b_r$, if and only if all measured bits are identical.
    If any two bits, $b_i \neq b_j$, disagree, then \EDM-$r$ aborts in failure.
    \label{def:EDM}
\end{definition}

\begin{definition}
    An $r$-round {\em error-corrected measurement} (\ECM-$r$) employs $r$ cat-based measurements of a given Pauli operator $\Pauli$ to obtain bits $b_1$, $b_2$, ..., $b_r$.
    It returns an estimate of the outcome of $\Pauli$ as $\hat b = \text{Majority}\{b_1,b_2,...,b_r\}$--- that is, it returns a majority vote of the $r$ measured bits.
    \label{def:ECM}
\end{definition}

\begin{figure}[h]
    \centering
    \includegraphics[width=0.8\linewidth]{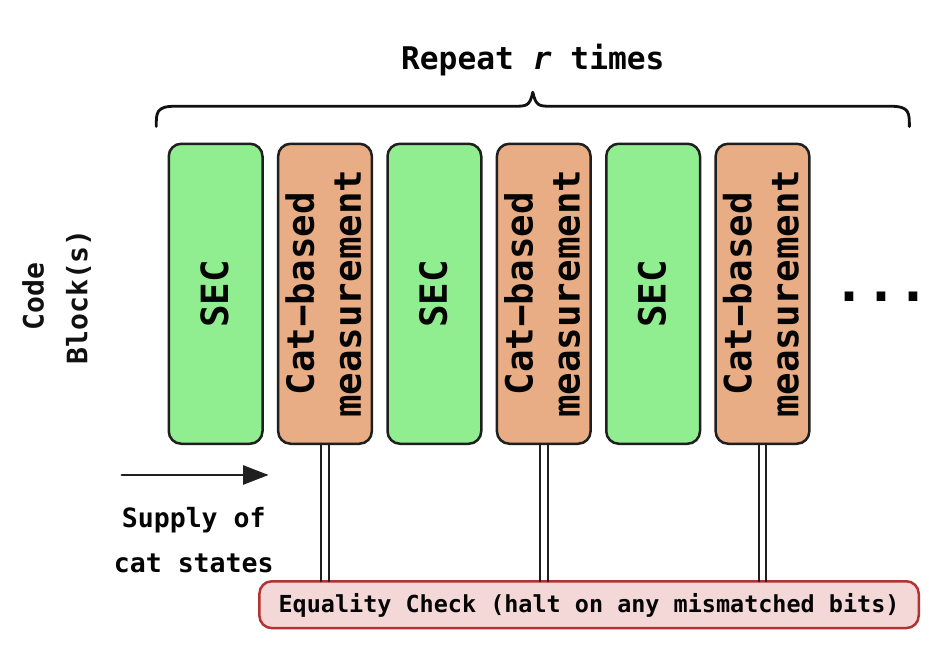}
    \caption{An error-detected logical measurement (\EDM) based on cat-states.
    A cat-based measurement is repeated $r$ times for reliability.
    An \EDM is deemed to have failed if \emph{any} extracted bits disagree.}
    \label{fig:EDM}
\end{figure}

\begin{figure}[h]
    \centering
    \includegraphics[width=0.8\linewidth]{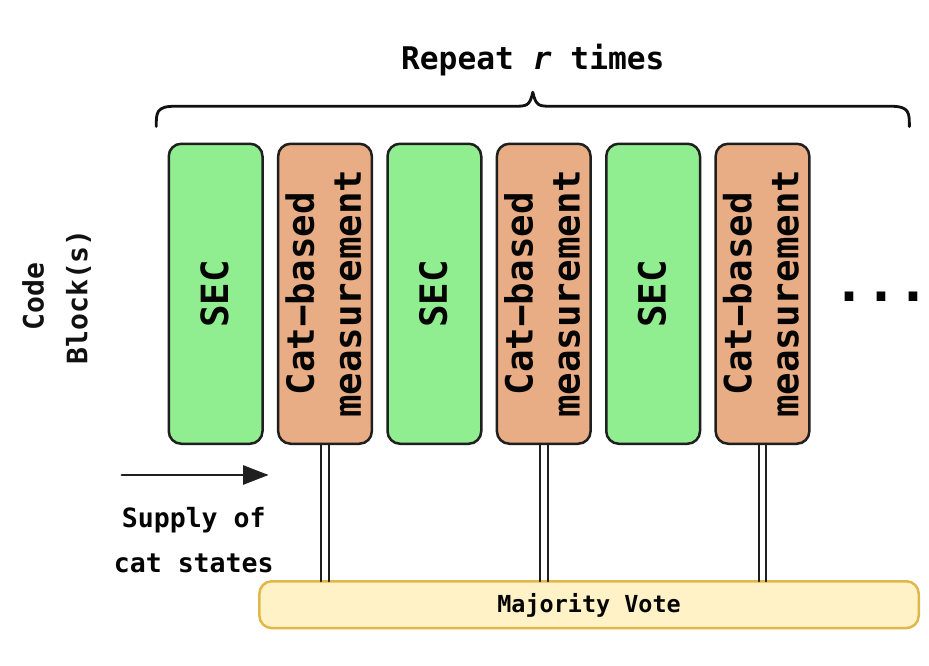}
    \caption{An error-corrected logical measurement (\ECM) based on cat states.
    A cat-based measurement is repeated $r$ times, and the value of the operator being measured is taken to be the majority vote on the $r$ measured bits.}
    \label{fig:ECM}
\end{figure}

For concreteness, we illustrate \EDM-$r$ and \ECM-$r$ of \cref{def:EDM,def:ECM} in \cref{fig:EDM,fig:ECM}.
Compared to a single cat-based measurement, \EDM (or \ECM) extracts a logical measurement outcome with lower logical error rate $\pEDM$ (or $\pECM$), \textit{i.e.}, the probability of a bit-flip error in $\hat b$.
If errors on each cat-based measurement outcome $b_1, ..., b_r$ are of $O(p)$ and independent, then $r$ repeated cat-based measurements suppresses $\pEDM$ (or $\pECM$) to $O(p^r)$ (or $O(p^{\lceil r/2\rceil})$) respectively.

Compared to \ECM, the \EDM procedure has the benefit of being faster, but at the cost of requiring an entire code block to be discarded upon failure.
Thus, \EDM is especially useful when a failure and discard occurs early in a computation; such is the case in a distillation block, for instance when injecting a physical magic state into code block (see~\cref{sec:magic_state_factory}).

To have independence of cat-based outcome flip, at least one syndrome extraction cycle (SEC) between consecutive cat-based measurements is necessary.
Otherwise a single fault that anti-commutes with $\Pauli$, introduced during/after the first cat-based measurement with probability $O(p)$, ensures all subsequent $r-1$ cat-based measurements are also incorrect (assuming no further faults).
Incidentally, we note that for the same reason, when a logical measurement is ``hybrid'' (\textit{i.e.}, involving both a code block and an unencoded physical qubit) only \EDM but not \ECM is possible.
How many SEC rounds are sufficient, on the other hand, is not {\em a priori} clear.
For surface codes or color codes, we expect to need many rounds because errors are detected in few locations only. 
In this work, we use quantum LDPC codes and we expect that their single-shot properties~\cite{bombin2015single, campbell2019theory, quintavalle2021single, gu2024single, lin2025single, jacob2025single, mian2026multivariate} could make consecutive cat-based outcome flips almost independent because the decoder can identify a fault within few SECs, before it flips many consecutive bits.
Here we conjecture that a single SEC round per cat-based measurement is sufficient, and we provide numerical evidence to that effect for specific instances of quantum LDPC codes selected for the walking cat architectures.

\begin{figure}[h]
    \centering
    \includegraphics[width=0.95\linewidth]{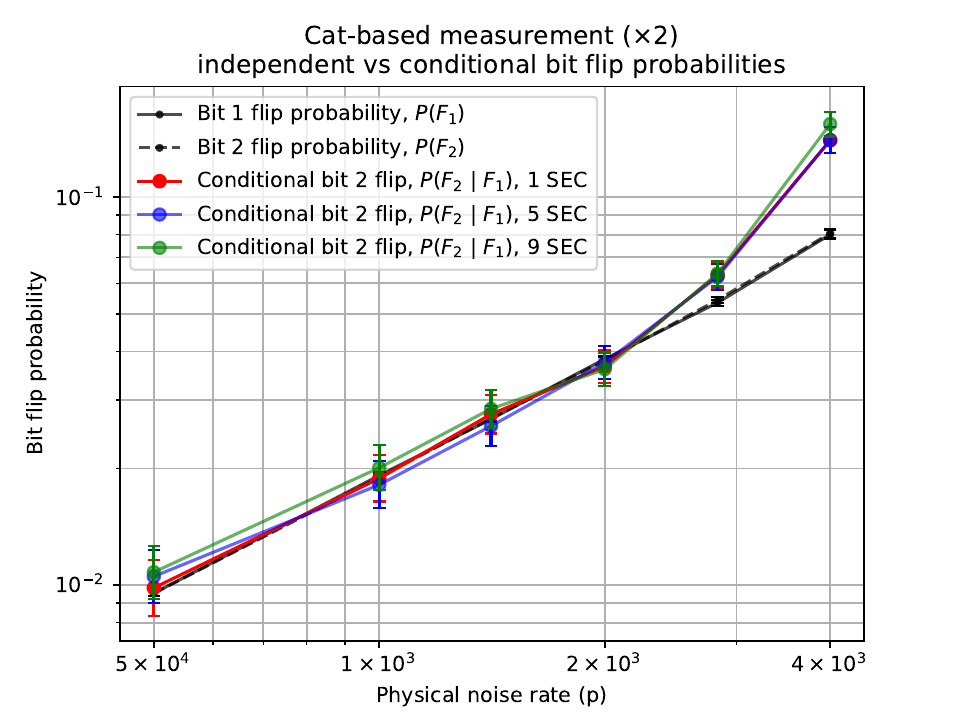}
    \caption{(\textbf{Black}) Probability of bit-flip $P(F_1)$ and $P(F_2)$ on each outcome bit, in $r=2$ rounds of cat-based measurements, on the \code{102} code.
    Also shown is the conditional probability $P(F_2~|~F_1)$, when the two cat-based measurements are separated by (\textbf{Red}) one, (\textbf{Blue}) five, and (\textbf{Green}) nine SEC rounds.
    The bit-flips are independent {\em iff} $P(F_2)=P(F_2~|~F_1)$.}
    \label{fig:Correlator}
\end{figure}

\Cref{fig:Correlator} shows numerical results for $r=2$ cat-based measurements, separated by 1,~5,~and~9 SEC rounds (see \cref{subsec:Numerics} for details).
Here, $\Pauli$ is a randomly chosen operator in a single \code{102} code block, whose value is initialized to $0$ ({\em i.e.}, it is the positive eigenstate of $\Pauli$).
Therein, $F_1$ (or $F_2$) denote events in which the cat-based measurement outcomes experience a bit-flip error; \textit{i.e.}, $b_1=1$ (or $b_2=1$) respectively.
The individual bit-flip error probabilities---$P(F_1)$ and $P(F_2)$---are shown.
Also shown is the conditional probability $P(F_2~|~F_1)$.
Observe that everywhere $p\leq 2\times 10^{-3}$, $P(F_2~|~F_1) \approxeq P(F_2)$, consistent with bit flip errors $F_1$ and $F_2$ being approximately independent.
At noise rates above $p>2\times 10^{-3}$, logical faults originating in the memory block {\em before} the first cat-based measurement occur with probability comparable to $P(F_2)$ (see~\cref{subsec:MemoryBlockPerf}), but causes {\em both} $F_1$ and $F_2$ to occur; we conjecture that this contributes to $P(F_2~|~F_1) > P(F_2)$ in that noise regime.
Moreover, $P(F_2~|~F_1)$ for 1,~5,~and~9 SEC rounds appear mutually indistinguishable everywhere.
We conclude from these results, that a single SEC round between consecutive cat-based measurements suffices.

\subsection{Heuristics}
\label{subsec:Heuristics}
To facilitate resource estimation, we now propose several heuristics, which we validate numerically in the next section.

We assume a source of cat states with independent noise with rate $p$. 
This is compatible with the heuristic estimate for the cat factory in \cref{tab:cat_heuristic}.
Based on this, we expect the cat factories to produce cat states suffering from $X$ errors with rate $p/2$ per qubit and $Z$ errors with rate $4p/5$ per qubit.
The cat state transport from the cat factory to the block where it is consumed requires in the worst case up to about 100 transport steps for the codes we consider because the largest code length we use is 102.
Based on the moving-qubit model, this transport induces depolarizing noise on the cat qubits with rate up to $5p/100$.
For simplicity, we replace the whole cat state noise by depolarizing noise with rate $p$, which is stronger in $X$ and weaker in $Z$, but which is close enough to make our performance estimate relevant.

\begin{heuristic}
The bit-flip error probability of a single cat-based measurement of weight $\physicalWidth$ is
\begin{align}
    \pflip = P(F)= C_1 \physicalWidth p \label{eq:heuristic_oneflip},
    \end{align}
    with $C_1$ a constant, and $p$ the physical noise rate.
    \label{heuristic:oneflip}
\end{heuristic}
\begin{proof}[Justification]
    A single $Z$ error on any qubit of a cat state results in a bit flip of that cat-based measurement outcome.
    Since the produced cat-state has a $Z$ error proportional to $\physicalWidth p$, and any one gate of \cref{fig:CatBasedMeasurement} can yield such a $Z$ error, the fact that there are $2 \physicalWidth$ such gates each with noise $\propto p$, naturally suggests \cref{eq:heuristic_oneflip}.
\end{proof}

We further propose heuristics for procedures that aggregate multiple cat-based measurements:
\newcommand{\combifact}{\binom{r}{\lceil r/2\rceil}}
\begin{heuristic}
When aggregating multiple cat-based measurements through EDM or ECM, the overall logical measurement error rate ({\em i.e.} error in $\hat b$) is
\begin{align}
\pEDM & = C_2 p_{\log} + \pflip^r\label{eq:heuristic_EDM},\\
\pECM & = C_2 p_{\log} + \combifact\pflip^{\lceil r/2\rceil}.\label{eq:heuristic_ECM}
\end{align}
Here, $C_2$ is a constant, $\combifact = \frac{r!}{\lceil r/2\rceil!(r-\lceil r/2\rceil)!}$, and $p_{\log}$ is the memory logical error rate per SEC---\textit{i.e.}, faults arising in the code block(s) absent any logical operations (see~\cref{subsec:MemoryBlockPerf}).
\end{heuristic}

\begin{remark}
    In \cref{eq:heuristic_EDM,eq:heuristic_ECM}, $C_2 p_{\log}$ approximates faults in the code block occurring before the first cat-based measurement (or early on) in an \EDM or \ECM procedure.
    Such faults, may result in highly correlated bit flips across successive cat-based measurements if they lead to a logical error which flips the logical outcome, and therefore set an error floor for \EDM-$r$ and \ECM-$r$.
    We will refer to such faults as {\em memory induced}, and distinguish them from {\em measurement induced} faults represented by the $p^r$ or $p^{\lceil r/2\rceil}$ terms.
    \label{remark:Heuristics}
\end{remark}

Even though we ascribe meaning to the terms of \cref{eq:heuristic_EDM,eq:heuristic_ECM} as being memory or measurement induced (see remark), note that we cannot distinguish the two types of fault in practice.
Instead, we merely draw attention to the fact that it is important to balance $r$ against memory performance of the target code block(s)---choosing an over-large $r$ well past the noise floor set by memory induced faults is wasteful.
In the next section, we establish specific values for constants $C_1$ and $C_2$ through numerical simulations.

\subsection{Numerical simulations}
\label{subsec:Numerics}

\begin{figure}[h]
    \centering
    \includegraphics[width=0.95\linewidth]{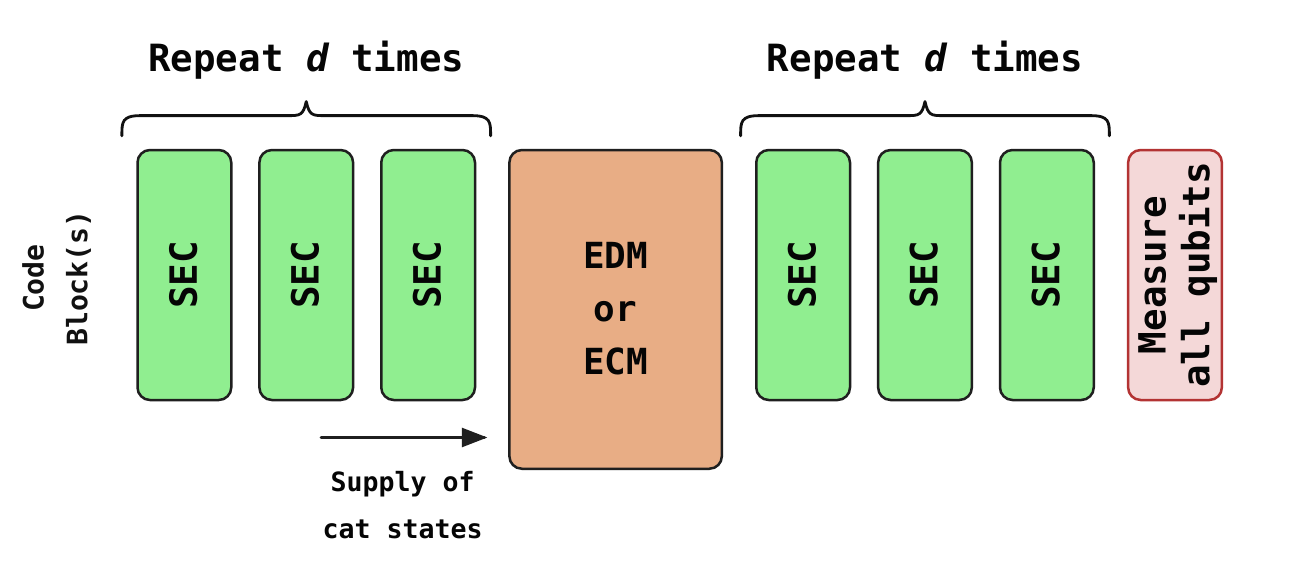}
    \caption{Block diagram of simulation circuits.}
    \label{fig:NumericsBlockDiagram}
\end{figure}

In all simulations of the \EDM or \ECM procedure, we use a circuit with the structure of \cref{fig:NumericsBlockDiagram}.
The circuit initializes all physical qubits of the code block in $\ket{+}$ (or $\ket{0}$), followed by $d$ SEC.
This prepares each $k$ logical qubits of the code in the logical state $\ket{\bar{+}}$ (or $\ket{\bar{0}}$) (we consider only CSS codes).
In the absence of faults, this sets $\hat b(\Pauli)=0$ for any $\Pauli$ containing only $I$ and $X$ (or $Z$).
Then, the block labeled ``\EDM or \ECM'' implements the procedure of \cref{fig:EDM} or \cref{fig:ECM}.
Another $d$ SECs are applied, followed by destructive measurement of all qubits and ancillae.
The noise model we use is the ionic circuit-level noise model of \cref{tab:fully connected model operations} for all circuit elements, and cat states are prepared with single-qubit depolarizing noise rate $p$ on each qubit in its support (see~\cref{subsec:CatHeuristics}).
Syndromes are decoded with Beam Search initialized with the {\texttt{beam32\_340iters}} configuration~\cite{ye2025beam}.
Finally, corrected cat-based measurement bits are aggregated via \EDM (or \ECM), and a logical measurement error is logged if the result is $\hat b = 0$.

The rest of this section discusses results from numerical simulations of: (i) the \code{102} code where only \ECM is considered; and (ii) the \code{70} code employing both \ECM and \EDM.
We consider these specific code as a representative example; generally, while we expect the validity of the heuristics we proposed in \cref{eq:heuristic_oneflip,eq:heuristic_EDM,eq:heuristic_ECM} to hold under different code choices, different values for constants $C_1,C_2$ may need re-fitting.

\subsubsection{Single Cat-Based Measurement}
\label{subsubsec:OneCat}
\begin{figure}[h]
    \centering
    \includegraphics[width=0.95\linewidth]{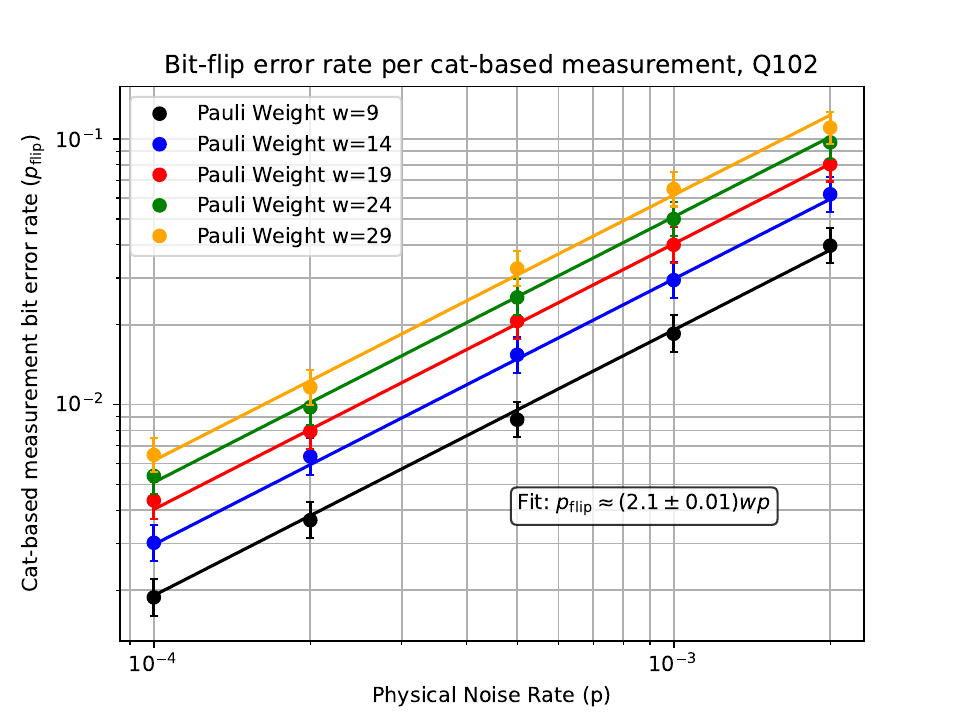}
    \caption{Bit flip probability $p_\text{flip}$ for \ECM-$1$, by physical noise rate ($p$) and physical weight ($\physicalWidth$) of the logical operator being measured.
    Also shown are fits to the ansatz $p_\text{flip} = C\physicalWidth p$, where the fit parameter is found to be $C_1 = 2.1\pm 0.01$.
    }
    \label{fig:ByWeight}
\end{figure}

\Cref{fig:ByWeight} shows bit-flip error rates for a single cat-based measurement (equivalently, \ECM-$1$), across a range of physical noise rates and Pauli weights, targeting a \code{102} memory code block.
Fitting the heuristic of \cref{eq:heuristic_oneflip} to this numerical data yields $C_1 = 2.1\pm 0.01$.
A similar numerical experiment on a \code{70} distillation code block yields identical results for $C_1$.

\subsubsection{\ECM for the \code{102} code}
\label{subsubsec:MemECM}
\begin{figure}[h]
    \centering
    \includegraphics[width=0.95\linewidth]{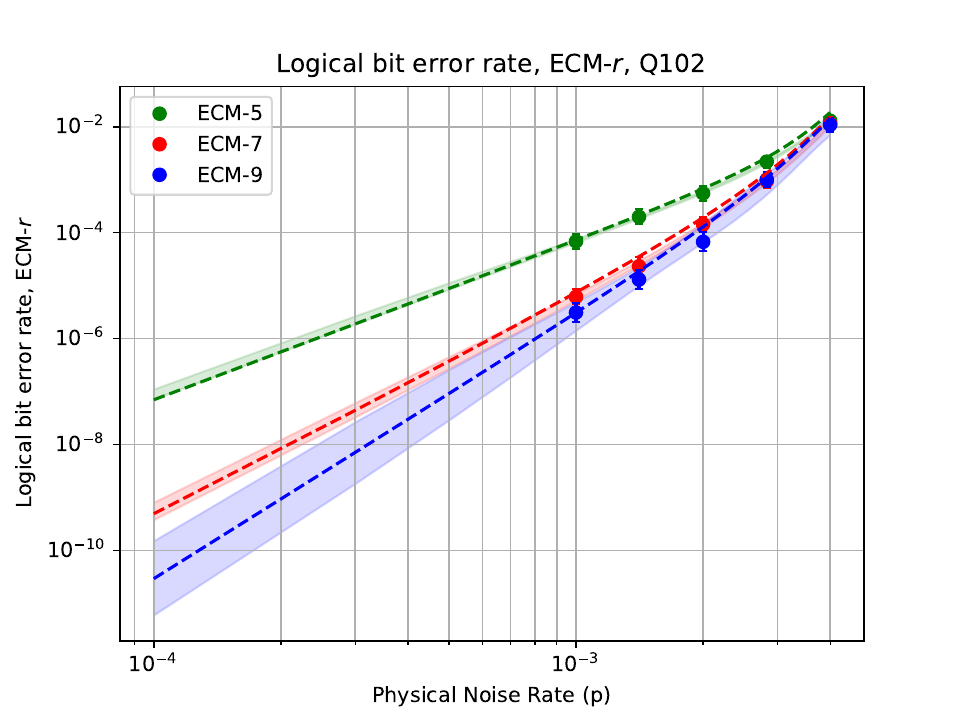}
    \caption{Logical bit error rate for \ECM-$r$, performed on the \code{102} code.
    Each \ECM procedure here extracts the value of a weight-9 Pauli operator.
    Dashed lines show heuristic \cref{eq:heuristic_ECM}, with $C_1 = 2.1$ and $C_2 = 3.4$.}
    \label{fig:ECM_Totoro}
\end{figure}

\Cref{fig:ECM_Totoro} shows the logical performance of \ECM-$5$, $7$, and $9$, applied to the memory code \code{102}.
Shaded regions depict 95\% confidence bands, derived from fitting and extrapolating simulation datapoints for each $r$ with the ansatz: $\pfitECM(r,p) = p^{\lceil r/2\rceil} \exp^{\alpha + \beta p + \gamma p^2}$, with fit parameters $\alpha$, $\beta$, $\gamma$.
Dashed lines show the heuristic of \cref{eq:heuristic_ECM}.
From these simulations, we determine $C_2 = 3.4\pm 0.4$.
Close agreement between our proposed heuristic and simulation data, support the validity of \cref{eq:heuristic_ECM}.

We do not contemplate \EDM on memory code, for efficiency's sake.
A memory code block targeted by $m$ rounds of \EDM can expect to experience at least one aborted \EDM---and therefore must be destroyed and re-initialized---with probability $O(p^m)$.
Moreover, aborted \EDM compounds upon the cost of an already long running computation, if it occurs late into its execution.

\subsubsection{\ECM and \EDM for the \code{70} code}
\label{subsubsec:DistECMEDM}
\begin{figure}[h]
    \centering
    \includegraphics[width=0.95\linewidth]{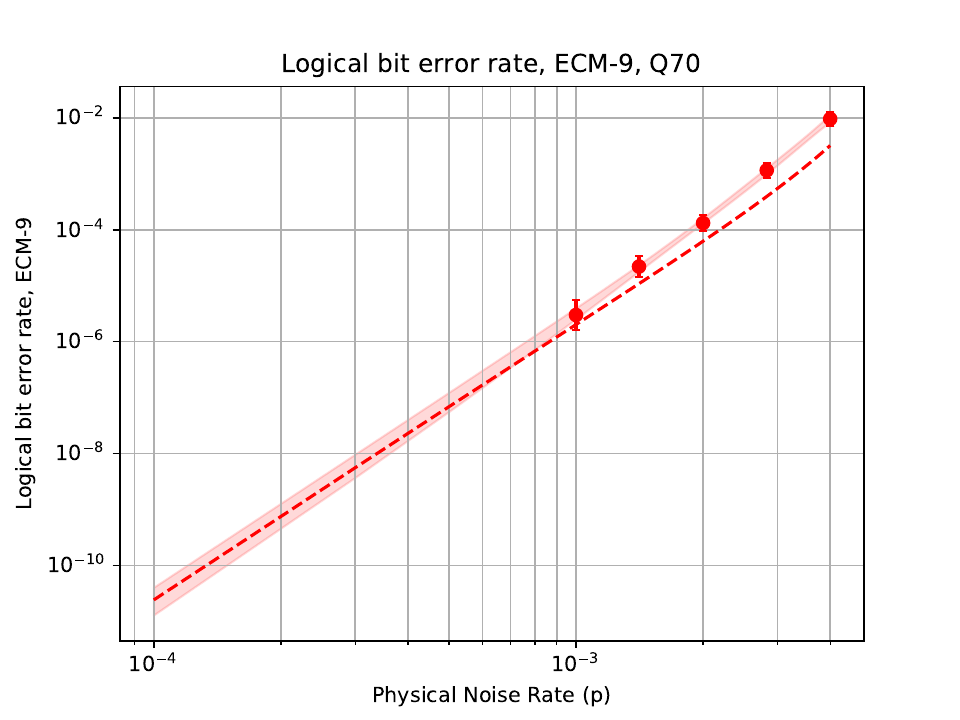}
    \caption{Logical bit error rate for \ECM-$9$, performed on the \code{70} code.
    Dashed line plots show heuristic \cref{eq:heuristic_ECM}, again with $C_1 = 2.1$ and $C_2 = 3.4$.}
    \label{fig:MEK_ECM}
\end{figure}

\Cref{fig:MEK_ECM} shows the logical performance of \ECM-$9$ applied to the \code{70} code.
The shaded region and dashed lines have the same meaning here as in \cref{fig:ECM_Totoro}.
The very same values for heuristic parameters $C_1=2.1$ and $C_2=3.4$ as in previous sections, were used here.

\begin{figure}[h]
    \centering
    \includegraphics[width=0.95\linewidth]{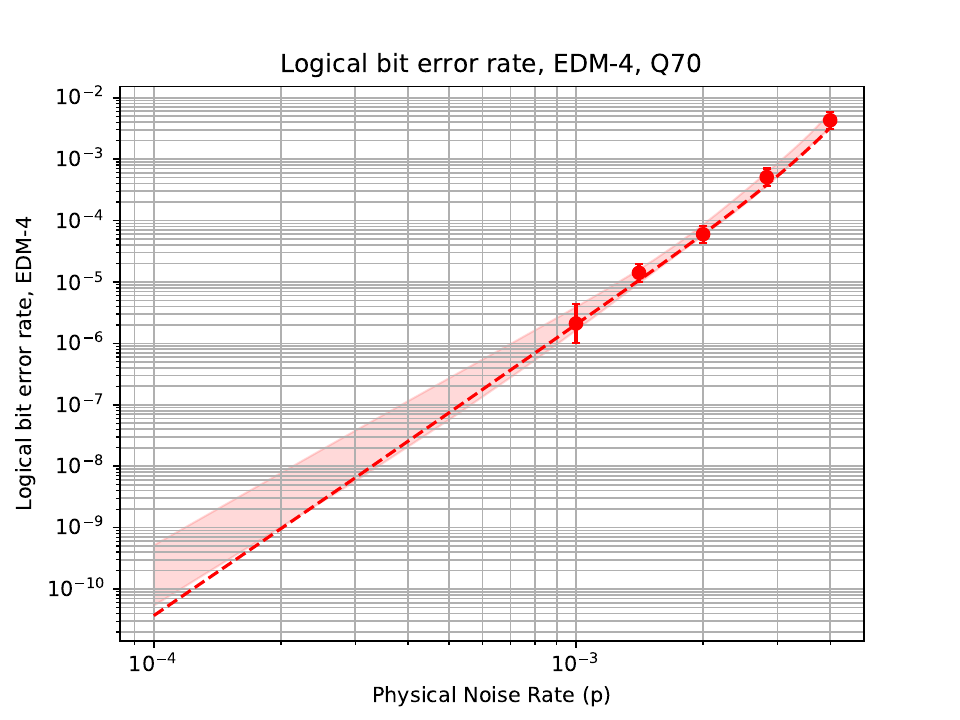}
    \caption{Logical bit error rate for \EDM-$4$, performed on the \code{70} code.}
    \label{fig:MEK_EDM}
\end{figure}

\begin{figure}[h]
    \centering
    \includegraphics[width=0.95\linewidth]{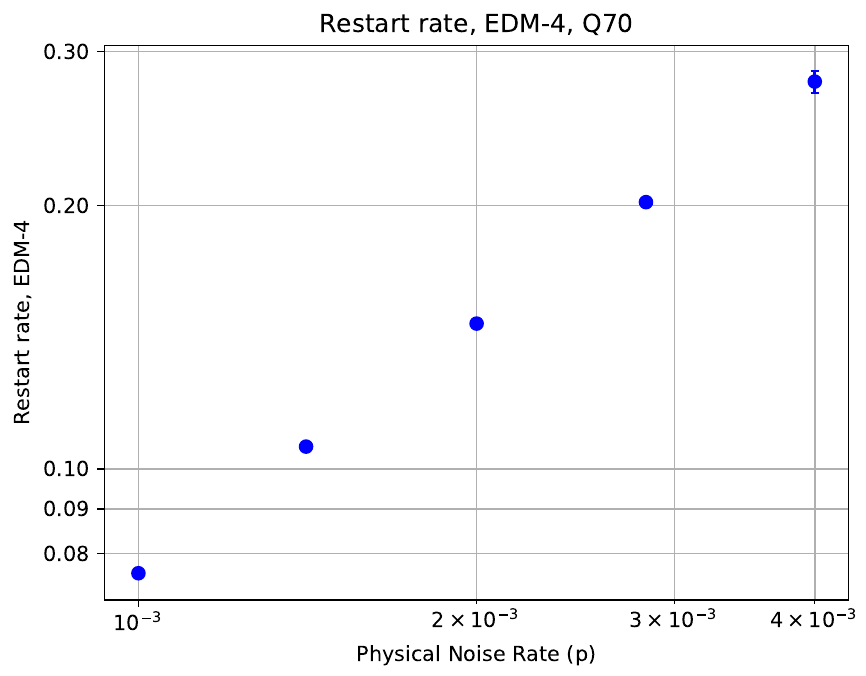}
    \caption{Restart rate for \EDM-$4$, performed on the \code{70} code.}
    \label{fig:EDM-restart}
\end{figure}

We use \EDM in a distillation code (for which \code{70} is a candidate), sparingly and early in the distillation process, to inject unverified magic states.
\Cref{fig:MEK_EDM} shows the logical performance of \EDM-$4$.
As before, the shaded region depicts 95\% confidence bands, albeit estimated by fitting simulation datapoints to the slightly different ansatz: $\pfitEDM(r,p) = p^{r} \exp^{\alpha + \beta p + \gamma p^2}$.
Dashed line shows the heuristic of \cref{eq:heuristic_EDM}, once more with the same values for $C_1, C_2$ already established above.
\Cref{fig:EDM-restart} plots the corresponding \EDM failure and restart rate.

\begin{remark}
    Ansatzae $\pfitECM$, $\pfitEDM$ used to extrapolate the confidence bands of \cref{fig:ECM_Totoro,fig:MEK_ECM,fig:MEK_EDM} neglect the error floor set by memory induced faults in the code (see~\cref{remark:Heuristics}).
    When estimating $\pECM$ (or $\pEDM$) with $d\neq r$ (or $\lceil d/2\rceil\neq r$), extrapolation to different physical noise rates $p$ can be especially inaccurate in regimes where the logical measurement error is mainly determined by memory error.
    We ascribe the overestimation of that extrapolated confidence band of \cref{fig:MEK_EDM} to such a mismatch (between \EDM $r=4$ vs $d=9$).
\end{remark}

\subsubsection{Inter-block \ECM}
\label{subsubsec:2BlockECM}
\begin{figure}[h]
    \centering
    \includegraphics[width=0.95\linewidth]{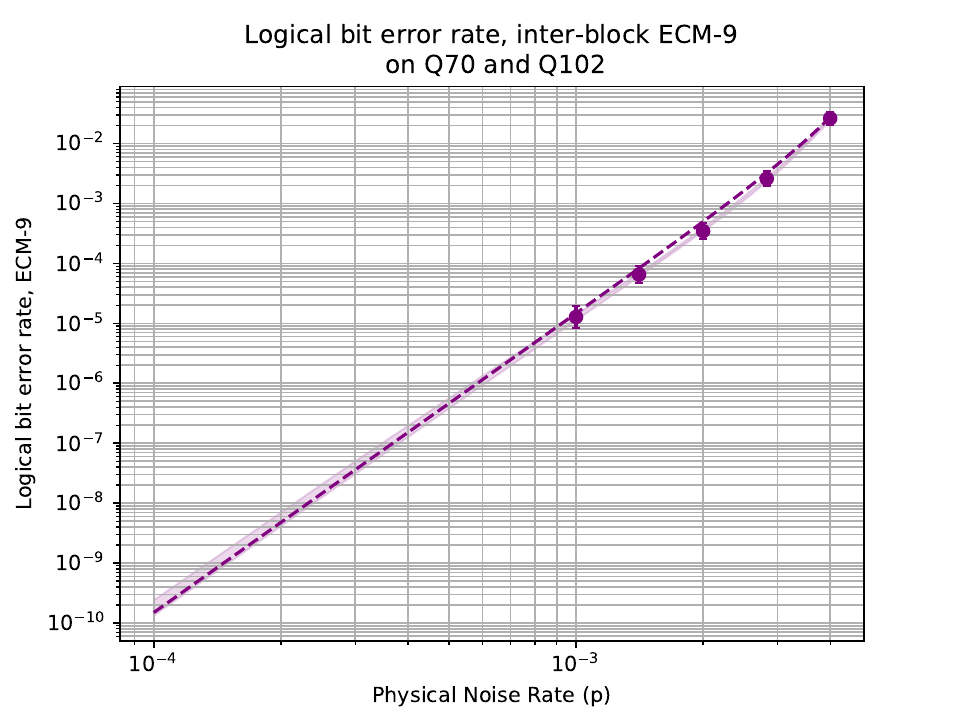}
    \caption{Logical bit error rate for \ECM-9 across code blocks, comprised of a \code{70} and a \code{102} codes.
    On each block, a weight 9 operator is measured on each block, with a cat state of weight 18.}
    \label{fig:TwoBlock}
\end{figure}

\Cref{fig:TwoBlock} shows a logical measurement across two code blocks, comprised of a \code{70} and a \code{102} code.
Such a logical measurement is used, for instance, to implement a logical instruction (LM2), to perform a logical $T$ gate in the \code{102} code by consuming a magic state stored in the \code{70} code, or to perform inter-block logical \CX gates.
In \cref{fig:TwoBlock} we measure a weight 9 logical operator in each block, using a cat state of weight 18.
The dashed line is the heuristic of \cref{eq:heuristic_ECM}, with $C_2 = 3.4$ as before, though with $p_{\log}$ therein being the sum of memory logical error rate of both code blocks.

\subsection{Viterbi measurements}
\label{subsec:AdaptiveMeas}
In this section we extend the \ECM fault-tolerant measurement scheme with an adaptive variant.
Adaptivity of cat-based measurements carries two advantages: (i) the cat factory of \cref{sec:Cat factories} has a small yet finite probability of failure, so a subset of the $r$ cat state used in \ECM-$r$ may occasionally be missing; and (ii) even without missing cat states, an adaptive measurement can be faster on average by halting earlier when sufficient information is available. For example, when the first $\lceil r/2\rceil$ measurements happen to all agree in \ECM-$r$.

\begin{definition}
    A {\em Viterbi measurement} with parameters $(\varepsilon, w, p)$ is an adaptive sequence of cat-based measurements that halts once a precision of $\varepsilon$ is reached, but continues with further cat-based measurements otherwise.
    \label{def:ViterbiMeas}
\end{definition}

In \cref{def:ViterbiMeas}, the halting condition is determined by Wald's sequential hypothesis test~\cite{viterbi2003error,wald1945SequentialTesting,wald2004SequentialAnalysis}.
We assume that all cat-based measurements undergo {\em i.i.d.} bit-flip error with probability $C_1 \physicalWidth p$ (see \cref{eq:heuristic_oneflip}).
Suppose that in $r$ existing repeated cat-based measurements, $m_b$ measured bits take value $b$ and the remaining $r-m_b$ take value $\neg b$.
Then, we halt if
\begin{align}
    \log\frac{1-\varepsilon}{\varepsilon}\leq |2m_b - r|\log\frac{1-\pflip}{\pflip}.
    \label{eq:ViterbiHalt}
\end{align}
Otherwise, we proceed with further cat-based measurements.
Here, we let $\pflip = C_1 \physicalWidth p$, as in \cref{eq:heuristic_oneflip}.
The idea behind \cref{eq:ViterbiHalt} is simply that when $\varepsilon$ is smaller and more stringent, we seek to observe events further into the tail (and therefore rarer) of an $r$-outcome binomial distribution in order to accept the logical outcome.
Crucially, if a cat state is unavailable, then we proceed with another SEC round without incrementing either $r$ or $m_b$.

\begin{figure}[h]
    \centering
    \includegraphics[width=0.95\linewidth]{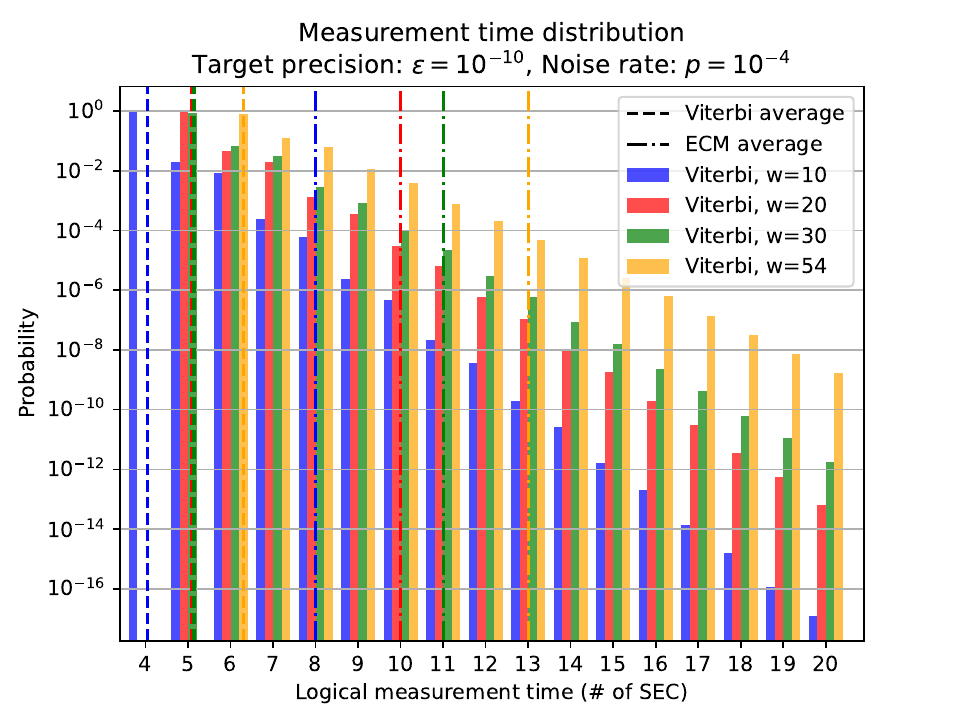}
    \caption{Duration of adaptive logical measurements versus \ECM, in terms of number of SEC rounds. Each cat-based measurement of weight $\physicalWidth$ is erroneous with probability $2.1\physicalWidth p$, where $p=10^{-4}$.
    Missing cat states occur with probability $5\physicalWidth p$.
    Dashed lines (‐ ‐) show average \# of SEC rounds needed for adaptive measurements.
    Dot-dashes (‐·‐) show the expected \# of SEC rounds for non-adaptive \ECM for comparison.}
    \label{fig:Viterbi}
\end{figure}

In \cref{fig:Viterbi} we compute logical measurement times using cat states of weights $\physicalWidth=10$, $20$, and $30$, denominated in terms of number of SECs, with physical noise rate $p=10^{-4}$.
Each individual cat-based measurement is assumed to be erroneous with probability $2.1\physicalWidth p$ (see \cref{eq:heuristic_oneflip}).
Whenever a cat state is missing, an SEC is counted, without a corresponding contribution towards precision $\varepsilon$; we let this occur with probability $5\physicalWidth p$ (see the rejection rate in \cref{tab:cat_heuristic}, with two verification rounds, {\em i.e.}, $m=2$ and with rejection rates due to leakage and loss that are negligible).
Therein, the dashed (‐ ‐) and dot-dashed (‐·‐) lines denote average number of SECs needed for Viterbi measurements vs standard \ECM respectively.
Note, that Viterbi measurements on average are about $2\times$ faster than \ECM with fixed $r$.

\section{Weight-Reduced Logical Pauli Operators}
\label{sec:Weight-Reduced Logical Pauli Operators}

\begin{table*}
    \begin{centering}
    \begin{tabular}{|c|c|c|c|c|}
    \hline
    \multirow{2}{*}{Code} & \multirow{2}{*}{\makecell{Block \\ width}} & \multirow{2}{*}{\makecell{Logical \\ width}} & \multicolumn{2}{c|}{Accessible}\tabularnewline
    \cline{4-5} \cline{5-5}
    &  &  & Pauli Operators & Clif. Gates\tabularnewline
    \hline
    \code{54} & 16 & 2 & All & All\tabularnewline
    \hline
    \code{70} & 18 & 6 & All & All\tabularnewline
    \hline
    \code{102} & 30 & $3$ & \makecell{Logical Paulis \\ with weight $\leq 3$} & \makecell{All single-qubit logical Cliffords \\ + all logical SWAPs}\tabularnewline
    \hline
    \end{tabular}
    \par\end{centering}
    \caption{Block widths, accessible operators, and accessible Clifford gates for \code{102}, \code{70}, and \code{54}.}
    \label{tab:LogicalConstraints}
\end{table*}

Based on our logical instruction set, the only logical operators that can be measured are accessible operators that admit low-weight physical representative(s). Otherwise, a cat state of size that is beyond the capacity of the cat factories available may be needed.

In \cref{subsec:Weight-Reduction}, we briefly discuss our strategy for finding low-weight physical representatives of various sets of logical Pauli operators, including weight-reduced symplectic bases.
We then proceed to report results for our candidate codes \code{102}, \code{70}, \code{54} in \cref{subsec:WeightReducedSpecificCodes}, and set reasonable logical widths for them (see~\cref{subsec:accessible_logical_operators}).

The results of this section are summarized in \cref{tab:LogicalConstraints}.

\subsection{Weight reduction}
\label{subsec:Weight-Reduction}

Our weight-reduction of the physical representative of Pauli operators is staged in two separate steps.
The first step finds {\em stabilizer-optimized representatives} of a set of Pauli operators, $\PauliOpt$.
Then, the second step uses $\PauliOpt$ to find a {\em weight-reduced symplectic basis}.
Using the notation $\PauliWeight(\bar P)$ to mean the weight of Pauli operator $\Pauli$, and the notion of symplectic basis introduced in \cref{subsec:Logical operators and their physical representatives}, we define $\PauliOpt$ and a weight-reduced symplectic basis thus:

\begin{definition}
    Given a logical Pauli operator $\Pauli$ of a code $\codeC$, its {\em stabilizer-optimized representative} is $\PauliOpt = \Pauli S$, where $S$ is a stabilizer of $\codeC$ chosen to minimize $\PauliWeight(\PauliOpt)$.
\end{definition}

\begin{definition}
    A {\em weight-reduced symplectic basis} is a symplectic basis $\logicalBasis = \{\bar X_1, \bar Z_1, \dots, \bar X_k, \bar Z_k\}$ such that $\max_{\bar P \in \logicalBasis} \PauliWeight(\bar P)$ is minimized.
    In other words, $\logicalBasis$ is a symplectic basis, wherein the highest stabilizer-optimized weight is minimized.
\end{definition}

Finding stabilizer-optimized representatives of a Pauli operator $\Pauli$ reduces to finding a vector on the lattice spanned by elements of the stabilizer group $\Stab$ of $\codeC$, $V=\sum_{S_i\in\Stab} v_i \rho(S_i)$, such that $||V - \rho(\Pauli)||_1$ is minimized.
Here, $v_i\in \mathbb{Z}_2$ are Boolean coefficients, and $\rho(S_i)$ denotes the binary vector representation of stabilizer / Pauli operators.
We use established off-the-shelf solvers for such closest vector problems, including the QDistEvol distance-finding algorithm~\cite{webster2026distancefindingalgorithmsquantumcodes}.

To find a weight-reduced symplectic basis, we turn to Tabu search~\cite{glover1986future, glover1989tabu}.
Given a valid (but possibly high-weight) symplectic basis, the metaheuristic search traverses successive sets of ``neighboring'' bases until a lower-weight one is found.
We define a {\em neighboring basis} as follows:

\begin{definition}
    \label{def:TabuNeighbor}
    Given a symplectic basis $\logicalBasis=\{(\bar X_i, \bar Z_i)~|~1\leq i\leq k\}$ for code $\codeC$, a {\em $(u,v)$-neighboring basis} $\logicalBasis^{'}=\{(\bar X_j^{'}, \bar Z_j^{'})~|~1\leq j\leq k\}$ is a symplectic basis that shares the same entries as $\logicalBasis$ except for two the following two substitutions: $\bar X_u\to \bar X_u \bar X_v$ and $\bar Z_v\to \bar Z_u \bar Z_v$.
\end{definition}

\begin{remark}
    When evaluating each $(u,v)$-neighboring basis, we evaluate the weight of each basis element's stabilizer-optimized representative.
    Otherwise, the weight of a naïve representative of products $\bar X_u \bar X_v$ and $\bar Z_u \bar Z_v$ will almost always be larger than $\bar X_u$, $\bar X_v$, $\bar Z_u$, or $\bar Z_v$ themselves.
\end{remark}

It is easy to verify that $\logicalBasis^{'}$ in \cref{def:TabuNeighbor} forms a valid symplectic basis for $\codeC$.
However in the event when $\codeC$ is self-orthogonal and admits a transversal $H$ gate, a condition for that transversality is self-similarity of its codewords: $\rho(\bar Z_i) = \rho(\bar X_i)$ for all $(\bar X_i, \bar Z_i)\in \logicalBasis$.
The transformation of \cref{def:TabuNeighbor} evidently does not preserve that self-similarity.
If transversality of $H$ must be preserved, we slightly modify \cref{def:TabuNeighbor}, and define a $(u,v,s,t)$-neighboring basis with the following substitution rules:
\begin{align}
    \Pauli_{j}^{'}=\begin{cases}
    \Pauli_{j} & j\notin\{u,v,s,t\}\\
    \prod_{i\in\{u,v,s,t\}\backslash j}\Pauli_{i} & j\in\{u,v,s,t\}
    \end{cases}
\end{align}
where $\Pauli_i$ (or $\Pauli_i^{'}$) are Pauli operators in $\logicalBasis$ (or $\logicalBasis^{'}$) respectively.
The $(u,v,s,t)$-neighboring bases preserve self-similarity, but at the cost of larger neighboring sets and therefore slower Tabu searches.

\subsection{Weight-reduced operators of specific code candidates}
\label{subsec:WeightReducedSpecificCodes}
We employ our two-step weight-reduction to codes \code{102}, \code{70}, and \code{54}.
For practical reasons, the sets of Pauli operators being considered in each case varies slightly.
In the case of \code{70} and \code{54}, $k=6,2$ respectively, which makes finding stabilizer-optimized representatives efficient in practice for all $4^k - 1$ logical Pauli operators.
By contrast with \code{102}, enumerating and optimizing all $4^{22}-1$ logical Pauli operators is challenging and remains an on-going endeavor.
Instead, we considered only $2^k-1 = 4,194,303$ logical Pauli operators in each CSS basis, alongside a small number of additional operators of low logical weights (see \cref{subsec:Accessible}).
We find weight-reduced symplectic bases for all three codes.

\begin{figure}[h]
    \centering
    \includegraphics[width=0.95\linewidth]{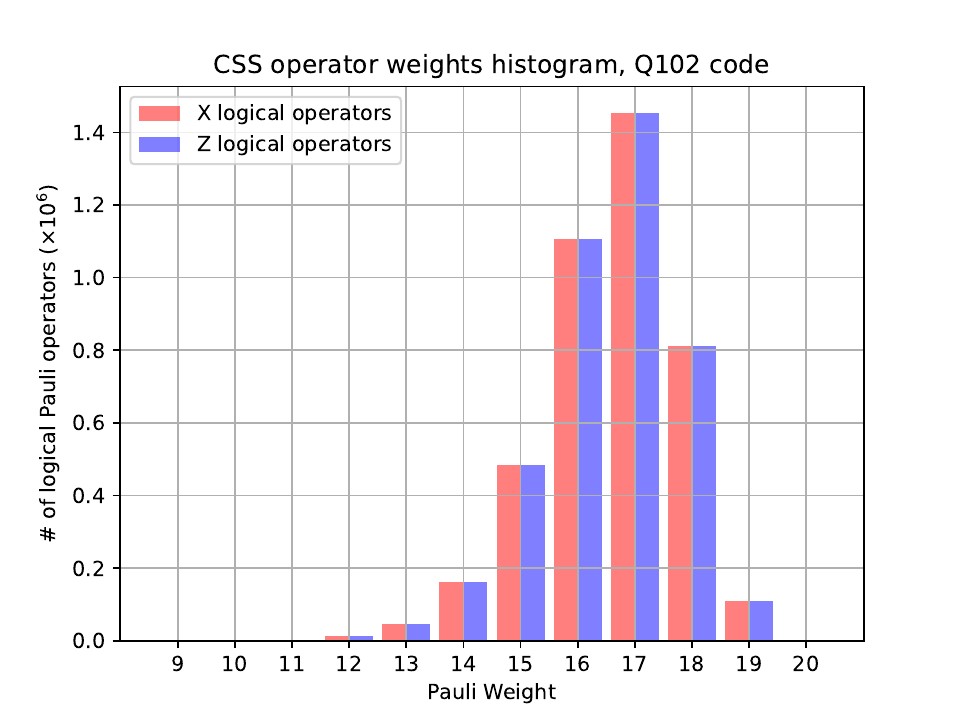}
    \caption{Distribution of weights of all $2^k-1=4,194,303$ operators in each CSS basis, for the \code{102} code.
    Not easily visible due to scale, are $612$ minimum weight $9$ operators; the maximum weight is $20$.}
    \label{fig:All_Totoro}
\end{figure}

\begin{figure}[h]
    \centering
    \includegraphics[width=0.95\linewidth]{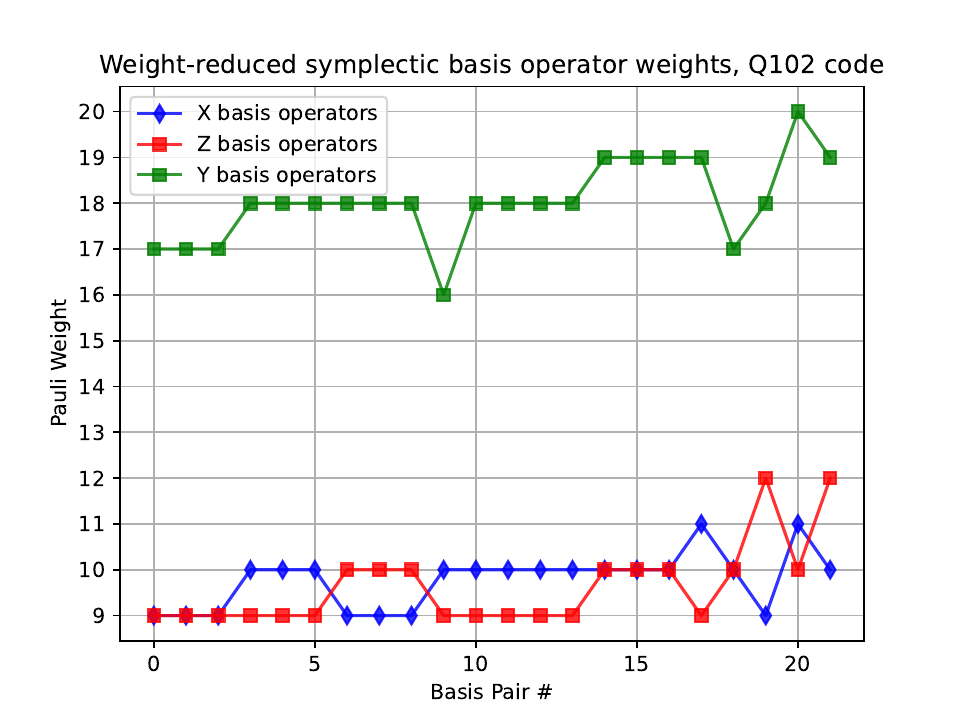}
    \caption{Weights of all $k=22$ weight-reduced symplectic basis operators for the \code{102} code.}
    \label{fig:Symp_Totoro}
\end{figure}

\Cref{fig:All_Totoro} shows weight distributions of $2^k-1$ stabilizer-optimized operators in each CSS basis for the \code{102} code.
We observe a maximum weight of $20$.
\Cref{fig:Symp_Totoro} shows weights of the resulting weight-reduced symplectic basis.
The maximum weight $\bar X_i$ or $\bar Z_i$ operator in that weight-reduced symplectic basis is $12$.
We also calculated corresponding $\bar Y_i$ operators, of which the maximum weight was $20$.
\Cref{tab:Symp_X_Totoro,tab:Symp_Z_Totoro} list the weight-reduced symplectic basis for \code{102} explicitly.

\begin{figure}[h]
    \centering
    \includegraphics[width=0.95\linewidth]{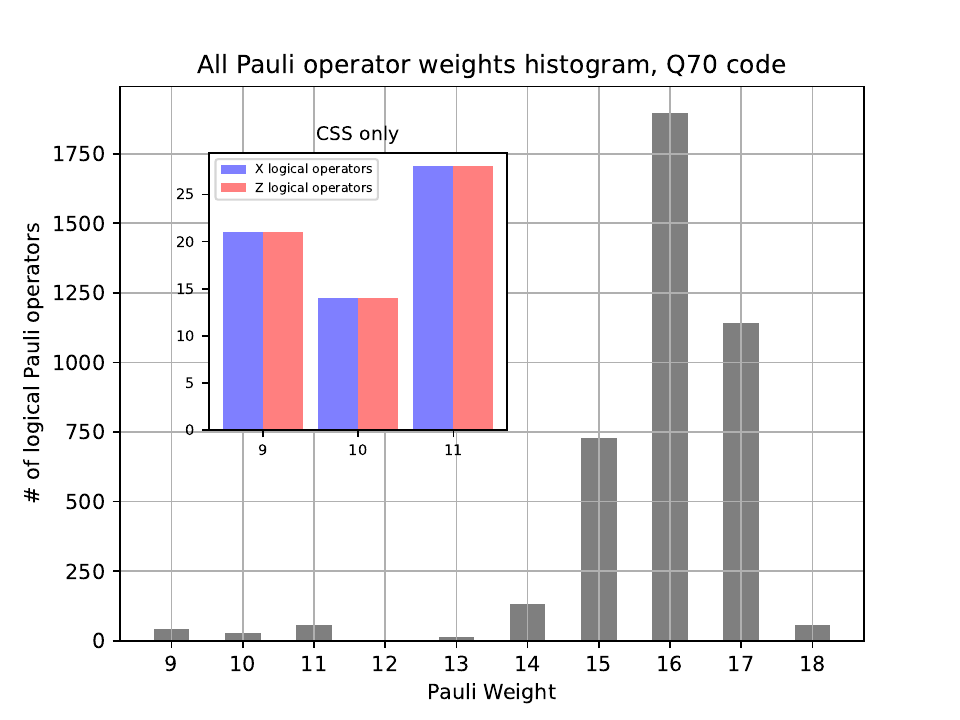}
    \caption{Distribution of weights of all $4^6-1=4,095$ Pauli operators for the \code{70} code.
    The inset shows the subset of $2^6-1=63$ operators in each CSS basis.}
    \label{fig:All_MEK}
\end{figure}

\begin{figure}[h]
    \centering
    \includegraphics[width=0.95\linewidth]{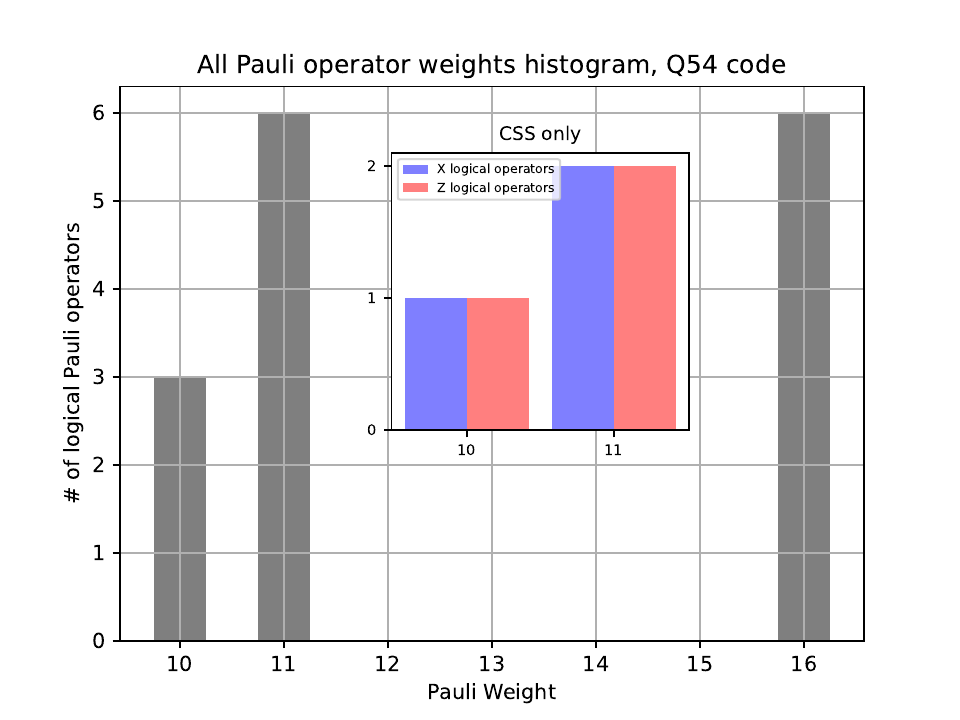}
    \caption{Distribution of weights  of all $4^{2}-1=15$ Pauli operators for the \code{54} code.
    The inset shows the subset of $2^2-1=3$ operators in each CSS basis.}
    \label{fig:All_TransH}
\end{figure}

\Cref{fig:All_MEK,fig:All_TransH} show weight distributions of all $4^k-1$ stabilizer-optimized operators for the \code{70} and \code{54} codes respectively.
We find the maximum weight operator to be $18$ for \code{70}, and $16$ for \code{54}.
The inset of each figure shows subsets of $2^k-1$ stabilizer-optimized operators in each CSS basis; in general these are of much lower weights, topping out at weight $11$.
For brevity, we do not show charts for the weight-reduced logical bases.
Instead, we simply state that \code{70} admits a minimum weight symplectic basis in which {\em all} operators have weight $9$, while for \code{54} all operators in its weight-reduced symplectic basis have weight $11$.
\Cref{tab:Symp_MEK,tab:Symp_TransH} list the weight-reduced logical bases for \code{70} and \code{54} explicitly.

\begin{figure}[h]
    \centering
    \includegraphics[width=0.95\linewidth]{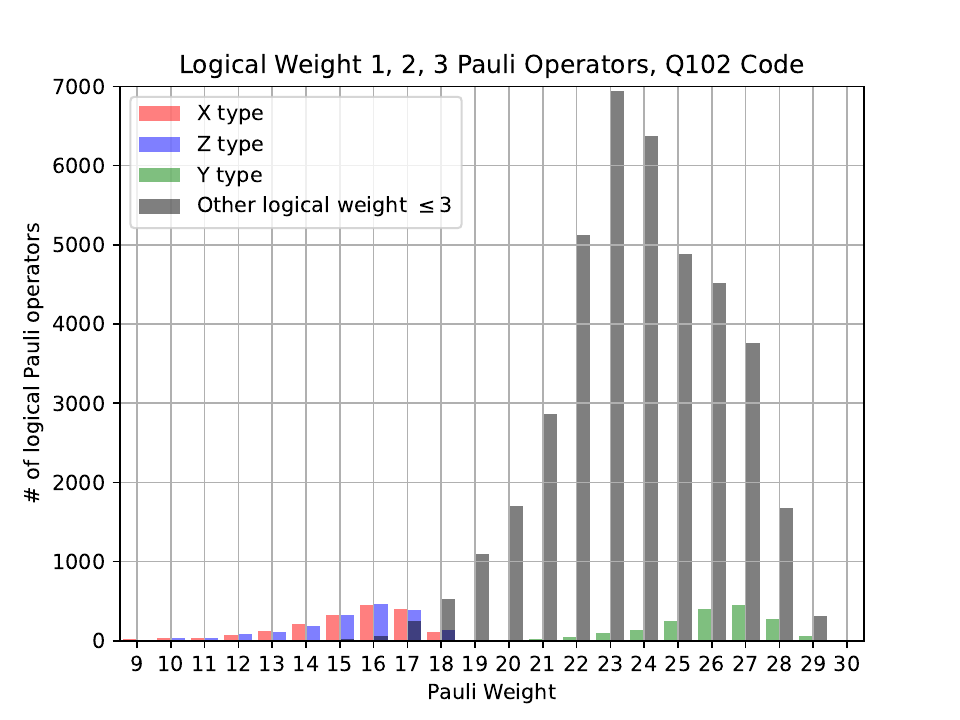}
    \caption{Distribution of stabilizer-optimized weights of all $43,725$ Pauli operators of form $P_1P_2P_3$ (logical weight 1, 2 or 3) for the \code{102} code.
    Here $P_1,P_2,P_3\in\{I,X,Y,Z\}$.
    Bars labeled $X$, $Y$, $Z$ type refer to operators wherein $P_1$, $P_2$, $P_3$ are only Pauli of the given type, or the identity.}
    \label{fig:LW3_Totoro}
\end{figure}

\subsection{Accessible logical gates}
\label{subsec:Accessible}

In this section, we describe accessible operators and accessible Clifford gates in the sense of \cref{subsec:accessible_logical_operators}.
Based on results from our weight-reduction of logical Pauli operators, we set specific block widths for each code in~\cref{tab:LogicalConstraints}.

From \cref{fig:All_MEK,fig:All_TransH}, we set block widths for \code{70} and \code{54} to 18 and 16 respectively, so that {\em all} Pauli operators in the block are accessible.
Consequently, all logical Clifford gates are also accessible.

For \code{102}, any logical Pauli of form $P_1 P_2 P_3$ with logical weight 1, 2 or 3 has a representative with weight at most 30 (\cref{fig:LW3_Totoro}).
For legibility, the same data is presented in table form in \cref{tab:LogicalWeight3Totoro}.
We set the block width to $\bar w = 30$, so that any Pauli operator up to logical weight $w=3$ is accessible.

Combining the accessible logical SWAP and single-qubit Clifford operations which can be implemented via frame tracking (see \cref{subsec:Accessible logical Clifford gates}) with logical \CX gates using logical measurements (as in \cref{fig:architecture_overview_Clifford_gates_by_meas}) allows us to implement all logical Clifford gates for $\code{102}$.

\section{The magic factory}
\label{sec:magic_state_factory}

As introduced in \cref{subsec:magic-factory-component}, we consider two protocols for producing logical $\ket{\bar{H}}$ states for the logical instruction set-supported implementation of $T$ gates shown in \cref{fig:architecture_overview_T_gates_by_meas}: the \CHfactory protocol and the \MEK protocol. Their physical realizations in code blocks are the \CHfactory factory embedded in the \code{54} code and the \MEK factory embedded in the \code{70} code. Throughout this section, for $F\in\{\mathrm{CH2},\mathrm{MEK}\}$, ``$F$ protocol'' refers to the abstract algorithm, while ``$F$ factory'' or simply ``$F$'' refers to its realization in a code block. The \MEK factory benefits from the fact that every logical Pauli operator on the six encoded qubits admits a representative of weight at most $18$; see \cref{fig:All_MEK}. This makes the \MEK factory efficient because all logical Clifford gates are accessible by frame tracking.

In \cref{subsec:magic_factory_error_model}, we state the factory-level error model used to determine the logical error rates of the output $\ket{\bar{H}}$ states.
In \cref{subsec:magic_factory_analysis}, we summarize the performance metrics used in subsequent analysis.
In \cref{subsec:t_factory_aux_ops}, we introduce the auxiliary operations used by these factories: controlled-$\bar{H}$ constructions, physical $H$-state injection, logical $\bar{H}^{\otimes k}$ measurements, and the transversal $H$ measurement primitive used by \CHfactory.
In \cref{subsec:ch2_factory}, we describe the \CHfactory protocol and analyze its realization as the \CHfactory factory (\cref{fig:CH2_circuit}), deriving its code-block schedule, output error, and average runtime.
In \cref{subsec:mek_factory}, we describe the \MEK protocol and analyze its realization as the \MEK factory (\cref{fig:logical_mek_circuit}) in the same way.

A summary of the essential properties of these magic factories is provided in \cref{tab:Magic factory parameters}.

\begin{table*}[t]
\centering
\begin{tabular*}{\textwidth}{@{\extracolsep{\fill}}|c|c|c|c|c|c|c|}
\hline
\textbf{Factory} & \makecell{\textbf{Physical} \\ \textbf{qubits}} & \makecell{\textbf{Width}} & \makecell{\textbf{Magic states /} \\ \textbf{success}} & \makecell{\textbf{Avg. SEC /} \\ \textbf{success}} & \makecell{\textbf{Per-attempt} \\ \textbf{failure rate}} & \makecell{\textbf{Output-state} \\ \textbf{error estimate}} \\
\hline
MEK in \code{70} & 221 & 18 & 2 & $47.6$ & $\approx 11.97\%$ & $3.6\times 10^{-7}$ \\ \hline
\CHfactory in \code{54} & 173 & 54 & 2 & $13.44$ & $\approx 2.28\%$ & $7.2\times 10^{-8}$ \\
\hline
\end{tabular*}
\caption{Magic factory parameters derived in this section; see \cref{subsec:ch2_factory} for information on the \CHfactory factory, and \cref{subsec:mek_factory} for information on the MEK factory. Physical qubit counts exclude the companion cat factory and use the component-level formula $3n+11$ from \cref{subsec:magic-factory-component}, where the extra qubit accounts for the physical ancilla used to inject a noisy $\ket{H}$ state. The width gives the maximum cat-state size required by the factory: for MEK this is the support-$18$ hybrid $Y\bar{Y}$ injection measurement in \code{70} (weight-$18$ logical $Y$ representative, with one cat state qubit reused to measure $Y$ on the physical ancilla; see \cref{fig:All_MEK,tab:LogicalConstraints,fig:h_state_injection}), while for \CHfactory it is the weight-$54$ transversal $\bar{H}^{\otimes 2}$ verification of \cref{fig:transversal_h_measurement,fig:CH2_circuit}. The average SEC count is the average number of SECs required to successfully produce a pair of $\ket{\bar{H}}$ states, the failure rate is the percentage of failed attempts due to detected errors, and the output-state error estimate is the rate of undetected errors.}
\label{tab:Magic factory parameters}
\end{table*}

\subsection{Magic factory error model}
\label{subsec:magic_factory_error_model}

Unless stated otherwise, all numerical estimates in this section are evaluated
at
\[
    p = 10^{-4}, \qquad \ploss = 10^{-7}, \qquad \pleak = 10^{-5},
\]
where $p$ denotes the physical two-qubit-gate error rate, $\ploss$ the
physical qubit-loss rate, and $\pleak$ the physical leakage rate. The
corresponding memory-block logical error rates and SEC depths are taken from
\cref{tab:loss_leakage_summary}.

For the factory-level output-error estimates, we use an
\emph{ideal-logical-operation approximation}: Clifford frame updates and
logical measurements, including Viterbi measurements, are treated as perfect,
so the only noisy inputs to the factory-level model are the injected logical
$\ket{\bar{H}}$ states. This is well motivated in the operating regime considered
here: the Viterbi logical measurements are run to target error
$\varepsilon = 10^{-10}$, and the memory blocks themselves have logical error
rates of $10^{-10}$ and $3\times 10^{-10}$ per SEC for \code{70} and \code{54}
(\cref{tab:loss_leakage_summary}), all orders of magnitude below the
injected-state and factory-output error rates estimated in this section.

We also assume that loss or leakage affecting a physical ancilla is detected
when that ancilla is measured, via the loss- and leakage-detection
measurements introduced in \cref{subsec:Loss and leakage measurement}.
Accordingly, loss and leakage on physical ancillas outside code blocks are
treated as detected faults that trigger a retry, rather than as undetected
Pauli errors.

Lastly, unless otherwise stated, for simplicity we consider only probabilities up to lowest-order in $p$.

\subsection{Magic factory analysis overview}
\label{subsec:magic_factory_analysis}

In the analyses below, for either factory protocol $F\in\{\mathrm{CH2},\mathrm{MEK}\}$, the primary output quantities are summarized in \cref{tab:magic_factory_analysis_outputs}, while the supporting notation is summarized in \cref{tab:magic_factory_analysis_notation}.

\begin{table}[t]
\centering
\small
\begin{adjustbox}{max width=\columnwidth}
\begin{tabular}{|l|l|}
\hline
\textbf{Quantity} & \textbf{Description} \\
\hline
$p_{F,\mathrm{fail}}$ & Per-attempt failure rate. \\
\hline
$p_{F,\mathrm{out}}$ & Output-state error estimate. \\
\hline
$N_{F,\mathrm{SEC}}^{\mathrm{avg}}$ & \makecell[l]{Average number of SECs required to successfully produce\\ one pair of $\ket{\bar{H}}$ states.} \\
\hline
\end{tabular}
\end{adjustbox}
\caption{Primary output quantities reported in the magic-factory performance analysis for either protocol $F\in\{\mathrm{CH2},\mathrm{MEK}\}$.}
\label{tab:magic_factory_analysis_outputs}
\end{table}

\begin{table}[t]
\centering
\small
\begin{adjustbox}{max width=\columnwidth}
\begin{tabular}{|l|l|}
\hline
\textbf{Quantity} & \textbf{Description} \\
\hline
$r_{F,\mathrm{inj}}$ & Number of rounds used for an \EDM \\
\hline
$p_{H,\mathrm{retry}}$ & Retry probability of a single $H$-state injection gadget. \\
\hline
$a_{H,\mathrm{acc}}$ & Acceptance probability of a single $H$-state injection gadget. \\
\hline
$p_{\mathrm{anc,det}}$ & Detected ancilla loss/leakage probability. \\
\hline
$q_{Y,F}$ & Factory-specific logical error rate of H-state injection. \\
\hline
$a_{F,\mathrm{ver}}$ & Acceptance probability of the verification stage. \\
\hline
$a_{F,\mathrm{inj}}$ & Probability that all injections in one attempt are accepted. \\
\hline
$\tau_{\mathrm{Vit},w}^{\mathrm{avg}}$ & Average SEC cost of a weight-$w$ Viterbi measurement. \\
\hline
$C_F$ & SEC cost of a complete attempt up to the verification stage. \\
\hline
$D_j$ & \makecell[l]{SEC cost accrued before restart if the $j$th injection\\ is the first one to fail.} \\
\hline
\end{tabular}
\end{adjustbox}
\caption{Supporting notation used in the magic-factory performance analysis for either protocol $F\in\{\mathrm{CH2},\mathrm{MEK}\}$.}
\label{tab:magic_factory_analysis_notation}
\end{table}

\begin{figure}[t]
    \centering
    \includegraphics[width=\linewidth]{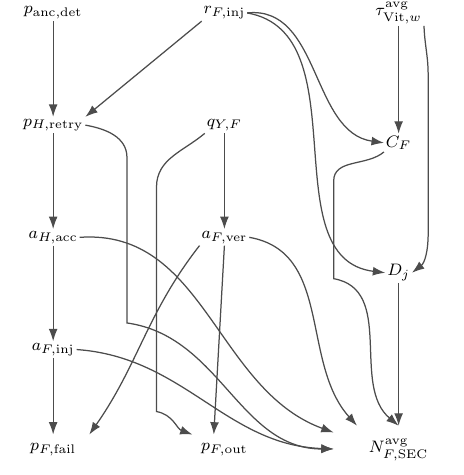}
    \caption{Dependency graph for the quantities used in the magic-factory performance analysis. The arrows indicate how the injection, verification, and runtime inputs feed into the derived quantities $p_{F,\mathrm{fail}}$, $p_{F,\mathrm{out}}$, and $N_{F,\mathrm{SEC}}^{\mathrm{avg}}$.}
    \label{fig:magic_factory_quantity_dependency_graph}
\end{figure}

\noindent The dependency structure among these quantities is summarized in \cref{fig:magic_factory_quantity_dependency_graph}.

\subsection{Magic factory auxiliary operations}
\label{subsec:t_factory_aux_ops}

The magic state factories require a number of auxiliary non-Clifford operations performed at the logical level. In this section we show how to implement these operations using the logical instruction set supplemented by physical non-Clifford operations.
Here the only non-Clifford ingredient is the preparation of a physical $\ket{H}$ state.
We refer to such constructions as \emph{IS+H-based}.

\subsubsection{\texorpdfstring{Logical $R_y(\pm \pi/4)$ gates}{Logical Ry(+/-pi/4) gates}}

This subsection describes IS+H-based implementations of the logical gates
$R_y(\pi/4)$ and $R_y(-\pi/4)$ used below in the magic-state factory
constructions.

\begin{figure}
    \centering
    \subfloat[]{\includegraphics[width=0.95\linewidth]{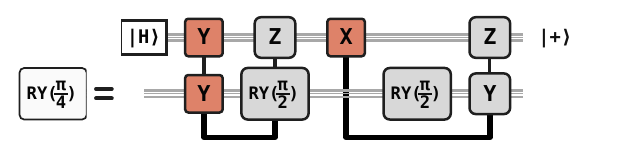}\label{fig:logical_ry_measurement_based}}

    \vspace{0.75em}
    \subfloat[]{\includegraphics[width=0.95\linewidth]{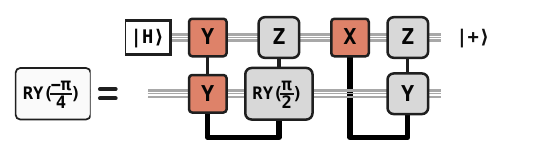}\label{fig:logical_ry_neg_measurement_based}}
    \caption{IS+H-based implementations of logical $R_y(\pm \pi/4)$. (a) A physical ancilla prepared in the $\ket{H}$ state, a joint $Y\bar{Y}$ measurement, and an $X$-basis ancilla readout realize logical $R_y(\pi/4)$ up to a Pauli frame update. (b) The corresponding IS+H-based circuit for logical $R_y(-\pi/4)$ derived from the $R_y(\pi/4)$ implementation.}
    \label{fig:logical_ry_measurement_based_pair}
\end{figure}

\Cref{fig:logical_ry_measurement_based_pair} shows the IS+H-based
implementations of logical $R_y(\pi/4)$ and $R_y(-\pi/4)$ used below:
\cref{fig:logical_ry_measurement_based} implements $R_y(\pi/4)$ and
\cref{fig:logical_ry_neg_measurement_based} implements $R_y(-\pi/4)$.
These are obtained from the
IS+H-based $T$-gate circuit of
\cref{fig:architecture_overview_T_gates_by_meas}(b) by conjugation with $SH$,
using the identity $R_y(\pi/4)=SHTH^\dag S^\dag$.

\subsubsection{Physical H-State injection}

Both of our magic state factories begin by preparing two $\ket{\bar{H}}$ states in a code block.
Preparing a single logical $\ket{\bar{H}}$ state can be done by implementing a logical $R_y(\pi/4)$ gate using a
physical ancilla prepared in the $\ket{H}$ state to provide the non-Clifford resource.
The circuit \cref{fig:h_state_injection} demonstrates the implementation
of the IS+H-based circuit for this in a code block.
Since $\ket{H}=R_y(\pi/4)\ket{0}$, we first prepare the physical
ancilla by applying a physical $R_y(\pi/4)$ rotation. We then measure the
joint $Y\bar{Y}$ operator between this physical ancilla and the target logical
qubit. Because this hybrid measurement couples an unencoded qubit to a
code block, it is implemented as an \EDM (see \cref{subsec:FTLM}); although
the code block can undergo SEC between repeated rounds, the unencoded ancilla
cannot, so a single ancilla fault can correlate the repeated $Y\bar{Y}$ outcomes
and invalidate an \ECM majority vote or Viterbi-measurement protocol run. Finally, the ancilla is measured
in the $X$
basis, and the resulting Pauli and Clifford byproduct corrections are absorbed
into the software frame via Clifford frame tracking rather than applied
physically; see \cref{subsec:Clifford frame-tracking}.

After preparing the logical $\ket{\bar{H}}$ state via this injection circuit, we
$\bar{H}$-twirl it by implementing the logical Hadamard with probability $1/2$ via frame tracking.
The $\bar{H}$-twirl leaves the injected state in a stochastic mixture of $\bar{H}$ and $\bar{Y}\bar{H}$ (Ref.~\cite{meier2013magic} for proof), so that the relevant first-order logical fault afterwards is an undetected logical $\bar{Y}$.

Even without $\bar{H}$-twirling, errors that do not correspond to an undetected logical $\bar{Y}$ fault are disregarded under the ideal-logical-operation approximation of \cref{subsec:magic_factory_error_model}, or contribute only through the hybrid-\EDM logical-outcome error $\pEDM$, which is not first order in $p$. This is relevant later when studying faults in EDM-based $R_y(\pm \pi/4)$ gates inside controlled-$\bar{H}$ constructions, as these gates cannot be twirled easily.

\begin{definition}[Injected $\bar{Y}$ fault]
    Consider the \EDM-$r$ $H$-state injection gadget of \cref{fig:h_state_injection}. An {\em injected $\bar{Y}$ fault} is an undetected logical $\bar{Y}$ fault on the prepared logical $\ket{\bar{H}}$ state induced by a fault in the $H$-state injection circuit. We denote its probability by $q_Y$.
    \label{def:injected_y}
\end{definition}

\begin{heuristic}[H-state injection $\bar{Y}$ fault rate]
    For the \EDM-$r$ $H$-state injection gadget, we estimate the
    injected-$\bar{Y}$-fault probability by
    \begin{equation}
        \begin{aligned}
            q_Y
            \approx{}& \left(\frac{8}{15}+\frac{7r}{15}\right)p \\
            &+ (r+1)\,\mathrm{Depth}(\mathrm{SEC})\frac{p}{150}.
        \end{aligned}
        \label{eq:h_injection_qy}
    \end{equation}
    Under the magic-factory error model of
    \cref{subsec:magic_factory_error_model}, we disregard memory errors on the
    target code block when applying this heuristic.
    We refer to \cref{eq:h_injection_qy} as the \emph{$H$-state injection
    $\bar{Y}$ fault rate heuristic}.
    \label{heur:h_injection_y_fault_rate}
\end{heuristic}

\begin{figure*}[t]
    \centering
    \includegraphics[width=\textwidth]{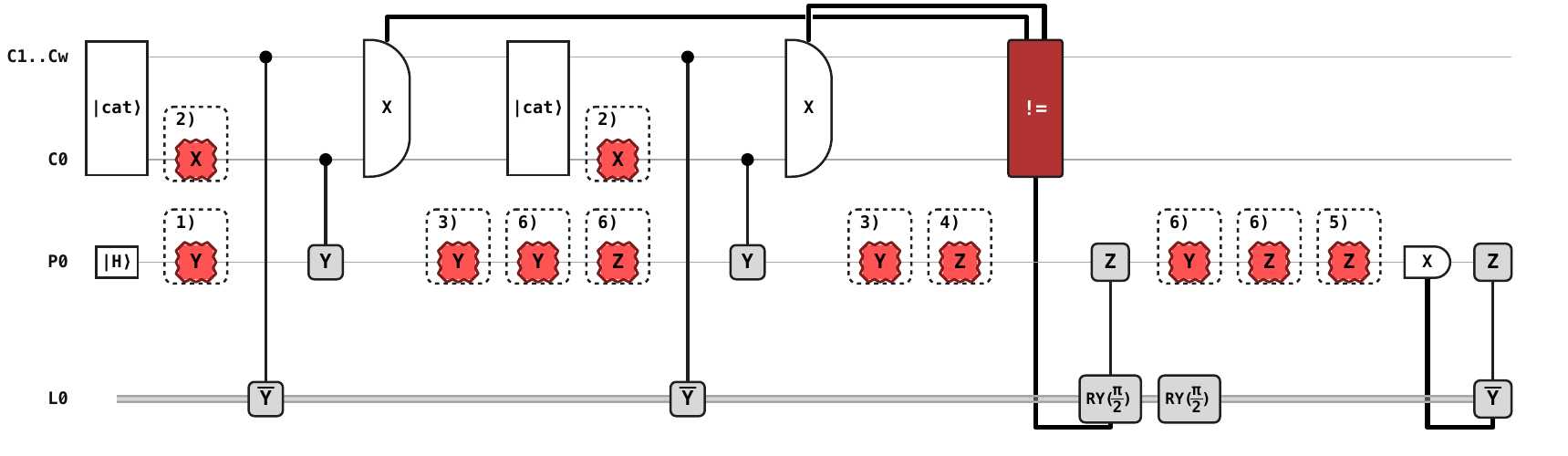}
    \caption{Fault locations contributing to an injected $\bar{Y}$ fault in the $H$-state injection gadget. Here, the joint $Y\bar{Y}$ measurement is implemented as \EDM-$2$ (that is, $r=2$). The figure represents $H$-state injection from the perspective of the single cat-state qubit \texttt{C0} aligned with the physical ancilla \texttt{P0}. \texttt{C1..Cw} represents the rest of the cat state carrying out the logical measurement of $\bar{Y}$ on the logical qubit target (\texttt{L0}). The bright red Pauli operators mark the location of all effective Pauli errors that lead to an injected $\bar{Y}$ fault for each source of root Pauli errors in items~1) through~6) of the justification of \cref{heur:h_injection_y_fault_rate}.}
    \label{fig:h_state_injection_faults}
\end{figure*}

\begin{proof}[Justification]
    The heuristic is obtained by accounting for first-order faults that can leave an
    undetected $\bar{Y}$ on the injected logical state. Under the ideal-logical-operation approximation
    of \cref{subsec:magic_factory_error_model}, we disregard memory faults on
    the target code block and also neglect the separate logical-outcome error of the hybrid $Y\bar{Y}$ \EDM itself.
    For the $r=3$ injection \EDM used below
    (\cref{heur:ch2_edm_rounds,heur:mek_edm_rounds}), the heuristic
    \cref{eq:heuristic_EDM} gives a residual accepted-outcome error of order
    $10^{-8}$ for both factories, far below the first-order ancilla-fault
    terms retained below, so to leading order it suffices to track only the
    effective Pauli left on the physical ancilla.
    In the injection gadget of \cref{fig:h_state_injection}, the final ancilla
    $X$ measurement determines whether a logical $\bar{Y}$ frame update is applied to
    the target logical qubit. First-order faults that anti-commute with the repeated $Y\bar{Y}$ check are
    rejected by the \EDM, and faults on the cat qubit or on the ancilla-cat
    entangling gate matter only through the ancilla Pauli they induce after
    propagation. Therefore, the undetected first-order faults contributing to an undetected
    logical $\bar{Y}$ fault are exactly those whose net effect is an accepted $Y$ or $Z$ fault on
    the physical ancilla immediately before the final ancilla $X$ readout,
    because either fault flips that readout and hence triggers the wrong
    logical-$\bar{Y}$ frame update.

    The fault locations and coefficients below were obtained by an exhaustive
    computational search over the moving-qubit noise model of
    \cref{subsec:noise-model,tab:moving qubit model operations}. For each
    elementary noise source, we inserted each Pauli fault, propagated it
    through the remainder of the gadget to the final measurements, and counted
    it as an undetected injected-$\bar{Y}$ fault exactly when it left all
    repeated $Y\bar{Y}$ outcomes unchanged but flipped the final $X$ measurement of
    the physical ancilla.
    In the moving-qubit model (\cref{tab:moving qubit model operations}),
    single-qubit preparation, one-qubit gates, and measurement readout faults
    occur with rate $p/10$, idle Pauli faults occur with rate $p/100$, and the
    cat-state entangling gate in one cat-based $Y\bar{Y}$ measurement is a
    two-qubit gate with depolarizing rate $p$. We also use
    the cat-state noise model of \cref{subsec:Heuristics}, which treats the
    relevant cat qubit as depolarized with rate $p$. Under the loss/leakage
    convention of \cref{subsec:magic_factory_error_model}, loss or leakage on
    the physical ancilla during the idle interval before the final $X$ readout
    causes a detected restart rather than an injected $\bar{Y}$ fault. Hence these
    events do not contribute to $q_Y$; instead, they are included in the retry
    rate of the injection gadget, as stated in \cref{heur:h_injection_retry}.

    We now list all fault locations that can lead to an injected $\bar{Y}$ fault.
    \Cref{fig:h_state_injection_faults} summarizes the effective fault locations
    for the six Pauli error sources counted below.
    \begin{enumerate}
        \item an effective $Y$ fault on the physical ancilla before the first
        cat-based $Y\bar{Y}$ measurement, arising either from the noisy $\ket{0}$
        preparation or from the subsequent physical $R_y(\pi/4)$ gate, with
        total first-order probability $p/15 + p/30 = p/10$. For the
        preparation term, the moving-qubit noise model gives a single-qubit
        depolarizing fault of rate $p/10$, so each Pauli branch occurs with
        probability $p/30$. Of these three branches, $Z$ acts trivially on
        $\ket{0}$, while $X\ket{0}$ and $Y\ket{0}$ both equal $\ket{1}$ up to
        phase, so after the subsequent physical $R_y(\pi/4)$ they yield the
        same effective ancilla-$Y$ fault. Thus preparation
        contributes $2(p/30)=p/15$. By contrast, for the post-gate
        depolarizing fault after the physical $R_y(\pi/4)$, only the literal
        $Y$ branch contributes to this item, giving $p/30$. This ancilla $Y$
        fault commutes with every repeated $Y\bar{Y}$ check, so the \EDM accepts, but
        its $Z$ component anticommutes with the final ancilla $X$ readout and
        flips that bit;
        \item an $X$ fault on the cat-state qubit that couples to the physical
        $\ket{H}$ ancilla in any one of the $r$ cat-based $Y\bar{Y}$ measurements,
        contributing at most $rp/3$. Here the cat qubit is modeled as a
        single-qubit depolarizing channel of rate $p$, and only the $X$
        component can produce an undetected injected-$\bar{Y}$ fault: a
        $Y=iXZ$ fault or $Z$ fault has a $Z$ component that flips the extracted bit and
        causes the \EDM to be rejected;
        \item a $Y$ fault on the physical $\ket{H}$ state induced by the
        cat-based entangling gate in any one of the $r$ $Y\bar{Y}$ measurements,
        contributing at most $2rp/15$. The entangling-gate fault is uniformly
        distributed over the $15$ non-identity two-qubit Pauli errors. For
        this count, we place the fault immediately after the ancilla-cat
        entangling gate of a given $Y\bar{Y}$ round and propagate it through the
        remainder of that round, including the cat-qubit $X$ readout and the
        conditioned Clifford updates in \cref{fig:h_state_injection}. Writing
        the post-gate Pauli fault in ancilla-cat order, the four faults $YI$,
        $YX$, $YY$, and $YZ$ yield an effective $Y$ on the physical
        $\ket{H}$ ancilla after that propagation. Of these, only $YI$ and
        $YX$ are accepted in that round: $YY$ and $YZ$ flip the extracted
        $Y\bar{Y}$ bit and are therefore rejected by the repeated measurement. The
        accepted branches are still harmful because, once the fault has
        propagated to an ancilla $Y$, it commutes with every subsequent $Y\bar{Y}$
        check and so is not caught by the \EDM, while its $Z$ component
        anticommutes with the final ancilla $X$ readout and flips that
        outcome, producing the injected-$\bar{Y}$ fault;
        \item a $Z$ fault on the physical $\ket{H}$ state after the last
        cat-based $Y\bar{Y}$ measurement, with probability $5p/15 = p/3$. Here we
        focus on propagated ancilla-$Z$ faults because the corresponding
        propagated ancilla-$Y$ faults were already counted in item~3. An effective $Z$ on the physical ancilla flips the final
        ancilla $X$ readout, but if it occurs in the last $Y\bar{Y}$ round there is
        no later repeated measurement through which the \EDM can catch it on
        the physical ancilla; at that stage, the \EDM can still reject only
        residual faults on the cat state before that cat qubit is measured.
        After placing the two-qubit Pauli fault immediately after the ancilla-cat entangling gate
        and propagating it through the rest of that round, $11$ non-identity
        Pauli terms are equivalent to an effective $Z$ on the physical ancilla
        before the final ancilla $X$ readout. Of these, only the five branches
        with cat Pauli $I$ or $X$ are accepted in the final \EDM round; the other
        six have cat Pauli $Y$ or $Z$, flip the final cat-qubit $X$ readout,
        and are therefore rejected by the \EDM. The surviving effective $Z$
        flips the final ancilla $X$ readout. If the same accepted $Z$ fault
        arises in any non-final $Y\bar{Y}$ round, it is caught by a subsequent
        repeated $Y\bar{Y}$ measurement; therefore only a fault in the last $Y\bar{Y}$
        round contributes here, giving $5p/15$;
        \item a flip of the final $X$-basis readout, with probability $p/10$;
        \item a $Y$ or $Z$ fault on the idle physical $\ket{H}$ state before the final
        $X$ readout. Since the moving-qubit idle noise is a single-qubit
        depolarizing channel of rate $p/100$, only the $Y$ and $Z$ components
        flip the final ancilla $X$ readout, so the relevant first-order rate is
        $(2/3)(p/100)=p/150$. This contributes at most
        $(r+1)\,\mathrm{Depth}(\mathrm{SEC})\,p/150$;
    \end{enumerate}
    Summing the non-idle contributions in items 1--5 gives
    \[
        \frac{p}{10}
        + r\frac{p}{3}
        + r\frac{2p}{15}
        + \frac{p}{3}
        + \frac{p}{10}
        = \left(\frac{8}{15}+\frac{7r}{15}\right)p.
    \]
    Adding the idle-noise contribution from item~6,
    \[
        (r+1)\,\mathrm{Depth}(\mathrm{SEC})\frac{p}{150},
    \]
    to this non-idle sum yields
    \cref{eq:h_injection_qy} as the resulting first-order estimate.
\end{proof}

\begin{heuristic}[Retry rate of $H$-state injection]
    Let $p_{H,\mathrm{retry}}$ denote the probability that the \EDM-$r$ $H$-state injection gadget aborts and must be restarted. Under the magic-factory error model of \cref{subsec:magic_factory_error_model}, we estimate the retry rate by
    \begin{equation}
        \begin{aligned}
            p_{H,\mathrm{retry}}
            &\approx
            1-\bigl((1-\pflip)^r+\pflip^r\bigr)(1-p_{\mathrm{anc,det}}), \\
            \pflip &\approx C_1 w p, \\
            p_{\mathrm{anc,det}}
            &\approx
            (r+1)\,\mathrm{Depth}(\mathrm{SEC})\,(\ploss+\pleak).
        \end{aligned}
        \label{eq:h_injection_retry}
    \end{equation}
    Here, $\pflip$ is the single-round cat-based outcome-flip probability from
    \cref{eq:heuristic_oneflip}, applied to the hybrid
    $Y\bar{Y}$ measurement with weight $w$.
    \label{heur:h_injection_retry}
\end{heuristic}

\begin{proof}[Justification]
    By \cref{def:EDM}, an \EDM-$r$ accepts only if all $r$ extracted bits agree.
    Therefore the \EDM outcomes are accepted in exactly two disjoint cases:
    either all $r$ outcomes are correct, with probability $(1-\pflip)^r$, or
    all $r$ outcomes are flipped, with probability $\pflip^r$.

    By the ancilla loss/leakage convention of
    \cref{subsec:magic_factory_error_model}, any loss or leakage on the
    physical ancilla during the idle interval before the final $X$ readout
    causes a detected restart rather than an injected $\bar{Y}$ fault. Using the
    idle loss and leakage rates from \cref{tab:moving qubit model operations},
    the probability of such a detected loss/leakage event is
    \[
        p_{\mathrm{anc,det}}
        \approx
        (r+1)\,\mathrm{Depth}(\mathrm{SEC})\,(\ploss+\pleak).
    \]
    Approximating this ancilla-detection event as independent of the cat-based
    outcome flips, the gadget is accepted if and only if the \EDM outcomes
    agree and no detected loss/leakage event occurs. Hence the total acceptance
    probability is estimated by
    \[
        \bigl((1-\pflip)^r+\pflip^r\bigr)(1-p_{\mathrm{anc,det}}),
    \]
    and taking the complement gives the heuristic estimate
    \cref{eq:h_injection_retry}.
\end{proof}

\begin{figure}
    \centering
    \includegraphics[width=0.95\linewidth]{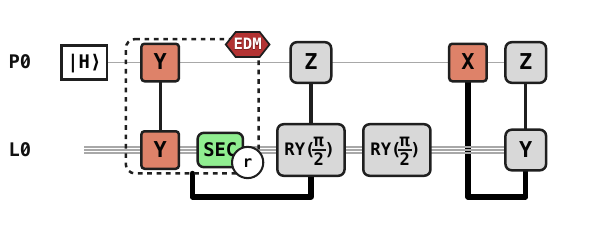}
    \caption{$H$-state injection from a physical ancilla \texttt{P0} onto a logical qubit \texttt{L0}, implemented in a code block with SEC-level scheduling. This realizes the IS+H-based circuit for logical $R_y(\pi/4)$, with the hybrid joint $Y\bar{Y}$ measurement implemented as an \EDM; the $r$ marker in the diagram denotes that the \EDM uses a variable number $r$ of rounds.}
    \label{fig:h_state_injection}
\end{figure}

\subsubsection{\texorpdfstring{Logical Controlled-$\bar{H}$ gate}{Logical Controlled-H gate}}
In this subsection, we discuss the IS+H-based implementation of a logical controlled-$\bar{H}$ gate, which is used in the verification stage of the \CHfactory protocol.
The construction is based on the standard decomposition of a controlled-Hadamard gate into a controlled-$\bar{Z}$ gate and the $R_y(\pm \pi/4)$ gates discussed above.

\begin{figure}
    \centering
    \includegraphics[width=0.95\linewidth]{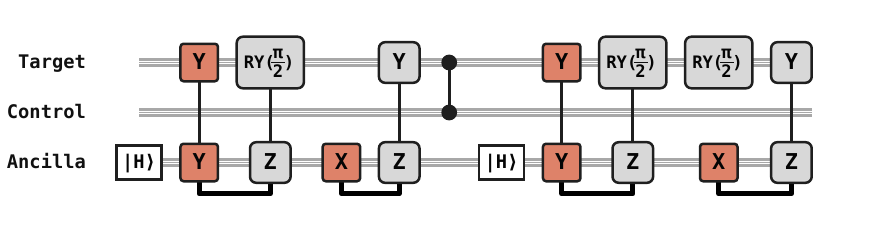}
    \caption{A logical controlled-$\bar{H}$ gate constructed from the decomposition $\mathrm{C}\bar{H} = (I \otimes R_y(\pi/4))\,\CZ\,(I \otimes R_y(-\pi/4))$. The logical $R_y(\pm \pi/4)$ gates are implemented using IS+H-based circuits. The intermediate logical $\CZ$ is left unexpanded, since as a Clifford gate it can be implemented via the logical instruction set or by frame tracking.}
    \label{fig:logical_h_measurement_based}
\end{figure}

\Cref{fig:logical_h_measurement_based} shows an IS+H-based implementation of a logical controlled-$\bar{H}$ gate.
The construction uses the standard identity
\[
    \mathrm{C}H = (I \otimes R_y(\pi/4))\,\CZ\,(I \otimes R_y(-\pi/4)),
\]
as in Fig.~3 of Ref.~\cite{dasu2026computing}. In code blocks where all logical
Pauli operators are accessible, such as the \MEK factory in the \code{70}
code, the logical controlled-$\bar{Z}$ is an accessible Clifford gate and can therefore be
implemented by frame tracking rather than expanded into instruction-set
operations. By contrast, the logical $R_y(\pm \pi/4)$ gates are non-Clifford
and here require injected $\ket{H}$ states, so they must be expanded into
their IS+H-based implementations
(\cref{fig:logical_ry_measurement_based} and
\cref{fig:logical_ry_neg_measurement_based}), yielding
\cref{fig:logical_h_measurement_based}.

\begin{figure*}[t]
    \centering
    \includegraphics[width=\textwidth]{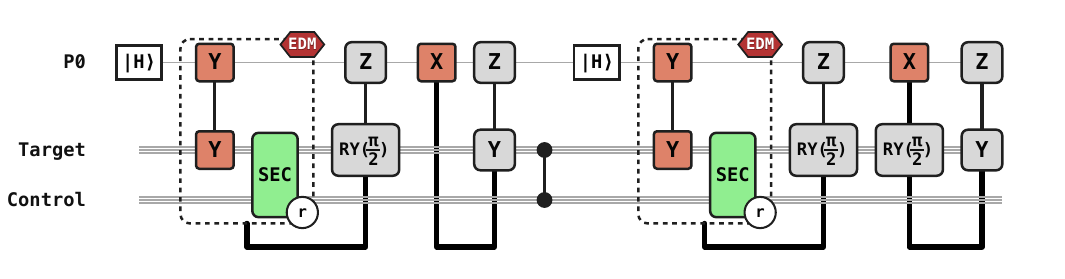}
    \caption{Implementation of the logical controlled-$\bar{H}$ gate in a code block with SEC-level scheduling. Each logical $R_y(\pm \pi/4)$ subcircuit is realized by the $H$-state injection gadget of \cref{fig:h_state_injection}, which uses a physical ancilla prepared in $\ket{H}$ and an \EDM for the hybrid $Y\bar{Y}$ measurement.}
    \label{fig:logical_ch_code_block}
\end{figure*}

Combining the logical decomposition of
\cref{fig:logical_h_measurement_based} with the IS+H-based implementations of
$R_y(\pm \pi/4)$ gives the corresponding code-block implementation of the
logical controlled-$\bar{H}$ gate with SEC-level scheduling, shown in
\cref{fig:logical_ch_code_block}. Each logical $R_y(\pm \pi/4)$ subcircuit uses a physical ancilla prepared in $\ket{H}$
and an \EDM for the hybrid $Y\bar{Y}$ measurement. The $R_y(\pi/4)$ code-block implementation can be seen in \cref{fig:h_state_injection}.

\subsubsection{Transversal H measurement}

\begin{figure}[b!]
    \centering
    \includegraphics[width=0.75\linewidth]{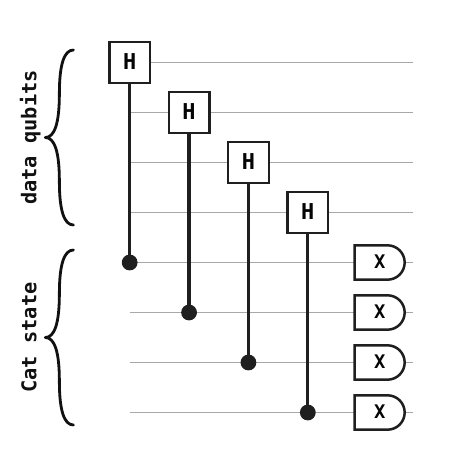}
    \caption{Cat-based measurement of the physical Hermitian unitary $H^{\otimes n}$ in a code with strongly transversal Hadamard, yielding a measurement of the joint logical observable $\bar{H}^{\otimes k}$. Every data qubit in the code block is targeted by a physical controlled-H gate controlled by a separate cat state qubit.}
    \label{fig:transversal_h_measurement}
\end{figure}

For a code with strongly transversal Hadamard, the cat-based measurement shown in \cref{fig:transversal_h_measurement} realizes the controlled-$M$ measurement gadget of~\cite{gottesman1999teleportation} for the Hermitian unitary $M=H^{\otimes n}$. Each physical controlled-$H$ in this layer can be implemented by the physical gate sequence: $R_y(\pi/4, \text{target}) \CZ(\text{control}, \text{target}) R_y(-\pi/4, \text{target})$. Starting from a cat ancilla and a data state $\ket{\psi}$, the transversal controlled-$H$ layer prepares
\[
    \frac{1}{\sqrt{2}}\left(\ket{0}^{\otimes n}\ket{\psi} + \ket{1}^{\otimes n}H^{\otimes n}\ket{\psi}\right).
\]
Measuring the cat state in the $X$ basis then projects the data onto the $\pm 1$ eigenspaces of $M$, equivalently applying the projector $(I \pm M)/2$. Since $H^{\otimes n}=\bar{H}^{\otimes k}$ on the code space, this gadget measures the joint logical observable $\bar{H}^{\otimes k}$. This is the verification primitive used by the \CHfactory protocol to verify the encoded $\ket{\bar{H}}\ket{\bar{H}}$ state.

\subsection{The CH2 factory}
\label{subsec:ch2_factory}

\subsubsection{The CH2 factory circuit}

The \CHfactory factory realizes the \CHfactory protocol in the \code{54} code. The protocol prepares a logical $\ket{\bar{H}}\ket{\bar{H}}$ state and then verifies that state by measuring the joint logical observable $\bar{H}^{\otimes 2}$. Because the code admits transversal Hadamard, the factory can implement this verification fault-tolerantly with the cat-based measurement of \cref{fig:transversal_h_measurement}. This is considerably cheaper than realizing the same check through a logical controlled-$\bar{H}$ measurement, and does not require additional physical $\ket{H}$ states.

\begin{figure*}[t]
    \centering
    \includegraphics[width=\textwidth]{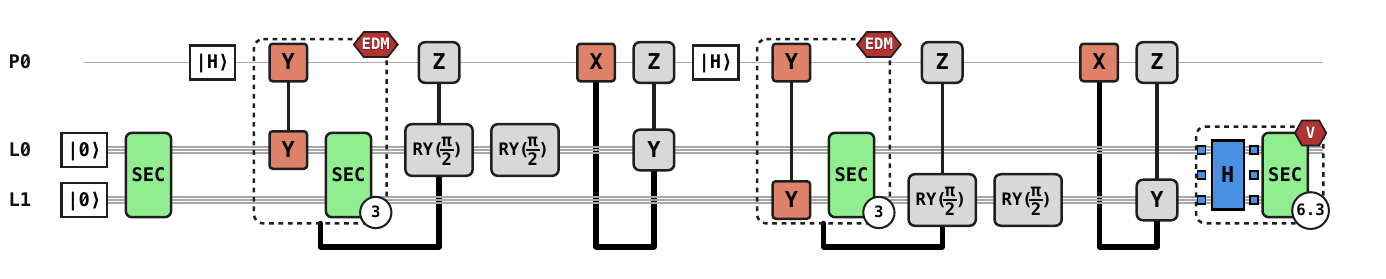}
    \caption{\CHfactory factory in the \code{54} code, implemented in a code block with SEC-level scheduling. The circuit first applies the $H$-state injection gadget of \cref{fig:h_state_injection} to logical qubit $0$, then repeats the same \EDM-$3$ injection on logical qubit $1$, where the required \EDM round count is estimated in \cref{heur:ch2_edm_rounds}. It finally verifies the resulting candidate state by a transversal measurement of the joint logical observable $\bar{H}^{\otimes 2}$ using \cref{fig:transversal_h_measurement}, implemented as a Viterbi measurement with average runtime $6.31~\mathrm{SEC}$ as estimated in \cref{heur:ch2_viterbi_runtime}.}
    \label{fig:CH2_circuit}
\end{figure*}

The code-block schedule, shown in \cref{fig:CH2_circuit}, first applies the $H$-state injection gadget of \cref{fig:h_state_injection} to logical qubit $0$, and then applies the same injection circuit to logical qubit $1$. These injections are implemented with \EDM-$3$ measurements, as estimated in \cref{heur:ch2_edm_rounds}. The factory then verifies the candidate $\ket{\bar{H}}\ket{\bar{H}}$ state by measuring the joint logical observable $\bar{H}^{\otimes 2}$ with the transversal gadget of \cref{fig:transversal_h_measurement}, implemented as a weight-$54$ Viterbi measurement with average runtime $6.31~\mathrm{SEC}$ from \cref{heur:ch2_viterbi_runtime}.
Depending on the outcome of EDM at the injection step, a Clifford $R_y(\pm \pi/2)$ correction on the first logical qubit may be required.
We account for this by measuring $\bar{H}^{'} \otimes \bar{H}$, $\bar{H} \otimes \bar{H}^{'}$, or $\bar{H}^{'} \otimes \bar{H}^{'}$ instead of $\bar{H}\otimes \bar{H}$, where $H^{'} = R_y(-\pi/2)H R_y(\pi/2)$. This is possible since $H^{'}=ZXH$.
On the symplectic basis of \cref{tab:Symp_TransH}, the physical support of $\bar X_i$ and $\bar Z_i$ are identical, so it follows that to measure $\bar{H}^{'}$ on one or both logical qubits of \code{54}, we replace the physical controlled-$H$ of \cref{fig:transversal_h_measurement} with controlled-$H^{'}$ on the support of their respective $\bar X$ (and $\bar Z$) representatives.

\subsubsection{Logical measurements}
\label{subsubsec:ch2_logical_measurements}

The \CHfactory schedule uses two measurement primitives: the hybrid $Y\bar{Y}$ \EDM inside each
$H$-state injection gadget, and the final logical $\bar{H}^{\otimes 2}$ Viterbi
measurement. For the hybrid \EDM we use logical-measurement error target
$\varepsilon_{\mathrm{inj}} = 10^{-5}$ to choose the round count; the resulting
choice $r=3$ already gives an estimated accepted-outcome error of order
$10^{-8}$, numerically about $3.90\times 10^{-8}$ by
\cref{heur:ch2_edm_rounds}. Larger \EDM round counts would only increase
retries. The purely logical Viterbi verification is instead run at the tighter
target $\varepsilon = 10^{-10}$ from \cref{subsec:magic_factory_error_model}.

\begin{heuristic}[Injection-\EDM round count for \CHfactory/\code{54}]
    For the \CHfactory/\code{54} factory at the operating point
    $p = 10^{-4}$ and $p_{\log,\mathrm{\code{54}}} = 3\times 10^{-10}$,
    choose the hybrid-injection \EDM round count
    \begin{equation}
        \begin{aligned}
            r_{\mathrm{CH2,inj}}
            &=
            \min\bigl\{
                r \geq 1 : \\
            &\qquad
                C_2 p_{\log,\mathrm{\code{54}}}
                + (C_1 w_{\mathrm{CH2,inj}} p)^r
                \leq 10^{-5}
            \bigr\} \\
            &= 3,
        \end{aligned}
        \label{eq:ch2_injection_rounds}
    \end{equation}
    with $w_{\mathrm{CH2,inj}} = 16$.
    \label{heur:ch2_edm_rounds}
\end{heuristic}

\begin{proof}[Justification]
    By \cref{fig:All_TransH,tab:LogicalConstraints}, the injected logical $Y$
    can be represented with weight $16$ in \code{54}, so we take
    $w_{\mathrm{CH2,inj}} = 16$ for the hybrid $Y\bar{Y}$ measurement in this
    analysis. Using the \EDM heuristic
    \cref{eq:heuristic_EDM} with the fitted constants $C_1 = 2.1$ and
    $C_2 = 3.4$ from \cref{subsubsec:OneCat,subsubsec:MemECM},
    \[
        \begin{aligned}
            C_2 p_{\log,\mathrm{\code{54}}}
            + (C_1 w_{\mathrm{CH2,inj}} p)^2
            &\approx
            3.4 \times 3\times 10^{-10} \\
            &\quad + (2.1\times 16\times 10^{-4})^2 \\
            &\approx 1.13\times 10^{-5}
            > 10^{-5},
        \end{aligned}
    \]
    whereas
    \[
        \begin{aligned}
            C_2 p_{\log,\mathrm{\code{54}}}
            + (C_1 w_{\mathrm{CH2,inj}} p)^3
            &\approx
            3.4 \times 3\times 10^{-10} \\
            &\quad + (2.1\times 16\times 10^{-4})^3 \\
            &\approx 3.90\times 10^{-8}
            < 10^{-5}.
        \end{aligned}
    \]
    Hence the smallest admissible \EDM round count is
    $r_{\mathrm{CH2,inj}} = 3$.
\end{proof}

\begin{heuristic}[Viterbi runtime for \CHfactory/\code{54}]
    For the final transversal verification in the \CHfactory/\code{54} factory, use the
    average Viterbi runtime
    \begin{equation}
        \tau_{\mathrm{Vit},54}^{\mathrm{avg}}
        \approx 6.31~\mathrm{SEC},
        \label{eq:ch2_viterbi_runtime}
    \end{equation}
    corresponding to target logical-measurement error $\varepsilon = 10^{-10}$.
    \label{heur:ch2_viterbi_runtime}
\end{heuristic}

\begin{proof}[Justification]
    For the final verification, the measured observable is the transversal
    $\bar{H}^{\otimes 2}$ of weight $54$, so the weight-$54$,
    $\varepsilon = 10^{-10}$ entry of \cref{tab:LMtime}, obtained from the
    Viterbi logical-measurement model of \cref{subsec:AdaptiveMeas}, gives the runtime estimate
    \cref{eq:ch2_viterbi_runtime}.
\end{proof}

\subsubsection{Error analysis}

Under the ideal-logical-operation assumption of \cref{subsec:magic_factory_error_model}, the only error sources for each \CHfactory attempt are the two injected
logical $\ket{\bar{H}}$ states. The rest of the logical circuit is treated as perfect.

To produce a standard output pair of logical $\ket{\bar{H}}$ states, we reject the $-1$ outcome of the logical $\bar{H}^{\otimes 2}$ measurement,
which corresponds to $\ket{\bar{Y}\bar{H}}\ket{\bar{H}}$ or $\ket{\bar{H}}\ket{\bar{Y}\bar{H}}$.
We accept either $\ket{\bar{H}}\ket{\bar{H}}$ or $\ket{\bar{Y}\bar{H}}\ket{\bar{Y}\bar{H}}$; the latter is an undetected logical error.
The \CHfactory error analysis therefore reduces to the chance that the two \EDM-$3$
$H$-state preparations from \cref{heur:ch2_edm_rounds} each produce an
independent injected-$\bar{Y}$ fault.

\begin{heuristic}[Injected-state error estimate for \CHfactory/\code{54}]
    For the \CHfactory factory implemented in the \code{54} block, taking $p=10^{-4}$,
    $r = r_{\mathrm{CH2,inj}} = 3$ from \cref{heur:ch2_edm_rounds},
    and the \code{54} SEC depth from \cref{tab:loss_leakage_summary}, the injected-$\bar{Y}$-fault
    probability per accepted logical $\ket{\bar{H}}$ preparation is estimated by
    \begin{equation}
        \begin{aligned}
            q_{Y,\mathrm{CH2}}
            &\approx
            1.93\times 10^{-4}
            + \frac{4 \times 28.15 \times 10^{-4}}{150} \\
            &\approx
            2.68\times 10^{-4}.
        \end{aligned}
        \label{eq:ch2_qy}
    \end{equation}
    \label{heur:ch2_qy}
\end{heuristic}

\begin{proof}[Justification]
    Applying \cref{heur:h_injection_y_fault_rate} with
    $p = 10^{-4}$ and
    $r = r_{\mathrm{CH2,inj}} = 3$ from
    \cref{heur:ch2_edm_rounds}, together with
    \[
        \mathrm{Depth}(\mathrm{SEC})
        =
        d_{\mathrm{POC}}
        =
        1 + 8 + 10.10 + 8.05 + 1
        =
        28.15
    \]
    from the \code{54} SEC time budget in \cref{tab:loss_leakage_summary},
    gives the heuristic estimate \cref{eq:ch2_qy}.
\end{proof}

\begin{heuristic}[Verification acceptance probability for \CHfactory/\code{54}]
    Assuming the two injection gadgets fail independently and the remaining
    logical operations in the factory are perfect, we estimate the \CHfactory
    verification/postselection acceptance probability by
    \begin{equation}
        a_{\mathrm{CH2,ver}}
        \approx
        (1-q_{Y,\mathrm{CH2}})^2 + q_{Y,\mathrm{CH2}}^2
        \approx 0.99946.
        \label{eq:ch2_acceptance_probability}
    \end{equation}
    \label{heur:ch2_acceptance_probability}
\end{heuristic}

\begin{proof}[Justification]
    We assume i.i.d. logical-$\bar{Y}$ faults on each injected logical
    $\ket{\bar{H}}$ state; this is consistent with
    \cref{heur:ch2_qy}, where $q_{Y,\mathrm{CH2}}$ is obtained by specializing
    \cref{heur:h_injection_y_fault_rate} to the \CHfactory/\code{54} injection
    gadget under the ideal-logical-operation approximation. The \CHfactory factory accepts
    when there are either none,
    or exactly two, logical-$\bar{Y}$ faults. This gives the heuristic estimate
    \[
        a_{\mathrm{CH2,ver}}
        \approx
        (1-q_{Y,\mathrm{CH2}})^2 + q_{Y,\mathrm{CH2}}^2.
    \]
    Evaluating at the injected-state estimate \cref{eq:ch2_qy} gives the quoted
    numerical value.
\end{proof}

\begin{heuristic}[Output error estimate for the \CHfactory factory]
    Conditioned on even-parity postselection, the \CHfactory output pair has estimated logical 
    error rate of 
    \begin{equation}
        p_{\mathrm{CH2,out}}
        \approx
        \frac{q_{Y,\mathrm{CH2}}^2}{a_{\mathrm{CH2,ver}}}
        \approx 7.2\times 10^{-8}.
        \label{eq:ch2_output_error}
    \end{equation}
    \label{heur:ch2_output_error}
\end{heuristic}

\begin{proof}[Justification]
    The quantity $q_{Y,\mathrm{CH2}}^2$ estimates the probability of
    the bad double-fault event before acceptance. However,
    $p_{\mathrm{CH2,out}}$ is the error rate
    after we keep only the runs that pass the final $\bar{H}^{\otimes 2}$
    check. That check rejects the odd-parity cases, so among accepted runs only
    two branches remain: the good branch
    $\ket{\bar{H}}\ket{\bar{H}}$ and the bad branch
    $\ket{\bar{Y}\bar{H}}\ket{\bar{Y}\bar{H}}$. The bad branch arises precisely when both
    injections produce an injected $\bar{Y}$ fault, which is estimated to occur with
    probability $q_{Y,\mathrm{CH2}}^2$.
    Therefore the output error is the fraction of accepted runs that lie in
    this bad branch,
    \[
        p_{\mathrm{CH2,out}}
        \approx
        \frac{\Pr[\text{bad and accepted}]}{\Pr[\text{accepted}]}
        \approx
        \frac{q_{Y,\mathrm{CH2}}^2}{a_{\mathrm{CH2,ver}}},
    \]
    where $\Pr[\text{accepted}] \approx a_{\mathrm{CH2,ver}}$ by
    \cref{heur:ch2_acceptance_probability}. Evaluating at \cref{eq:ch2_qy}
    yields the stated numerical estimate.
\end{proof}

\subsubsection{Time analysis}
\label{subsubsec:ch2_time_analysis}

We next convert the \CHfactory schedule shown in \cref{fig:CH2_circuit} into an
average SEC cost. Apart from the initial logical $\ket{0}$ preparation counted
below, one \CHfactory attempt consists exactly of the two \EDM-based injections and the
final transversal verification.

\begin{heuristic}[Per-attempt failure rate estimate for the \CHfactory factory]
    For the \CHfactory/\code{54} implementation, the estimated per-attempt
    failure rate is
    \begin{equation}
        p_{\mathrm{CH2,fail}}
        \approx
        1-a_{\mathrm{CH2,inj}}a_{\mathrm{CH2,ver}}
        \approx 2.28\%.
        \label{eq:ch2_failure_rate}
    \end{equation}
    \label{heur:ch2_failure_rate}
\end{heuristic}

\begin{proof}[Justification]
    For the \CHfactory/\code{54} injection,
    \cref{fig:All_TransH,tab:LogicalConstraints} show that the logical $Y$
    representative can be taken to have weight $16$, so we use $w=16$ in the
    retry estimate. Therefore, \cref{heur:h_injection_retry} together with the fitted
    value $C_1=2.1$ from \cref{subsubsec:OneCat},
    $r = r_{\mathrm{CH2,inj}} = 3$ from \cref{heur:ch2_edm_rounds},
    $p=10^{-4}$, $\ploss=10^{-7}$, and $\pleak=10^{-5}$ gives
        \[
        \begin{aligned}
            \pflip
            &\approx C_1 w p
            = 2.1\times 16 \times 10^{-4}
            = 3.36\times 10^{-3}, \\
            p_{\mathrm{anc,det}}
            &\approx
            4\times 28.15 \times (10^{-7}+10^{-5}) \\
            &= 1.13726\times 10^{-3}
            \approx 1.14\times 10^{-3}, \\
            p_{H,\mathrm{retry}}
            &\approx
            1 \\
            &\quad -
            \Bigl((1-3.36\times 10^{-3})^3 \\
            &\qquad\quad + (3.36\times 10^{-3})^3\Bigr) \\
            &\qquad \times (1-1.14\times 10^{-3}) \\
            &\approx 1.12\times 10^{-2}.
        \end{aligned}
    \]
    Hence, one injection is accepted with probability
    $a_{H,\mathrm{acc}} = 1-p_{H,\mathrm{retry}} \approx 0.9888$, so the
    probability that both injections are accepted is
    \[
        a_{\mathrm{CH2,inj}}
        =
        a_{H,\mathrm{acc}}^2
        \approx 0.9778.
    \]
    By \cref{heur:ch2_acceptance_probability}, the \CHfactory
    verification/postselection acceptance probability is
    $a_{\mathrm{CH2,ver}} \approx 0.99946$, so the per-attempt failure rate is
    $p_{\mathrm{CH2,fail}} \approx 1-a_{\mathrm{CH2,inj}}a_{\mathrm{CH2,ver}} \approx 2.28\%$.
\end{proof}

\begin{heuristic}[Average runtime estimate for the \CHfactory factory]
    For the \CHfactory/\code{54} implementation, the expected SEC count per successful \CHfactory
    output pair is
    \begin{equation}
        N_{\mathrm{CH2,SEC}}^{\mathrm{avg}}
        \approx 13.44~\mathrm{SEC},
        \label{eq:ch2_avg_runtime}
    \end{equation}
    or about $6.72~\mathrm{SEC}$ per logical $\ket{\bar{H}}$ state.
    \label{heur:ch2_runtime}
\end{heuristic}

\begin{proof}[Justification]
    Write $N_{\mathrm{CH2,SEC}}^{\mathrm{avg}}$ for the expected SEC count
    until one successful \CHfactory output pair. A first-step decomposition over the
    four possible outcomes of one \CHfactory attempt gives the recursive relation 
    \[
        \begin{aligned}
            N_{\mathrm{CH2,SEC}}^{\mathrm{avg}}
            &\approx
            p_{H,\mathrm{retry}}
            \left(D_1 + N_{\mathrm{CH2,SEC}}^{\mathrm{avg}}\right) \\
            &\quad +
            a_{H,\mathrm{acc}} p_{H,\mathrm{retry}}
            \left(D_2 + N_{\mathrm{CH2,SEC}}^{\mathrm{avg}}\right) \\
            &\quad +
            a_{\mathrm{CH2,inj}} (1-a_{\mathrm{CH2,ver}}) \\
            &\qquad \times
            \left(C_{\mathrm{CH2}} + N_{\mathrm{CH2,SEC}}^{\mathrm{avg}}\right) \\
            &\quad +
            a_{\mathrm{CH2,inj}} a_{\mathrm{CH2,ver}} C_{\mathrm{CH2}}.
        \end{aligned}
    \]
    Here $D_1$ is the SEC cost if the first injection fails. Likewise,
    $D_2$ is the cost if the second injection fails after the first succeeds.
    Finally, $C_{\mathrm{CH2}}$ is the SEC cost of a full \CHfactory schedule
    through the final verification.

    As shown in \cref{fig:CH2_circuit}, after the initial logical $\ket{0}$
    preparation the factory consists only of two uses of the
    $H$-state injection gadget of \cref{fig:h_state_injection} followed by the
    final transversal $\bar{H}^{\otimes 2}$ verification. Each logical
    $\ket{\bar{H}}$ preparation is implemented with an \EDM-$r_{\mathrm{CH2,inj}}$
    measurement from \cref{heur:ch2_edm_rounds}, so each injection attempt costs
    \[
        N_{H\text{-inj}} = r_{\mathrm{CH2,inj}} = 3~\mathrm{SEC}.
    \]
    The final verification is the weight-$54$ Viterbi measurement from
    \cref{fig:CH2_circuit}. By \cref{heur:ch2_viterbi_runtime}, its
    average runtime is
    $\tau_{\mathrm{Vit},54}^{\mathrm{avg}} = 6.31~\mathrm{SEC}$. Since the
    initial logical $\ket{0}$ preparation costs $1~\mathrm{SEC}$, we have
    \[
        \begin{aligned}
            D_1 &= 1 + N_{H\text{-inj}} = 4~\mathrm{SEC}, \\
            D_2 &= 1 + 2N_{H\text{-inj}} = 7~\mathrm{SEC}, \\
            C_{\mathrm{CH2}}
            &= 1 + 2N_{H\text{-inj}} + \tau_{\mathrm{Vit},54}^{\mathrm{avg}} \\
            &\approx 13.31~\mathrm{SEC},
        \end{aligned}
    \]
    and from the justification of \cref{heur:ch2_failure_rate},
    \[
        \begin{aligned}
            p_{H,\mathrm{retry}} &\approx 1.12\times 10^{-2}, \\
            a_{H,\mathrm{acc}}
            &= 1-p_{H,\mathrm{retry}}
            \approx 0.9888, \\
            a_{\mathrm{CH2,inj}}
            &= a_{H,\mathrm{acc}}^2
            \approx 0.9778, \\
            a_{\mathrm{CH2,ver}} &\approx 0.99946.
        \end{aligned}
    \]
    Substituting these values into the recursion, first collect the terms
    proportional to $N_{\mathrm{CH2,SEC}}^{\mathrm{avg}}$:
    \[
        \begin{aligned}
            N_{\mathrm{CH2,SEC}}^{\mathrm{avg}}
            &\approx
            p_{H,\mathrm{retry}} D_1 \\
            &\quad +
            a_{H,\mathrm{acc}} p_{H,\mathrm{retry}} D_2 \\
            &\quad +
            a_{\mathrm{CH2,inj}} C_{\mathrm{CH2}} \\
            &\quad +
            \Bigl(
                p_{H,\mathrm{retry}}
                + a_{H,\mathrm{acc}} p_{H,\mathrm{retry}} \\
            &\qquad\qquad
                + a_{\mathrm{CH2,inj}}(1-a_{\mathrm{CH2,ver}})
            \Bigr)
            N_{\mathrm{CH2,SEC}}^{\mathrm{avg}}.
        \end{aligned}
    \]
    Since
    \[
        \begin{aligned}
            &p_{H,\mathrm{retry}}
            + a_{H,\mathrm{acc}} p_{H,\mathrm{retry}} \\
            &\quad
            + a_{\mathrm{CH2,inj}}(1-a_{\mathrm{CH2,ver}}) \\
            &\quad
            = 1-a_{\mathrm{CH2,inj}}a_{\mathrm{CH2,ver}},
        \end{aligned}
    \]
    this becomes
    \[
        \begin{aligned}
            N_{\mathrm{CH2,SEC}}^{\mathrm{avg}}
            &\approx
            p_{H,\mathrm{retry}} D_1 \\
            &\quad +
            a_{H,\mathrm{acc}} p_{H,\mathrm{retry}} D_2 \\
            &\quad +
            a_{\mathrm{CH2,inj}} C_{\mathrm{CH2}} \\
            &\quad + (1-a_{\mathrm{CH2,inj}}a_{\mathrm{CH2,ver}})
            N_{\mathrm{CH2,SEC}}^{\mathrm{avg}}.
        \end{aligned}
    \]
    Rearranging gives
    \[
        \begin{aligned}
            S_{\mathrm{CH2}}
            &=
            p_{H,\mathrm{retry}} D_1 \\
            &\quad +
            a_{H,\mathrm{acc}} p_{H,\mathrm{retry}} D_2 \\
            &\quad +
            a_{\mathrm{CH2,inj}} C_{\mathrm{CH2}},
        \end{aligned}
    \]
    so that
    \[
        \begin{aligned}
            a_{\mathrm{CH2,inj}}a_{\mathrm{CH2,ver}}
            N_{\mathrm{CH2,SEC}}^{\mathrm{avg}}
            &\approx S_{\mathrm{CH2}}, \\
            N_{\mathrm{CH2,SEC}}^{\mathrm{avg}}
            &\approx
            \frac{S_{\mathrm{CH2}}}
            {a_{\mathrm{CH2,inj}}a_{\mathrm{CH2,ver}}}.
        \end{aligned}
    \]
    Numerically,
    \[
        \begin{aligned}
            S_{\mathrm{CH2}}
            &\approx
            (1.12\times 10^{-2})(4) \\
            &\quad +
            (0.9888)(1.12\times 10^{-2})(7) \\
            &\quad +
            (0.9778)(13.31) \\
            &\approx 13.14,
        \end{aligned}
    \]
    so
    \[
        \begin{aligned}
            N_{\mathrm{CH2,SEC}}^{\mathrm{avg}}
            &\approx
            \frac{13.14}{0.9773} \\
            &\approx 13.44~\mathrm{SEC}.
        \end{aligned}
    \]
    Dividing by the two logical $\ket{\bar{H}}$ outputs gives $6.72~\mathrm{SEC}$ per
    output state.
\end{proof}

\subsection{The MEK factory}
\label{subsec:mek_factory}

The \MEK protocol~\cite{meier2013magic} uses a $[[4,2,2]]$ code to distill two injected $\ket{\bar{H}}$ states.
The $[[4,2,2]]$ code does not have a strongly transversal Hadamard, but does have a transversal Hadamard up to a swap of the two logical qubits.
The protocol therefore realizes the $\bar{H}^{\otimes 2}$ verification through the more complex circuit of \cref{fig:mek_circuit}.

\subsubsection{The MEK factory circuit}

\begin{figure}[t]
    \centering
    \includegraphics[width=\linewidth]{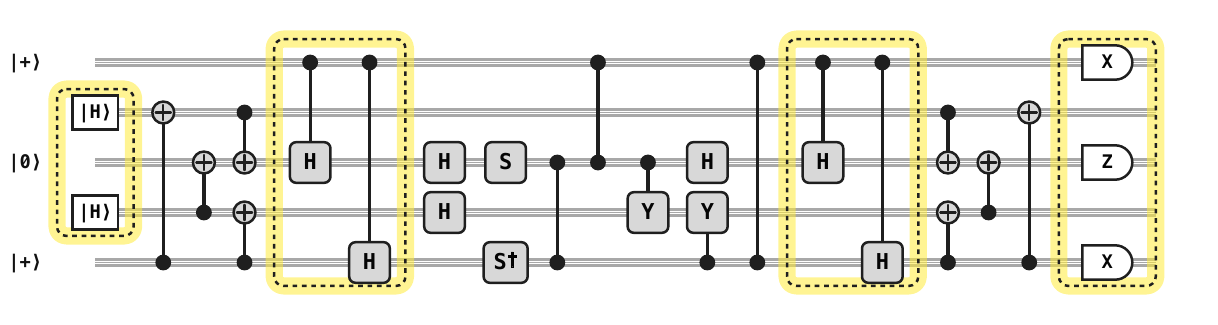}
    \caption{Circuit for the \MEK protocol from Ref.~\cite{meier2013magic}. The highlighted operations are the only steps that remain to be implemented as actual logical measurements once the Clifford part of the protocol is absorbed into frame updates.}
    \label{fig:mek_circuit}
\end{figure}

The \MEK protocol leverages a transversal implementation of logical $\overline{\mathrm{SWAP}}\cdot \bar{H}^{\otimes 2}$ in the $[[4,2,2]]$ error-detection code to verify two encoded logical $\ket{\bar H}$ states.
\Cref{fig:mek_circuit} shows the corresponding protocol circuit; therein, unitary encoding of two logical $H$ states is followed by a verification measurement via a sequence of logical $CH$ gates (see Ref.~\cite{meier2013magic} and \cref{fig:logical_h_measurement_based}).
The distillation succeeds only if both the verification measurement as well as syndromes of the $[[4,2,2]]$ indicate no error(s).
A unitary decoding then yields two verified $H$ states. In our architecture, the \MEK factory hosts the entire \MEK protocol and its $[[4,2,2]]$ code within a \code{70} code block, so that the resulting logical $\bar H$ states reside within a high-performance qLDPC memory.

This choice of code is also convenient for logical measurements: in the \code{70} block, every logical Pauli observable on the six encoded qubits admits a representative of weight at most $18$; see \cref{fig:All_MEK,tab:LogicalConstraints}. Hence all logical Clifford operations can be absorbed into Clifford frame tracking, so the highlighted operations in \cref{fig:mek_circuit} are the only steps that require physical implementation.

\begin{figure*}[t]
    \centering
    \includegraphics[width=\textwidth]{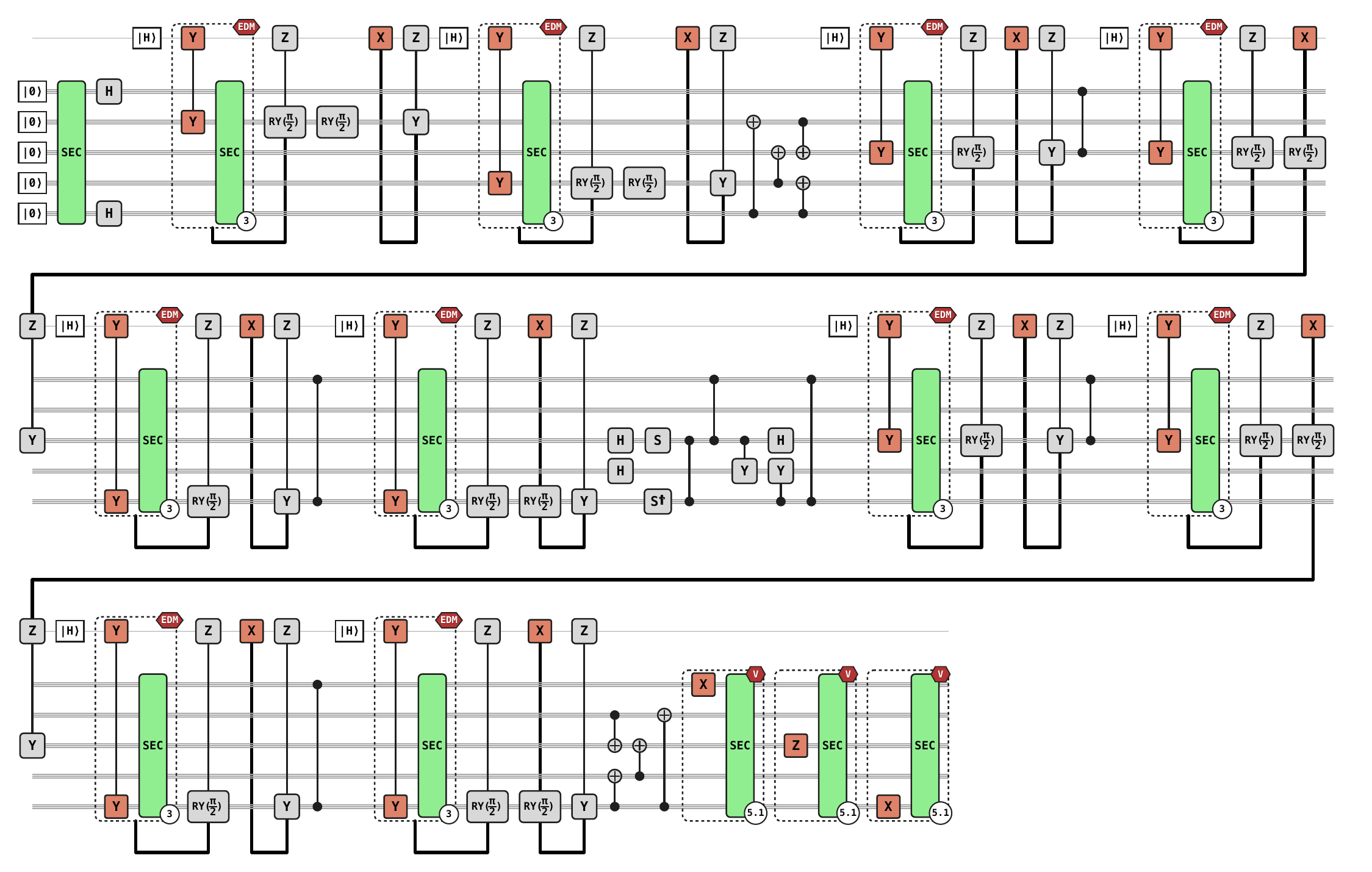}
    \caption{\MEK factory in the \code{70} code, implemented in a code block with SEC-level scheduling. This schedule is obtained from \cref{fig:mek_circuit} by expanding each highlighted operation and the final logical qubit measurements into their IS+H-based implementation. Only five of the six logical qubits of \code{70} are needed to host the \MEK protocol, leaving one spare logical qubit. This spare qubit is a consequence of the BB/GB two-block group-algebra construction, which forces the number of encoded logical qubits to be even.}
    \label{fig:logical_mek_circuit}
\end{figure*}

\Cref{fig:logical_mek_circuit} expands those operations, together with the final logical-qubit readout, into an IS+H-based schedule. Only five of the six logical qubits encoded by \code{70} are needed to host the \MEK protocol, so the block effectively has one spare logical qubit. This spare qubit arises because BB and GB codes are built from a two-block group-algebra structure, which forces the number of encoded logical qubits to be even. The initial ${\ket{\bar{H}}}$ states are prepared using the \EDM-$3$ injection circuit of \cref{fig:h_state_injection}. Each logical controlled-$\bar{H}$ gadget is decomposed into two \EDM-based logical $R_y(\pm \pi/4)$ injections plus Clifford frame updates, and the intermediate logical $\CZ$ is likewise absorbed into the frame. The only remaining purely logical Viterbi measurements are the three terminal readouts, each of weight at most $18$, timed using \cref{heur:mek_viterbi_runtime}. In total, the schedule uses ten injected states: two to prepare the initial logical $\ket{\bar{H}}$ states and eight for the four controlled-$\bar{H}$ gadgets, since each gadget contains two $R_y$ gates.

\subsubsection{Logical measurements}
\label{subsubsec:mek_logical_measurements}

As in \CHfactory, the MEK schedule uses two measurement primitives: the hybrid
$Y\bar{Y}$ {\EDM}s inside the ten $H$-state injection gadgets, and the three
terminal purely logical Viterbi readouts. The $R_y(\pm \pi/4)$ gates inside the
controlled-$\bar{H}$ gadgets are therefore compiled entirely into \EDM-based
injections and Clifford frame updates. We again use
$\varepsilon_{\mathrm{inj}} = 10^{-5}$ as the round-count target for the hybrid
{\EDM}s; the resulting choice $r=3$ gives an estimated accepted-outcome error
of order $10^{-8}$, numerically about $5.44\times 10^{-8}$ by
\cref{heur:mek_edm_rounds}. The terminal Viterbi measurements are run at the
target $\varepsilon = 10^{-10}$ of \cref{subsec:magic_factory_error_model}.

\begin{heuristic}[Injection-\EDM round count for \MEK/\code{70}]
    For the \MEK/\code{70} factory at the operating point
    $p = 10^{-4}$ and $p_{\log,\mathrm{\code{70}}} = 10^{-10}$,
    choose the hybrid-injection \EDM round count
    \begin{equation}
        \begin{aligned}
            r_{\mathrm{MEK,inj}}
            &=
            \min\bigl\{
                r \geq 1 : \\
            &\qquad
                C_2 p_{\log,\mathrm{\code{70}}}
                + (C_1 w_{\mathrm{MEK,inj}} p)^r
                \leq 10^{-5}
            \bigr\} \\
            &= 3,
        \end{aligned}
        \label{eq:mek_injection_rounds}
    \end{equation}
    with $w_{\mathrm{MEK,inj}} = 18$.
    \label{heur:mek_edm_rounds}
\end{heuristic}

\begin{proof}[Justification]
    By \cref{fig:All_MEK,tab:LogicalConstraints}, the injected logical $Y$ can
    be represented with weight $18$ in \code{70}. Thus
    $w_{\mathrm{MEK,inj}} = 18$. Using \cref{eq:heuristic_EDM} together with the fitted constants
    $C_1 = 2.1$ and $C_2 = 3.4$ from
    \cref{subsubsec:OneCat,subsubsec:MemECM},
    \[
        \begin{aligned}
            C_2 p_{\log,\mathrm{\code{70}}}
            + (C_1 w_{\mathrm{MEK,inj}} p)^2
            &\approx
            3.4 \times 10^{-10} \\
            &\quad + (2.1\times 18\times 10^{-4})^2 \\
            &\approx 1.43\times 10^{-5}
            > 10^{-5},
        \end{aligned}
    \]
    whereas
    \[
        \begin{aligned}
            C_2 p_{\log,\mathrm{Q70}}
            + (C_1 w_{\mathrm{MEK,inj}} p)^3
            &\approx
            3.4 \times 10^{-10} \\
            &\quad + (2.1\times 18\times 10^{-4})^3 \\
            &\approx 5.44\times 10^{-8}
            < 10^{-5}.
        \end{aligned}
    \]
    Hence the smallest admissible \EDM round count is
    $r_{\mathrm{MEK,inj}} = 3$.
\end{proof}

\begin{heuristic}[Viterbi runtime for \MEK/\code{70}]
    For the three terminal logical readouts in the \MEK circuit, use the
    conservative Viterbi runtime estimate
    \begin{equation}
        \tau_{\mathrm{Vit},20}^{\mathrm{avg}}
        \approx 5.10~\mathrm{SEC},
        \label{eq:mek_viterbi_runtime}
    \end{equation}
    which upper bounds the average runtime of the weight-at-most-$18$ Viterbi
    measurements appearing at the end of \cref{fig:logical_mek_circuit}.
    \label{heur:mek_viterbi_runtime}
\end{heuristic}

\begin{proof}[Justification]
    Each of the three terminal logical Pauli readouts in
    \cref{fig:logical_mek_circuit} has weight at most $18$. The timing data in
    \cref{tab:LMtime}, obtained from the Viterbi model of
    \cref{subsec:AdaptiveMeas}, are tabulated for weights $10$, $20$, $30$, and
    $54$. We therefore use the weight-$20$, $\varepsilon = 10^{-10}$ entry as
    a conservative upper bound for the average runtime of the weight-at-most-$18$
    measurements in \MEK.
    giving \cref{eq:mek_viterbi_runtime}.
\end{proof}

\subsubsection{Error analysis}

Under the ideal-logical-operation approximation of \cref{subsec:magic_factory_error_model}, the ten noisy
inputs to one execution of the \MEK protocol are the injected logical $\ket{\bar{H}}$
states, while the rest of the logical circuit is treated as perfect.

\begin{heuristic}[Injected-state error estimate for \MEK/\code{70}]
    For the \MEK factory implemented in the \code{70} block, taking $p=10^{-4}$,
    $r = r_{\mathrm{MEK,inj}} = 3$ from \cref{heur:mek_edm_rounds},
    and the \code{70} SEC depth from \cref{tab:loss_leakage_summary}, the injected-$\bar{Y}$-fault
    probability per accepted logical $\ket{\bar{H}}$ preparation is estimated by
    \begin{equation}
        \begin{aligned}
            q_{Y,\mathrm{MEK}}
            &\approx
            1.93\times 10^{-4}
            + \frac{4 \times 27.70 \times 10^{-4}}{150} \\
            &\approx
            2.67\times 10^{-4}.
        \end{aligned}
        \label{eq:mek_qy}
    \end{equation}
    \label{heur:mek_qy}
\end{heuristic}

\begin{proof}[Justification]
    Applying \cref{heur:h_injection_y_fault_rate} with
    $p = 10^{-4}$ and
    $r = r_{\mathrm{MEK,inj}} = 3$ from
    \cref{heur:mek_edm_rounds}, together with
    \[
        \mathrm{Depth}(\mathrm{SEC})
        =
        d_{\mathrm{POC}}
        =
        1 + 7 + 11.65 + 7.05 + 1
        =
        27.70
    \]
    from the \code{70} SEC time budget in \cref{tab:loss_leakage_summary},
    gives the heuristic estimate \cref{eq:mek_qy}.
\end{proof}

\begin{heuristic}[Verification acceptance probability for \MEK/\code{70}]
    Model the ten logical $\ket{\bar{H}}$ inputs to one \MEK attempt as independent
    twirled resource states, each with stochastic $\bar{Y}$ error probability $q$,
    and assume the remaining logical operations in the factory are perfect.
    Then, upon substituting the heuristic input value
    $q_{Y,\mathrm{MEK}}\approx 2.67\times 10^{-4}$, the
    verification/postselection stage succeeds with probability
    \begin{equation}
        a_{\mathrm{MEK,ver}}(q_{Y,\mathrm{MEK}})
        \approx 0.99733.
        \label{eq:mek_postselection_value}
    \end{equation}
    \label{heur:mek_postselection}
\end{heuristic}

\begin{proof}[Justification]
    Sections~II--III of Ref.~\cite{meier2013magic} define the standard \MEK
    protocol acceptance event: the protocol decodes the outer $[[4,2,2]]$ block and
    accepts only if neither the encoded $\bar{H}^{\otimes 2}$ measurement nor
    the final syndrome indicates an error. Section~III then gives the exact
    expressions
    \begin{equation}
        \begin{aligned}
            a(q)
            &= 1 - 10q + 58q^2 - 192q^3 + 400q^4 \\
            &\quad - 544q^5 + 480q^6 - 256q^7 + 64q^8,
        \end{aligned}
        \label{eq:mek_acceptance_poly}
    \end{equation}
    \begin{equation}
        \begin{aligned}
            u(q)
            &= 9q^2 - 56q^3 + 160q^4 - 256q^5 \\
            &\quad + 240q^6 - 128q^7 + 32q^8,
        \end{aligned}
        \label{eq:mek_marginal_poly}
    \end{equation}
    \begin{equation}
        \begin{aligned}
            u_2(q)
            &= 13q^2 - 80q^3 + 228q^4 - 368q^5 \\
            &\quad + 352q^6 - 192q^7 + 48q^8,
        \end{aligned}
        \label{eq:mek_pair_poly}
    \end{equation}
    Here $a(q)$ is the total acceptance probability of the \MEK protocol,
    $u(q)$ is the accepted probability that a specified output state is bad,
    and $u_2(q)$ is the accepted probability that at least one of the two
    output states is bad.
    Section~III also shows that every single input $\bar{Y}$ fault is rejected. Let
    $P_1$ denote the accepted probability that only output $1$ is bad, let
    $P_2$ denote the same event for output $2$, and let $P_{12}$ denote the
    accepted probability that both outputs are bad. By symmetry,
    \[
        u(q) = P_1 + P_{12} = P_2 + P_{12},
        \qquad
        u_2(q) = P_1 + P_2 + P_{12}.
    \]
    Therefore the odd-parity accepted branch, in which exactly one output is in
    $\ket{\bar{Y}\bar{H}}$, has probability
    \begin{equation}
        \begin{aligned}
            2\bigl(u_2(q)-u(q)\bigr)
            &=
            8q^2 - 48q^3 + 136q^4 - 224q^5 \\
            &\quad + 224q^6 - 128q^7 + 32q^8,
        \end{aligned}
        \label{eq:mek_odd_poly}
    \end{equation}
    so the branch that yields a standard pair of logical $\ket{\bar{H}}$ states
    has probability
    \begin{equation}
        a_{\mathrm{MEK,ver}}(q)
        =
        a(q)-2\bigl(u_2(q)-u(q)\bigr).
        \label{eq:mek_postselection}
    \end{equation}
    Evaluating at the injected-state estimate \cref{eq:mek_qy} gives
    $2\bigl(u_2(q_{Y,\mathrm{MEK}})-u(q_{Y,\mathrm{MEK}})\bigr)
    \approx 5.70\times 10^{-7}$, together with
    \cref{eq:mek_postselection_value}.
\end{proof}

\begin{heuristic}[Output error estimate for the \MEK factory]
    Conditioned on the successful branch counted by
    \cref{heur:mek_postselection}, each \MEK output state has estimated error
    \begin{equation}
        p_{\mathrm{MEK,out}} \approx e_{\mathrm{even}}(q_{Y,\mathrm{MEK}})
        \approx 3.6\times 10^{-7}.
        \label{eq:mek_output_error}
    \end{equation}
    \label{heur:mek_output_error}
\end{heuristic}

\begin{proof}[Justification]
    In the same idealized model, the even-parity bad branch has probability
    $P_{12} = 2u(q)-u_2(q)$, so the conditional per-output-state error is
    \begin{equation}
        e_{\mathrm{even}}(q)
        =
        \frac{2u(q)-u_2(q)}{a_{\mathrm{MEK,ver}}(q)}.
        \label{eq:mek_conditional_error}
    \end{equation}
    Evaluating \cref{eq:mek_conditional_error} at the heuristic input value
    from \cref{eq:mek_qy}, and using \cref{heur:mek_postselection}, gives
    \cref{eq:mek_output_error}.
\end{proof}

\subsubsection{Time analysis}
\label{subsubsec:mek_time_analysis}

We now translate the expanded logical schedule of
\cref{fig:logical_mek_circuit} into an SEC cost.

\begin{heuristic}[Per-attempt failure rate estimate for the \MEK factory]
    Assuming the ten injection \EDM outcomes are independent, the
    \MEK/\code{70} implementation has estimated per-attempt failure rate
    \begin{equation}
        p_{\mathrm{MEK,fail}}
        \approx
        1-a_{\mathrm{MEK,inj}}\,a_{\mathrm{MEK,ver}}
        \approx 11.97\%.
        \label{eq:mek_failure_rate}
    \end{equation}
    \label{heur:mek_failure_rate}
\end{heuristic}

\begin{proof}[Justification]
    The first source of failure is the ten \EDM-$3$ $H$-state injections from
    \cref{heur:mek_edm_rounds}. As in
    the \CHfactory analysis, we model each injection restart with
    \cref{heur:h_injection_retry}. For the \code{70} implementation,
    \cref{fig:All_MEK,tab:LogicalConstraints} show that logical $Y$
    representatives have weight at most $18$.
    Therefore, using the fitted value $C_1=2.1$ from
    \cref{subsubsec:OneCat}, together with
    $r = r_{\mathrm{MEK,inj}} = 3$ from \cref{heur:mek_edm_rounds},
    $p=10^{-4}$, $\ploss=10^{-7}$, and $\pleak=10^{-5}$,
    \[
        \begin{aligned}
            \pflip
            &\approx C_1 w p
            = 2.1\times 18 \times 10^{-4}
            = 3.78\times 10^{-3}, \\
            p_{\mathrm{anc,det}}
            &\approx
            4\times 27.70 \times (10^{-7}+10^{-5}) \\
            &= 1.11908\times 10^{-3}
            \approx 1.12\times 10^{-3}, \\
            p_{H,\mathrm{retry}}
            &\approx
            1
            - \Bigl((1-\pflip)^3 \\
            &\qquad\quad + \pflip^3\Bigr) \\
            &\quad \times (1-p_{\mathrm{anc,det}}) \\
            &\approx 1.24\times 10^{-2}.
        \end{aligned}
    \]
    Hence one injection is accepted with probability
    \[
        a_{H,\mathrm{acc}}
        =
        1-p_{H,\mathrm{retry}}
        \approx 0.9876,
    \]
    and therefore all ten injections succeed with probability
    \[
        a_{\mathrm{MEK,inj}}
        =
        a_{H,\mathrm{acc}}^{10}
        \approx
        0.883.
    \]

    The second source of failure is rejection at the \MEK
    verification/postselection stage. By
    \cref{heur:mek_postselection},
    \[
        a_{\mathrm{MEK,ver}}
        \approx
        a_{\mathrm{MEK,ver}}(q_{Y,\mathrm{MEK}})
        \approx
        0.99733.
    \]
    Combining the ten injection acceptances with this verification/postselection
    success probability gives
    $p_{\mathrm{MEK,fail}} \approx 1-a_{\mathrm{MEK,inj}}a_{\mathrm{MEK,ver}} \approx 11.97\%$,
    which is the heuristic estimate \cref{eq:mek_failure_rate}.
\end{proof}

\begin{heuristic}[Average runtime estimate for the \MEK factory]
    For the \MEK/\code{70} implementation, taking the logical-measurement parameters
    from \cref{heur:mek_edm_rounds,heur:mek_viterbi_runtime}, the expected SEC count per
    accepted \MEK output pair is
    \begin{equation}
        N_{\mathrm{MEK,SEC}}^{\mathrm{avg}}
        \approx
        47.6~\mathrm{SEC},
        \label{eq:mek_avg_runtime}
    \end{equation}
    or about $23.8~\mathrm{SEC}$ per logical $\ket{\bar{H}}$ state.
    \label{heur:mek_avg_runtime}
\end{heuristic}

\begin{proof}[Justification]
    Write $N_{\mathrm{MEK,SEC}}^{\mathrm{avg}}$ for the expected SEC count
    until one successful \MEK output pair. A first-step decomposition over the
    three possible outcomes of one \MEK attempt gives
    \[
        \begin{aligned}
            N_{\mathrm{MEK,SEC}}^{\mathrm{avg}}
            &\approx
            \sum_{j=1}^{10}
            a_{H,\mathrm{acc}}^{j-1} p_{H,\mathrm{retry}} \\
            &\qquad \times
            \left(D_j + N_{\mathrm{MEK,SEC}}^{\mathrm{avg}}\right) \\
            &\quad +
            a_{\mathrm{MEK,inj}} (1-a_{\mathrm{MEK,ver}}) \\
            &\qquad \times
            \left(C_{\mathrm{MEK}} + N_{\mathrm{MEK,SEC}}^{\mathrm{avg}}\right) \\
            &\quad +
            a_{\mathrm{MEK,inj}} a_{\mathrm{MEK,ver}} C_{\mathrm{MEK}}.
        \end{aligned}
    \]
    Here $D_j$ is the accrued SEC count if the $j$th injection is the first one
    to fail, and $C_{\mathrm{MEK}}$ is the SEC cost of a complete \MEK schedule
    up to and including the final logical readouts.

    Each logical $\ket{\bar{H}}$ preparation is implemented with an
    \EDM-$r_{\mathrm{MEK,inj}}$ measurement from
    \cref{heur:mek_edm_rounds}, so
    \[
        I = r_{\mathrm{MEK,inj}} = 3~\mathrm{SEC}.
    \]
    For the terminal Viterbi measurements in \cref{fig:logical_mek_circuit}, we use the
    conservative runtime estimate from \cref{heur:mek_viterbi_runtime} and
    set
    \[
        V = \tau_{\mathrm{Vit},20}^{\mathrm{avg}} = 5.10~\mathrm{SEC}.
    \]
    One successful \MEK schedule uses ten logical $\ket{\bar{H}}$
    preparations in total: two standalone injections, plus two injections
    inside each of the four logical controlled-$\bar{H}$ gadgets.
    Only the three terminal logical readouts require Viterbi measurements.
    Therefore
    \[
        C_{\mathrm{MEK}} = 10I + 3V \approx 45.3~\mathrm{SEC}.
    \]

    Index the ten logical $\ket{\bar{H}}$ preparations in
    \cref{fig:logical_mek_circuit} by the order in which their $H$-state
    injection gadgets are executed. Since all ten injections occur before the
    three terminal readouts, if the $j$th injection is the first one to fail
    then the accrued SEC cost is simply $jI$. Therefore
    \[
        \begin{aligned}
            D_1 &= I, &
            D_2 &= 2I, &
            D_3 &= 3I, \\
            D_4 &= 4I, &
            D_5 &= 5I, \\
            D_6 &= 6I, &
            D_7 &= 7I, \\
            D_8 &= 8I, &
            D_9 &= 9I, \\
            D_{10} &= 10I.
        \end{aligned}
    \]
    and from the justification of \cref{heur:mek_failure_rate},
    \[
        \begin{aligned}
            p_{H,\mathrm{retry}} &\approx 1.24\times 10^{-2}, \\
            a_{H,\mathrm{acc}} &= 1-p_{H,\mathrm{retry}} \approx 0.9876, \\
            a_{\mathrm{MEK,inj}} &\approx 0.883, \\
            a_{\mathrm{MEK,ver}} &\approx 0.99733.
        \end{aligned}
    \]
    Substituting these values into the recursion, first collect the terms
    proportional to $N_{\mathrm{MEK,SEC}}^{\mathrm{avg}}$:
    \[
        \begin{aligned}
            N_{\mathrm{MEK,SEC}}^{\mathrm{avg}}
            &\approx
            p_{H,\mathrm{retry}}
            \sum_{j=1}^{10} a_{H,\mathrm{acc}}^{j-1} D_j \\
            &\quad +
            a_{\mathrm{MEK,inj}} C_{\mathrm{MEK}} \\
            &\quad +
            p_{H,\mathrm{retry}}
            \sum_{j=1}^{10} a_{H,\mathrm{acc}}^{j-1} \\
            &\qquad \times
            N_{\mathrm{MEK,SEC}}^{\mathrm{avg}} \\
            &\quad +
            a_{\mathrm{MEK,inj}} (1-a_{\mathrm{MEK,ver}})
            N_{\mathrm{MEK,SEC}}^{\mathrm{avg}}.
        \end{aligned}
    \]
    Since
    \[
        \begin{aligned}
            p_{H,\mathrm{retry}}
            \sum_{j=1}^{10} a_{H,\mathrm{acc}}^{j-1}
            &= 1-a_{H,\mathrm{acc}}^{10}
            = 1-a_{\mathrm{MEK,inj}},
        \end{aligned}
    \]
    \[
        \begin{aligned}
            &p_{H,\mathrm{retry}}
            \sum_{j=1}^{10} a_{H,\mathrm{acc}}^{j-1} \\
            &\quad + a_{\mathrm{MEK,inj}}(1-a_{\mathrm{MEK,ver}}) \\
            &\quad =
            (1-a_{\mathrm{MEK,inj}}) \\
            &\qquad +
            a_{\mathrm{MEK,inj}}(1-a_{\mathrm{MEK,ver}}) \\
            &\qquad = 1-a_{\mathrm{MEK,inj}}a_{\mathrm{MEK,ver}},
        \end{aligned}
    \]
    this becomes
    \[
        \begin{aligned}
            N_{\mathrm{MEK,SEC}}^{\mathrm{avg}}
            &\approx
            p_{H,\mathrm{retry}}
            \sum_{j=1}^{10} a_{H,\mathrm{acc}}^{j-1} D_j \\
            &\quad +
            a_{\mathrm{MEK,inj}} C_{\mathrm{MEK}} \\
            &\quad +
            (1-a_{\mathrm{MEK,inj}}a_{\mathrm{MEK,ver}})
            N_{\mathrm{MEK,SEC}}^{\mathrm{avg}}.
        \end{aligned}
    \]
    Rearranging gives
    \[
        \begin{aligned}
            S_{\mathrm{MEK}}
            &=
            p_{H,\mathrm{retry}}
            \sum_{j=1}^{10} a_{H,\mathrm{acc}}^{j-1} D_j \\
            &\quad +
            a_{\mathrm{MEK,inj}} C_{\mathrm{MEK}},
        \end{aligned}
    \]
    so that
    \[
        \begin{aligned}
            a_{\mathrm{MEK,inj}}a_{\mathrm{MEK,ver}}
            N_{\mathrm{MEK,SEC}}^{\mathrm{avg}}
            &\approx S_{\mathrm{MEK}}, \\
            N_{\mathrm{MEK,SEC}}^{\mathrm{avg}}
            &\approx
            \frac{S_{\mathrm{MEK}}}
            {a_{\mathrm{MEK,inj}}a_{\mathrm{MEK,ver}}}.
        \end{aligned}
    \]
    Numerically,
    \[
        \begin{aligned}
            S_{\mathrm{MEK}}
            &\approx
            (1.24\times 10^{-2})
            \sum_{j=1}^{10} (0.9876)^{j-1} D_j \\
            &\quad +
            (0.883)(45.3) \\
            &\approx 41.88,
        \end{aligned}
    \]
    so
    \[
        \begin{aligned}
            N_{\mathrm{MEK,SEC}}^{\mathrm{avg}}
            &\approx
            \frac{41.88}{(0.883)(0.99733)} \\
            &\approx
            47.6~\mathrm{SEC}.
        \end{aligned}
    \]
    Dividing by the two logical $\ket{\bar{H}}$ outputs gives
    $23.8~\mathrm{SEC}$ per output state.
\end{proof}

\section{The qubit factory}
\label{sec:The qubit factory}

The walking cat architecture comprises multiple copies each of four distinct types of subsystems, denoted M (memory blocks), T (magic factories), C (Cat factories), and B (Bell factories).
During each SEC, every subsystem independently experiences qubit losses. 

To maintain continuous operation and ease the burden of reloading qubits, we utilize a global reservoir, referred to in this section simply as the reservoir, of maximum capacity $R$ that replenishes the qubits lost across all copies of M, T, C, and B. Concurrently, $L$ loading zones each produce a new qubit at an average rate of one per second to refill the reservoir. We assume that a critical system failure occurs if the aggregate qubit loss across all subsystems during an SEC exceeds the current available inventory in the reservoir, implying that at least one subsystem will be unable to replace its lost qubits. 

Our goal in this section is to determine a minimum reservoir size that guarantees the long-term probability of such a system failure remains below a chosen threshold $\epsilon = 10^{-10}$, given a specific number of loading zones and an exact allocation of the subsystems. 
In \cref{tab: qubit factory operating points}, we report values of $L$ and $R$ for which the reservoir will operate at this threshold for two different allocations when using \code{102} for memory and \code{54} for the magic factory.
These values are chosen from \cref{fig: lr_curves}; the particular values we report in \cref{tab: qubit factory operating points} represent a point of diminishing returns where increasing the number of loading zones offers a minimal reduction in required reservoir size. This demonstrates that continuous system operation can be maintained in the presence of qubit loss, with a reasonable reservoir size and number of loading zones.

In \cref{subsec:reservoir DTMC}, we introduce a discrete-time Markov chain model for the reservoir, the qubit loss distributions we assume, and the method for determining operating values of $L$ and $R$ needed to maintain at most $\epsilon$ failure probability.

\begin{table}[b!]
    \centering
    \begin{tabular}{*{6}{>{\centering\arraybackslash}p{0.3in}}}
        \hline
        \textbf{\#M}&
        \textbf{\#T} &
        \textbf{\#C} &
        \textbf{\#B} &
        $L$ &
        $R$\\
        \hline
        20& 20& 40& 5& 28& 188\\
        5& 5& 10& 2& 15& 139\\
        \hline
    \end{tabular}
    \caption{Choices for number of qubit loading zones $L$ and reservoir size $R$ to operate below the critical system failure probability $\epsilon=10^{-10}$ for two different  allocations of the numbers of memory blocks (\textbf{\#M}), $T$ factories (\textbf{\#T}), cat factories (\textbf{\#C}), and Bell factories (\textbf{\#B}). We assume here that the memory code is \code{102} and that the magic factory code is \code{54}.}\label{tab: qubit factory operating points}
\end{table}

\subsection{Discrete-time Markov chain model}\label{subsec:reservoir DTMC}

\begin{figure}[t!]
    \centering
    \includegraphics[width=\linewidth]{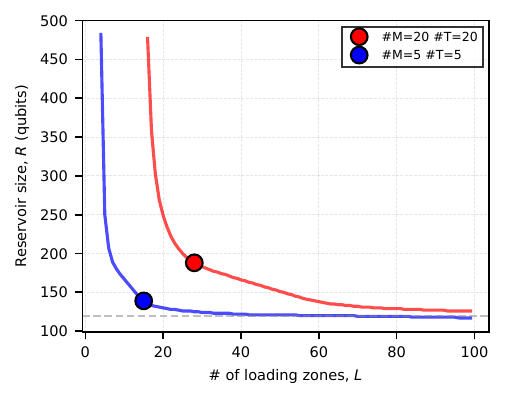}
    \caption{\textbf{Reservoir Capacity ($R$) vs. Number of Loading Zones ($L$).} This figure displays the minimum required reservoir capacity $R$ to maintain a system failure probability of $\le 10^{-10}$ across a varying number of loading zones $L$, and for different allocations of M and T blocks. We assume that each M and T block has an associated C block, so \#C$=$\#M+\#T. For simplicity, we take \#B$=\lceil$\#C/9$\rceil$.
    The asymptotic vertical behavior indicates the minimum number of loading zones required to maintain operation above the system failure threshold; asymptotic horizontal behavior shows that a minimum of 120 qubits are needed to avoid system failure, consistent with the case where 2 C components (containing 60 qubits each) each lose all of their qubits in a given SEC.}
    \label{fig: lr_curves}
\end{figure}

\begin{figure}[t!]
    \centering
    \includegraphics[width=\linewidth]{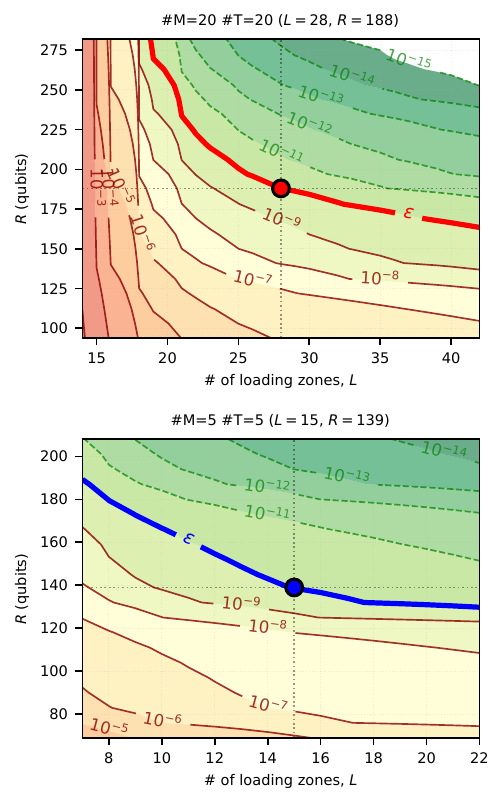}
    \caption{For each of the subsystem allocations from \cref{fig: lr_curves}, we consider how the probability of system failure changes as the chosen operating point values of $L$ loading zones and reservoir size $R$ are perturbed by up to $50\%$ each. Isolines on the 2D plots correspond to configurations of $L$ and $R$ that achieve an order of magnitude deviation from $\epsilon=10^{-10}$.}
    \label{fig: topographical failure}
\end{figure}

To determine a reservoir size and number of loading zones needed to operate below critical failure, we model the occupancy of the reservoir after each SEC as a discrete-time Markov chain. The state space is defined as the set of integers from $0$ to $R$, representing the current number of available qubits.
In the context of our system, the steady state distribution of this Markov chain represents the equilibrium behavior of the reservoir after an infinite number of SECs. Given a configuration of $L$ and $R$, we approximate system failure as the steady-state probability that the reservoir is empty.

To model the system, we assume the following:
\begin{itemize}
    \item \textbf{Discrete Time Steps:} The system dynamics are evaluated at discrete intervals corresponding to the fixed SEC time, $\Delta t$. All loss and reloading events are aggregated at these time steps. Losses in the components are governed by subsystem-specific probability mass functions, $p^{(S)}_k$, representing the probability that a single subsystem of type $S \in \{\text{M, T, C, B}\}$ loses exactly $k$ qubits during one SEC.
    \item \textbf{Aggregated Loss Distribution:} The total number of qubits lost in a single SEC is the sum of the losses from all individual subsystems. Because the individual subsystems operate independently, the aggregate loss probability distribution, $P_{\text{total loss}}$, can be computed as the convolution of the individual subsystem distributions.
    \item \textbf{Binary Loading:} After each SEC, the $L$ loading zones collectively add either zero or one qubit. As loading zones add approximately 1 qubit per second each to the reservoir, in the model we assume that after a single SEC a single qubit is added with probability $P_{\mathrm{add}}(1) = L\cdot \Delta t$. The total stored qubits cannot exceed the reservoir capacity $R$.
\end{itemize}

Denote the random variable of the state of the reservoir at time step $t$ by $X_t$. During a single SEC, let $k$ be the number of qubits lost across all subsystems, governed by the convolved probability distribution $P_{\mathrm{loss}}(k)$. Let $a$ be the number of qubits added, governed by the binary distribution $P_{\mathrm{add}}(a)$ where $a\in\{0,1\}$. The value of the system at time step $t+1$ is given by
$$X_{t+1} = \max(0, \min(R, X_t - k + a)).$$

The transition matrix $P$ of size $(R+1) \times (R+1)$ is constructed by marginalizing over all possible loss and addition events; the transition probability from state $i$ to state $j$ is given by
$$P_{i,j} = \sum_{k,a \colon X_{i+1} = j} P_{\text{total loss}}(k) \cdot P_{\mathrm{add}}(a).$$

The steady-state distribution vector $\pi$ describes the long-term, time-independent probability of finding the reservoir in any given state, and is given by the normalized left eigenvector of the transition matrix $P$ corresponding to the eigenvalue $\lambda = 1$.
The probability of critical system failure corresponds to the first element of the distribution vector, $\pi_0$. We require this failure probability to be strictly less than the critical system failure threshold, $\pi_0<\epsilon=10^{-10}$.

For the numerical simulations of this section, we make the following choices:
\begin{itemize}
    \item The loss distribution $p_k^{(M)}$ for the memory blocks is taken to be the distribution for \code{102} reported in \cref{tab:loss_leakage_summary}.
    \item The loss distribution $p_k^{(T)}$ for the T factories is taken to be the distribution for \code{54} reported in \cref{tab:loss_leakage_summary}.
    \item The loss distribution $p_k^{(C)}$ for the cat factories is taken to be the distribution from \cref{tab:cat_heuristic} with weight $w=30$, $m=2$ verification rounds, and $p_{\mathrm{loss}}=10^{-7}$. The cat factories associated with the magic factories could, in principle, produce smaller-weight cat states, so this choice is pessimistic.
    \item The loss distribution $p_k^{(B)}$ for the Bell factories is taken to have only two non-trivial contributions $p_2^{(B)}=3.6\times 10^{-6}$ and $p_0^{(B)}=1-p_2^{(B)}$, \emph{i.e.}, we suppose that with probability $p_2^{(B)}$ the Bell factory fails to produce a Bell state.
    \item We assume an SEC time of $\Delta t=6\times 10^{-3}$, so the probability of successfully adding a qubit at the end of a single SEC is taken to be $P_{\mathrm{add}}(1)=0.006 L$.
\end{itemize}

Because the probability of an empty reservoir strictly decreases when increasing the maximum reservoir capacity $R$ for a fixed value of $L$, we compute the minimal required capacity for a given number of loading zones using binary search. We define a search space between a minimum capacity of $1$ and a predefined maximum practical limit $R_{\max}=500$. By fixing an $L$ and iteratively computing the failure probability for the value of $R$ at the midpoint of the search space, we converge on the smallest reservoir size for which the system failure probability remains below $\epsilon$ for the given number of loading zones.

In \cref{fig: lr_curves}, we report the relationship between $L$ and $R$ operating at this critical threshold $\epsilon$ for different allocation and loss rate scenarios. For each of these scenarios, we choose the operating values of $L$ and $R$ for which increasing $L$ by one only decreases the required reservoir size by two qubits. This choice is simply meant to represent a point at which adding more loading zones does not result in a drastic reduction of reservoir size; any point on the curve would maintain continuous operation below the $\epsilon$ threshold. In \cref{fig: topographical failure}, we examine the effect on the total system failure probability when perturbing the values for this operating point.

\begin{figure}[t!]
    \centering
    \includegraphics[width=\linewidth]{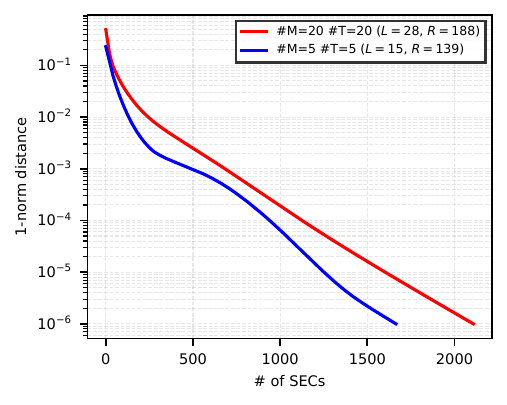}
    \caption{\textbf{1-norm distance to the steady-state distribution vs. the number of SECs.} By $\approx250$ SECs, both distributions are within $1\%$ of the steady state.}
    \label{fig: mixing curves}
\end{figure}

To validate our DTMC model, we demonstrate rapid convergence of the reservoir state to the steady-state distribution in \cref{fig: mixing curves}.

\section{The decoder}
\label{sec:decoder}

\begin{table*}[]
    \centering
    \begin{tabular}{|c|c|c|c|c|}
    \hline
        Code & Latency Measure & Mean (ms) & 99 percentile (ms) & 99.9 percentile (ms) \\
        \hline
        Q70  & reaction time  & 0.3484 & 0.5400 & 0.7934 \\
        \hline
        Q70  & decoding time per SEC  & 0.1468 & 0.2443 & 0.3257 \\
        \hline
        Q102  & reaction time  & 0.8462 & 1.2859 & 1.6521 \\
        \hline
        Q102 & decoding time per SEC  & 0.3969 & 0.7084 & 1.4936 \\
        \hline
    \end{tabular}
    \caption{Statistics of decoding time per SEC and reaction time at physical error rate $10^{-4}$ using a $(5,3)$ sliding window beam search decoder. Reaction time is the interval between the final qubit measurement and the acquisition of the final decoding result. Decoding time statistics are derived from a single run of $1,000,008$ SECs ($333,336$ windows). Reaction time statistics are generated from $10,000$ independent runs of $1,000,008$ SECs each.}
    \label{tab:time_stats_summary_1e-4}
\end{table*}

In this section, we describe the streaming version of the beam search decoder proposed in \cite{ye2025beam}. The beam search decoder and the widely used BP-OSD decoder are both built upon Belief Propagation (BP). Beam search improves upon the BP-OSD framework by significantly reducing worst-case decoding latency while simultaneously achieving a lower logical error rate \cite{ye2025beam}. We use the streaming beam search decoder to decode the Q70 code and the Q102 code. To provide a comprehensive assessment, we analyze the logical error rate per SEC alongside key decoding latency metrics. The statistics of these latency metrics, derived from memory experiments spanning one million SECs, are summarized in \cref{tab:time_stats_summary_1e-4}. Our simulation results show that the streaming decoder remains within $2\times$ of the global decoder in logical error rate across both codes and all tested physical error rates; see \cref{fig:stream_global_compare_logical_error_rate}.

\subsection{Background on streaming decoder}

A common benchmark for evaluating new decoders is the simulation of the logical error rate over $d$ SECs, where $d$ is the code distance. In this global decoding approach, the decoder processes a linear combination of all syndrome measurements as a single batch to produce a correction. While effective for benchmarking, this presents a significant scalability challenge for practical quantum computing. Real-world algorithms may span thousands or millions of SECs. Waiting for the entire sequence of SECs to finish before initiating a global decoding would introduce prohibitive latency and demand massive memory overhead to store the accumulated syndrome data. Furthermore, fault-tolerant execution of non-Clifford operations often requires active feedback, where subsequent gates depend on the results of intermediate logical measurements. To obtain these logical results reliably, the system must decode syndromes in real time to resolve physical errors before they propagate.
A prominent example is the measurement-based $R_y(\pi/4)$ gate via gate teleportation with the magic state $\ket{H}$; see \cref{fig:logical_ry_measurement_based}. The conditional Clifford correction $R_y(\pi/2)$ affects future logical measurements on the same qubit because it is implemented through Clifford frame tracking.
For these two reasons, a transition from global decoding to continuous, real-time streaming decoding is essential for viable fault-tolerant quantum systems.

Streaming decoders, also referred to as sliding window decoders, were initially proposed only for surface codes \cite{dennis2002topological, tan2023scalable, skoric2023parallel}. More recently, these techniques have been generalized to support the broader class of quantum LDPC codes \cite{huang2024increasing, gong2024toward}.

\subsection{Structure of the parity-check matrix}

The sliding window decoder exploits the fact that the parity-check matrix for repetitive syndrome extraction naturally adopts a staircase structure.  By leveraging the localized dependency in the staircase structure, we can utilize a limited temporal window of syndrome data to resolve errors in the earliest part of that window. The decoder then ``slides" forward by a fixed number of SECs to process subsequent data. An inner decoder is applied to each window to identify a minimum-weight correction, ensuring that any performance degradation relative to global decoding remains within acceptable limits.

Let $\mathbf{s}_0, \mathbf{s}_1, \mathbf{s}_2, \dots$ represent the syndrome vectors measured from successive SECs. Rather than processing these raw syndromes directly, we transform them into detector outcomes, defined as $\mathbf{d}_0 = \mathbf{s}_0$ and $\mathbf{d}_i = \mathbf{s}_i \oplus \mathbf{s}_{i-1}$ for $i \ge 1$, where $\oplus$ denotes the bitwise XOR operation. This transformation is critical as it yields a parity-check matrix with a sparse, staircase structure. As an illustration, the parity-check matrix for six SECs exhibits the following banded staircase structure:
\begin{equation} \label{eq:pcm_6cycles}
\begin{bmatrix}
H_0' & H_1 \\
 & H_2 & H_0 & H_1    \\
 & & & H_2 & H_0 & H_1  \\
 & & & & & H_2 & H_0 & H_1 \\
 & & & & & & & H_2 & H_0 & H_1 \\
 & & & & & & & & & H_2 & H_0'' \\
\end{bmatrix}
\end{equation}
Each column of this matrix corresponds to a potential error source, and the statistical likelihood of these errors is captured by the prior probability vector. The length of this vector matches the number of columns in the parity-check matrix, with the $i$-th coordinate representing the prior probability that the $i$-th error source is active. For the six-SEC case, the prior probability vector follows a block structure corresponding to the 11 block columns in \cref{eq:pcm_6cycles}:
\begin{equation} \label{eq:priorp_6cycles}
(\mathbf{p}_0', \mathbf{p}_1, \mathbf{p}_0, \mathbf{p}_1, \mathbf{p}_0, \mathbf{p}_1, \mathbf{p}_0, \mathbf{p}_1, \mathbf{p}_0, \mathbf{p}_1, \mathbf{p}_0'').
\end{equation}
The dimensions of these sub-vectors are consistent with their associated block columns: the lengths of $\mathbf{p}_0'$, $\mathbf{p}_0''$, and $\mathbf{p}_0$ match the number of columns in $H_0'$, $H_0''$, and $H_0$, respectively, while the length of $\mathbf{p}_1$ matches the number of columns in $H_1$ (and by extension, $H_2$).

The following proposition characterizes the general structure of the parity-check matrix and the prior probability vector.

\begin{proposition} \label{prop:pcm_staircase}
The parity-check matrix corresponding to $r$ repeated SECs can be arranged into an $r \times (2r-1)$ block matrix form. Both block column indices and block row indices are assumed to start from 0. The matrix structure is defined as follows:
\begin{itemize}
\item \textbf{Even Block Columns:} For $0 \le i \le r-1$, block column $2i$ contains a single non-zero block located in the $i$-th block row.
\item \textbf{Odd Block Columns:} For $0 \le i \le r-2$, block column $2i+1$ contains two non-zero blocks, situated in the $i$-th and $(i+1)$-th block rows.
\item \textbf{Uniformity:} The non-zero blocks in the intermediate odd block columns ($1, 3, \dots, 2r-3$) are identical, as are the non-zero blocks in the intermediate even block columns ($2, 4, \dots, 2r-4$).
\item \textbf{Boundary Conditions:} The non-zero blocks in the initial and final block columns (indices $0$ and $2r-2$) differ from the internal blocks due to temporal boundary effects.
\end{itemize}
The prior probability vector can be partitioned into $2r-1$ sub-vectors $(\mathbf{q}_0, \mathbf{q}_1, \dots, \mathbf{q}_{2r-2})$, where the length of each $\mathbf{q}_j$ matches the number of columns in block column $j$ of the parity-check matrix.
All sub-vectors with odd indices ($j \in \{1, 3, \dots, 2r-3\}$) are identical.
All intermediate sub-vectors with even indices $j \in \{2, 4, \dots, 2r-4\}$ are identical.
\end{proposition}

While the staircase structure and the translational invariance of the intermediate block columns were established in \cite{gong2024toward}, the specific boundary conditions that break this uniformity have not been addressed in prior literature. Specifically, in \cite{gong2024toward}, both $H_0'$ in the initial block column and $H_0''$ in the last block column were written as $H_0$. Accounting for these boundary effects is essential for the correct initialization of the first and terminal decoding windows.

To motivate the structure described in this proposition, we classify error events into four types:
\begin{itemize}
\item Type I (Transient Errors): These errors occur during the $i$-th SEC and trigger non-zero syndrome entries only in $\mathbf{s}_i$. Upon transformation to detector outcomes, they manifest as non-zero entries in both $\mathbf{d}_i$ and $\mathbf{d}_{i+1}$. A primary example is a measurement error on an ancilla qubit, which incorrectly reports a parity value that is immediately ``cleared" in the next SEC.

\item Type II (Early-Cycle Persistent Errors): These errors occur in the $i$-th SEC and trigger identical non-zero entries in all syndrome vectors $\mathbf{s}_j$ for $j \ge i$. The detector transformation significantly sparsifies the parity-check matrix by localizing the fault to non-zero entries within a single detector vector $\mathbf{d}_i$. A typical example is a Pauli error on a data qubit occurring before it undergoes any two-qubit gates with ancilla qubits during the $i$-th SEC.

\item Type III (Mid-Cycle Persistent Errors): These errors occur during the $i$-th SEC and trigger non-zero entries in all $\mathbf{s}_j$ for $j \ge i$. While the syndrome vectors are uniform for all $j > i$, the initial syndrome $\mathbf{s}_i$ differs from this steady state. This occurs when a data qubit error arises between its interactions with different ancilla qubits. For example, if a data qubit participates in four stabilizer checks, a mid-cycle error may flip only a subset of the four corresponding ancilla qubits in $\mathbf{s}_i$, whereas all four will be flipped in all subsequent $\mathbf{s}_j$ ($j > i$). In the detector basis, this event is localized to $\mathbf{d}_i$ and $\mathbf{d}_{i+1}$.

\item Type IV (Late-Cycle Persistent Errors): These errors occur in the $i$-th SEC after the data qubit has completed all interactions with ancilla qubits. Consequently, the error remains ``invisible" to the $i$-th syndrome extraction, triggering identical non-zero entries in $\mathbf{s}_j$ only for $j \ge i+1$. After transformation, this fault is localized solely to $\mathbf{d}_{i+1}$.
\end{itemize}

The block structure defined in \cref{prop:pcm_staircase} is a direct consequence of these four error types. Type II and Type IV errors—which manifest as single-detector events—populate the even block columns, while Type I and Type III errors—which trigger two adjacent detectors—generate the odd block columns. The observed uniformity of the internal blocks stems from the periodic nature of the syndrome extraction circuit; since each SEC is identical, the resulting spatiotemporal sub-matrices repeat consistently.
The deviations from this uniformity occur at the temporal boundaries:
\begin{itemize}
    \item The Initial Block Column: The first even block column (index $0$) differs from the internal even columns because the latter incorporate contributions from both Type II and Type IV errors. A Type IV error manifesting in the $i$-th detector vector actually originates from a fault in the preceding $(i-1)$-th SEC. Since there is no SEC prior to the initial measurement, Type IV errors are absent from the first column. Consequently, the initial block column is constituted solely of Type II errors, resulting in a distinct structural profile compared to the internal columns.
    
    \item The Final Block Column: The terminal even block column (index $2r-2$) differs from its internal counterparts due to the truncation of the syndrome extraction sequence. Specifically, a Type I error (such as a measurement error) occurring in the final SEC can only trigger the terminal detector vector $\mathbf{d}_{r-1}$. Because the sequence terminates, there is no subsequent $\mathbf{d}_r$ to capture the second half of the transient error signature. Consequently, while this error event would normally populate an odd block column (triggering two detectors) in any intermediate SEC, it is compressed into the terminal even block column (triggering only one detector) when it occurs at the final boundary.
\end{itemize}

The sliding window decoder is governed by two key parameters, denoted as the tuple $(w, c)$. The first parameter, window size $w$, defines the number of SECs the decoder considers at once. The second parameter, commit size $c$, specifies how many of the oldest SECs are finalized before the window slides forward.

\subsection{An illustrative example}

As an illustrative example, consider a $(3, 1)$ sliding window decoder processing six SECs. The parity-check matrix takes the form shown in \cref{eq:pcm_6cycles}. We partition the error vector into 11 sub-vectors $(e_0, e_1, \dots, e_{10})$, where each sub-vector corresponds to a block column in \cref{eq:pcm_6cycles}.
The decoder receives six detector vectors $\mathbf{d}_0, \mathbf{d}_1, \dots, \mathbf{d}_5$ as input. In the first window, the decoder processes only the first three detector vectors $(\mathbf{d}_0, \mathbf{d}_1, \mathbf{d}_2)$ and utilizes the first three block rows of \cref{eq:pcm_6cycles} to form the active sub-parity-check matrix
\begin{equation} \label{eq:pcm_sub1}
H_{\text{win}1} = \begin{bmatrix}
H_0' & H_1 \\
 & H_2 & H_0 & H_1    \\
 & & & H_2 & H_0 & H_1  \\
\end{bmatrix}.
\end{equation}
In this example, $H_{\text{win}1}$ is a $3\times 6$ block matrix because we set $w=3$. In general, $H_{\text{win}1}$ is a $w\times (2w)$ block matrix.
The decoder also extracts the corresponding sub-vectors from the global prior probability vector \cref{eq:priorp_6cycles} to construct the active prior probability vector for the window:
$$
\mathbf{p}_{\text{win}1} = (\mathbf{p}_0', \mathbf{p}_1, \mathbf{p}_0, \mathbf{p}_1, \mathbf{p}_0, \mathbf{p}_1) .
$$
These prior probabilities serve as the inputs for the inner decoder, matching the dimensionality of the six block columns present in $H_{\text{win}1}$.
Ideally, the decoder would utilize $H_{\text{win}1}, \mathbf{p}_{\text{win}1}$, and the detectors $(\mathbf{d}_0, \mathbf{d}_1, \mathbf{d}_2)$ to infer the six error sub-vectors $(e_0, e_1, \dots, e_5)$ corresponding to the block columns in $H_{\text{win}1}$. However, there is a structural issue within this truncated window—specifically regarding the final block column—which requires us to further modify this submatrix before proceeding with the decoding process.

Without loss of generality, we assume the original global parity-check matrix \cref{eq:pcm_6cycles} contains no identical columns. Indeed, standard tools like Stim \cite{gidney2021stim} automatically merge identical columns to produce a simplified parity-check matrix. This implies that the combined vertical block
$$
\begin{bmatrix}
H_1 \\
H_2
\end{bmatrix}
$$
contains only unique columns.

However, when considering the truncated submatrix $H_{\text{win}1}$ in \cref{eq:pcm_sub1}, the matrix $H_1$ in the final block column may contain identical columns because it is no longer supported by the $H_2$ block below it. Since identical columns significantly degrade the performance of BP-based decoders, we must merge these columns before decoding. Specifically, the subblock $H_1^{\text{merge}}$ is obtained from $H_1$ by merging all of its identical columns. Thus, the modified submatrix used for the first window becomes
$$
H_{\text{win}1}^{\text{merge}} = \begin{bmatrix}
H_0' & H_1 \\
 & H_2 & H_0 & H_1    \\
 & & & H_2 & H_0 & H_1^{\text{merge}}  \\
\end{bmatrix} ,
$$
and the modified prior probability vector becomes
$$
\mathbf{p}_{\text{win}1}^{\text{merge}} = (\mathbf{p}_0', \mathbf{p}_1, \mathbf{p}_0, \mathbf{p}_1, \mathbf{p}_0, \mathbf{p}_1^{\text{merge}}) .
$$
Here, $\mathbf{p}_1^{\text{merge}}$ represents the merged prior probability vector corresponding to $H_1^{\text{merge}}$. If two identical columns have independent prior probabilities $p_i$ and $p_j$, their combined prior probability is calculated as $p_i(1-p_j) + p_j(1-p_i)$. This combination rule is applied iteratively to all pairs of identical columns in $H_1$ to produce the vector $\mathbf{p}_1^{\text{merge}}$.
The sliding window decoder then employs an inner decoder—typically based on Belief Propagation—to infer the error sub-vectors $(e_0, e_1, \dots, e_5^{\text{merge}})$ using $H_{\text{win}1}^{\text{merge}}, \mathbf{p}_{\text{win}1}^{\text{merge}}$, and $(\mathbf{d}_0, \mathbf{d}_1, \mathbf{d}_2)$. Note that the final error block is denoted as $e_5^{\text{merge}}$ to reflect the column merging performed on the last block column of $H_{\text{win}1}$.

With a commit size of $c=1$, the decoder only finalizes and ``commits" the values for $e_0$ and $e_1$ to the global decoding result. These two specific sub-vectors are chosen because they constitute the complete error support required to satisfy the parity-check constraints of $\mathbf{d}_0$, the detector for the initial SEC. Once these values are committed, the window slides forward, treating the committed errors as known offsets that modify the syndromes for subsequent windows.

In the second window, the decoder shifts its focus to the detector vectors $(\mathbf{d}_1, \mathbf{d}_2, \mathbf{d}_3)$ and utilizes the second, third, and fourth block rows of \cref{eq:pcm_6cycles}. Since the error sub-vectors $e_0$ and $e_1$ have already been finalized, their corresponding block columns are removed from the active submatrix, resulting in the sub-parity-check matrix $H_{\text{win}2}$:
$$
H_{\text{win}2} = \begin{bmatrix}
H_0 & H_1 \\
 & H_2 & H_0 & H_1    \\
 & & & H_2 & H_0 & H_1  \\
\end{bmatrix} .
$$
The active prior probability vector for this window is
$$
\mathbf{p}_{\text{win}2} = (\mathbf{p}_0, \mathbf{p}_1, \mathbf{p}_0, \mathbf{p}_1, \mathbf{p}_0, \mathbf{p}_1) .
$$
The removal of the finalized columns necessitates a corresponding update to the syndrome. Specifically, in the global parity-check matrix \cref{eq:pcm_6cycles}, the second block row corresponds to the constraint $H_2 e_1 + H_0 e_2 + H_1 e_3 = \mathbf{d}_1$. Since the block column $H_2$ is excluded from the active submatrix $H_{\text{win}2}$, we must account for the influence of the committed error. Thus, we modify the detector value to $\mathbf{d}_1' = \mathbf{d}_1 - H_2 \hat{e}_1$, where $\hat{e}_1$ represents the error value determined and finalized during the first decoding window.

As in the first window, we must merge identical columns in the final $H_1$ block to prevent performance degradation in the inner decoder. The resulting modified submatrix for the second window is
$$
H_{\text{win}2}^{\text{merge}} = \begin{bmatrix}
H_0 & H_1 \\
 & H_2 & H_0 & H_1    \\
 & & & H_2 & H_0 & H_1^{\text{merge}}  \\
\end{bmatrix} ,
$$
and the modified prior probability vector is
$$
\mathbf{p}_{\text{win}2}^{\text{merge}} = (\mathbf{p}_0, \mathbf{p}_1, \mathbf{p}_0, \mathbf{p}_1, \mathbf{p}_0, \mathbf{p}_1^{\text{merge}}) .
$$
The inner decoder then utilizes $H_{\text{win}2}^{\text{merge}},\mathbf{p}_{\text{win}2}^{\text{merge}}$, and the updated syndrome $(\mathbf{d}_1', \mathbf{d}_2, \mathbf{d}_3)$ to infer the error sub-vectors $(e_2, e_3, \dots, e_7^{\text{merge}})$. Finally, the values for $e_2$ and $e_3$ are committed to the global decoding result, as these sub-vectors constitute the complete support for the modified detector $\mathbf{d}_1'$.

The third window proceeds in a manner essentially identical to the second. The inner decoder utilizes the merged submatrix $H_{\text{win}3}^{\text{merge}} = H_{\text{win}2}^{\text{merge}}$, the merged prior probability vector $\mathbf{p}_{\text{win}3}^{\text{merge}}=\mathbf{p}_{\text{win}2}^{\text{merge}}$, and the syndrome segment $(\mathbf{d}_2', \mathbf{d}_3, \mathbf{d}_4)$ to infer the error sub-vectors $(e_4, e_5, \dots, e_9^{\text{merge}})$.
As in the previous step, the first detector in the window is updated to $\mathbf{d}_2' = \mathbf{d}_2 - H_2 \hat{e}_3$, where $\hat{e}_3$ is the finalized error value from the second window. Following the decoding process, $e_4$ and $e_5$ are committed to the global decoding result.

The active submatrix for the last window is
$$
H_{\text{win}4} = \begin{bmatrix}
H_0 & H_1 \\
 & H_2 & H_0 & H_1    \\
 & & & H_2 & H_0''  \\
\end{bmatrix} ,
$$
and the active prior probability vector is
$$
\mathbf{p}_{\text{win}4} = (\mathbf{p}_0, \mathbf{p}_1, \mathbf{p}_0, \mathbf{p}_1, \mathbf{p}_0'') .
$$
In this concluding step, column merging is no longer required. This is because $H_0''$ represents the final block of the original parity-check matrix \cref{eq:pcm_6cycles} which, by our earlier assumption, contains no identical columns.
As with previous windows, the first detector in the sequence must be updated to account for the finalized error from the preceding window: $\mathbf{d}_3' = \mathbf{d}_3 - H_2 \hat{e}_5$, where $\hat{e}_5$ is the value committed during the third window. The inner decoder then utilizes $H_{\text{win}4}, \mathbf{p}_{\text{win}4}$, and the syndrome $(\mathbf{d}_3', \mathbf{d}_4, \mathbf{d}_5)$ to decode the remaining error sub-vectors $(e_6, e_7, \dots, e_{10})$. Since this is the final window in the sequence, all decoded values are committed to the global result, completing the decoding process for all six SECs.

\subsection{Number of windows and size of the last window}

In the specific case of a $(3, 1)$ decoder, the window size—defined as the number of detector vectors processed by the inner decoder—remains constant throughout the process. However, for general parameters $(w, c)$, the size of the terminal window may differ from that of the preceding windows.
Suppose there are $r$ total SECs. After the first $t$ windows have been decoded, $r - tc$ SECs remain unfinalized. The next window is determined as follows:
If $r - tc > w$, the subsequent window is not the terminal one and maintains a size of $w$.
If $r - tc \le w$, the subsequent window is the terminal window, and its size is exactly $r - tc$.

The size of the terminal window, $w_{\text{last}}$, necessarily falls within the range $[w - c + 1, w]$. This can be shown by contradiction: if we assume $w_{\text{last}} \le w - c$, then the previous window would have contained only $(w_{\text{last}} + c) \le w$ remaining SECs. By definition, that previous window would have already satisfied the condition to be the terminal window, contradicting the assumption that a subsequent window exists.

For a total of $r$ SECs and decoding parameters $(w, c)$, the total number of windows, $N_{\text{win}}$, is given by
$$
N_{\text{win}} = \left\lfloor \frac{r - w - 1}{c} \right\rfloor + 2 .
$$
This expression can be derived as follows. The first $t = \lfloor (r - w - 1)/c \rfloor$ windows commit a total of $tc$ SECs to the global decoding result. This leaves a remainder of $R$ SECs to be processed, where
$$
R = r - tc = w + 1 + [(r - w - 1) \pmod c] .
$$
Since the properties of the modulo operator ensure that $0 \le (r - w - 1) \pmod c \le c - 1$, the number of remaining SECs is bounded by
$$
w + 1 \le R \le w + c .
$$
Because $R$ is strictly greater than the window size $w$, but no greater than $w + c$, these remaining SECs are processed in exactly two additional windows: one window of size $w$ and a final terminal window of size $R - c$. Adding these two to the initial $t$ windows yields the total $N_{\text{win}}$.
The first $(N_{\text{win}} - 1)$ windows commit $(N_{\text{win}} - 1)c$ SECs, so the size of the terminal window is
$$
w_{\text{last}} = r - (N_{\text{win}} - 1)c .
$$

\subsection{Pseudocode}

As demonstrated in the toy example, the sliding window process for four windows utilizes four corresponding submatrices $H_{\text{win}1}^{\text{merge}}$, $H_{\text{win}2}^{\text{merge}}$, $H_{\text{win}3}^{\text{merge}}$, and $H_{\text{win}4}$, together with four prior probability vectors $\mathbf{p}_{\text{win}1}^{\text{merge}},\mathbf{p}_{\text{win}2}^{\text{merge}},\mathbf{p}_{\text{win}3}^{\text{merge}}$, and $\mathbf{p}_{\text{win}4}$. Notably, the two intermediate windows share an identical structure, such that $H_{\text{win}2}^{\text{merge}} = H_{\text{win}3}^{\text{merge}}$ and $\mathbf{p}_{\text{win}2}^{\text{merge}} = \mathbf{p}_{\text{win}3}^{\text{merge}}$. This uniformity holds for arbitrary parameters $(w, c)$: the submatrices and prior probability vectors required for all ``middle" windows are invariant, with only the first and the last windows requiring distinct configurations.
Consequently, the entire decoding procedure can be executed using only three unique pairs of sub-parity-check matrices and prior probability vectors. We denote these as $(H_{\text{first}}, \mathbf{p}_{\text{first}})$, $(H_{\text{mid}}, \mathbf{p}_{\text{mid}})$, and $(H_{\text{last}}, \mathbf{p}_{\text{last}})$, representing the first, middle, and last windows, respectively.
\cref{algorithm:init_inner_decoder} outlines the construction of these three pairs, generalizing the structural logic established in the previous toy example. Furthermore, \cref{algorithm:stream_decoder} details the online execution of the streaming decoder, demonstrating how these pre-initialized matrices are deployed across sliding windows and how detector outcomes are iteratively adjusted to subtract the influence of finalized errors.

In a practical implementation, the performance of the decoder relies heavily on the efficiency of the syndrome updates. Specifically, the matrix-vector multiplications between $H_2$ and the committed error vectors (Lines 9 and 13 of \cref{algorithm:stream_decoder}) must be implemented using sparse-matrix operations. Given the high sparsity of $H_2$, failing to exploit this structure would result in significant computational overhead, substantially deteriorating the real-time decoding throughput.

\begin{algorithm}
\caption{Calculate parity-check matrices and prior probability vectors for inner decoder}
\label{algorithm:init_inner_decoder}
\KwData{\\
$H_0', H_1, H_2, H_0, H_0''$: submatrices in the global parity-check matrix; \\
$\mathbf{p}_0', \mathbf{p}_1, \mathbf{p}_0,\mathbf{p}_0''$: sub-vectors in the global prior probability vector; \\
$(w,c)$: window size and commit size; \\
$r$: total number of SECs
}
\SetKwProg{Fn}{Function}{}{}
\Fn{\textsc{Init\_First\_Window}$()$}{
  $(H_1^{\text{merge}}, \mathbf{p}_1^{\text{merge}})\gets \textsc{Merge}(H_1, \mathbf{p}_1)$\;
  $H_{\text{first}} \gets$ a $w\times (2w)$ block matrix\;
  $H_{\text{first}}(i,j)$ denotes the block at the $i$th row and $j$th column. Indices start from $0$\;
  $H_{\text{first}}(0,0) \gets H_0'$\;
  $H_{\text{first}}(i,2i) \gets H_0$ for $i=1,2,\dots,w-1$\;
  $H_{\text{first}}(i,2i+1) \gets H_1$ for $i=0,1,\dots,w-2$\;
  $H_{\text{first}}(i,2i-1) \gets H_2$ for $i=1,2,\dots,w-1$\;
  $H_{\text{first}}(w-1,2w-1) \gets H_1^{\text{merge}}$\;
  $\mathbf{p}_{\text{first}} \gets (\mathbf{p}_0', \underbrace{\mathbf{p}_1, \mathbf{p}_0, \dots, \mathbf{p}_1, \mathbf{p}_0}_{\text{repeat } w-1 \text{ times}}, \mathbf{p}_1^{\text{merge}})$\;
  \KwRet{$(H_{\text{first}}, \mathbf{p}_{\text{first}})$}\;
}
\Fn{\textsc{Init\_Middle\_Window}$()$}{
  $(H_1^{\text{merge}}, \mathbf{p}_1^{\text{merge}})\gets \textsc{Merge}(H_1, \mathbf{p}_1)$\;
  $H_{\text{mid}} \gets$ a $w\times (2w)$ block matrix\;
  $H_{\text{mid}}(i,2i) \gets H_0$ for $i=0,1,\dots,w-1$\;
  $H_{\text{mid}}(i,2i+1) \gets H_1$ for $i=0,1,\dots,w-2$\;
  $H_{\text{mid}}(i,2i-1) \gets H_2$ for $i=1,2,\dots,w-1$\;
  $H_{\text{mid}}(w-1,2w-1) \gets H_1^{\text{merge}}$\;
  $\mathbf{p}_{\text{mid}} \gets (\mathbf{p}_0, \underbrace{\mathbf{p}_1, \mathbf{p}_0, \dots, \mathbf{p}_1, \mathbf{p}_0}_{\text{repeat } w-1 \text{ times}}, \mathbf{p}_1^{\text{merge}})$\;
  \KwRet{$(H_{\text{mid}}, \mathbf{p}_{\text{mid}})$}\;
}
\Fn{\textsc{Init\_Last\_Window}$()$}{
  $w_{\text{last}} \gets r - (\lfloor \frac{r - w - 1}{c} \rfloor + 1)c$\;
  $H_{\text{last}} \gets$ a $w_{\text{last}}\times (2w_{\text{last}}-1)$ block matrix\;
  $H_{\text{last}}(i,2i) \gets H_0$ for $i=0,1,\dots,w_{\text{last}}-2$\;
  $H_{\text{last}}(i,2i+1) \gets H_1$ for $i=0,1,\dots,w_{\text{last}}-2$\;
  $H_{\text{last}}(i,2i-1) \gets H_2$ for $i=1,2,\dots,w_{\text{last}}-1$\;
  $H_{\text{last}}(w_{\text{last}}-1,2w_{\text{last}}-2) \gets H_0''$\;
  $\mathbf{p}_{\text{last}} \gets (\underbrace{\mathbf{p}_0, \mathbf{p}_1, \dots, \mathbf{p}_0, \mathbf{p}_1}_{\text{repeat } w_{\text{last}}-1 \text{ times}}, \mathbf{p}_0'')$\;
  \KwRet{$(H_{\text{last}}, \mathbf{p}_{\text{last}})$}\;
}
\Fn{\textsc{Merge}$(H, \mathbf{p})$}{
    \tcc{Requirement: Length of vector $\mathbf{p}$ equals number of columns in matrix $H$}
  Initialize $H^{\text{merge}}$ as an empty matrix and initialize $\mathbf{p}^{\text{merge}}$ as an empty vector\;
  Denote $i$th column of $H$ as $H(i)$. Indices start from $0$\;
  $m\gets$ length of $\mathbf{p}$\;
  \For{$i=0,1,\dots,m-1$}
  {\If{$H(i)=H^{\text{merge}}(j)$ \emph{for some} $j$}{
  $\mathbf{p}^{\text{merge}}(j) \gets \mathbf{p}(i)(1-\mathbf{p}^{\text{merge}}(j)) + \mathbf{p}^{\text{merge}}(j)(1-\mathbf{p}(i))$
  }
  \Else{
  Append $H(i)$ as the last column of $H^{\text{merge}}$. Append $\mathbf{p}(i)$ as the last entry of $\mathbf{p}^{\text{merge}}$
  }}
  \KwRet{$(H^{\text{merge}}, \mathbf{p}^{\text{merge}})$}\;
}
\end{algorithm}

\begin{algorithm}
\caption{Streaming decoder}
\label{algorithm:stream_decoder}
\KwData{\\
$H_0', H_1, H_2, H_0, H_0''$: submatrices in the global parity-check matrix; \\
$\mathbf{p}_0', \mathbf{p}_1, \mathbf{p}_0,\mathbf{p}_0''$: sub-vectors in the global prior probability vector; \\
$(w,c)$: window size and commit size; \\
$r$: total number of SECs
}
\SetKwInOut{KwStreamIn}{Streaming Input}
\KwStreamIn{$\mathbf{d}_0,\mathbf{d}_1,\dots,\mathbf{d}_{r-1}$}
\SetKwInOut{KwStreamOut}{Streaming Output}
\KwStreamOut{$\hat{e}_0, \hat{e}_1, \dots, \hat{e}_{2r-2}$}
$(H_{\text{first}}, \mathbf{p}_{\text{first}})\gets \textsc{Init\_First\_Window}()$\;
Configure \textsc{First\_Window\_Decoder} with $(H_{\text{first}}, \mathbf{p}_{\text{first}})$\;
$(H_{\text{mid}}, \mathbf{p}_{\text{mid}})\gets \textsc{Init\_Middle\_Window}()$\;
Configure \textsc{Middle\_Window\_Decoder} with $(H_{\text{mid}}, \mathbf{p}_{\text{mid}})$\;
$(H_{\text{last}}, \mathbf{p}_{\text{last}})\gets \textsc{Init\_Last\_Window}()$\;
Configure \textsc{Last\_Window\_Decoder} with $(H_{\text{last}}, \mathbf{p}_{\text{last}})$\;
$(\hat{e}_0, \hat{e}_1, \dots, \hat{e}_{2w-2}, \hat{e}_{2w-1}^{\text{merge}}) \gets \textsc{First\_Window\_Decoder}(\mathbf{d}_0,\mathbf{d}_1,\dots,\mathbf{d}_{w-1})$\;
\textbf{streaming output} $\hat{e}_0, \hat{e}_1, \dots, \hat{e}_{2c-1}$\;
$\mathbf{d}_c\gets \mathbf{d}_c- H_2 \hat{e}_{2c-1}$\;
\For{$i=1,2,\dots,\lfloor \frac{r - w - 1}{c} \rfloor$}{
$(\hat{e}_{2ic}, \hat{e}_{2ic+1}, \dots, \hat{e}_{2ic+2w-2}, \hat{e}_{2ic+2w-1}^{\text{merge}})\gets \textsc{Middle\_Window\_Decoder}(\mathbf{d}_{ic},\dots,\mathbf{d}_{ic+w-1})$\;
\textbf{streaming output} $\hat{e}_{2ic}, \hat{e}_{2ic+1}, \dots, \hat{e}_{2ic+2c-1}$\;
$\mathbf{d}_{ic+c}\gets \mathbf{d}_{ic+c}- H_2 \hat{e}_{2ic+2c-1}$\;
}
$w_{\text{last}} \gets r - (\lfloor \frac{r - w - 1}{c} \rfloor + 1)c$\;
$(\hat{e}_{2r-2w_{\text{last}}}, \hat{e}_{2r-2w_{\text{last}}+1}, \dots, \hat{e}_{2r-2})\gets \textsc{Last\_Window\_Decoder}(\mathbf{d}_{r-w_{\text{last}}},\dots,\mathbf{d}_{r-1})$\;
\textbf{streaming output} $\hat{e}_{2r-2w_{\text{last}}}, \hat{e}_{2r-2w_{\text{last}}+1}, \dots, \hat{e}_{2r-2}$\;
\end{algorithm}

\subsection{Simulation results}

In this subsection, we present the numerical simulation results for the sliding-window decoder, focusing on two primary objectives. First, we demonstrate that the logical error rates of the sliding-window decoder remain highly competitive with those of the global decoder. Across our tested code instances, the logical error rate of the sliding-window approach is at most $2\times$ that of the global baseline. In certain cases, the streaming version even achieves a lower error rate.
Second, we assess the decoder's real-time viability by processing one million continuous SECs. By recording comprehensive decoding time statistics, we demonstrate that the streaming approach is fast enough to keep pace with ion-trap hardware.

As mentioned at the beginning of this section, we utilize the Q70 and Q102 codes as our benchmarks. For these two code instances, we observe that a $(5, 3)$ sliding-window configuration—defined by a window size of 5 and a commit size of 3—provides an optimal balance between error suppression and decoding latency. Consequently, all sliding-window simulation results reported hereafter utilize these $(5, 3)$ parameters.
Both the global baseline and the inner decoder within the streaming framework employ the beam search decoder with the \texttt{beam32\_340iters} configuration \cite{ye2025beam}. Detailed specifications for this configuration are available in Table~I of \cite{ye2025beam}.

\begin{figure}[b!]
    \centering
    \includegraphics[width=0.95\linewidth]{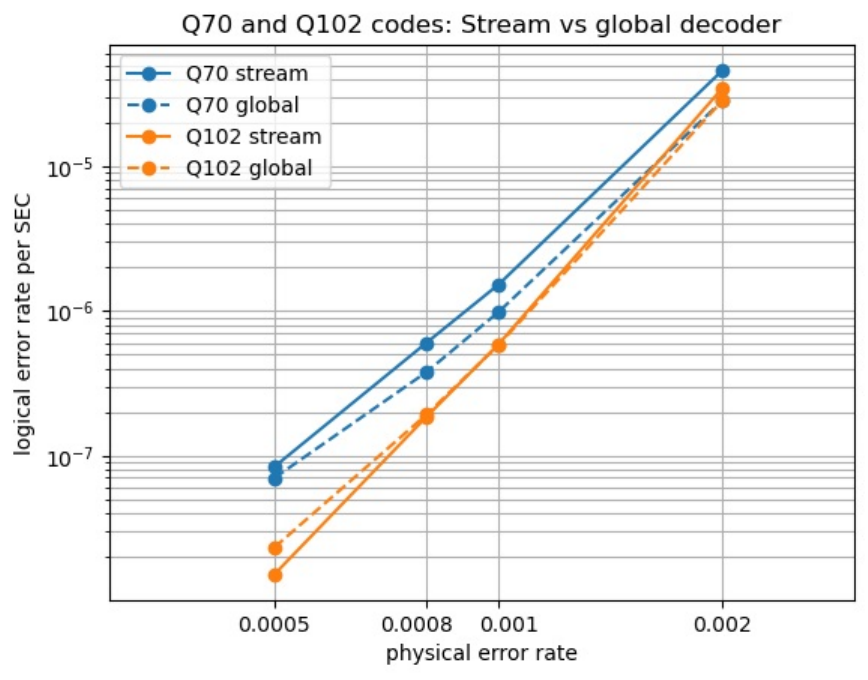}
    \caption{Logical error rate comparison between the $(5,3)$ sliding-window decoder and the global baseline. To ensure a consistent comparison, both decoders process 9 SECs for each code, matching the code distance $d=9$.}
\label{fig:stream_global_compare_logical_error_rate}
\end{figure}

\cref{fig:stream_global_compare_logical_error_rate} presents a comparison of logical error rates between the streaming and global decoders. To ensure a consistent baseline, we performed nine SECs for both codes, reflecting their code distance of 9. We evaluated both code instances across four physical error rates: $0.0005, 0.0008, 0.001$, and $0.002$.
Across all eight data points, the streaming decoder consistently maintained a logical error rate within $2\times$ that of the global decoder. Notably, this performance gap narrows as the physical error rate decreases. In the case of the Q102 code at $p=0.0005$, the streaming decoder actually achieved a lower logical error rate than its global counterpart.
In \cref{app:sec_additional_simulation_stream_decoder}, we provide further simulation results demonstrating that the logical error rate per SEC shows only a minor increase or remains nearly constant as the number of SECs grows under the streaming decoder.

Next, we evaluate the streaming decoder by processing $1,000,008$ consecutive SECs for both the Q70 and Q102 codes at physical error rate $10^{-4}$. We selected $1,000,008$ SECs to ensure the simulation exceeds one million rounds while remaining a multiple of the code distance, $d=9$.

\begin{figure}[b!]
    \centering
    \includegraphics[width=0.95\linewidth]{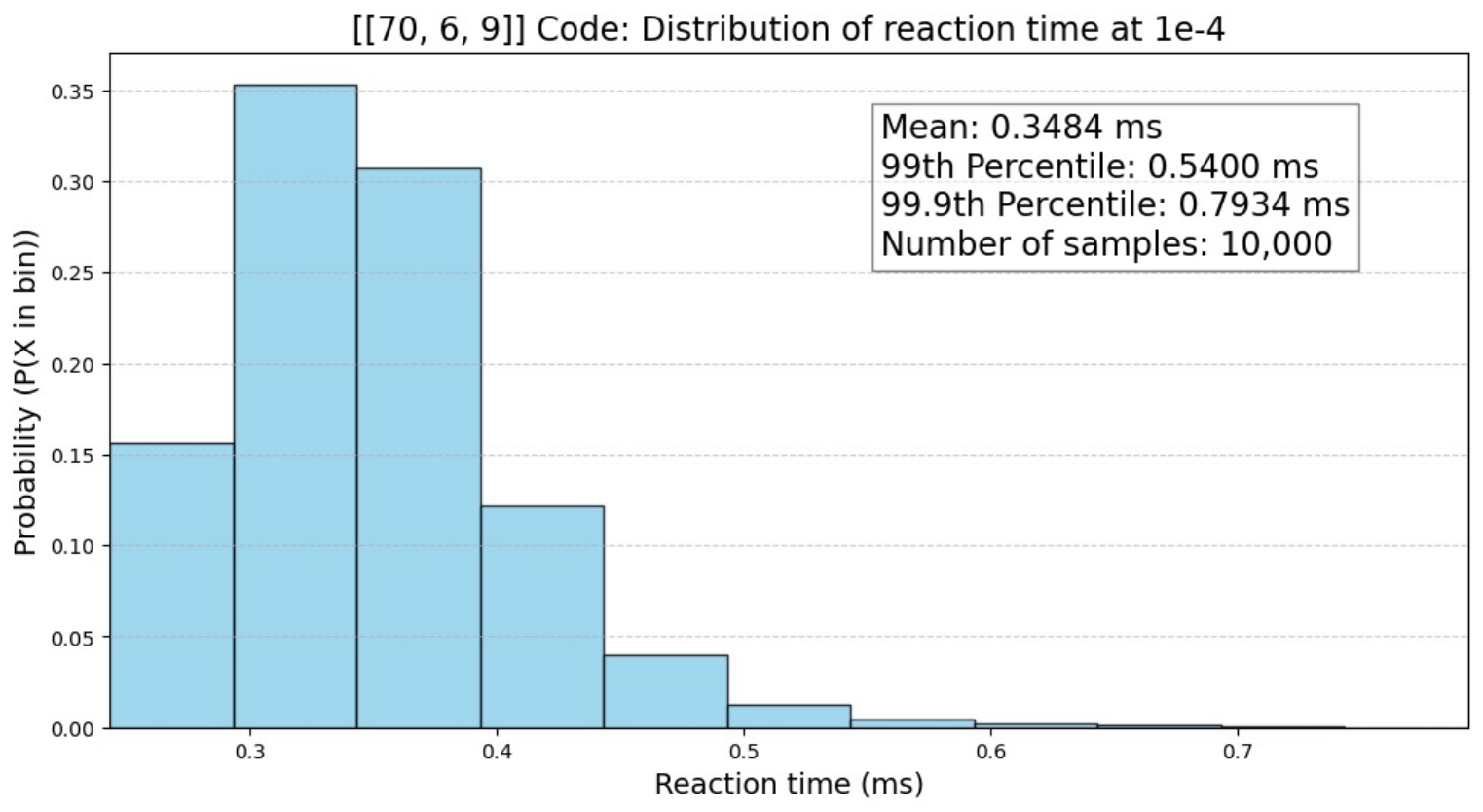}
    \vspace*{0.05in}

    \includegraphics[width=0.95\linewidth]{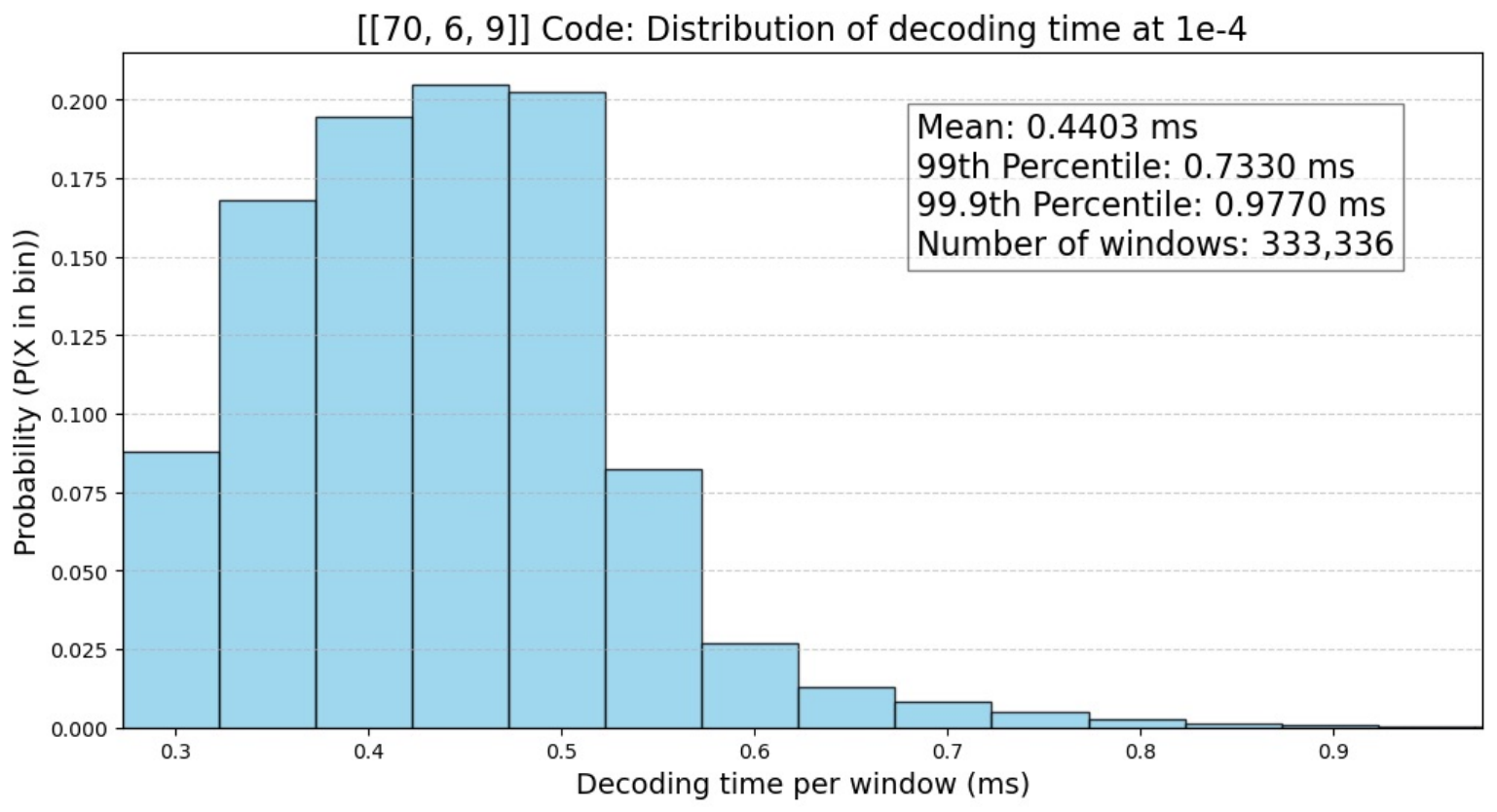}
    \vspace*{0.05in}

    \includegraphics[width=0.95\linewidth]{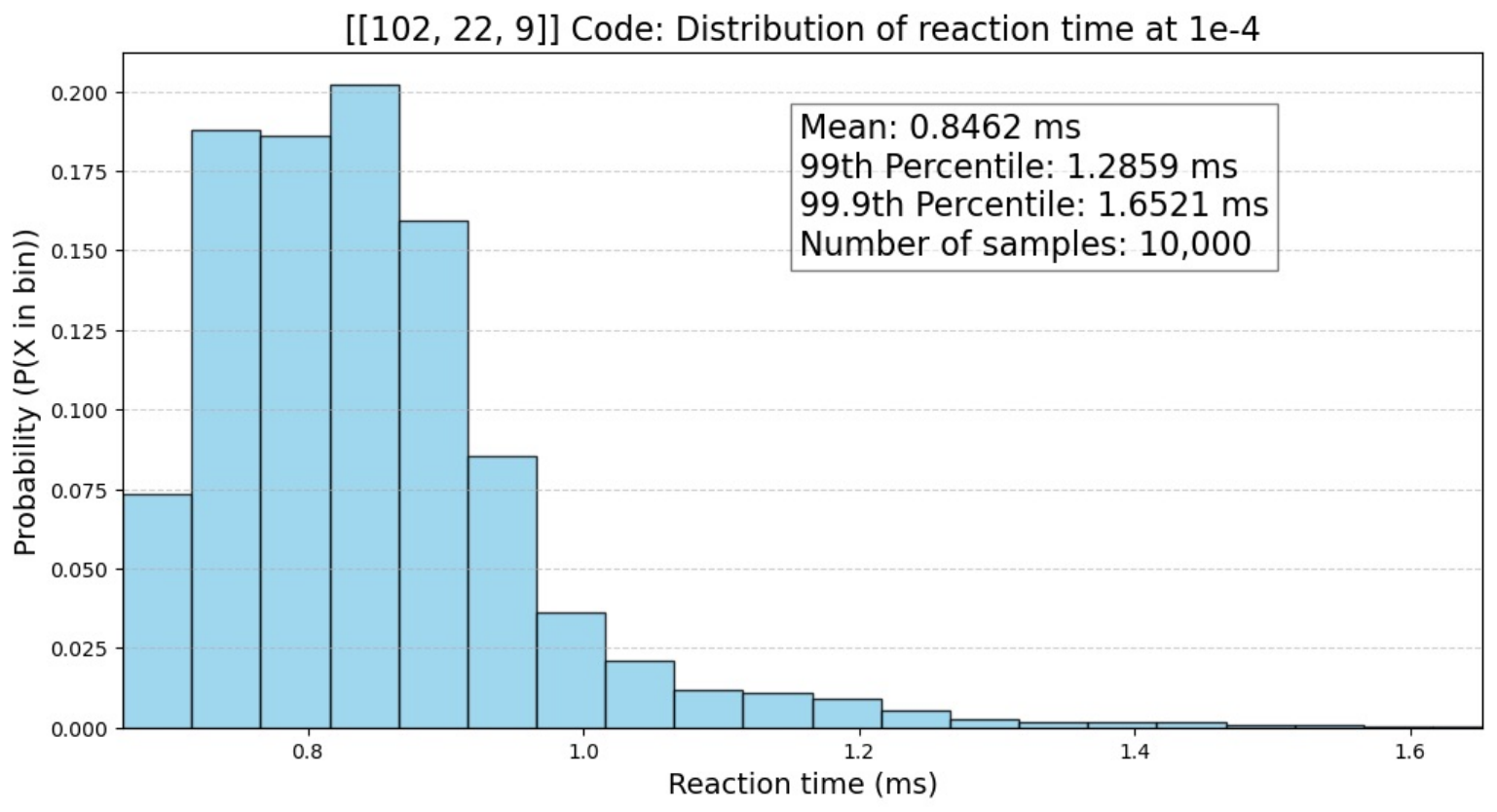}
    \vspace*{0.05in}

    \includegraphics[width=0.95\linewidth]{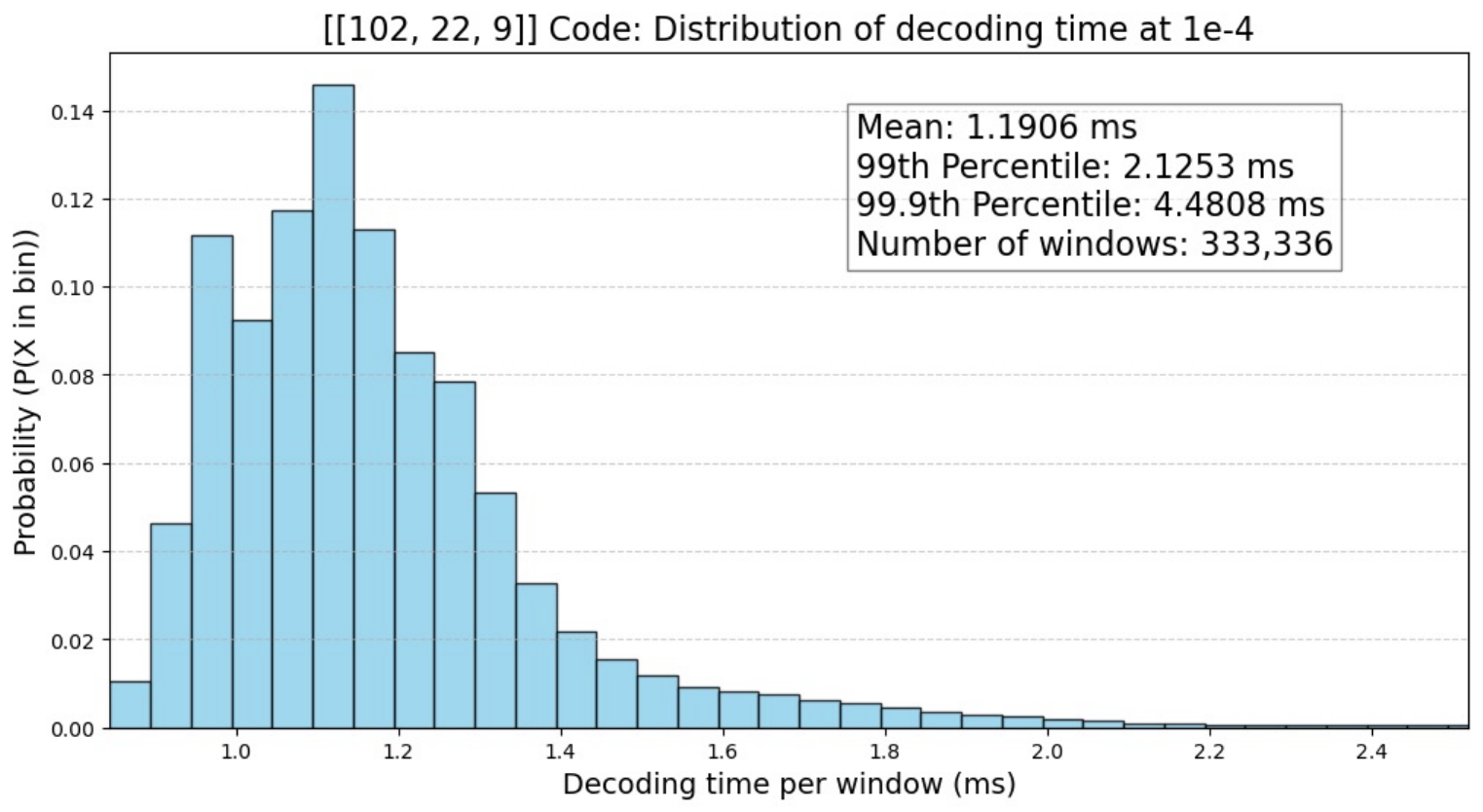}
    \caption{Probability distribution histograms for decoding time per window and reaction time at physical error rate $10^{-4}$ using a $(5,3)$ sliding window beam search decoder. Decoding time statistics are derived from a single run of $1,000,008$ SECs ($333,336$ windows). Reaction time statistics are generated from $10,000$ independent runs of $1,000,008$ SECs each.}
    \label{fig:decoding_and_reaction_time_1e-4}
\end{figure}

\cref{fig:decoding_and_reaction_time_1e-4} illustrates the simulation results for two critical decoding latency measures. The first metric, decoding time per window, represents the processing duration for each sliding-window instance. Using a $(5,3)$ configuration, a single run of $1,000,008$ SECs results in $333,336$ decoding windows. The probability distribution histograms for these windows are shown in \cref{fig:decoding_and_reaction_time_1e-4}. Because the $(5,3)$ decoder commits three SECs per window, real-time processing is achieved if the decoding time per window remains below three times the hardware's SEC time. Given a high-performance estimate of $1\text{ ms}$ per SEC in trapped-ion architectures, the threshold for real-time operation is $3\text{ ms}$ per window. For both codes, the mean and 99th percentile decoding times fall safely below this $3\text{ ms}$ limit.

The second metric is the reaction time following destructive measurements. This is defined as the interval between the final qubit measurement and the acquisition of the corresponding decoding solution. Minimizing this latency is paramount because these final results often determine the selection of gates or operations in subsequent computational stages. In our streaming context, the reaction time is equivalent to the duration required to process the final window.

Because a single run of $1,000,008$ SECs yields only one reaction time sample, we performed $10,000$ independent runs to generate reliable statistics. The resulting probability distribution histograms, shown in \cref{fig:decoding_and_reaction_time_1e-4}, demonstrate that the reaction times for both codes are small, with average values consistently below $1\text{ ms}$.

Key statistical metrics—including the mean, 99th percentile, and 99.9th percentile values for both decoding time per SEC and reaction time—are summarized for the Q70 and Q102 codes in \cref{tab:time_stats_summary_1e-4}. While raw data records the decoding time per window, the decoding time per SEC is derived by dividing the per-window duration by three. This reflects the $(5,3)$ sliding-window configuration, where each processed window commits three SECs to the final result.

Finally, while this section illustrates the latency distributions at physical error rate $10^{-4}$ in \cref{fig:decoding_and_reaction_time_1e-4}, the corresponding histograms for physical error rate $5 \times 10^{-4}$ are provided in \cref{fig:decoding_and_reaction_time_5e-4} within \cref{app:sec_additional_simulation_stream_decoder}. At this higher noise level, the average decoding time per SEC and average reaction time remain below $1\text{ ms}$, though the distribution exhibits a heavier tail as decoding complexity increases.

\section{Enriched walking cat architecture}
\label{sec:Enriched walking cat architecture}
The logical instruction set of the walking cat architecture (see \cref{sec:Logical instruction set architecture}) was designed to be simple, yet universal for quantum computation. In this section, we show that the set of logical instructions can be enriched by exploiting physical Clifford operators that implement logical operations on the chosen memory block codes. In particular, we can use permutations implemented by physically moving the qubits along cyclic shifts and transversal Clifford gate operations to perform unitary gates.
Applying such a gate before a destructive measurement (DMX) or (DMZ) gives us access to new sets of fast logical measurements.

In \cref{subsec: parity measurements}, we discuss joint parity measurements and using Clifford gates to enrich the set of available parity measurements. In \cref{subsec: cyclic gates}, we define permutation-based gates for BB codes that are natural for our architecture. In \cref{subsec: transversal}, we discuss transversal Clifford gates for BB codes.

\subsection{Parity measurements}
\label{subsec: parity measurements}
Consider two active logical code blocks $A$ and $D$, where block $D$ is slated for destructive measurement. 
To extract joint parity measurements, we couple the two blocks via a transversal \CX gate across all qubits. When implementing the instruction (DMX), we use block $D$ as the control and block $A$ as the target so that the destructive measurement of the $D$ block yields the parity measurements $\{\bar X^{A}_i\otimes \bar X^{D}_j \mid i,j\in[k]\}$. For the purposes of this section, we will assume that the operators $\bar X_i^A$ and $\bar X_i^D$ are always the logical $X$ operators of the chosen symplectic bases without the Clifford frame update. That is, each $\bar X_i\in L_X$ as given in \cref{subsec:Destructive measurements}. Similarly, we reverse the control and target for instruction (DMZ) to obtain the measurements $\{\bar Z^{A}_i\otimes \bar Z^{D}_j \mid i,j\in[k]\}$.

\begin{figure}[tb!]
    \centering
    \includegraphics[width=0.7\linewidth]{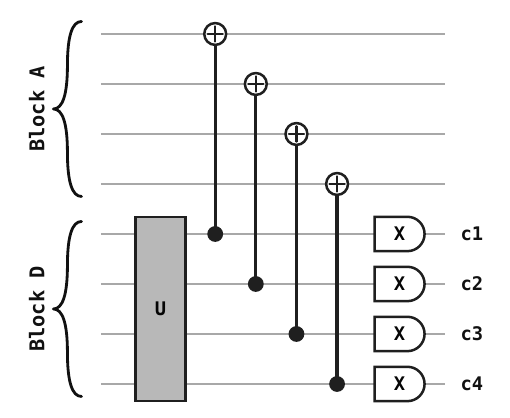}
    \caption{Logical instruction (DMX) enriched by applying a physical Clifford unitary $U$ to code block $D$ prior to a round of \CX and destructive $X$ measurements.}
    \label{fig: enriched DMX}
\end{figure}

We can enrich this primitive by introducing a collection of Clifford operators $\mathcal{E}$ that will be physically applied to block $D$ prior to coupling (\cref{fig: enriched DMX}), so that the observables measured on $D$ are conjugated by $U \in \mathcal{E}$. This yields a larger collection of available joint parity measurements to expose to the compiler, allowing for mixed-basis measurements between the blocks. The available parity measurements of the destructive $X$ and $Z$ protocols, respectively, become
\begin{align*}
    &\{\bar X^{A}_i\otimes U\bar X^{D}_j U^\dagger \mid i,j\in[k], U\in\mathcal{E}\},\\
    &\{\bar Z^{A}_i\otimes U\bar Z^{D}_j U^\dagger \mid i,j\in[k], U\in\mathcal{E}\}.
\end{align*} 
Assuming that each $U\in\mathcal{E}$ is a logical Clifford operation for the code on block $D$, we can increase the number of available joint parity measurements from 2 to a maximum of $2|\mathcal{E}|$ as each $U\bar X^{D}_j U^\dagger$ and $U\bar Z^{D}_j U^\dagger$ will be representatives of logical Pauli operators.

\subsection{Cyclic gates}
\label{subsec: cyclic gates}
We first consider permutation-based gates \cite{calderbank1997errorcorrection} that preserve the memory block code space. As physical swap operations do not propagate errors between qubits, we are fault-tolerantly enriching our space of available joint parity measurements.
Of particular interest are permutation gates that exploit the three-ring structure used for ancilla routing during syndrome extraction (see  \cref{sec:Memory block}). Given $(r,s,t)\in\mathbb{Z}_a\times \mathbb{Z}_b\times \mathbb{Z}_c$, the $(r,s,t)$ \emph{three-ring cyclic gate}, or \emph{cyclic gate}, for short, is a permutation-based gate implemented via the following shift operation on the data qubits (as opposed to the ancilla qubits during syndrome extraction):
\[
\delta(u,v,w) \mapsto \delta(u \oplus r,\, v \oplus s,\, w \oplus t).
\]

For BB codes, whose qubits are indexed by elements of $\mathbb{Z}_2\times \mathbb{Z}_\ell\times \mathbb{Z}_m$, the $(0,i,j)$ cyclic gates have a particularly nice interpretation: they correspond precisely to multiplication by a monomial $x^i y^j$ \cite{eberhardt2024logicaloperatorsfoldtransversalgates, kim2026timedynamiccircuitsfaulttolerantshift}. The action of the $(0,i,j)$ cyclic gate on the $X$ stabilizer matrix of a BB code corresponds to right multiplication by the $2\ell m\times 2\ell m$ block permutation matrix
\[
\begin{bmatrix}
x^iy^j &0_{\ell m\times \ell m}\\ 0_{\ell m\times \ell m}&x^i y^j
\end{bmatrix}.
\]
Given that $x$ and $y$ commute and that $A$ and $B$ are sums of monomial terms, we can see that
\begin{align*}
    H_X \begin{bmatrix}
x^iy^j &0\\ 0&x^i y^j
\end{bmatrix} &= [A\mid B]\begin{bmatrix}
x^iy^j &0\\ 0&x^i y^j
\end{bmatrix} \\
&= [Ax^iy^j\mid B x^i y^j] \\
&= x^iy^j H_X,
\end{align*}
and likewise for $H_Z$. As left multiplication by $x^i y^j$ simply permutes rows of both $H_X$ and $H_Z$, we have demonstrated that for every $i\in\mathbb{Z}_\ell$ and $j\in\mathbb{Z}_m$, the $(0,i,j)$ cyclic gate, which requires $im+j$ transport steps, preserves the stabilizers of any BB code, and hence is a logical operation. 

The logical operation implemented by a $(0,i,j)$ cyclic gate is equivalent to a circuit of logical \CX gates \footnote{When $\ell$ and $m$ are both odd, it is known that the $(0,1,0)$ and $(0,0,1)$ gates act as multiplication by roots of unity in a particular symplectic basis \cite{eberhardt2024logicaloperatorsfoldtransversalgates}.}. Suppose that $L_X, L_Z\in\mathbb{F}_2^{k\times n}$ are two matrices representing a symplectic basis for the code. Applying the $(0,i,j)$ shift implements the transformations
\begin{align}
    L'_X &= G L_X \pmod 2, \label{eq: X transformation under shift}\\
    L'_Z &= (G^{-1})^T L_Z \pmod 2,\label{eq: Z transformation under shift}
\end{align}
on the logical $X$ and $Z$ bases, respectively, where $G\in \mathbb{F}_2^{k\times n}$ is an invertible matrix. In particular, the $2k\times 2k$ logical symplectic matrix implemented by the physical $(0,i,j)$ shift is given by
\begin{equation*}
    \bar C = 
    \begin{bmatrix}
        G & 0_{k\times k}\\
        0_{k\times k}& ({G}^{-1})^T
    \end{bmatrix}.
\end{equation*}

This $G$ can be determined from $L_X$, $L_Z$, and the $(0,i,j)$ shift: 
The CSS condition guarantees that $L_X L_Z^T = L_Z L_X^T=I_k$, so multiplying \cref{eq: X transformation under shift} on the right by $L_Z^T$ and \cref{eq: Z transformation under shift} on the right by $L_X^T$ shows
\begin{align}
    G &= L'_X L_Z^T \pmod 2,\\
    (G^{-1})^T &= L'_Z L_X^T  \pmod 2,
\end{align}
where $L'_X$ and $L'_Z$ are determined by applying the $(0,i,j)$ shift to the columns of $L_X$ and $L_Z$, respectively. 

For the $[[70, 6, 9]]$ BB7 memory code, the symplectic representation of the logical action implemented by the $(0,1,0)$ shift for the basis in \cref{tab:Symp_MEK} is given by the $12\times 12$ block matrix
\begin{equation*}
    \bar C^{(70)} = 
    \begin{bmatrix}
        G_1 & 0_{6\times 6}\\
        0_{6\times 6}& ({G_1}^{-1})^T
    \end{bmatrix},
\end{equation*}
where $G_1$ is the $6\times 6$ matrix:
\begin{equation}\label{eq: 70 memory x shift}
G_1 = 
\begin{bmatrix}
0 & 0 & 0 & 1 & 0 & 1 \\
1 & 1 & 0 & 0 & 1 & 1 \\
0 & 1 & 1 & 1 & 1 & 0 \\
1 & 0 & 0 & 1 & 0 & 1 \\
0 & 1 & 1 & 1 & 0 & 0 \\
1 & 0 & 0 & 0 & 0 & 1
\end{bmatrix}.
\end{equation}
The $(0,1,0)$ shift in this BB code is an order-7 logical \CX circuit; the $(0,0,1)$ shift is a logical identity.

For the $[[102, 22, 9]]$ GB8 memory code, the symplectic representation of the logical action implemented by the $(0,1,0)$ shift for the basis in \cref{tab:Symp_X_Totoro,tab:Symp_Z_Totoro} is given by the $44\times 44$ block matrix
\begin{equation*}\label{eq: 102 memory x shift}
    \bar C^{(102)} = 
    \begin{bmatrix}
        G_2 & 0_{22\times 22}\\
        0_{22\times 22}& ({G_2}^{-1})^T
    \end{bmatrix},
\end{equation*}
where $G_2$ is the $22\times 22$ matrix:
{\fontsize{8}{4}
\begin{equation}
G_2=
{\setlength{\arraycolsep}{2.5pt}
    \begin{bmatrix}
0 & 1 & 0 & 1 & 1 & 1 & 0 & 1 & 1 & 1 & 1 & 1 & 1 & 0 & 1 & 0 & 1 & 1 & 0 & 0 & 1 & 0 \\
1 & 1 & 1 & 0 & 0 & 0 & 1 & 1 & 1 & 1 & 1 & 0 & 1 & 1 & 0 & 1 & 1 & 0 & 1 & 0 & 1 & 1 \\
1 & 1 & 0 & 0 & 0 & 0 & 1 & 1 & 0 & 0 & 0 & 1 & 1 & 0 & 1 & 1 & 0 & 0 & 0 & 0 & 1 & 1 \\
0 & 0 & 1 & 1 & 1 & 0 & 0 & 1 & 1 & 1 & 0 & 0 & 1 & 0 & 0 & 0 & 0 & 0 & 1 & 1 & 0 & 1 \\
0 & 1 & 1 & 1 & 1 & 1 & 1 & 1 & 0 & 1 & 0 & 0 & 0 & 0 & 1 & 0 & 0 & 1 & 0 & 0 & 1 & 0 \\
0 & 0 & 1 & 1 & 1 & 0 & 1 & 1 & 0 & 0 & 1 & 0 & 1 & 1 & 0 & 0 & 1 & 0 & 0 & 1 & 0 & 1 \\
0 & 1 & 0 & 1 & 0 & 1 & 0 & 0 & 0 & 0 & 1 & 1 & 0 & 0 & 1 & 1 & 0 & 1 & 1 & 1 & 1 & 0 \\
0 & 0 & 0 & 1 & 1 & 0 & 1 & 0 & 0 & 0 & 0 & 1 & 1 & 0 & 1 & 0 & 1 & 0 & 0 & 1 & 1 & 0 \\
0 & 1 & 1 & 1 & 1 & 0 & 1 & 0 & 1 & 1 & 1 & 0 & 0 & 0 & 1 & 0 & 1 & 0 & 1 & 1 & 1 & 0 \\
0 & 1 & 0 & 1 & 0 & 1 & 0 & 0 & 0 & 1 & 1 & 1 & 1 & 1 & 0 & 1 & 1 & 0 & 1 & 1 & 1 & 1 \\
0 & 0 & 1 & 0 & 0 & 1 & 0 & 0 & 1 & 0 & 0 & 0 & 1 & 0 & 1 & 1 & 0 & 0 & 0 & 0 & 1 & 1 \\
1 & 0 & 0 & 0 & 1 & 1 & 0 & 1 & 0 & 0 & 0 & 1 & 1 & 0 & 0 & 0 & 1 & 0 & 0 & 1 & 0 & 1 \\
1 & 0 & 1 & 0 & 1 & 1 & 1 & 0 & 1 & 1 & 0 & 1 & 1 & 0 & 1 & 0 & 1 & 1 & 1 & 0 & 0 & 1 \\
1 & 1 & 1 & 0 & 1 & 1 & 0 & 1 & 0 & 1 & 0 & 1 & 0 & 1 & 0 & 1 & 1 & 0 & 1 & 0 & 1 & 1 \\
0 & 1 & 0 & 0 & 1 & 1 & 0 & 1 & 1 & 0 & 1 & 1 & 0 & 0 & 0 & 0 & 1 & 0 & 1 & 1 & 1 & 0 \\
0 & 1 & 1 & 0 & 1 & 0 & 0 & 1 & 1 & 0 & 1 & 1 & 0 & 1 & 1 & 0 & 1 & 0 & 0 & 1 & 0 & 0 \\
0 & 1 & 1 & 0 & 1 & 1 & 0 & 0 & 1 & 0 & 1 & 1 & 0 & 1 & 1 & 1 & 1 & 1 & 1 & 1 & 1 & 1 \\
1 & 1 & 0 & 0 & 0 & 1 & 1 & 1 & 0 & 0 & 1 & 0 & 1 & 1 & 1 & 0 & 1 & 1 & 1 & 0 & 1 & 0 \\
0 & 1 & 0 & 0 & 1 & 1 & 0 & 1 & 1 & 1 & 1 & 0 & 0 & 1 & 0 & 0 & 0 & 1 & 0 & 1 & 1 & 1 \\
0 & 1 & 1 & 1 & 1 & 0 & 1 & 1 & 1 & 0 & 0 & 0 & 1 & 0 & 0 & 1 & 0 & 1 & 0 & 0 & 1 & 1 \\
1 & 0 & 0 & 0 & 1 & 1 & 1 & 0 & 1 & 1 & 0 & 1 & 1 & 1 & 1 & 0 & 0 & 1 & 0 & 0 & 1 & 0 \\
0 & 0 & 1 & 0 & 1 & 1 & 1 & 0 & 0 & 0 & 0 & 0 & 0 & 1 & 1 & 0 & 1 & 1 & 0 & 0 & 1 & 1
\end{bmatrix}.
}
\end{equation}
}
The $(0,1,0)$ shift in this GB code is an order-51 logical \CX circuit.

\subsection{Transversal gates}
\label{subsec: transversal}
We can also consider transversal Clifford operators, which likewise do not propagate faults between qubits. We take the convention that \emph{transversal Clifford gates} are operators of the form $U=\bigotimes_{i=1}^n U_i$ where each $U_i$ is a (potentially different) single-qubit Clifford gate. If a BB code is \emph{self-orthogonal}---the $X$ and $Z$ stabilizer spaces are isomorphic---then the transversal $H^{\otimes n}$ gate is guaranteed to be a logical operator, and for some choices of $b,c\in\{1,-1\}^n$ the operators $\bigotimes_{i=1}^n S^{b_i}$, and $\bigotimes_{i=1}^n \sqrt{X}^{c_i}$ are also logical operators \cite{Tansuwannont2025selfdual}. 

While the algebraic structure of BB codes enables a simplified search for transversal Clifford logical gates, we believe that only self-orthogonal BB codes admit such logical gates.
Thus, we consider only self-orthogonal BB codes for the remainder of this section.
We leave the study of fold-transversal gates that fit within the three-ring framework, which, in principle, should exist, for future work.

Given polynomials $a(x,y)$ and $b(x,y)$, the corresponding BB code with $A=a(x,y)$ and $B=b(x,y)$, as defined in \cref{sec:Memory block}, is guaranteed to be self-orthogonal if there exists a monomial $x^i y^j$ for which $x^iy^j a(x,y) = b(x^{-1},y^{-1})$ and $x^iy^j b(x,y) = a(x^{-1},y^{-1})$. Indeed, by considering $x^iy^j$ as an $\ell m\times \ell m$ matrix, we see that $x^i y^j[A\mid B] = [x^i y^jA\mid x^i y^jB] = [B^T\mid A^T]$; since left multiplication by $x^i y^j$ simply permutes the rows of $H_X=[A\mid B]$, the $X$ and $Z$ stabilizer spaces are isomorphic. 

While restricting to self-orthogonality may impact the achievable distance for BB and GB codes, there are choices of $n$ and $k$ for which self-orthogonal GB8 codes achieve near-optimal distance when compared to the best-known BB8 codes, in general. In \cref{tab: self-orthogonal codes} we give examples of such codes, which are, in fact, bicycle codes. Self-orthogonal codes with the same parameters were achieved in \cite{Liang2025selfdual}, where they considered rotated versions of BB8 codes. 
Given that the constructions we found are simple bicycle codes, their syndrome extraction can be highly optimized when compared to rotated BB codes. Finally, we note that self-orthogonal BB codes have been proposed in other fault-tolerant architectures \cite{xu2025batchedhighratelogicaloperations}, demonstrating that they retain practical interest despite a possible decrease in performance.

\begin{table}[t]
    \centering
    \begin{tabular}{lcccc}
        \hline
        \makecell[l]{$[[n,k,d]]$}&
        \makecell[c]{$\ell$} &
        \makecell[c]{$m$} &
        \makecell[c]{$a(x)$}\\
        \hline
        $[[66,6,8]]$& $33$ & $1$ & $1+x+x^3+x^{10}$ \\
        $[[72,12,6]]$& $36$ & $1$ & $1+x+x^4+x^9$ \\
        $[[100,12,8]]$& $50$ & $1$  & $1+x+x^5+x^{16}$ \\
        \hline
    \end{tabular}
    \caption{Three examples of self-orthogonal GB8 codes, with distances near or at the best known distance for BB8 codes of the same length and rate.  For these codes, we define $b(x) = a(x^{-1})$, \textit{i.e.}, these codes are all bicycle codes.}\label{tab: self-orthogonal codes}
\end{table}

Beyond identifying transversal Clifford gates of a self-orthogonal BB code, it is natural to also consider the logical operations they perform. For an arbitrary symplectic basis, the corresponding logical circuit can be found by conjugating all of the basis operators and directly computing the logical symplectic matrix. In the case where a self-orthogonal code admits even a single odd-weight logical operator, Theorem 1 of \cite{Tansuwannont2025selfdual} constructs a symplectic basis of the code for which $L_X=L_Z$. In particular, the physical $H^{\otimes n}$ operation implements $\bar H^{\otimes k}$. The authors also establish that for any $(a_1,\dots,a_k)\in\{-1,1\}^k$, the logical $\bigotimes_{j=1}^k \bar S_j^{a_j}$ operator can be implemented in the same symplectic basis by a physical $\bigotimes_{i=1}^n S_i^{b_i}$ for some choice of $(b_1,\dots,b_n)\in\{-1,1\}^n$.

\clearpage

\part{The micro-architecture}
\label{part:The Micro-architecture}
\section{Physical implementation of the walking cat architecture}\label{sec:micro-architecture}

The walking cat architecture was described using the moving-qubit model in the prior Parts. That coarse-grained model is abstracted to capture the features of the device that matter for the design of quantum error correction protocols. In this section, we provide a specific micro-architectural implementation of the walking cat architecture as it will be implemented using trapped ions with electronic qubit control (EQC)~\cite{malinowski2023wire}. 

Ions, which are electrically charged particles in which our qubits are encoded, are attracted to local minima in an electric potential. Suitably shaped electric fields generated by electrodes on a quantum charge-coupled device (QCCD) chip can trap ions near the chip's surface and also move them with changing fields~\cite{kielpinski2002architecture, malinowski2023wire}. QCCD devices have been shown to do so with control voltages well within standard digital-to-analog output ranges. Slow-varying electric fields used for ion transport only weakly perturb the internal states of the ion that encode a qubit, so that ions can preserve the encoded quantum information while moving at a wide range of speeds. Experiments have demonstrated such transport with low motional excitation (less than one motional quantum) at speeds of several tens of $\mathrm{m/s}$~\cite{sterk2022closed} and approaching $100~\mathrm{m/s}$~\cite{PhysRevA.107.043119}. 

While electric fields are used for precise motional control of ions, magnetic fields are required for control of internal states of qubits, in a similar way nuclear magnetic resonance experiments control nuclear spins of samples~\cite{vandersypen2004nmr}. Two ion-based qubits can be entangled when they vibrate in a collective manner such that their internal states change collectively in response to a spatially varying magnetic field. So far, existing realizations of QCCD architecture rely on using lasers to impose an effective magnetic field to rotate between qubit states or generate entanglement between two ions~\cite{moses2023race, ransford2025helios}, but their scalability is hindered by the high laser power per ion required to operate at large detunings to suppress photon scattering error~\cite{ozeri2007errors}. While employing an efficient configuration of output grating couplers with integrated photonics can partially alleviate the challenge~\cite{kolhatkar2026efficient}, the fundamental hurdle of evenly distributing high-power laser light across many locations remains. On the other hand, the EQC architecture uses a real magnetic field from near-field radio-frequency (RF) signals generated by antennas on a QCCD chip~\cite{loschnauer2025scalable}. Moreover, a single active antenna can drive parallel operations across the ions trapped above it, supporting either exclusively single-qubit gates or exclusively two-qubit gates simultaneously, though not a mixture of both. Thus, power dissipated per qubit can be greatly reduced compared to laser-based schemes with one grating coupler per qubit. While RF fields cannot be focused sharply like laser beams, antennas with optimized shape can significantly suppress field strength outside the gate zones (to be defined later in the section) and precise ion positioning can be used to achieve excellent crosstalk error (estimated to be $\leq 1 \times 10^{-6}$ per single-qubit gate in Ref.~\cite{loschnauer2025scalable}) during zone-selective operations. We design our micro-architecture to take advantage of the scalability and hardware efficiency of EQC.

\subsection{Functions of a QCCD chip}
A strategy for building a scalable quantum hardware architecture is tiling high-fidelity unit cells across a planar grid. These cells can host multiple qubits and are often specialized into distinct types with specific functions, such as quantum gates or qubit measurement and reset. Control signal density overhead is managed as the system scales by leveraging signal multiplexing and co-wiring among the unit cells. In contrast with hardware where qubits are fixed in place, a crucial element of our architecture is the ability to move/shuttle qubits within and between unit cells to where they are needed.
QCCD provides an ideal platform for the required routing of mobile qubits with managed control overhead~\cite{malinowski2023wire}.  

\begin{figure}[b!]
    \centering
    \includegraphics[width=\linewidth]{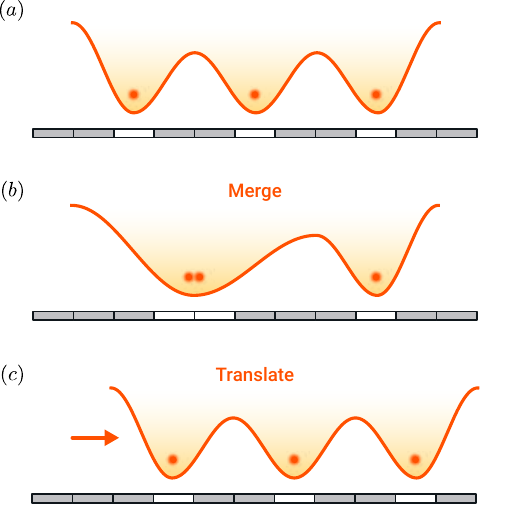}
    \caption{Simplified illustration of potential well configuration changes on a QCCD chip. The orange curves qualitatively represent the axial trapping potential landscape seen by positively charged ions. The small rectangles correspond to an array of segmented DC voltage electrodes, with the color on each segment qualitatively representing a voltage value (gray = high, white = low). The axial positions of the ions follow the local minima of the axial trapping potential created by the electrodes. With appropriate voltage ramps, one can transition from a) initial configuration to b) configuration with a merged well, where the balance between external field and Coulomb repulsion determines the final positions of the merged ions, or c) linearly translated configuration. In general, a voltage update on a DC voltage electrode affects both axial and radial confinement (not depicted here), where the radial confinement primarily comes from the pseudopotential created by RF electrodes of the trap.
    }
    \label{fig:merge_and_transport_example}
\end{figure}

At its core, a QCCD realizes an electric potential landscape in which ions can be trapped and shuttled. The potential landscape can be dynamically adjusted with control voltages.
A QCCD is generally manufactured on a substrate using planar processes and contains different functional layers, with the top layer featuring patterned electrodes that define potential well configurations for holding multiple ions at specific locations. A {\em potential well} is a local region of positive trapping curvature (confinement strength) that can hold a non-negative number of ions, where Coulomb repulsion creates small spacing between ions within the well; a {\em potential well configuration} refers to the spatial distribution of potential wells across the grid, where some wells can have different curvatures from others and well spacing is generally much larger than ion-ion spacing within a well (tens to hundreds of micrometers vs. less than ten micrometers). Changes in voltages applied to the electrodes lead to a transition from one potential well configuration to another, resulting in different levels of confinement, spacings between wells, and even the number of wells (cf. \cref{fig:merge_and_transport_example}). Functional layers below the top layer of the QCCD are generally designed to implement specific functions supported by the unit cells. For example, scalable QCCD architectures generally include photonic layers for distributing laser power through a network of nanophotonic waveguides terminated by output grating couplers directed at trapped ions above. These distributed lasers operate at various wavelengths for specific purposes such as ion cooling, state preparation, and measurement. For EQC, which has demonstrated state-of-the-art gate fidelities~\cite{loschnauer2025scalable, hughes2025trapped}, the QCCD includes an antenna layer with embedded RF antennas that deliver gate waveforms to the trapped ions above.

\subsection{Extensions of the moving-qubit model}
\label{subsec:extension_of_the_moving_qubit_model}

In this subsection we refine the moving-qubit model introduced in \ref{sec:The moving qubit model} with hardware constraints relevant to QCCD systems. We note several key areas of refinement.

\begin{figure}[b!]
    \centering
    \includegraphics[width=\linewidth]{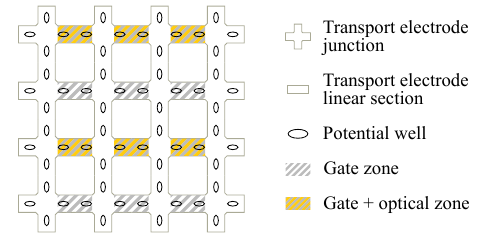}
    \caption{Embedding of a ``moving qubit'' grid in the top metal of our trap. The transport electrodes represent the two-dimensional network of junctions along which ions can be shuttled; shuttling operations facilitate transport of ions to establish proximity between ion pairs required for entangling gates. The ellipses denote potential wells at which ions can be held; here we show a possible well configuration used for gating. Gate zones are indicated by gray hatching. As emphasized in the main text, we assume a dense embedding of gating antenna in every row. Optical zones, indicated by yellow hatching, are available on every other horizontal row. 
    }
    \label{fig:uarch_grid_embedding}
\end{figure}

\textit{Ion-specific implementation} - Abstract qubit models neglect details regarding the explicit physical implementation of fundamental operations such as gating, measurement, state preparation. Further physical-level considerations, such as heating, cooling, and loading, are important in ion-specific implementations of qubit systems. An ion has many internal states, two of which are chosen to define a qubit. In particular, in an ``optical, metastable, ground state'' (OMG) encoding, the qubit levels correspond to two particular energy levels in a metastable orbital \cite{allcock2021omg}. Gating involves driving transitions between those two internal states, while avoiding transitioning to spectator ({\em i.e.} non-qubit) states. It is also possible to map one or both of the qubit states into other internal states for various purposes. We refer to this as \emph{shelving} the state. Measurement involves a sequence of laser pulses to shelve a target qubit state to ``bright'' states and induce fluorescence that is measurable on a photodetector or camera. The specific preparation of the computational zero state $\ket{0}$ involves a specific sequence of photonic operations. Furthermore, physical-level effects like ion heating must be mitigated with cooling techniques to prevent buildup of non-Markovian error, which must be built explicitly into a micro-architecture design. The exact nature of each of these operations depends strongly on the particular ion species an architecture is built around. 

\textit{Explicit shuttling constraints} - The moving-qubit model assumes a fixed two-dimensional grid of sites, and it does not specify particular intermediary states and transitions that permit re-arrangement within the model. In QCCD systems, one similarly starts with a fixed two-dimensional grid of junctions connected by linear sections, depicted in \cref{fig:uarch_grid_embedding} as corners and edges of squares respectively. Ions can be transported from one linear section to another through shuttling sequences. Both the junctions and the linear sections between the junctions can support trapping sites, and the number of trapping sites (potential wells) can change between different potential well configurations by changing control voltages on the surface electrodes. The dynamic reconfiguration is not arbitrary in the sense that the detailed geometry of the electrode layout determines the types of potential well configurations that can be implemented. 

\textit{Weighted transport cost} - Transport operations are implemented as a sequence of primitive shuttling steps.
The cost of each primitive shuttling  step may differ, and depends on the transport time and motional excitation (which necessitates cooling) incurred under the given control voltage budget.
We reduce the cost of transport operations by co-designing the logical architecture and QCCD micro-architecture to ensure that low-cost primitives are most often used for implementing the necessary qubit alignment between data and ancilla qubits. 

\textit{Zonal operation} - In order to perform a two-qubit gate, the participating ions must be located in the same linear section above a gate antenna and their potential wells merged into a single well. This requires us to refine the moving-qubit model assumption that two-qubit gates are available for all nearest-neighbors pairs of qubits. However, we note that the same wire for a gate antenna can be routed under multiple linear sections, such that a large number of two-qubit gates can be performed in parallel. This extensive parallelism based on simple classical control is one of the key advantages of trapped ion architecture based on electronic qubit control~\cite{malinowski2023wire}. In this work, we assume that every horizontal section has a gate RF antenna underneath - we call those horizontal sections the gate zones. Because single-qubit gates also require near-field RF, they are only available in the gate zones as well. Similarly, qubit reset and readout become available when the participating ion is at an optical zone - a section with output grating couplers for various laser wavelengths and a collecting lens that relays photons to a photodetector. We assume optical zone density is sparser than the stated gate zone density because of the engineering challenge of photonic waveguide routing with small bend radii and optical crosstalk between neighboring optical zones. In this work, optical zones are available in every alternating row.

We conclude the subsection with a note on the difference in accounting of transport cost between the moving-qubit model and the micro-architecture model. The moving-qubit model does not assume knowledge of transport junctions and hence an ion can increment its column index within a row in one shuttling step. On the other hand, because the micro-architecture model assumes the ion has to go through transitions between allowed potential well configurations, it requires two shuttling steps to increment its column index (effectively junction index). Also, as stated, the moving-qubit model assumes availability of two-qubit gates between any nearest neighbors, whereas the micro-architecture incurs extra shuttling steps to bring the control and target ions into the same linear section. Thus, the micro-architecture model will always have a strictly larger count of shuttling steps.

\subsection{Physical implementation of the moving-qubit operations}

\begin{table*}
\centering
\begin{adjustbox}{max width=\textwidth}
\begin{tabular}{|p{0.75in}|p{2in}|p{2in}|p{2in}|c|c|}
\hline
\textbf{Operation} & \textbf{Purpose} & \makecell{\textbf{Location} } & \makecell{\textbf{Error} \\ \textbf{mechanisms}} & \makecell{\textbf{Timings}} & \makecell{\textbf{Reference}}  \\
\hline
Single-qubit drives & Implementing SU(2) rotations   & Any gate zone under single-ion wells    &   Coherent and incoherent noise    & 5-10 \unit{\micro s} & \cite{loschnauer2025scalable}       \\ \hline
Entangling drives     & Implement a specific entangling interaction & Any gate zone under two-ion wells   & Coherent noise, incoherent noise, and leakage    & 100-300 \unit{\micro s} & \cite{loschnauer2025scalable, hughes2025trapped}    \\
\hline
Measurement laser pulses     & Measurement of $\ket{0}$, $\ket{1}$, and possibly other leakage states.  & Optical zones     & Measurement error from stray photons & 350~\unit{\micro s} per image &  \cite{an2022high} \\\hline
Preparation laser pulses  & Preparation of $\ket{0}$  & Optical zones & Polarization error, intensity fluctuation, optical crosstalk  & 450~\unit{\micro s}; $>90\%$ post-selection rate & \cite{sotirova2024high}\\ \hline
Cooling laser pulses  & Cooling   & Optical zones & Beam misalignment from ions, polarization error, intensity fluctuation  & 0.3~phonon/\unit{\micro s} & \cite{clements2026sub} \\ \hline
Split        & Split two potential wells to separate 2-ion crystals        & Zones with supporting control electrodes     & Heating, loss, incoherent error  & 50-100 \unit{\micro s}   & \cite{PhysRevLett.109.080502}       \\ \hline
Merge      & Merge two 1-ion crystals into a 2-ion crystal           & Zones with supporting control electrodes    & two-qubit gate error increase from motional heating, loss, incoherent error  & 50-100 \unit{\micro s}   & \cite{PhysRevLett.109.080502}    \\ \hline
Idle                & Do nothing             & Any ion anywhere         & Heating, incoherent memory error, leakage, loss  & n/a    & n/a  \\ \hline
Shuttling              & Move an ion across a junction  & Anywhere in the trap  & Incoherent memory error, phase tracking errors, heating  & 5-15 \unit{\micro s}  & \cite{sterk2022closed, PhysRevA.107.043119} \\
\hline
\end{tabular}
\end{adjustbox}
\caption{Operations in the micro-architecture model. Error sources list the expected dominant error sources and are not exhaustive. Timings are ranges demonstrated in state of the art results in associated references.}
\label{tab:uarch_model}
\end{table*}

In \cref{sec:The moving qubit model} we introduced the high-level model of moving qubits, including a tabulation of all basic operations in \cref{tab:moving qubit model operations}. In the micro-architecture model, each of these operations further decomposes into refined physical operation sequences. This mapping is summarized below. 

\textit{State preparation} - To prepare a cooled ion into a qubit, its internal state must be initialized using optical pumping and then mapped to a specific internal state that serves as one of the two levels comprising a qubit. Optical pumping is achieved by illuminating the ion with a laser of a specific wavelength and polarization such that the internal state population performs a random walk within the energy level manifold as a sequence of stimulated excitation followed by spontaneous emission, until most of the population falls into a dark state from which the population cannot escape. Physical and engineering constraints generally limit the steady-state population in the dark state, namely the fidelity of optical pumping. At the end of optical pumping, shelving pulses from laser and/or RF can be used to transfer the dark state population to a target qubit state. Following these mapping pulses, a heralding process may be employed to detect failed state preparation \cite{sotirova2024high}. 

\textit{One-qubit \& two-qubit gates} - Both single-qubit and two-qubit gates are implemented with high fidelity using EQC, which combines electronic position control of ionic qubits with spatially structured RF magnetic fields generated by RF currents. An RF signal can induce rotation between two hyperfine levels of an ion that define a Bloch sphere. We control site-selective single-qubit operations by locally addressable offset operations that control the interaction of individual ions with the driving field. This allows us to apply massively parallel layers of single-qubit gates with locally controlled addressing. For two-qubit gates, two ions are first merged to form an ion crystal inside a common potential well, in which the motions of the ions hybridize to form collective motional modes. An RF gradient can strongly couple to a particular collective mode and induce the two ions to follow a state-dependent trajectory in the motional phase space, making the qubits acquire state-dependent geometric phase.
Like other motional mode mediated gates, that phase space trajectory must begin and end in the same state so that the ion's internal state remains unentangled with that motional mode. Perturbations to the phase space trajectory by phonons and leakage outside the qubit subspace due to off-resonant driving of unwanted transitions are the dominant sources of remaining error after calibration. As with the single-qubit gates, we can achieve massively parallel two-qubit gate layers with additional site-selection by exploiting the fact that the entangling gate drive, when coupled with sufficient dynamical decoupling, acts like an identity operation (up to single-qubit corrections) on ions that have not been merged. Low occupation of motional modes is important for reducing sensitivity to temperature-dependent physics that can leave residual spin-motion entanglement or deviation from the target geometric phase after a two-qubit gate~\cite{sutherland2022one}, although it should be emphasized that RF gate techniques with a judicious gate drive ramp can be made relatively robust to thermal mode occupation while maintaining state-of-the-art fidelity~\cite{hughes2025trapped}. 

\textit{Leakage reset} - Leakage reset works analogously to state preparation. However, there are two physically distinct leakage channels: ground state and metastable state leakages---these are distinguished by which specific non-qubit states in the ion a qubit might have leaked to. For ground state leakage, we can reset by directly applying the preparation sequence on an ion in the ground state. For metastable leakage, the ion can be routed to specialized optical zones in the loading chip where one can optically pump the ion out of the metastable state without inducing optical crosstalk on qubits encoded in metastable states~\cite{allcock2021omg}.

\textit{Measurement} - Qubit state measurement works by performing shelving pulses necessary to map the qubit back to a ground state of the ion where it can continuously fluoresce under illumination by excitation laser(s). Collection optics focused on the ions can collect fluorescent photons and relay them onto photodetectors; each detector signal is then binarized based on whether the photon count within the exposure time is below or above a threshold count. This imaging procedure can be sequentially applied to each of the two qubit states, so that the two classical bits obtained can be used to inform if the qubit was in a first state, a second state, or neither, which would indicate a potential qubit leakage or loss. Ground state leakage can be directly measured by applying imaging without shelving pulses and checking if the image is bright, before the two images for qubit state readout are taken. Metastable leakage cannot be immediately distinguished from loss as both appear as dark in all three images, but it can be routed to the aforementioned special optical zones with optical pumping out of the metastable orbital, which can then distinguish metastable leakage from loss.

\textit{Idle} - Idling is a similarly simple process in the micro-architecture model as in the moving-qubit model. The primary distinction is the precession of a physical qubit's frame under Z-type memory errors due to spatially varying magnetic fields. The presence of spatially varying AC magnetic fields from the trap RF electrodes and active gate RF antennas can cause dephasing if the inhomogeneous AC Zeeman shifts are not corrected via phase tracking, if the phase tracking calibration drifts substantially, or if the fields fluctuate in time. Technical electric field noise from the electrodes and anomalous electric field noise from the surfaces can also cause motional heating, which, if untreated by scheduled cooling steps, can degrade gate fidelities. 

\textit{Transport} - Low-excitation ion transport can be achieved by quasistatically updating electrode voltages to transition from one potential well configuration into another, such that the potential well centered at one location adiabatically shuttles to another nearby potential well location, similar to how signal electrons are transferred in CCD cameras. Following the formalism in Ref.~\cite{blakestad2010transport}, time-dependent control waveforms are obtained by framing each shuttling step as a constrained optimization problem, matching the desired moving potential to a linear combination of basis potentials generated from transport electrodes.

\subsection{Lower-level operations}

\textit{Loading} - A high flux of ions can be generated by photo-ionizing a beam of neutral atoms generated by a variety of sources, such as by ablation, ovens, or a magneto-optical trap. The flux can then be photoionized inside of the trapping region and be laser-cooled to remove motional quanta (phonons), a necessary step for high-fidelity ion transport and gate control. As loading is fundamentally a non-deterministic operation, loading rates in trapped ion systems will generally need to be kept higher than loss rates, with a mechanism for ejecting or storing excess loaded ions. Single-zone loading rates in excess of 400 ions per second have been experimentally demonstrated~\cite{bruzewicz2016scalable}, although loading rates vary by ion species~\cite{shi2023ablation}. This can be further increased by loading in multiple zones in parallel. 

\textit{Qubit loss and replenishment} - A QCCD is generally operated at cryogenic temperature such that ion heating and loss due to experimental imperfection are minimized. At cryogenic temperature, background gas pressure contributing to ion collision with background neutral molecule is strongly suppressed. Similarly, cryogenic temperature strongly suppresses anomalous ion heating rate from electrode surfaces~\cite{labaziewicz2008suppression}. Ions can be trapped above a QCCD for a time scale that is many orders of magnitude longer than the slowest physical operation timescale of the system, but ion loss is generally inevitable and thus fresh ions must be transported from the loading zones to replenish lost qubits.

\textit{Cooling} - Laser cooling removes motional entropy from ions and dissipates it to the environment through spontaneous emission. The simplest and most robust cooling method is Doppler cooling, which provides frictional cooling force through velocity-dependent radiation pressure and is used for cooling hot ions freshly captured in a well. However, Doppler cooling yields a relatively high steady-state motional entropy with average phonon number well above 1 for typical radial mode frequencies. To reach motional ground state, additional cooling techniques that engineer higher phonon-removing scattering rates than non-phonon-removing rates are used, such as sideband cooling~\cite{monroe1995resolved} or EIT cooling~\cite{lechner2016electromagnetically}, but such techniques often operate at a lower cooling rate and/or increased experimental complexity than Doppler cooling. Polarization gradient cooling can provide a good balance of fast cooling rate and low steady-state phonon number ($\bar{n} \approx 1$)~\cite{clements2026sub}. We emphasize again that motional ground state is not strictly required for state-of-the-art gate fidelity in EQC, as it has been proved to work at steady-state Doppler cooling temperature (with average phonon number of $\bar{n}\approx 3.5$ in Ref.~\cite{hughes2025trapped} for ${}^{40}\mathrm{Ca}^{+}$) although initializing in such a state tolerates more heating steps between scheduled cooling steps. While data qubits cannot be laser-cooled, their motional entropies can be effectively reduced by sympathetic cooling~\cite{home2009memory}.

\section{Micro-architecture of the components}
\label{sec:Micro-architecture of the components}

In this section, we design a micro-architecture for the three most critical components of the walking cat architecture: the memory block, the magic factory and the cat factory.
Thanks to our unified design for the memory block and the magic factory, the magic factory micro-architecture can be immediately derived from the memory micro-architecture.
The design of these components is described in \cref{subsec:Memory and magic factory micro-architecture}.
Then, the cat factory micro-architecture is described in \cref{subsec:Cat factory micro-architecture}, which also discusses the implementation of cat qubit transport at the core of the cat-based measurement interface (see \cref{fig:architecture_subcomponents}).

\subsection{Memory and magic factory micro-architecture}
\label{subsec:Memory and magic factory micro-architecture}

Mapping the three-ring structure of \cref{fig:architecture_overview_memory} to a QCCD device involves mapping the abstract cyclic structure to physical shuttling of ions and implementing the required gating and measurement operations subject to the physical device constraints previously detailed. Explicitly, our memory block embedding must support: the required cyclic shifts, entangling operations between ancilla and data qubits, ancilla qubit measurement and resets, data qubit measurements and resets, and leakage and loss checks. Providing efficient embeddings and physical support for each operation is an interesting and complex trade space optimization problem. 

\begin{figure}
    \centering
     \includegraphics[width=\linewidth]{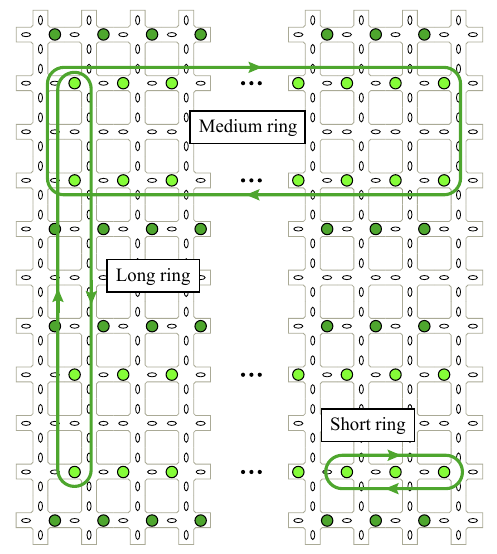}
    \caption{Mapping of the three-ring framework to a two-dimensional embedding. Here, the data (dark green) and ancilla (light green) ions are arranged in rows of the trap. Long rings are accomplished through swaps of the top and bottom blocks. Medium rings are accomplished through cyclic shifts of the ancilla within each block. Small rings are accomplished through embedded cycles, as detailed in \cref{fig:percolation_demo}.}
    \label{fig:three_loop_embedding}
\end{figure}

Beginning with the cyclic structure, we envision a prototypical two-block embedding, as shown in \cref{fig:three_loop_embedding}, which suffices for all codes considered in this work. The three-ring structure can be efficiently embedded in this implementation according to the mapping of: long ring~$\mapsto$~block swaps, medium ring~$\mapsto$~cyclic shift, short ring~$\mapsto$~embedded shift. Effectively, our embedding amounts to a twice folded version of the abstract embedding of the block on a line. By folding our embedding and implementing long shifts as stacked block swaps, we are able to significantly reduce the transport overhead associated with syndrome extraction removing the need for a highway used for fast transport along a linear block of qubits in \cref{fig:architecture_overview_memory} and \cref{sec:Memory block}.

\begin{figure}
    (a)\\
    \includegraphics[width=\linewidth]{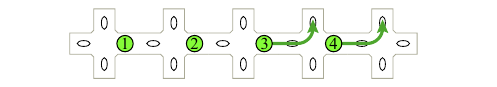} \\
    (b)\\
    \includegraphics[width=\linewidth]{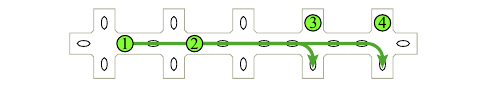} \\
    (c)\\
    \includegraphics[width=\linewidth]{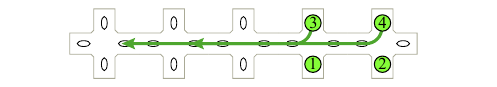} \\
    (d)\\
    \includegraphics[width=\linewidth]{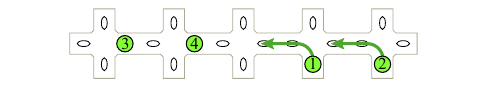} \\
    (e)\\
    \includegraphics[width=\linewidth]{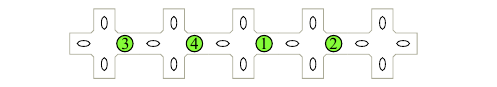} \\
    \caption{Example of embedded ring dynamics. Here, we implement a cyclic shift of ions in a single row embedding. (a)~First, the right set of ions are ``hidden'' in the vertical legs. (b)~Next, the left set of ions translate right and are hidden in the lower vertical legs. (c)--(e)~Finally, the right set of ions move left and we move the left set of ions out of the vertical legs. This operation is efficient for cycles that involve a small number of ions, but dense ring embeddings are more efficient when there is a large number of ions involved in the cyclic shift.}
    \label{fig:percolation_demo}
\end{figure}

Next, for gating, the ancilla qubits proceed according to the cyclic shifts described above. Once the required alignment between the ancilla and the data qubits is achieved, two-qubit gates can be performed by shuttling data qubits up/down one row to meet their associated ancilla qubits, merging the pairs, driving the entangling transition, and finally splitting the pairs and returning the data qubits to the original locations. Clearly, single-qubit operations can be performed on the data or ancilla qubits before the merge and after the split. This process of cyclic shifts followed by gating is repeated until all the required gates between data and ancilla qubits have been made.

To measure and reset the ancilla qubits, we must move the ancilla qubits to an associated photonic row. This can be accomplished by shuttling the ancilla up/down a row to the embedded photonic row contained in the middle of the code block. In practice, we expect that we will only support one optical zone per horizontal section, so we will have to measure the ancilla in two batches. A sparser embedding could achieve ancilla measurement in a single batch, highlighting a common space-time tradeoff in these design problems. After measurement, we must cool and reset the ancilla back to the metastable qubit state, which can also be accomplished in the optical zones. In practice, we may wish to employ a second batch of ancilla qubits so that we can pipeline the state preparation and measurement procedure, which would remove all contribution of ancilla measurement and reset to the logical clock cycle. We emphasize that, though more ions are actually employed in a pipelined protocol, the number of device zones employed remains constant. When dealing with physical embeddings at the micro-architecture level, the correct figure of merit to cost a protocol is the number of zones required. 

To measure and reset the data qubits, we proceed analogously to the ancilla qubits. In our embedding, the data qubits are already naturally localized on photonic rows, so minimal shuttling  is required. The only difference is the inclusion of beacon qubits in the layout, as discussed below, but the beacon ion can be moved out of the way as with the second ancilla ion. After measurement, the data ion must be reset. This can either be done directly in the optical zone after measurement, or we can replace the measured data ion with another freshly prepped qubit, which reduces the clock cycle through pipelining. 

Integrating beacon qubits within the code blocks for leakage and loss check requires introducing an additional ion with new functionality in the embedding. In our layout, we imagine placing the beacons in the same rows as the natural resting location of the data qubits. In this way, every data qubit has a ``partner'' beacon ion that can be used for leakage and loss checks. In fact, the beacon ions can additionally be used for cooling, naturally integrating the leakage, loss, and cooling of the data qubits in a single unit that we call a ``leakage and loss reduction unit'' or LLRU. 

To measure ion loss with a beacon ion, we may exploit the fact that a split/merge sequence that occurs between an ion and a hole (from a previously lost ion) will dramatically heat the beacon ion. As the trapping curvature has to change sign between positive and negative at the center, there is a moment during the sequence where there is very little restoring force where the lone beacon ion sits (note that Coulomb repulsion contributes to the restoring force for a splitting/merging ion pair)~\cite{kaufmann2014dynamics}. This exposes the beacon ion momentarily to potentially many orders of magnitude higher heating rate than what is felt by an ion pair, as the heating rate scales with $S_{E}(\omega)/\omega$ where $S_{E}(\omega)$ is the power spectral density of the electric field noise at the trapping frequency $\omega$ and generally becomes worse near zero frequency~\cite{brownnutt2015ion}. A discrimination procedure that leaves a cold beacon ion trapped and ejects a hot beacon ion out of the trap is followed by a standard imaging step. In this way, we can detect loss of data ions by periodically merging the data ion with a beacon ion, splitting the pair, and measuring the beacon ion. If the beacon ion is lost, then we infer that the data ion was also lost. We emphasize that the beacon ion need not be in the metastable state for this procedure to work.

To measure leakage with a beacon qubit, we may employ well-established techniques of leakage checks \cite{stricker2020experimental}. Experimentally, these protocols amount to performing a gating and measurement sequence that teleports the information from a data ion to a beacon ion and then measures the data ion, see \cref{subsec:atom-loss-gadget}. Physically, this check follows a similar sequence to loss checks with two additions: 1) an active gate is performed in the merged state, and 2) the beacon qubit must be allocated to $\ket{0}$ in the metastable manifold. We note that leakage checks also check for loss, but the leakage check requires a higher overhead because now there are additional qubit reset and gating operations that must be performed. 

Finally, we note that the beacon qubit naturally provides the capabilities for cooling data qubits. When set in the ground state, we may cool beacon qubits with integrated photonics by cycling a transition without affecting the data qubit. This opens the door to various cooling protocols, including sympathetic and exchange \cite{fallek2024rapid} cooling.  

\begin{figure*}
    \centering
    \includegraphics[width=\linewidth]{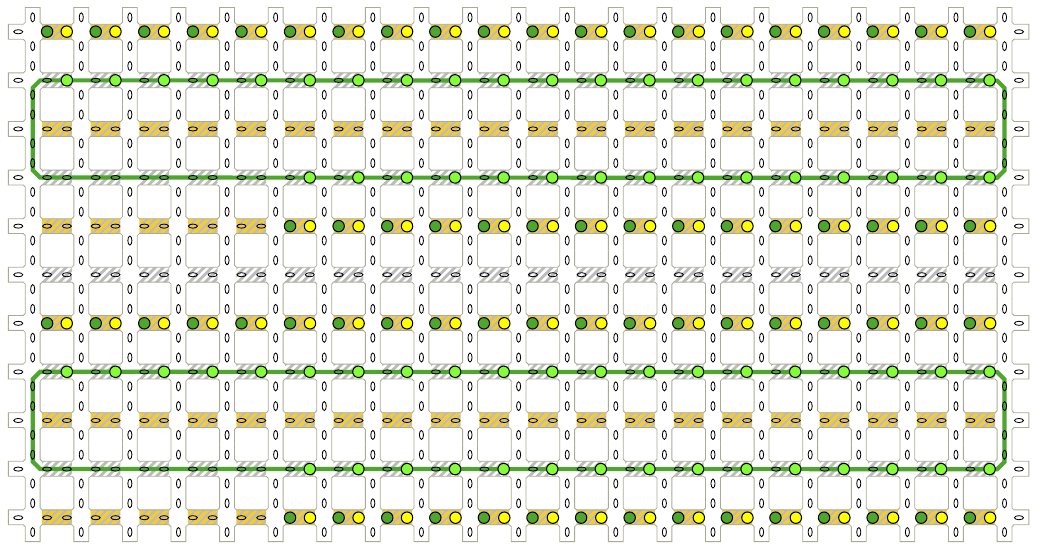}
    \caption{Embedding of the \code{70} code on the two-dimensional junction network. Dark green (yellow) circles represent data (beacon) qubits. Light green circles represent ancilla qubits. The number of ancilla qubits on each row is a multiple of five, which is the period of the short cyclic shift in this code, so that short cyclic shifts can be performed entirely within each row. The green lines indicate the path for the medium ring ancilla transport. Vertical legs are kept vacant so that they can be used for the block swap that implements the long ring ancilla transport. Vacant rows in the interior of the medium ring paths (rows 3 and 9) can be used for state preparation and measurement of a second set of ancilla qubits. The vacant middle row (row 6), as well as vacant rows to the top and bottom of the embedding shown here, are used to bring in a supply of fresh ions to replace leaked or lost qubits. The embedding for the \code{102} code is defined analogously, but less constrained: since the code only utilizes long and medium rings for ancilla transport, the number of ancilla qubits per row is not required to be a multiple of five, which allows for a slightly more space-efficient embedding.}
    \label{fig:uarch_memory_embedding}
\end{figure*}

\begin{table}
    \centering
    \begin{tabular}{|c|c|c|}
        \hline
         & Q70 & Q102\\
         \hline
         Transport & 424 & 387 \\
         \hline
         Merge / split & 16 & 18 \\
         \hline
         Parallel 1q gate layer & 2 & 2 \\
         \hline
         Parallel 2q gate layer & 8 & 9 \\
         \hline
         Readout & 3 & 3 \\
         \hline
         State preparation & 1 & 1 \\
         \hline
    \end{tabular}
    \caption{Primitive operation rounds for a single syndrome extraction cycle in the \code{70} and \code{102} codes. This includes initial state preparation of ancillas, transport and gating with data qubits, as well as the final readout and leakage check. Ion loss checks can be performed in parallel to ancilla transport sequence and are not included in this table. State preparation of a new set of ancilla qubits could also be performed in parallel in a separate part of the trap, \emph{cf.} \cref{fig:uarch_memory_embedding}.}
    \label{tab:uarch_memory_embedding_steps}
\end{table}

With these specifications, we may define a micro-architecture for our memory blocks that encompasses all codes in the three-cycle layout considered in this paper. In \cref{fig:uarch_memory_embedding}, we specify embeddings for the \code{102} code and the \code{70} code. We have performed detailed, device-level emulations of the syndrome extraction cycles of these codes under the micro-architecture specified here. Because device parameters can vary significantly subject to experimental trade spaces, we do not report actual syndrome extraction cycle estimates under a micro-architecture model. Instead, we report the detailed break-down of the syndrome extraction cycle into primitive operations for the \code{102} and for the \code{70} in \cref{tab:uarch_memory_embedding_steps}. 

\begin{figure*}
    \centering
    \includegraphics[width=\linewidth]{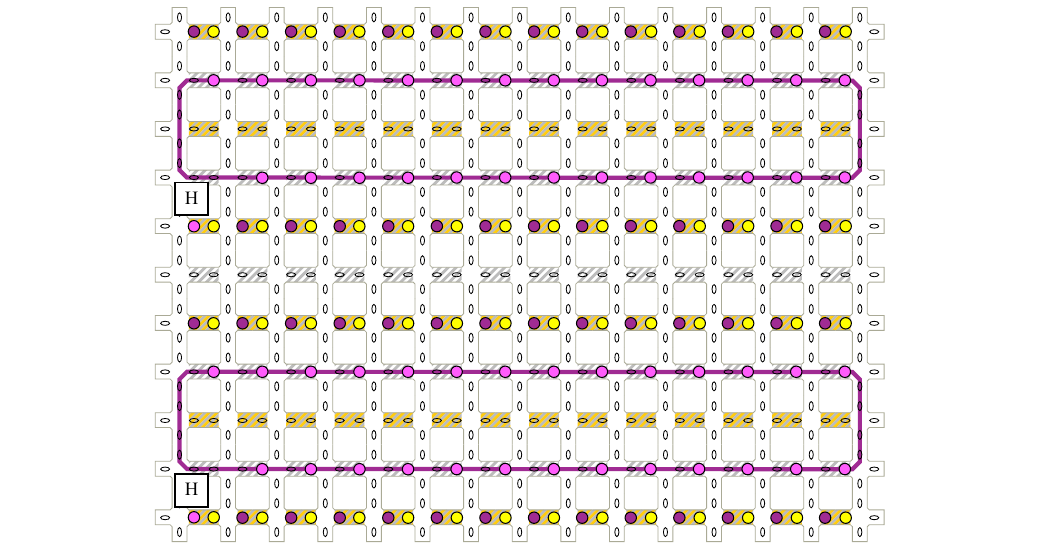}
    \caption{Embedding of the CH2 magic factory. Dark purple (yellow) circles represent data (beacon) qubits. Light purple circles represent ancilla qubits. In this embedding we use two H state resource qubits, aligned in the same rows as the data qubits. Operation with a single H state qubit is also possible.}
    \label{fig:uarch_magic_factory_embedding}
\end{figure*}

The micro-architecture for magic factories extends straightforwardly from the memory block micro-architecture. As was said in \cref{sec:magic_state_factory}, both the \MEK and the \CHfactory protocols can be achieved by including one additional physical qubit with the support for a single-qubit $R_y(\pi/4)$ operation. Coupled with cat-facilitated measurements of joint observables with this additional qubit and the code block, we may implement either protocol in a suitably sized code block. This additional resource qubit is low overhead and can be placed in any zone that supports gating. See \cref{fig:uarch_magic_factory_embedding} for an illustration of the micro-architecture embedding of the \CHfactory protocol.

\subsection{Cat factory micro-architecture}
\label{subsec:Cat factory micro-architecture}

\begin{figure*}[t]
    \centering

    \begin{tabular}{c@{\hspace{0.5cm}}c}
    \small (a)~Initial layout & \small (b)~Cyclical data shift \\
    \includegraphics[width=0.48\textwidth]{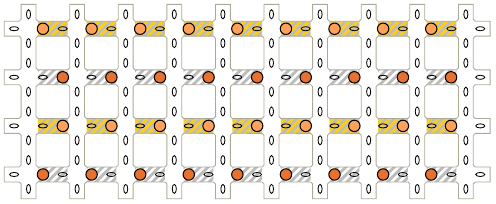} &
    \includegraphics[width=0.48\textwidth]{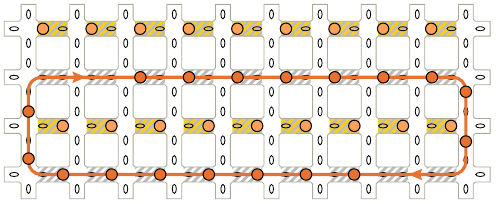} \\[0.5cm]

    \small (c)~Vertical data transport & \small (d)~\CX gate -- 8 pairs \\
    \includegraphics[width=0.48\textwidth]{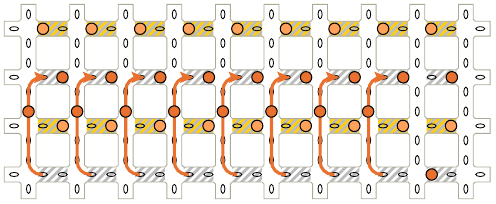} &
    \includegraphics[width=0.48\textwidth]{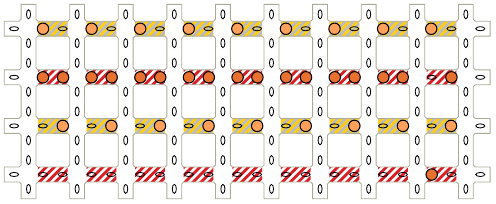} \\[0.5cm]

    \small (e)~Ancilla-data \CX & \small (f)~Ancilla measurement \\
    \includegraphics[width=0.48\textwidth]{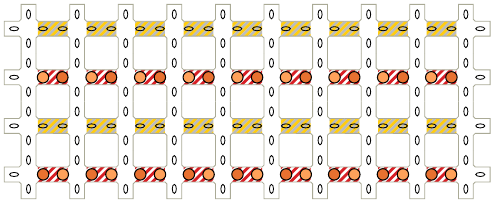} &
    \includegraphics[width=0.48\textwidth]{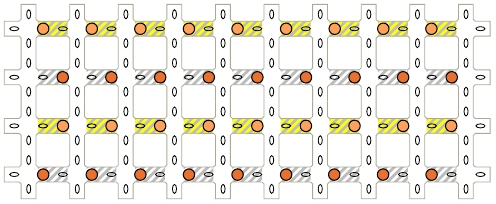}
    \end{tabular}

    \caption{Operations of weight-18 cat factory in a four-row layout. Dark (light) orange circles represent cat data (ancilla) qubits. In subpanels (d)--(f), the red tinted hatching of gate zones and bright yellow hatching of optical zones signal active use of the respective zones.}
    \label{fig:w18_cat_factory_uarch_phases}
\end{figure*}

Here we show how the abstract cat factory layout introduced in \cref{subsec:Design of cat factories} maps to rows in QCCD. \cref{algorithm:cat_verfication} shows how the ancilla and data qubits are arranged into four abstract rows and $w/2$ columns, where $w$ is the weight of the target cat state. It should first be pointed out that ions merge axially in a QCCD row to perform two-qubit gates, as opposed to merging radially (vertically) as it was depicted in \cref{fig: cat prep} for ease of visualizing \CX connections. Thus, a faithful representation of \cref{fig: cat prep} on the QCCD grid requires bringing ions on separate rows (which comprise the cyclical orbit) into a common row before a two-qubit gate layer. Ions can slide into nearby junctions and climb up or down the vertical sections to find their target row. Because of the assumed constraint that the photonic operations are only available at alternating rows, a simple mapping yields the following ordering of rows: ancilla row 1 (photonic row 1), data row 1 (gate row 1), ancilla row 2 (photonic row 2), and data row 2 (gate 2). We emphasize that there can be more space-efficient and/or time-efficient layouts although those deviate from the qubit arrangement sequence discussed and simulated in \cref{subsec:Design of cat factories} and require a deeper analysis of spacetime tradeoffs. The number of horizontal sections is still equal to the number of abstract columns, $w/2$. See \cref{fig:w18_cat_factory_uarch_phases} for an example $w=18$ cat factory layout. 

In the four-row layout, the data qubits can circulate around the perimeter of the second photonic row, while the ancilla qubits stay idle in the photonic rows. During the cat preparation stage, data qubits cyclically shift (\cref{fig:w18_cat_factory_uarch_phases}(b)), and then the data qubits in the bottom row selectively slide through the vertical sections to form two-ion pairs at the upper data row (\cref{fig:w18_cat_factory_uarch_phases}(c)-(d)). This means that in general there will be some gate zones where a data qubit does not have a partner qubit to merge with during the two-qubit gate waveform. Because we know the locations of such lone qubits, we can design a trap chip where a conditional electric field can be activated to bias lone qubit ions to stay at their current positions instead of merging with vacant wells and experiencing high heating rates. In the cat verification stage, the ancilla qubits in the photonic rows slide down the vertical sections to partner with data qubits at every gate zone (\cref{fig:w18_cat_factory_uarch_phases}e). After each ancilla qubit is entangled with two data qubits, its state is read out at the optical zone it was originally prepared in to obtain the parity of the ZZ stabilizer (\cref{fig:w18_cat_factory_uarch_phases}f). The ancilla ions can be cooled and reset at their respective optical zones, but at the cost of potentially long idle times for the data qubits, as multiple preparation pulses may be needed to reduce reset error. To facilitate prompt repeating of ZZ stabilizer measurements, the cat factory block can be augmented with an extra photonic row below the block, so that cooling and qubit reset operations can be pipelined (not shown in \cref{fig:w18_cat_factory_uarch_phases}). While a ZZ stabilizer measurement is going on, ions at the extra photonic row go through a high-fidelity but potentially slow qubit reset protocol, and then exchange places with the measured ancilla ions at the original photonic rows after each stabilizer measurement. Each horizontal section of the extra photonic row holds two ions, and while one ion undergoes photonic operations, the other ion is displaced from the optical zone within the same horizontal section. The ion transport during the final one-bit teleportation step proceeds similarly to the first transport step of the ZZ stabilizer measurement, except that after the data-ancilla entanglement, the old data qubits slide to the photonic rows to be measured out while the old ancilla (now data) qubits stay at the gate rows.

\begin{table*}[t!]
    \centering
    \begin{tabular}{|c|c|c|}\hline
        {} & {\bf Preparation} & {\bf m = 2 verification rounds} \\\hline
        Transport & $w -6 + 14\lceil \log_2 w \rceil$ & $18m + 6 = 42$ \\\hline
        Parallel 1q gate layer & 1 & 1  \\\hline
        Parallel 2q gate layer & $ \lceil \log_2 w \rceil$ & $2m + 1 = 5$ \\\hline
        Merge / split & $2\lceil \log_2 w \rceil$ & $4m + 2 = 10$ \\\hline
        Readout & 0 & $m+1=3$ \\\hline
        State preparation & 1 & $m = 2$ \\\hline
    \end{tabular}
    \caption{Primitive operation rounds required to prepare a $(\epsilon = 10^{-10}, p=10^{-4})$-independent, weight-$w$ cat state in the four-row layout of cat factory; see \cref{tab:cat_state_cost} for comparison and the end of \cref{subsec:extension_of_the_moving_qubit_model} for the explanation of the difference in counting. Note that the transport steps for verification rounds could incur additional counts if ancilla reset pipelining is used to bring fresh ancilla from an independent photonic row to the cat block.}
    \label{tab:cat_state_uarch_cost}
\end{table*}

We refine the timing estimate for producing high-quality cat state provided in \cref{tab:cat_state_cost} using the stated hardware constraints and assumptions and summarize the result in \cref{tab:cat_state_uarch_cost}. There is an extra transport overhead that scales with the two-qubit gate layer count $\lceil \log_2 w \rceil$, which comes from the combination of vertical sliding of data qubits to form two-qubit pairs and the extra transport steps taken around the corners of the cyclical shift trajectory. The linear scaling with respect to the cat weight $w$ has a factor of two compared to the moving-qubit model because the micro-architecture model requires two primitive steps to do a linear translation from one horizontal section to another (see \cref{subsec:extension_of_the_moving_qubit_model}). The overall time cost of producing a verified cat state of relevant weight size ($w=18$ for \code{70}, see \cref{fig:All_MEK}) is still dominated by the cost of slow two-qubit gate, readout and reset operations (see \cref{tab:uarch_model}) but the analysis highlights the importance of keeping adiabatic transport cost small relative to the gate operation time.

\subsection{Micro-architecture of the cat-based measurement interface}

\begin{figure*}
    \centering
    \includegraphics[width=\linewidth]{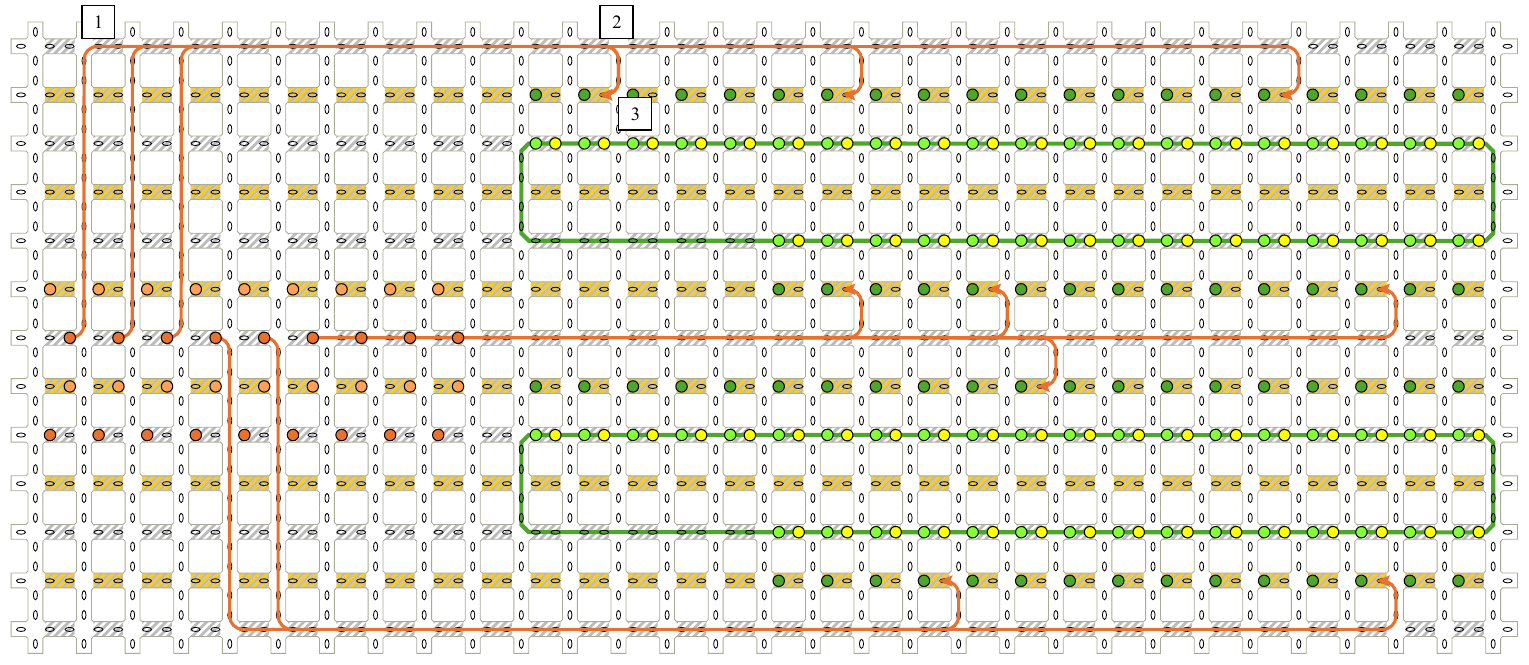}
    \caption{Cat-based measurement interface between a $w=18$ cat factory and the \code{70} code block shown in \cref{fig:uarch_memory_embedding}. Trajectories of the first batch of $w/2$ qubits from the cat factory to align with the data qubits of a logical operator of the code block are shown in orange; the second batch moves analogously. Each trajectory is realized in three phases. 1) the qubits of the cat state are vertically transported to match the target row indices. 2) They are translated horizontally while stretching their relative spacings, in what we call an accordion movement, until they are above or below the target data qubits. 3) A final short vertical transport merges each of the $w/2$ qubits of the cat state with the target data qubits for parallel two-qubit gates.}
    \label{fig:uarch_cat_state_interface_to_memory}
\end{figure*}

To analyze the transport cost of aligning a verified cat state against specific data qubits of a logical operator being measured, we consider placing cat factories to either the left or the right side of the memory block (cf. \cref{fig:uarch_cat_state_interface_to_memory}). The memory block has data qubits folded into four physical rows, so we need to re-shape the spatial extent of the cat state from two physical rows into up to four physical rows, depending on the support of the target logical operator. To avoid collision in the vertical sections during routing, which could happen if all $w$ qubits in the cat need to be aligned against single data row in the memory block, we send out the cat state in two separate rounds, where each round has at most $w/2$ qubits. For the first round, selectively stage at most $w/2$ qubits in the cat state onto the vertical section, and slide them up or down, until every selected qubit is at the nearest gate row above or below the target data qubit in the memory block. Then the selected qubits are linearly translated along the rows, possibly changing their relative horizontal spacings as they move in an accordion-like fashion, until their vertical offsets from the target data qubits are zeroed. The second round applies the same sequence of moves to the remaining qubits in the cat state. The said accordion-like movement in each round should approximately take $2\left(h-w/2\right)$ rounds of transport steps, where $h$ is the number of horizontal sections occupied by the memory block and each horizontal section requires two primitive transport steps to traverse as explained in \cref{subsec:extension_of_the_moving_qubit_model}; we assume digitally controlled electrodes can be used to stop the translation of qubits at appropriate positions while a broadcast linear transport waveform is applied to rows of interest. Thus, we expect cat state routing and alignment to require at least $4h - 2w$ transport steps plus additional overhead for vertical transports over two rounds, which we estimate to be $v$, the number of vertical sections spanned by the memory block. For the \code{70} memory block embedding shown in \cref{fig:uarch_memory_embedding}, we expect at least $4h-2w+v=80-36+10 = 54$ transport steps or approximately 2.7 POC in the moving-qubit model unit. If we add up the POC cost of transport, gates, and readout (state preparation is assumed to be pipelined; merge and splits are considered part of the two-qubit gate sequence), then the total POC cost of preparing and routing a verified $w=18$ cat state in the micro-architecture model is at least 23.9 POC. Same method of POC counting for the micro-architecture model of the \code{70} code in \cref{tab:uarch_memory_embedding_steps} yields at least 34.2 POC. Thus, the cat state preparation and routing time cost can still be below that of one SEC and thus theoretically allows us to pipeline cat state routing during SEC for immediate cat state consumption at the end of SEC. The cat state is consumed by bringing out the target data qubits in the memory block to merge with the qubits in the cat state in the gate rows above or below their original rows, and then routing the cat state to some photonic rows for readout, one option for such photonic rows being the original photonic rows in the cat factory. 

\clearpage

\part{Compilation and applications}
\label{part:Compilation and Applications}
\section{The logical compiler}

This section describes our logical compiler and the process of hierarchical synthesis of a given application in terms of logical instruction set.
We use the described synthesis process to produce quantum circuits for this FT architecture and estimate their execution time.  We chose two sample algorithms to illustrate our approach: Heisenberg Hamiltonian dynamics simulation and Shor's integer factorization.

We compared our fault-tolerant Hamiltonian-simulation implementation against state-of-the-art NISQ approaches and found that the walking-cat architecture should enable more than an order-of-magnitude improvement in both achievable circuit depth and algorithmic precision. This would push the computation into a regime that is classically intractable and out of reach for NISQ hardware.

We also show that Shor’s algorithm could factor a 30-bit integer, such as 1,071,514,531, in less than one day. Note that our focus here is on proof-of-concept, rather than pushing application performance as much as possible. In keeping with the general design principles of this paper, we prioritize generating concrete implementations rather than more speculative performance gains. 

\subsection{Hierarchical circuit synthesis}

\begin{figure}[b!]
    \centering
    \includegraphics[width=1\linewidth]{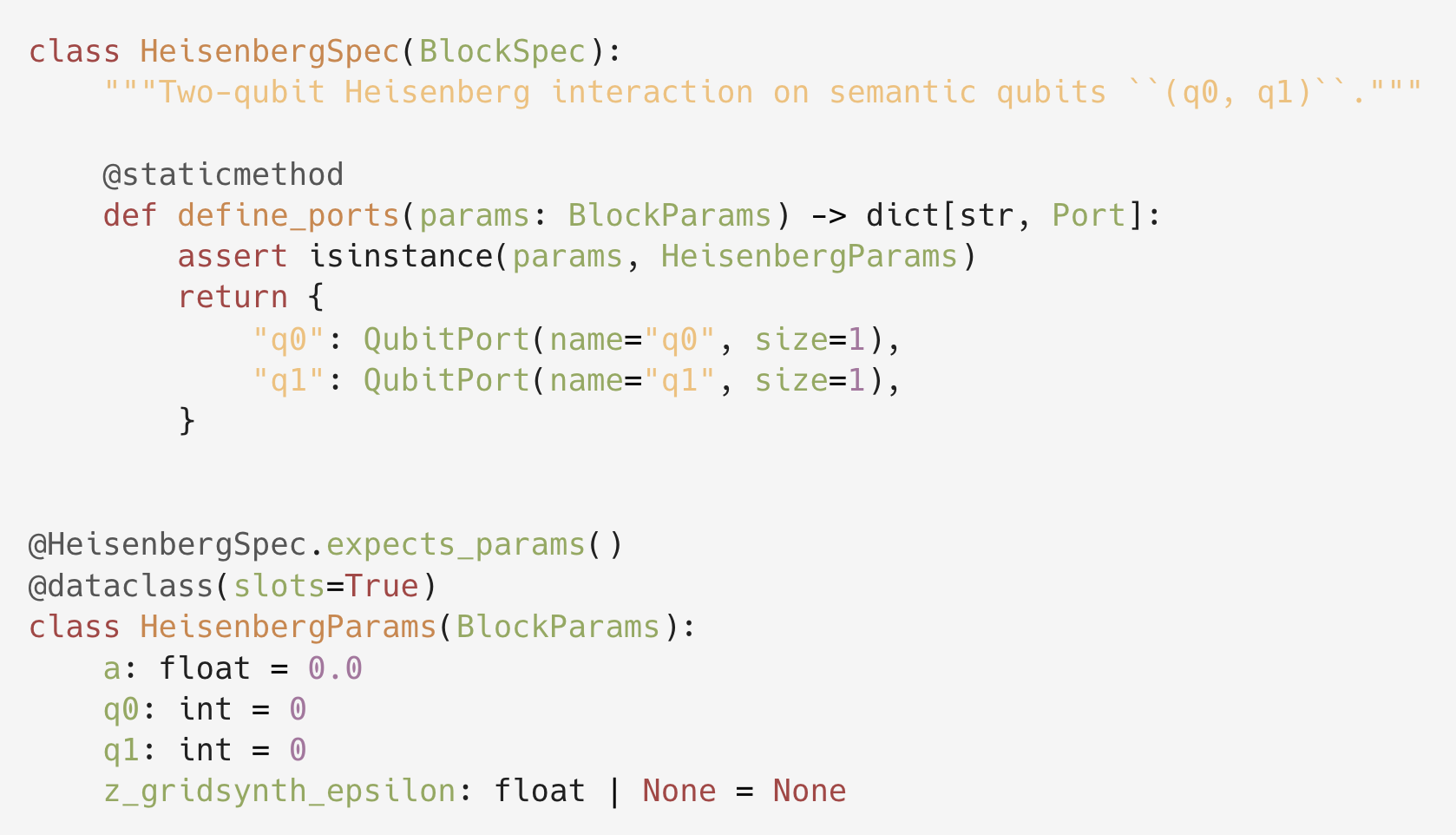}
    \caption{Heisenberg interaction term {\em spec}.}
    \label{fig:heisenberg_spec}
\end{figure}

\begin{figure}[t!]
    \centering
    \includegraphics[width=1\linewidth]{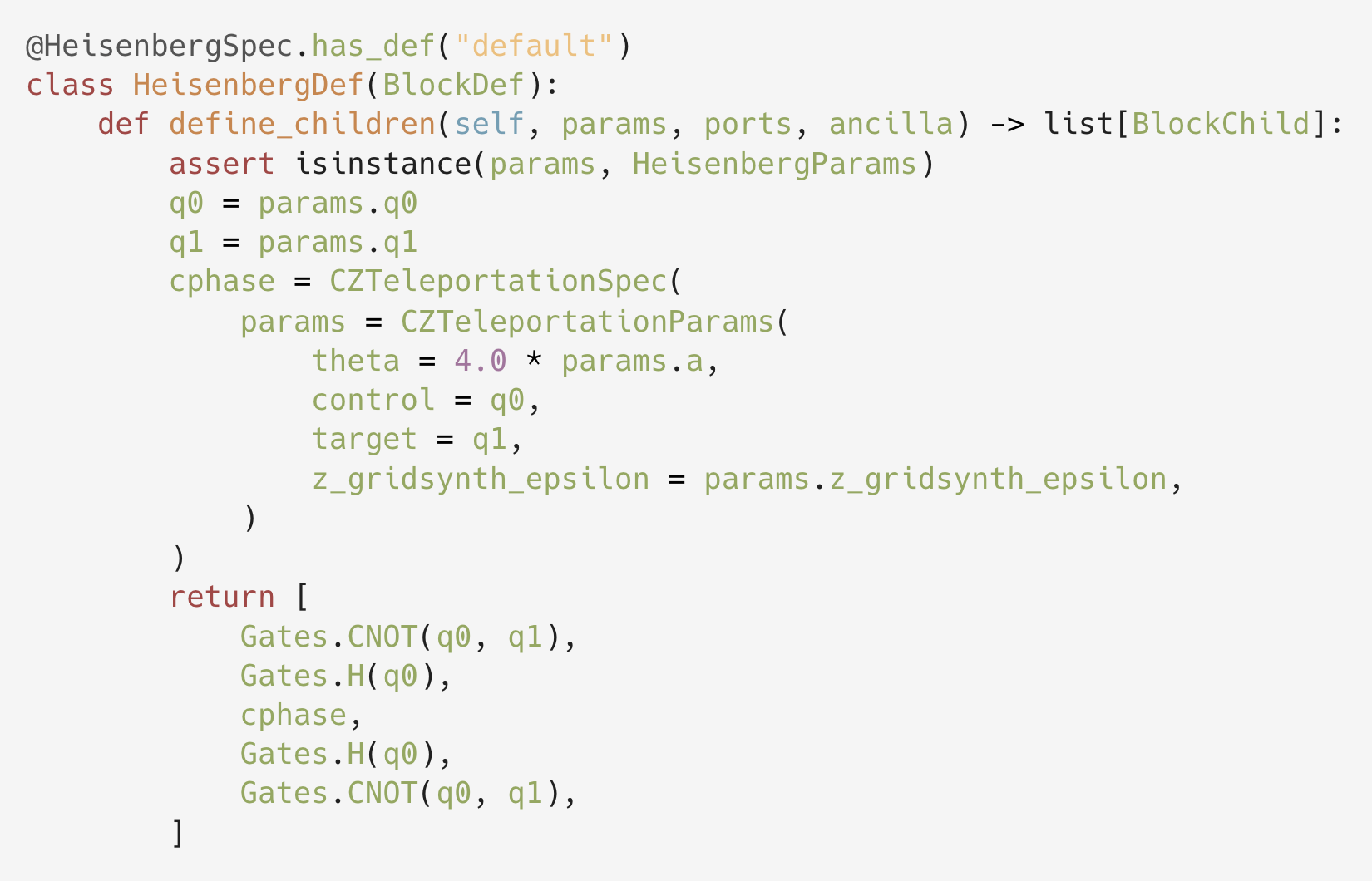}
    \caption{Heisenberg interaction term {\em def}.}
    \label{fig:heisenberg_def}
\end{figure}

We use a compiler instruction set similar to that of Qualtran~\cite{harrigan2024qualtran, qualtran2026} and QREF~\cite{qref2024, qref_format}.
It takes as an input a sequence of instruction from the compiler instruction set which contains Pauli measurements, single-qubit and two-qubit Clifford gates, single-qubit $Z$ rotation with arbitrary angle and custom-built instructions that extend the library of available operations. All these instructions can be conditioned on a classical outcome.
Each instruction of the compiler instruction set is defined by a {\em spec}, which specifies the input and output of the instruction and its parameters, and a {\em def}, which specifies the decomposition of the instruction into previously defined compiler instructions.
Examples of a spec and a def are provided in \cref{fig:heisenberg_spec} and \cref{fig:heisenberg_def} respectively.
The purpose of the compiler is to decompose the input sequence into instructions from the logical instruction set described in \cref{sec:Logical instruction set architecture}.

For example, consider an $n$-bit adder. 
The spec takes two $n$-qubit registers, $\ket{x}$ and $\ket{y}$, and semantically defines the output to be $\ket{x, x+y}$. A def encodes the actual implementation of that spec, for example, a textbook long adder, a carry-lookahead adder~\cite{DraperKutinRainsSvore2004QCLA}, or a temporary AND-based Gidney adder~\cite{Gidney2018}. Defs declare their own ancillae, allowing different realizations of a component to achieve various resource tradeoffs, be it depth, $T$-count, width, or other. This nested decomposition system carries down to the instructions of the underlying logical instruction set.

We leverage this system to implement and adapt different versions of our circuit blocks and select decompositions based on the constraints of our architecture and the broader circuit context. We may also construct hand-optimized implementations of important components, identifying and leveraging opportunities for architectural parallelism.

We compile in stages by recursively expanding child specs within a given definition, carefully selecting ancilla allocations for chosen defs, resulting in a fully decomposed synthesis tree which may then be processed to yield a time-ordered list of leaf operations on our target architecture---in our case, intra-memory-block free tracked Clifford gates, logical measurements within a memory block or touching two distinct memory blocks, and $T$ states prepared in magic factories.  We then group this exact circuit decomposition into operation layers that can be executed in parallel on the target architecture, yielding the final executable circuit with exact accounting.

\subsection{Hamiltonian simulation: Heisenberg model}

Here we review the current state-of-the-art results in Hamiltonian simulation across classical methods and NISQ experiments. We then compare these results to fault-tolerant instances of Hamiltonian simulation compiled for the walking cat architecture.

{\bf Hamiltonian simulation with classical methods.}
Current state-of-the-art classical methods, which rely on matrix-product-state (MPS)-based methods~\cite{google56mps2025, quantinuum56hsim2025ising} reach 56-qubit simulations of the 2D transverse-field Ising Hamiltonian on a rectangular lattice, achieving about 5\% relative accuracy on up to roughly 20 second-order Trotter steps, although robust MPS convergence in the most challenging regimes is typically limited to around 6–8 steps. Larger MPS-based simulations may be possible at the cost of reduced accuracy, assuming the entanglement growth remains sufficiently slow; this becomes harder as lattice connectivity increases.

\begin{figure}[b!]
    \centering
    \includegraphics[width=0.85\linewidth]{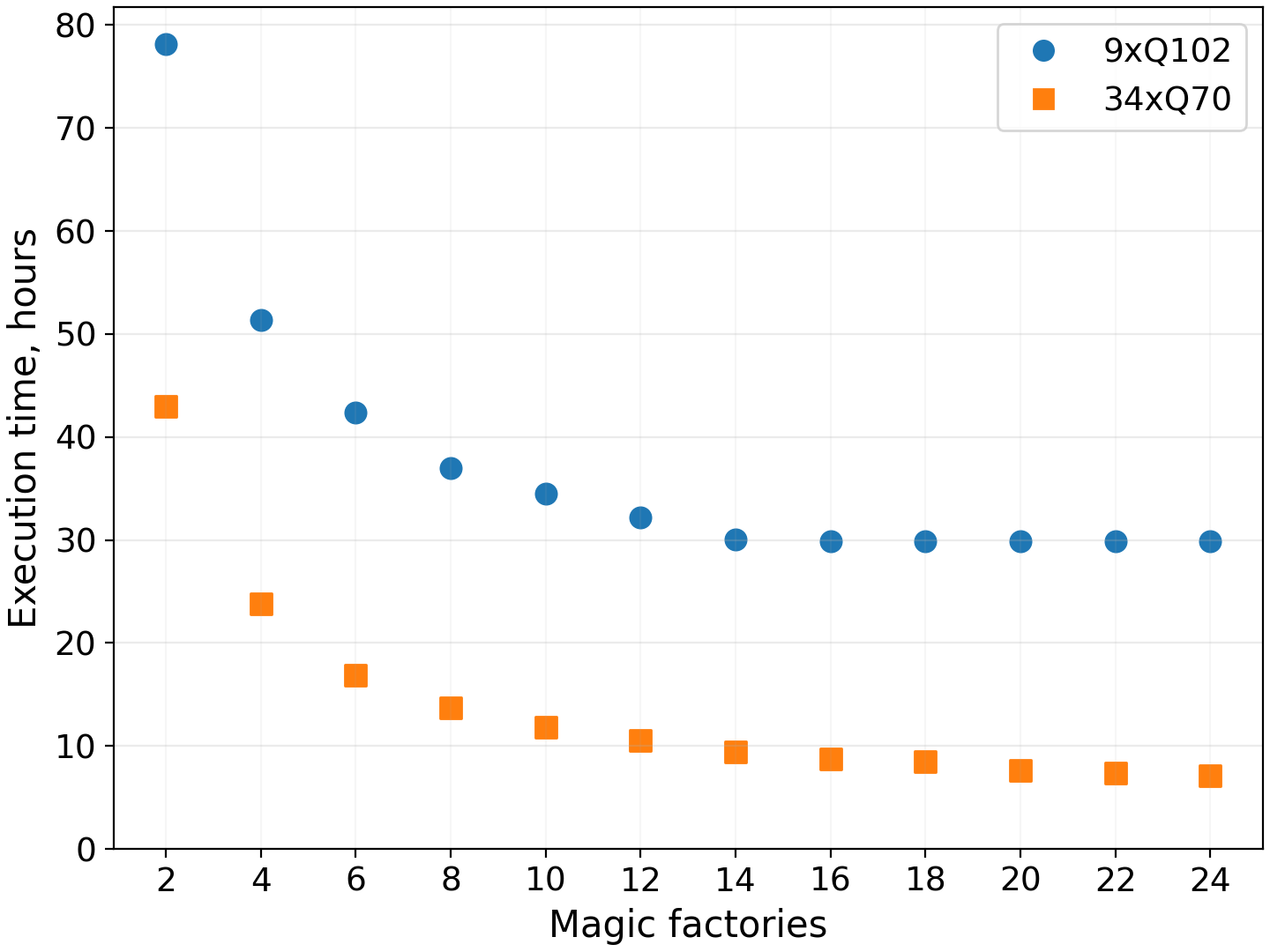}
    \caption{Single-shot execution time estimates of the Heisenberg Hamiltonian simulation for a random degree-seven regular graph of size 100 with 10 sixth-order Trotter steps (250 second order steps) shown for different walking cat architectures based on \code{102} and \code{70} as a function of the number of \CHfactory factories while keeping the total number of logical qubits at 198 for \code{102} and 204 for $\code{70}$.}
    \label{fig:hsim_tfact}
\end{figure}

{\bf Hamiltonian simulation with NISQ.} Recent digital Hamiltonian simulation experiments have reached: 2D square lattice Fermi-Hubbard dynamics on 72 qubits with up to 3 second-order Trotter steps on Google’s Willow processor~\cite{google72hsim2025fh}, 1D Fermi-Hubbard dynamics on 104 qubits with 10 optimized second-order Trotter steps on IBM hardware~\cite{ibm104hsim2026fh}, 1D Heisenberg spin-chain dynamics up to 100 qubits with second-order Trotterization on IBM hardware~\cite{ibm100hsim2024heisenberg}, 46-qubit 1D Heisenberg dynamics on Google hardware via Floquet cycling~\cite{google46hsim2024heisenberg}, and 56-qubit 2D transverse-field Ising dynamics on Quantinuum H2~\cite{quantinuum56hsim2025ising} with 40 second-order Trotter steps. Taken together, the strongest digital, Trotterized, benchmarked NISQ results are still mostly at 2D grid connectivity, at fewer than {$\sim$}100 qubits, with observable errors around 3-5\% in the best-controlled regimes, and with 40 second-order Trotter steps as the largest verified step count in this list.

\begin{figure}[t!]
    \centering
    \includegraphics[width=0.85\linewidth]{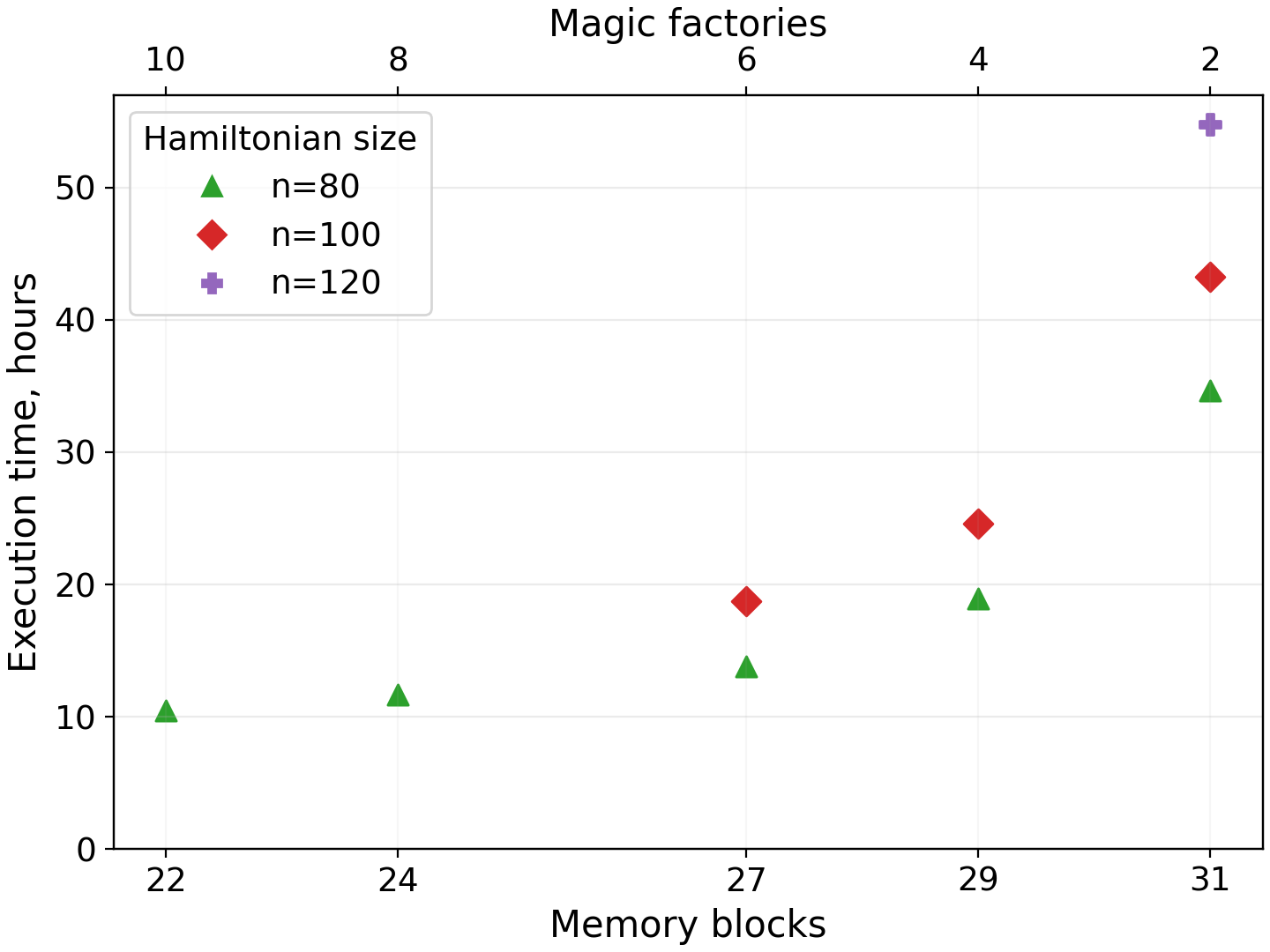}
    \caption{Single-shot execution time estimates of the Heisenberg Hamiltonian simulation for random degree-seven regular graphs with 80, 100, and 120 sites, using 10 sixth-order Trotter steps (250 second-order steps) shown for different walking cat architecture configurations $\conf{N_M}{70}{N_T}{\CHfactory}$ that fit on 10,000 physical qubits as a function of the number of memory blocks and magic factories.}
    \label{fig:hsim_q70}
\end{figure}

We study Heisenberg Hamiltonian simulation on 100 sites for degree-seven random graphs, a model that includes systems such as 3D spin glasses. Heisenberg Hamiltonian with random on disorder along the $Z$ axis $h_i \in [-1,1]$ on a regular graph $G$ is described as

\begin{align}
    H = \sum_{ (i,j) \in E(G)} \mathbf{S}_i \cdot \mathbf{S}_j + \sum_i h_i S_i^z.
\end{align}

Our aim is to achieve chemical accuracy in observables such as staggered magnetization, which can be used to study phase transitions. However,  moving beyond current NISQ capabilities to higher connectivity, longer evolution time, and a target error of $10^{-3}$ requires circuits roughly an order of magnitude deeper than are currently feasible, even with state-of-the-art error mitigation.

NISQ execution time estimates from~\cite{flasq2025} define
\begin{align}
T_{\mathrm{NISQ}}=(D_{\mathrm{2Q}}\cdot T_{\mathrm{2Q}})\times \frac{N_{\mathrm{samples}}}{N_{\mathrm{parallel}}},
\end{align}
where $D_{\mathrm{2Q}}$ is the number of 2Q-gate layers, $T_{\mathrm{2Q}}$ is the duration of one such layer, $N_{\mathrm{samples}} = \Gamma^2/\sigma^2$, $\Gamma^2=\gamma^{\,2\times N\times D_{\mathrm{2Q}}}$, and $\gamma=(1+\epsilon/2)/(1-\epsilon)$.  
For the Hamiltonian we considered, circuit depth $D_{\mathrm{2Q}}=4,000$, which is an order of magnitude larger than the circuit depths considered in the NISQ papers discussed above. Because $\Gamma^2$ scales exponentially, this leads to an astronomical NISQ time-to-solution estimate, even assuming a two-qubit-gate-layer time of 50ns, a physical gate fidelity of $99.99\%$, and $N_{\mathrm{parallel}}=100$, corresponding to 100 replicas of a 100-site Hamiltonian running on 10,000 physical qubits.

\begin{figure}[t!]
    \centering
    \includegraphics[width=0.85\linewidth]{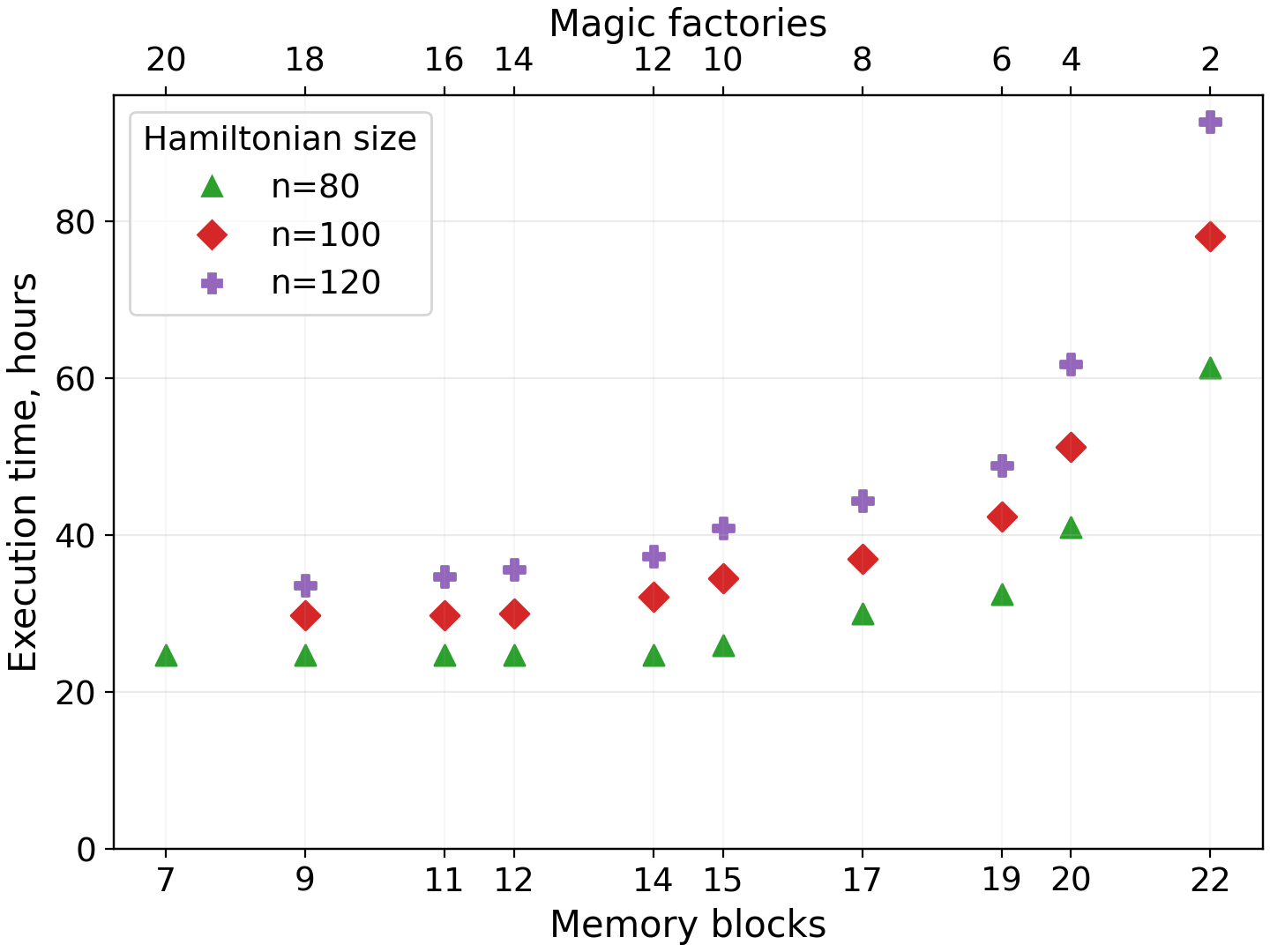}
    \caption{Single-shot execution time estimates of the Heisenberg Hamiltonian simulation for a random degree-seven regular graph with 80, 100, and 120 sites, using 10 sixth-order Trotter steps (250 second-order steps) shown for different for different walking cat architecture configurations $\conf{N_M}{102}{N_T}{\CHfactory}$ that fit on 10,000 physical qubits as a function of the number of memory blocks and magic factories.}
    \label{fig:hsim_q102}
\end{figure}

{\bf Fault-tolerant Hamiltonian simulation.} To reach chemical accuracy, the state preparation part of the circuit is synthesized using the recursive expansion of the higher-order product formulas to build sixth-order steps:
\begin{align}
\exp\!\left(-it\sum_{j=1}^{L}\alpha_j H_j\right)\approx \left[S_{2k}(\lambda)\right]^r,
\qquad \lambda:=-it/r
\end{align}
and 
\begin{align*}
    & S_2(\lambda):=\prod_{j=1}^{L}\exp(\alpha_j H_j\lambda/2)\prod_{j=L}^{1}\exp(\alpha_j H_j\lambda/2)\\
    & S_{2k}(\lambda):=
    \left[S_{2k-2}(p_k\lambda)\right]^2
S_{2k-2}((1-4p_k)\lambda)
\left[S_{2k-2}(p_k\lambda)\right]^2
\end{align*}
with $p_k = (4-4^{1/(2k-1)})^{-1}$.

To achieve chemical accuracy $10^{-3}$, \cite{Maslov2019heisenberg} predicts that roughly 80 sixth-order Trotter steps, or 2,000 second-order Trotter steps, are needed. Randomized compilation~\cite{Robertson2025} can reduce this to roughly 10 sixth-order Trotter steps, or 250 second-order Trotter steps. This is still about an order of magnitude higher than the 10-20 second-order Trotter steps explored in NISQ regimes for $10^{-2}$ error.

\begin{figure}[b!]
    \centering
    \includegraphics[width=0.95\linewidth]{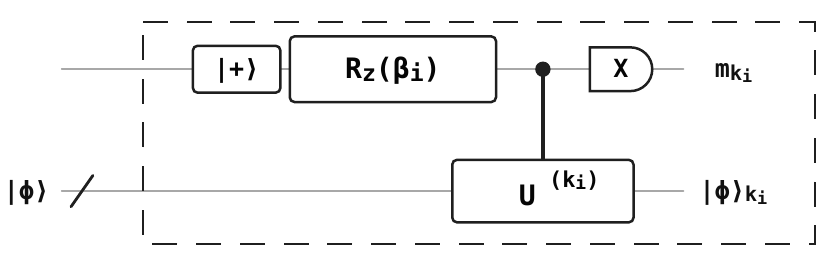}
    \caption{Iterative QPE on the prepared state $\left| \Phi \right>$. The block inside the dashed rectangle is repeated multiple times for different exponents $k_1, k_2, \dots$}
    \label{fig:hsim_iqpe}
\end{figure}

{\bf Resource estimation.}
We compiled Heisenberg Hamiltonian simulation circuits for a random degree-seven regular graphs of sizes 80, 100, 120 targeting instances of the walking cat architecture based on \code{102} and \code{70} with varying numbers of memory blocks and magic factories. The compiled sequence of logical instructions is used to estimate the runtime based on the logical instruction time reported in \cref{tab:example_logical_op_times}.

\begin{figure}[t]
    \centering
    \subfloat[]{\includegraphics[width=0.95\linewidth]{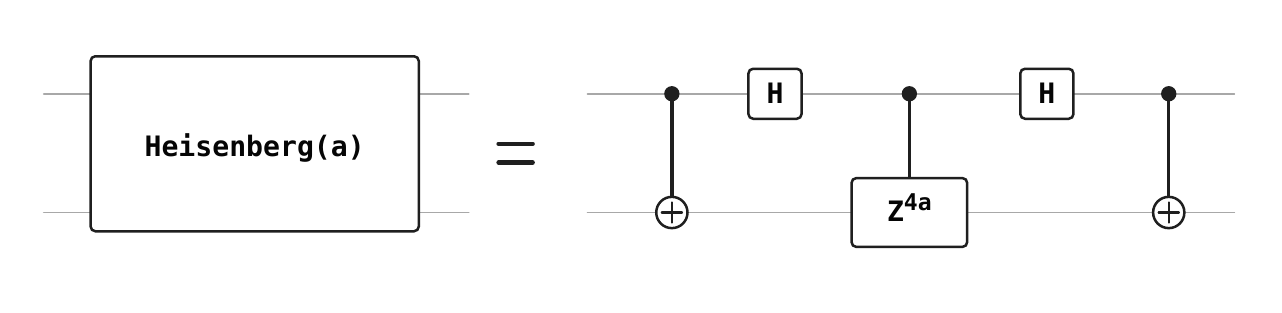}\label{fig:heisenberg_decomposition}}

    \vspace{0.75em}
    \subfloat[]{\includegraphics[width=0.95\linewidth]{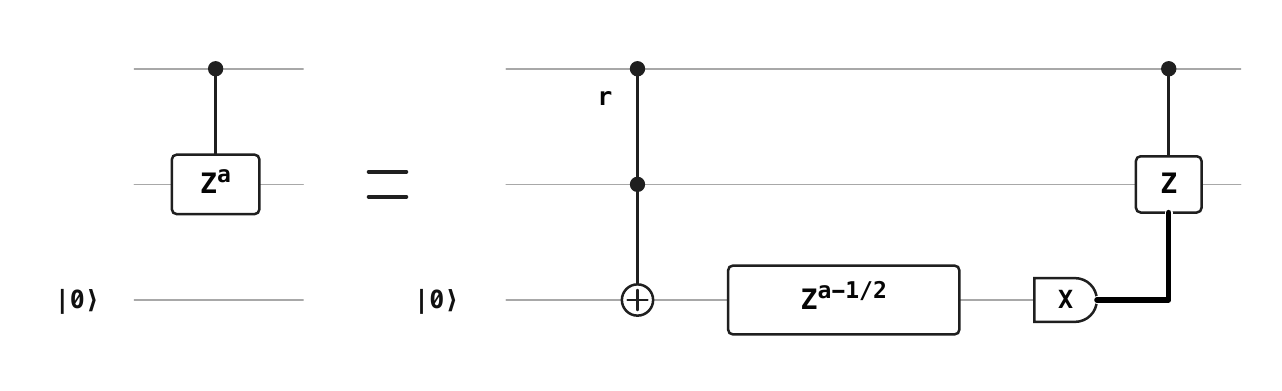}\label{fig:controlled_z_power_measurement_decomposition}}
    \caption{Ancilla-aided decomposition used for the Heisenberg-simulation compilation. (a) Decomposition of a Heisenberg interaction block into basis changes and a controlled-$Z^{4a}$. (b) Controlled-$Z^a$ power-measurement gadget used within that decomposition. The X-basis readout conditionally executes a final controlled-$Z$ gate.}
    \label{fig:heisenberg_decomposition_pair}
\end{figure}

We use an ancilla-aided construction for the Heisenberg interaction terms~\cite{Maslov2019heisenberg}, which can reuse $N/2$ ancillae between the interaction layers (see~\cref{fig:heisenberg_decomposition_pair}). This allows for parallel decomposition of single-qubit $Z$ rotations to T gates, which requires $\sim 10\cdot \log_{10} \frac{1}{\varepsilon}$ T layers using the Gridsynth algorithm~\cite{Selinger2014gridsynth}, but can get a 2.5x reduction in T counts using the repeat-until-success algorithm~\cite{Bocharov2015rus} and a number of ancillae proportional to the number of magic factories to make them non-blocking.

\begin{table}[h]
          \centering
          \begin{tabular}{|c|c|c|c|c|c|c|}\hline
              & \multicolumn{3}{c|}{ $ N_M \times \text{Q}70 + N_T\times \text{CH}2$} & \multicolumn{3}{c|}{$ N_M \times \text{Q}102 + N_T\times \text{CH}2$}\\\hline
              {\bf Size} &{\bf $N_M$} &{\bf $N_T$} &\makecell{\bf Time \\ \bf(hours)} & {\bf $N_M$} &{\bf $N_T$} &\makecell{\bf Time \\ \bf(hours)}\\\hline
              80 &  22& 10& 10& 6& 12& 24\\\hline
              100 &  27& 6& 18& 9& 16& 29\\\hline
              120 &  31& 2& 53& 9& 18& 33\\\hline
          \end{tabular}
          \caption{Best single-shot execution time estimates of the Heisenberg Hamiltonian simulation for a random degree-seven regular graph with 80, 100, and 120 sites, using 10 sixth-order Trotter steps (250 second-order steps) selected out of different walking cat architecture configurations that fit on 10,000 physical qubits.}
          \label{tab:compare_q70_q102}
      \end{table}
      
Both instances of the walking cat architecture benefit from larger numbers of magic factories (see \cref{fig:hsim_q70,,fig:hsim_q102,,fig:hsim_tfact}). Increasing the number of magic factories allows for the parallel implementation of single-qubit $Z$ rotations, but the improvement saturates when the number of magic factories is no longer the main bottleneck, as it is the case for $\code{102}$ with more than 12 magic factories (see~\cref{fig:hsim_tfact}). On the other hand, increasing the number of magic factories past 12 still improves the parallelism for $\code{70}$ in case of a 100-site Hamiltonian. At 24 magic factories the estimated single-shot execution time for the 100-site Heisenberg Hamiltonian simulation with 10 sixth-order Trotter steps on the \code{70} architecture is predicted to be around seven hours. 

With the total number of physical qubits fixed at 10,000, the 80- and 100-site instances run faster on \code{70} than on \code{102} (see \cref{fig:hsim_q70,,fig:hsim_q102}). The 120-site instance, however, runs faster on \code{102}. This is because \code{102} can fit the larger instance into fewer memory blocks and benefit more from the increased parallelism enabled by its larger number of magic factories (see~\cref{tab:compare_q70_q102}).

To measure our observable of interest (\emph{e.g.}, staggered magnetization) we use iterative quantum phase estimation (QPE), see \cref{fig:hsim_iqpe}. For our resource estimation, we assume the measurement budget of 50-100 shots based on the following approaches:
\begin{itemize}
   \item Bayesian QPE or rejection filtering phase estimation~\cite{Wang2019avqe} gives an estimate of ~40 shots for the error budget of $10^{-3}$.
   \item Random walk phase estimation~\cite{Lubinski2022,Cassandra2022randIPE} requires about 30 iterations for the error budget of $10^{-3}$. 
   \item Based on the Bayesian phase estimation with priors~\cite{vandenBerg2021qpe}, the expected number of rounds is at order of 100 for the error budget of $10^{-3}$. 
\end{itemize}

For the staggered magnetization observable, all controlled-$U^k$ gates can be implemented with one layer of parallel phases. Thus, the most expensive part of the circuit remains state preparation implemented through Trotterization. Using the operation times from \cref{tab:example_logical_op_times} for \code{70} on 33 memory blocks with more than 15 magic factories, we estimate that the total execution time to measure the staggered magnetization observable with chemical accuracy would take around one month.

\subsection{Shor's period finding}

Shor’s integer factoring can be interpreted as an instance of the quantum phase estimation (QPE) algorithm for the operator of modular multiplication by a fixed constant~\cite{shor1994algorithms, kitaev1995quantum}. To implement this algorithm, we again rely on an iterative QPE construction~\cref{fig:hsim_iqpe}, which requires a loop over $2n$ repeated controlled modular multiplications by precomputed constants, where $n$ is the number of bits of the integer being factorized. 
There is a considerable body of existing work on implementing the underlying arithmetic for Shor's algorithm, see e.g.,
\cite{Cuccaro2004RippleCarryAdder, DraperKutinRainsSvore2004QCLA, Haener2020ECDL, Parent2017Karatsuba, gidney2021factor}
The specific choice of arithmetic depends on multiple factors such as the bit-size $n$, optimization criteria (number of qubits vs time-to-solution), and testability of the circuits. 

In the implementation chosen in this paper, each controlled $U$ block in the iterative QPE construction~\cref{fig:hsim_iqpe} implements a controlled modular multiplier with a dual set of $n$ forward and backward doubly controlled constant modular adders, with a controlled swap block in the middle. The doubly controlled constant modular adders are themselves constructed in the style of \cite{Luongo2024}, with a doubly controlled adder, an uncontrolled constant comparator, a singly controlled subtractor, and a final doubly controlled comparator to uncompute the comparator's flag bit. The adders are implemented as Gidney-style temporary-AND-based register construction \cite{Gidney2018}, and the comparators are implemented similarly, as borrow-chain comparators in the style of \cite{Luongo2024}. This implementation is aligned with our principle to prioritize generating concrete implementations that can be used to perform end-to-end device level performance simulations. We leave further optimizations of the trade-space for future work. 

As an example, we compile four instances of Shor's period finding circuits for integers with 10, 20, 30, and 40 bits targeting architecture configurations $\conf{N_M}{70}{9}{CH2}$ with $N_M=17,34$ (see~\cref{tab:q70_synth_shor}). Our execution time estimates are based on the operation times shown in the~\cref{tab:example_logical_op_times}.

\begin{table}[h!]
    \centering
    \begin{tabular}{|c|c|c|c|c|}
    \hline
          {\bf Size} &\makecell{\bf Memory \\ \bf blocks} &{\bf T gates} &{\bf SEC }&\makecell{\bf Time \\ \bf(hours)}\\
    \hline
           10& 17 & 71,536& 598,520& 1\\ \hline
           20& 17 & 532,692& 4,805,861& 7.5\\ \hline
           30& 34 & 1,760,876& 14,748,419& 23\\ \hline
           40& 34 & 4,132,048& 34,733,136& 53.5\\
    \hline
      \end{tabular}
    \caption{Circuit synthesis results for Shor's factorization circuits on architecture configurations $\conf{N_M}{70}{9}{CH2}$ with $N_M$ memory blocks.}
    \label{tab:q70_synth_shor}
\end{table}

\clearpage
\part{Conclusion}
\section{Conclusion}

In this work, we proposed a complete architecture for fault-tolerant quantum computing with trapped ions, including a compiler, a thorough description of all components of the logical architecture, and a detailed micro-architecture for the most critical components.
We anticipate that this blueprint provides opportunities for optimization and co-design as well as future improvements at all levels, from the compiler to the error-correcting codes, the logical gates, the layout, and the micro-architecture.
We also expect the co-design of adjacent layers to bring significant performance improvements, as in the case of our design of the micro-architecture of entire components of the logical architecture.

Moving from this theoretical blueprint to the realization of such a machine would be a major breakthrough.
Although key hardware components such as high-fidelity two-qubit gates achieved with EQC~\cite{loschnauer2025scalable, hughes2025trapped} or ion transport~\cite{pino2021demonstration, ransford2025helios} have been demonstrated experimentally on small devices, scaling to the regime of thousands of physical qubits is a significant challenge.
A core design principle in our architecture is simplicity. This strategy extends across the whole architecture, from memory and magic factory design to the selection of cat-based measurement as the main subroutine for all logical operations.
This concept is witnessed at the micro-architecture level, relying extensively on simple cyclic moves of the qubits instead of more complex permutations.
It holds at the hardware level with the scalable control electronics proposed in~\cite{malinowski2023wire} and the gate design that removes the need for laser-based gates~\cite{hughes2025trapped}.
We believe that the simplicity at all levels of the architecture makes it viable for scaling FTQCs.

Finally, we note that many relevant candidates for applications have been proposed \emph{in theory} by the community~\cite{babbush2025grand}. However, history teaches us that, until a FTQC capable of running millions of logical operations is in the hands of the broad community of scientists, we will only scratch the surface of what is possible.

\section{Acknowledgments}

The authors would like to thank the whole IonQ team, particularly Jason Amini, Andrew Arrasmith, Chris Ballance, Tom Harty, Jeremy Sage, Curtis Volin, and Dave Wecker.

\bibliography{references}

\clearpage
\part{Appendices}
\appendix
\section{Glossary}
\label{app:Glossary}

\begin{figure}[b!]
    \centering
    \includegraphics[width=\linewidth]{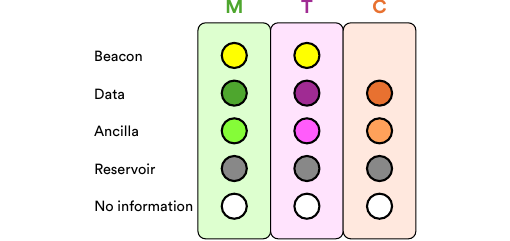}
    \caption{\textbf{Summary of the color conventions:} The memory block is green, the magic factory is purple, and the cat factory is orange. There are five types of qubits that can be present within a block. Yellow always represents beacon qubits, gray represents reservoir qubits, and white represents qubits that carry no information/are reset. Data qubits are always given a dark version of that block's color, and ancilla qubits are given a light version.}
    \label{fig: color conventions}
\end{figure}

\begin{itemize}
    \item Logical qubit: Error-corrected qubit encoded into multiple physical qubits.
    \item Physical qubit: Qubit in the moving-qubit model. We form logical qubits using blocks of multiple physical qubits.
    \item Data qubits: Physical qubits storing logical information.
    \item Ancilla qubit: Physical qubit used to perform an error correction protocol but does not store logical information.
    \item Beacon qubit: Physical qubit used for loss detection.
    \item POC: The physical operation cycle. Time for one round of physical operations in the moving-qubit model.
    \item SEC: The syndrome extraction cycle. Time for one round of measurement of all the stabilizer generators of the code.
    \item Logical error rate: Probability that at least one of the qubit suffers from a logical error after correction, normalized by the number of SEC.
    \item Memory block: Component of the walking cat architecture that stores the logical information. It is equipped with correction of circuit-level Pauli faults, leakage and loss.
    \item Magic factory: Component of the walking cat architecture that produces magic states and store them as a memory. It is equipped with correction of circuit-level Pauli faults, leakage and loss.
    \item Block: Memory block or magic factory.
    \item Cat factory: Component of the walking cat architecture producing cat states.
    \item Bell factory: Component of the walking cat architecture producing Bell states.
    \item Qubit factory: Component of the walking cat architecture producing new qubit to replace lost qubits.
    \item Local reservoir: Qubit reservoir located inside a component of the walking cat architecture.
    \item Global reservoir: Qubit reservoir located in the qubit factory whose qubits are used to replace lost qubit in the whole architecture and to refill local reservoirs.
    \item Symplectic basis: Basis $\bar X_1, \bar Z_1, \dots, \bar X_k, \bar Z_k$ for the set of $n$-qubit Pauli operators commuting with the stabilizers of a $[[n, k, d]]$ codes. These operators satisfy the same commutation and product relations as the standard Pauli operators $X_1, Z_1, \dots, X_k, Z_k$ on $k$ qubits.
    \item Compiler instruction set: Instructions available to describe the compiler input.
    \item Logical instruction set: Instructions available on logical qubits.
    \item Physical instruction set: Instructions available on physical qubits, that is within the moving-qubit model.
    \item QCCD: Quantum coupled-charged device. Architecture for trapped-ion quantum computer allowing to move ions and implement gates and measurements.    
    \item Device instruction set: Instructions available within a QCCD device.
    \item Logical width: Maximum weight of a logical operator that can be measured within the logical instruction set.
    \item Block width: Maximum weight of a physical representative of a logical operator.
    \item Cat-based measurement: Measurement of a Pauli operators with weight $w$ using a $w$-qubit cat state. 
    \item Accessible logical Pauli operator: Logical operator that can be measured.
    \item  Accessible logical Clifford gate: Logical Clifford gate that can be implemented by frame-tracking.
    \item In-block logical measurement: Measurement of a logical operator supported on a single code block.
    \item Inter-block logical measurement: Measurement of a logical operator supported on a pair of code blocks.
    \item LZ / LP: Preparation of all the logical qubits of a block in the logical zero states / logical plus state. 
    \item LT: Preparation of two logical magic states.
    \item LM1: In-block logical measurement of an accessible logical Pauli operator.
    \item LM2: Inter-block logical measurement of a product of two accessible logical Pauli operators.
    \item CLIF: Accessible Clifford gate implemented by frame-tracking.
    \item DMX / DMZ: Destructive measurement extracting many logical measurement outcomes in parallel by measuring the data qubits of a code in the $X$ and $Z$ basis.
    \item Three-ring framework: Formalism unifying the circuit and layout for the GB codes, BB codes and cyclic GPH codes.
    \item Quantum LDPC: Low-density parity check codes~\cite{mackay2004sparse}. Stabilizer codes defined by low-weight stabilizer generators.        
    \item GB codes: Generalized bicycle codes~\cite{kovalev2013quantum}.
    \item BB codes: Bivariate bicycle codes~\cite{bravyi2024high}.
    \item HGP codes: Hypergraph product codes~\cite{tillich2013quantum}
    \item Cyclic HGP codes: cyclic hypergraph product codes~\cite{aydin2025cyclic}.
    \item \code{102}: GB8 code with parameters $[[102, 22, 9]]$.
    \item \code{70}: BB7 code with parameters $[[70, 6, 9]]$. 
    \item \code{54}: GB8 code with parameters $[[54, 2, 10]]$ and strongly transversal $H$ gate.
    \item Cat state preparation: Circuit producing an unverified cat state that may contains correlated errors.
    \item Cat state verification: Circuit testing whether a state contains correlated errors, leakages or losses so that a cat state containing such errors can be rejected.
    \item Cat stitching: Circuit merging two cat states into a single cat state.    
    \item \EDM: Error-detected measurement. Logical measurement based on multiple consecutive cat-based measurement, with rejection when the cat-based measurement outcomes differ. 
    \item \ECM: Error-corrected measurement. Logical measurement based on multiple consecutive cat-based measurement. The logical outcome returned is obtained by majority vote.
    \item Viterbi measurement: Logical measurement obtained by performing an adaptive sequence of cat-based measurements. The likelihood of each of the two possible logical outcomes is updated after each new cat-based measurement outcome using Viterbi algorithm~\cite{viterbi2003error}.
    The measurement sequence is interrupted as soon as the likelihood ratio reaches a threshold value.
    \item \CHfactory factory: Magic factory producing two $H$ states per attempt based on the measurement of the logical operator $H^{\otimes 2}$.
    \item \MEK factory: Magic factory producing two $H$ states per attempt based on the \MEK distillation scheme~\cite{meier2013magic} implemented in a quantum LDPC code.
    \item $H$ injection: Circuit injecting a physical $H$ state into a logical qubit.
    \item Decoder: The classical subroutine that takes as an input the measurement data extracted by a circuit and returns a correction.
    \item Streaming decoder: Decoder that takes as an input a stream of measurement data.
    \item Sliding window decoder: A type of streaming decoder.
    \item Beam: A quantum LDPC code decoder~\cite{ye2025beam}.
    \item \texttt{beam32\_340iters}: Beam decoder with width 32 and a total of 340 iterations.
    \item Decoder reaction time: Time it takes for the decoder to extract all logical measurement outcomes after a destructive measurement (DMX) or (DMZ).
    \item Three-ring cyclic gate: Physical permutation of the qubits along three families of rings.
    \item Transversal Clifford gate: Tensor products of (possibly distinct) single-qubit Clifford gates.
\end{itemize}

\section{Background on stabilizer codes}
\label{app:Background on stabilizer codes}

This subsection reviews the necessary background on quantum error correction~\cite{nielsen2010quantum, preskill1998lecture}.
All the quantum error correction codes discussed in this paper are stabilizer codes~\cite{gottesman1997stabilizer}.
A {\em stabilizer code} with length $n$ is defined by a set of commuting $n$-qubit Pauli operators $S_1, \dots, S_r$, called the {\em stabilizer generators}.
We refer to the group they generate as the {\em stabilizer group}, which we denote $\Stab$.
An element of the stabilizer group is called a {\em stabilizer} of the code.

The {\em code space} is defined to be the set of $n$-qubit quantum states $\ket \psi$ fixed by the stabilizers, that is 
\begin{align}
    Q = \{ \ket \psi \in (\C^2)^{\otimes n} \ | \ \forall S \in \Stab, \ S \ket\psi = \ket\psi \}
    \cdot
\end{align}
Its dimension is given by $2^k$ where $k = n-\rank(\Stab)$,
and $\rank(\Stab)$ is the minimum number of generators of the group $\Stab$.
One can define a bijection from the $k$-qubit space $(\C^2)^{\otimes k}$ onto the code space $Q$ included in the $n$-qubit space.
Such a map can be interpreted as encoding $k$ {\em logical qubits} into $n$ {\em physical qubits}.

In this work, we focus on CSS codes~\cite{calderbank1996good, steane1996multiple}, which are stabilizer codes defined by stabilizer generators in $\{I, X\}^{\otimes n}$ or $\{I, Z\}^{\otimes n}$.
Moreover, we restrict ourselves to {\em quantum LDPC codes}~\cite{mackay2004sparse}, which are defined by low-weight stabilizer generators, making them typically easier to implement in practice.

Quantum error correction is performed by executing the so-called {\em syndrome extraction circuit}, which performs the measurement of the stabilizer generators of the code.
The outcome extracted is called the {\em syndrome}.
The syndrome extraction circuit is generally executed using additional qubits that we call {\em ancilla qubits}.
Some syndrome extraction circuits such as the color-based circuit~\cite{tremblay2022constant} are available for all quantum LDPC codes.
For a specific code, one may prefer a fine-tuned syndrome extraction circuit as in \cite{bravyi2024high}.

To avoid the accumulation of errors on unchecked qubits, the syndrome extraction circuit is executed at regular intervals.
In what follows, we refer to each run of the syndrome extraction circuit as a {\em syndrome extraction cycle} or SEC.
One can think of the SEC time as the logical clock cycle time of a fault-tolerant quantum computer.

The syndrome provides information about errors that occur on the data qubits. This information is fed to the {\em decoder}, which is the classical subroutine in charge of determining which correction to apply.
The topic of quantum LDPC decoding is a fast-moving research area. The BP-OSD decoder~\cite{panteleev2021degenerate, Roffe_LDPC_Python_tools_2022} is one of the most popular decoders. Recent progress led to the design of more efficient decoders such as the BP-GDC decoder~\cite{gong2024toward}, the BP-relay decoder~\cite{muller2025improved}.
In this work, we use the recent beam search decoder, which significantly outperforms BP-OSD both in terms of speed and accuracy~\cite{ye2025beam}.

The error-correction capability of a stabilizer code depends on its {\em minimum distance}, which is defined to be the minimum weight of a Pauli operator commuting with all the stabilizer generators and which is not a stabilizer.
The design of quantum LDPC codes with a large minimum distance is a highly non-trivial theoretical problem.
For two decades, it remained unclear whether quantum LDPC codes with minimum distance growing linearly with $n$ exist.
This is particularly surprising as building classical LDPC codes with linear minimum distance is easy~\cite{gallager1963low}.
This question was resolved in the asymptotic regime~\cite{panteleev2022asymptotically, leverrier2022quantum}.
In practice, small quantum LDPC codes with sublinear minimum distance are sufficient to outperform surface codes~\cite{tremblay2022constant, bravyi2024high}.

The performance of a quantum error correction protocol is estimated through Monte Carlo simulations of multiple SECs with noise inserted at the circuit level.
We perform numerical simulations using Stim~\cite{gidney2021stim} and our own extensions of this simulator.
We report estimates for the {\em logical error rate}, which in this work refers to the probability of a logical error per SEC.
The logical error rate is estimated by simulating $d$ SECs where $d$ is the minimum distance of the code.

\clearpage
\section{Code Database Table}
\label{app:sec_code_database_table}

The walking cat architecture supports many codes. Once fabricated, a device has a fixed micro-architectural configuration of magic-state, T-state, and cat-state factories, plus memory blocks. Within that fixed configuration, codes can still be selected dynamically at runtime, provided the code block size matches the region block size. Changing the number of factories may require hardware changes, but not a micro-architectural redesign; this makes it relatively straightforward to commission a device that uses a different code family for memory or magic-state factories. We therefore consider a large database of codes compatible with a single micro-architecture. \cref{tab:code_database} lists several promising three-ring codes from our search.
\begin{table*}[t]
\centering
\scriptsize
\setlength{\tabcolsep}{4pt}
\renewcommand{\arraystretch}{1.15}
\begin{tabular}{|c|c|c|c|c|c|c|c|c|c|}
\hline
\textbf{Family} & \textbf{$w$} & \textbf{$\ell$} & \textbf{$m$} & \textbf{$A$} & \textbf{$B$} & \textbf{$n$} & \textbf{$k$} & \textbf{$d$} & \textbf{$kd^2/n$} \\
\hline
BB & 8 & 16 & 4 & $1 + y + x + x^{3}$ & $y^{3} + x^{5}y^{3} + x^{7}y^{2} + x^{10}$ & 128 & 14 & $\le 14$ & 21.44 \\
GB & 8 & 63 & 1 & $1 + x + x^{4} + x^{36}$ & $x^{5} + x^{7} + x^{22} + x^{29}$ & 126 & 18 & $\le 12$ & 20.57 \\
GB & 8 & 63 & 1 & $1 + x + x^{3} + x^{41}$ & $x^{19} + x^{29} + x^{52} + x^{58}$ & 126 & 20 & $\le 11$ & 19.21 \\
BB & 8 & 9 & 7 & $1 + xy + x^{2}y^{4} + x^{5}y^{6}$ & $x + x^{2}y + x^{4}y^{6} + x^{7}y^{4}$ & 126 & 20 & $\le 11$ & 19.21 \\
BB & 8 & 12 & 6 & $1 + y + x + x^{5}y^{4}$ & $1 + x^{2}y^{3} + x^{6}y^{2} + x^{7}y^{5}$ & 144 & 14 & $\le 14$ & 19.06 \\
GB & 8 & 62 & 1 & $1 + x + x^{4} + x^{38}$ & $x^{8} + x^{18} + x^{34} + x^{45}$ & 124 & 12 & $\le 14$ & 18.97 \\
GB & 8 & 51 & 1 & $x^{22}+x^{26}+x^{37}+x^{50}$ & $x^{19}+x^{28}+x^{29}+x^{35}$ & $102^{*}$ & 22 & 9 & 17.47 \\
GB & 8 & 70 & 1 & $1+x+x^{4}+x^{13}$ & $1+x^{20}+x^{24}+x^{26}$ & 140 & 10 & 15 & 16.07 \\
BB & 8 & 12 & 4 & $1 + y + x + x^{4}$ & $x^{2} + x^{7}y^{2} + x^{8}y^{3} + x^{10}y^{3}$ & 96 & 10 & $\le 12$ & 15 \\
GB & 8 & 48 & 1 & $1 + x + x^{3} + x^{22}$ & $x^{8} + x^{15} + x^{24} + x^{35}$ & 96 & 10 & $\le 12$ & 15 \\
BB & 8 & 12 & 4 & $1 + y + x + x^{7}y$ & $xy + x^{3}y^{2} + x^{3}y^{3} + x^{7}$ & 96 & 22 & 8 & 14.67 \\
GB & 8 & 49 & 1 & $1 + x + x^{3} + x^{41}$ & $x^{5} + x^{22} + x^{32} + x^{48}$ & 98 & 8 & $\le 13$ & 13.8 \\
BB & 8 & 7 & 6 & $1 + x + x^{2} + x^{4}y$ & $y^{4} + xy^{5} + x^{2}y^{4} + x^{4}y^{2}$ & 84 & 8 & $\le 12$ & 13.71 \\
GB & 7 & 63 & 1 & $1 + x + x^{8} + x^{18}$ & $1 + x^{23} + x^{28}$ & 126 & 12 & $\le 12$ & 13.71 \\
GB & 7 & 63 & 1 & $1 + x + x^{3} + x^{19}$ & $x^{13} + x^{18} + x^{53}$ & 126 & 16 & $\le 10$ & 12.7 \\
GB & 8 & 42 & 1 & $1 + x + x^{3} + x^{8}$ & $1 + x^{20} + x^{30} + x^{34}$ & 84 & 16 & 8 & 12.19 \\
BB & 6 & 12 & 6 & $x^{3} + y + y^{2}$ & $y^{3} + x + x^{2}$ & $144^{\times}$ & 12 & 12 & 12 \\
BB & 6 & 12 & 6 & $1 + y + x^{3}y^{2}$ & $x + x^{2} + x^{3}y^{3}$ & 144 & 12 & $\le 12$ & 12 \\
GB & 7 & 45 & 1 & $1 + x + x^{4} + x^{40}$ & $x^{8} + x^{22} + x^{34}$ & 90 & 8 & $\le 11$ & 10.76 \\
BB & 8 & 12 & 2 & $1 + y + x + x^{4}$ & $y + xy + x^{3}y + x^{7}$ & 48 & 8 & 8 & 10.67 \\
GB & 7 & 42 & 1 & $1 + x + x^{4} + x^{9}$ & $x^{4} + x^{17} + x^{36}$ & 84 & 10 & 9 & 9.64 \\
GB & 6 & 60 & 1 & $1 + x + x^{4}$ & $x^{22} + x^{35} + x^{58}$ & 120 & 8 & $\le 12$ & 9.6 \\
BB & 6 & 15 & 3 & $x^{9} + y + y^{2}$ & $1 + x^{2} + x^{7}$ & $90^{\times}$ & 8 & 10 & 8.89 \\
BB & 6 & 15 & 3 & $1 + y + x^{3}y^{2}$ & $1 + x^{4} + x^{14}$ & 90 & 8 & $\le 10$ & 8.89 \\
GB & 6 & 49 & 1 & $1 + x + x^{3}$ & $x^{6} + x^{18} + x^{45}$ & 98 & 6 & $\le 12$ & 8.82 \\
GB & 8 & 18 & 1 & $1 + x + x^{3} + x^{10}$ & $1 + x^{13} + x^{15} + x^{16}$ & 36 & 8 & 6 & 8 \\
GB & 8 & 24 & 1 & $1 + x + x^{3} + x^{6}$ & $x^{7} + x^{14} + x^{16} + x^{17}$ & 48 & 6 & 8 & 8 \\
BB & 8 & 8 & 3 & $1 + x + x^{2}y + x^{3}y^{2}$ & $y^{2} + xy + x^{3} + x^{6}$ & 48 & 6 & 8 & 8 \\
GB & 7 & 31 & 1 & $1 + x + x^{3} + x^{27}$ & $1 + x^{17} + x^{30}$ & 62 & 10 & 7 & 7.9 \\
BB & 6 & 12 & 3 & $1 + y + x^{3}y^{2}$ & $y^{2} + x + x^{8}y$ & 72 & 8 & 8 & 7.11 \\
GB & 7 & 21 & 1 & $1 + x + x^{2} + x^{5}$ & $1 + x + x^{10}$ & 42 & 6 & 7 & 7 \\
BB & 7 & 7 & 5 & $y^{2}+x^{2}+x^{3}+x^{4}$ & $y+x+x^{3}$ & $70^{*}$ & 6 & 9 & 6.94 \\
BB & 7 & 7 & 5 & $y^{3} + xy + x^{2}y^{2} + x^{5}$ & $xy^{3} + x^{5} + x^{6}y^{2}$ & 70 & 6 & 9 & 6.94 \\
GB & 6 & 28 & 1 & $1 + x + x^{5}$ & $1 + x^{12} + x^{22}$ & 56 & 6 & 8 & 6.86 \\
BB & 6 & 6 & 6 & $x^{3} + y + y^{2}$ & $y^{3} + x + x^{2}$ & $72^{\times}$ & 12 & 6 & 6 \\
BB & 6 & 6 & 6 & $1 + y + x^{3}y^{2}$ & $1 + x + x^{2}y^{3}$ & 72 & 12 & 6 & 6 \\
GB & 7 & 21 & 1 & $1 + x + x^{3} + x^{13}$ & $x^{3} + x^{19} + x^{20}$ & 42 & 10 & 5 & 5.95 \\
GB & 8 & 22 & 1 & $1 + x + x^{3} + x^{12}$ & $1 + x^{4} + x^{9} + x^{15}$ & 44 & 4 & 8 & 5.82 \\
BB & 8 & 11 & 2 & $1 + x + x^{2} + x^{4}$ & $y + xy + x^{3} + x^{7}$ & 44 & 4 & 8 & 5.82 \\
GB & 6 & 31 & 1 & $1 + x + x^{12}$ & $1 + x^{3} + x^{8}$ & 62 & 10 & 6 & 5.81 \\
BB & 6 & 8 & 3 & $1 + y + x^{2}y^{2}$ & $x + x^{4}y^{2} + x^{5}y$ & 48 & 4 & 8 & 5.33 \\
GB & 6 & 24 & 1 & $1 + x + x^{8}$ & $x^{2} + x^{19} + x^{21}$ & 48 & 4 & 8 & 5.33 \\
GB & 7 & 24 & 1 & $1 + x + x^{3} + x^{19}$ & $x^{9} + x^{11} + x^{16}$ & 48 & 4 & 8 & 5.33 \\
BB & 6 & 7 & 3 & $1 + x + x^{3}y$ & $x + x^{5}y^{2} + x^{6}y$ & 42 & 6 & 6 & 5.14 \\
GB & 6 & 21 & 1 & $1 + x + x^{3}$ & $x + x^{12} + x^{20}$ & 42 & 6 & 6 & 5.14 \\
BB & 6 & 5 & 3 & $1 + x + x^{3}y$ & $y + xy^{2} + x^{4}y^{2}$ & 30 & 8 & 4 & 4.27 \\
GB & 6 & 15 & 1 & $1 + x + x^{4}$ & $x^{2} + x^{9} + x^{11}$ & 30 & 8 & 4 & 4.27 \\
BB & 5 & 15 & 2 & $1 + x + x^{5}$ & $x^{4} + x^{10}y$ & 60 & 4 & 8 & 4.27 \\
GB & 5 & 30 & 1 & $1 + x^{2} + x^{10}$ & $x^{17} + x^{20}$ & 60 & 4 & 8 & 4.27 \\
BB & 5 & 8 & 3 & $1 + y + xy^{2}$ & $x^{3}y^{2} + x^{6}y^{2}$ & 48 & 4 & 7 & 4.08 \\
GB & 5 & 24 & 1 & $1 + x + x^{8}$ & $x^{2} + x^{23}$ & 48 & 4 & 7 & 4.08 \\
BB & 5 & 6 & 3 & $1 + y + x^{2}$ & $y^{2} + xy$ & 36 & 4 & 6 & 4 \\
BB & 7 & 7 & 5 & $1 + x + x^{2} + x^{5}$ & $1 + x^{4} + x^{5}$ & 70 & 30 & 3 & 3.86 \\
GB & 8 & 27 & 1 & $x^{8}+x^{13}+x^{15}+x^{16}$ & $x^{7}+x^{8}+x^{10}+x^{15}$ & $54^{*}$ & 2 & 10 & 3.7 \\
GB & 5 & 15 & 1 & $1 + x + x^{5}$ & $x + x^{4}$ & 30 & 4 & 5 & 3.33 \\
GB & 5 & 45 & 1 & $1 + x^{3} + x^{15}$ & $x^{30} + x^{39}$ & 90 & 12 & 5 & 3.33 \\
GB & 5 & 60 & 1 & $1 + x^{4} + x^{20}$ & $x^{18} + x^{30}$ & 120 & 16 & 5 & 3.33 \\
BB & 5 & 6 & 4 & $1 + x^{2} + x^{4}y^{2}$ & $xy + x^{4}y^{2}$ & 48 & 8 & 4 & 2.67 \\
GB & 5 & 24 & 1 & $1 + x^{2} + x^{10}$ & $1 + x^{18}$ & 48 & 8 & 4 & 2.67 \\
BB & 5 & 12 & 6 & $1 + y^{2} + x^{3}y$ & $x^{3}y^{3} + x^{6}$ & 144 & 24 & 4 & 2.67 \\
GB & 5 & 72 & 1 & $1 + x^{12} + x^{24}$ & $x^{30} + x^{48}$ & 144 & 24 & 4 & 2.67 \\
\hline
\end{tabular}
\caption{
Code database with code family (GB = Generalized Bicycle, BB = Bivariate Bicycle), check weight $w$, matrix dimensions $(\ell,m)$, and defining $(A,B)$ from Bravyi et al.~\cite{bravyi2024high}, where $x=S_{\ell}\otimes I_m$, $y=I_{\ell}\otimes S_m$, and $A,B$ are binary matrix-polynomials in $x,y$. Entries marked with $^*$ correspond to the codes \code{102}, \code{54}, and \code{70} used elsewhere in the paper, and entries marked with $^\times$ correspond to the IBM BB codes cited in the memory section. Among codes with identical values of $kd^2/n$, we omit entries whose $n$ and $k$ are both exactly twice those of another listed code with the same value of $kd^2/n$.
When the distance is reported as $d \le d_{\mathrm{ub}}$, the upper bound $d_{\mathrm{ub}}$ is estimated with the qdist RND algorithm using two rounds of $10^6$ iterations; according to the probability estimate in Ref.~\cite{pryadko2023qdist}, this bound is probabilistically tight with probability $>99.999999\%$. When the distance is reported without the inequality, it was computed exactly by exhaustive enumeration.
}
\label{tab:code_database}
\end{table*}

\section{Additional simulation results for the streaming decoder} \label{app:sec_additional_simulation_stream_decoder}

\begin{figure}[t!]
    \centering
    \includegraphics[width=0.95\linewidth]{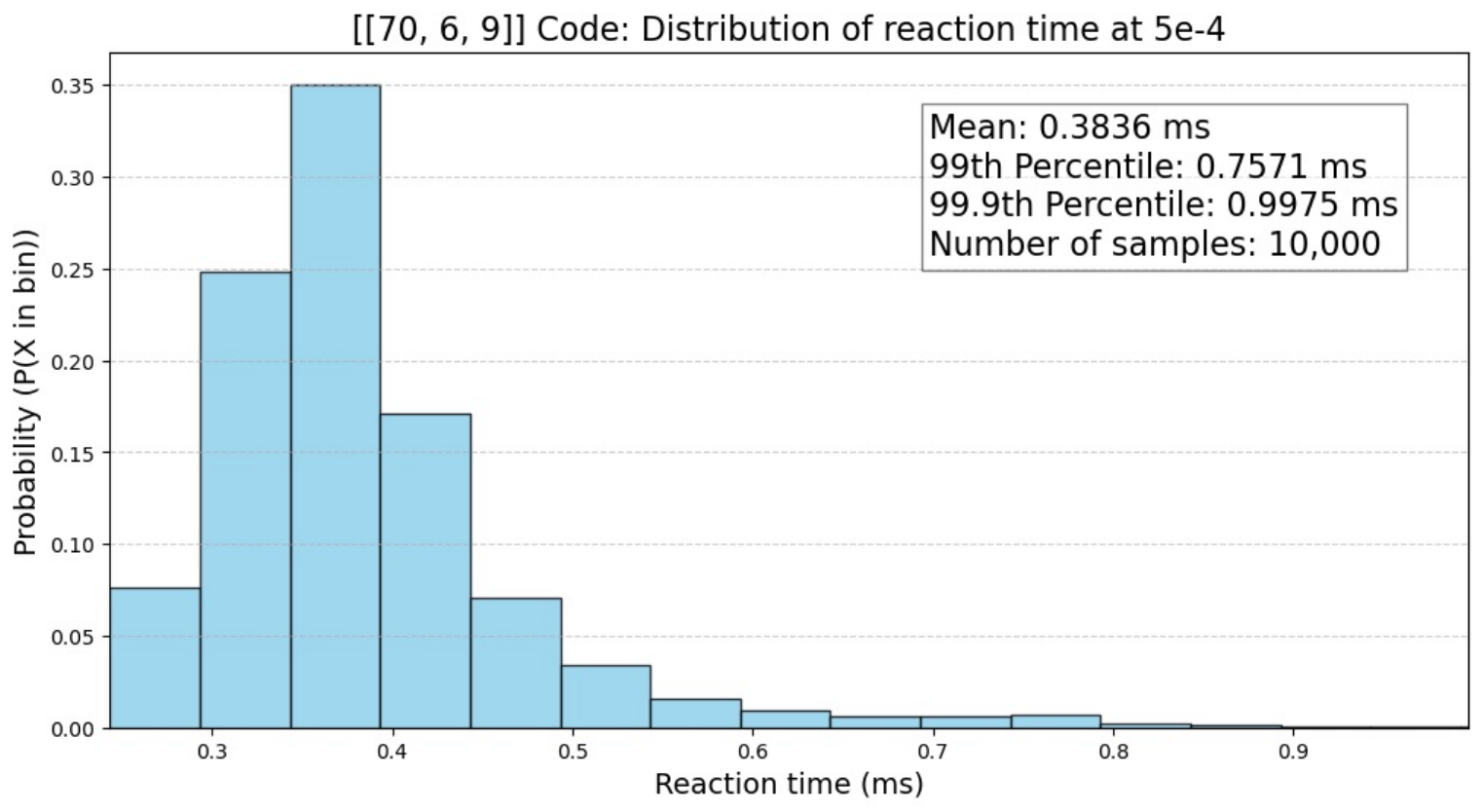}
    \vspace*{0.05in}

    \includegraphics[width=0.95\linewidth]{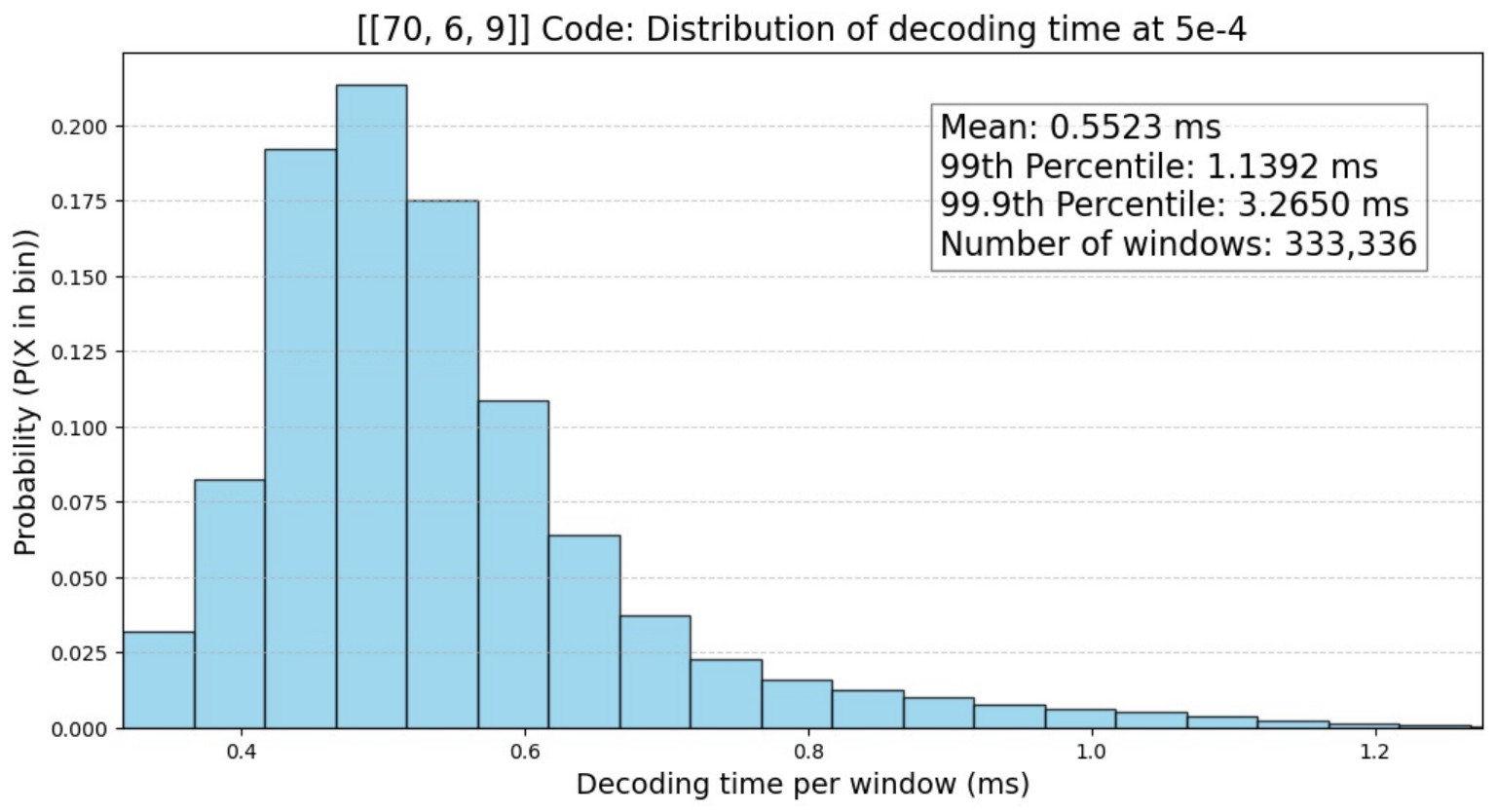}
    \vspace*{0.05in}

    \includegraphics[width=0.95\linewidth]{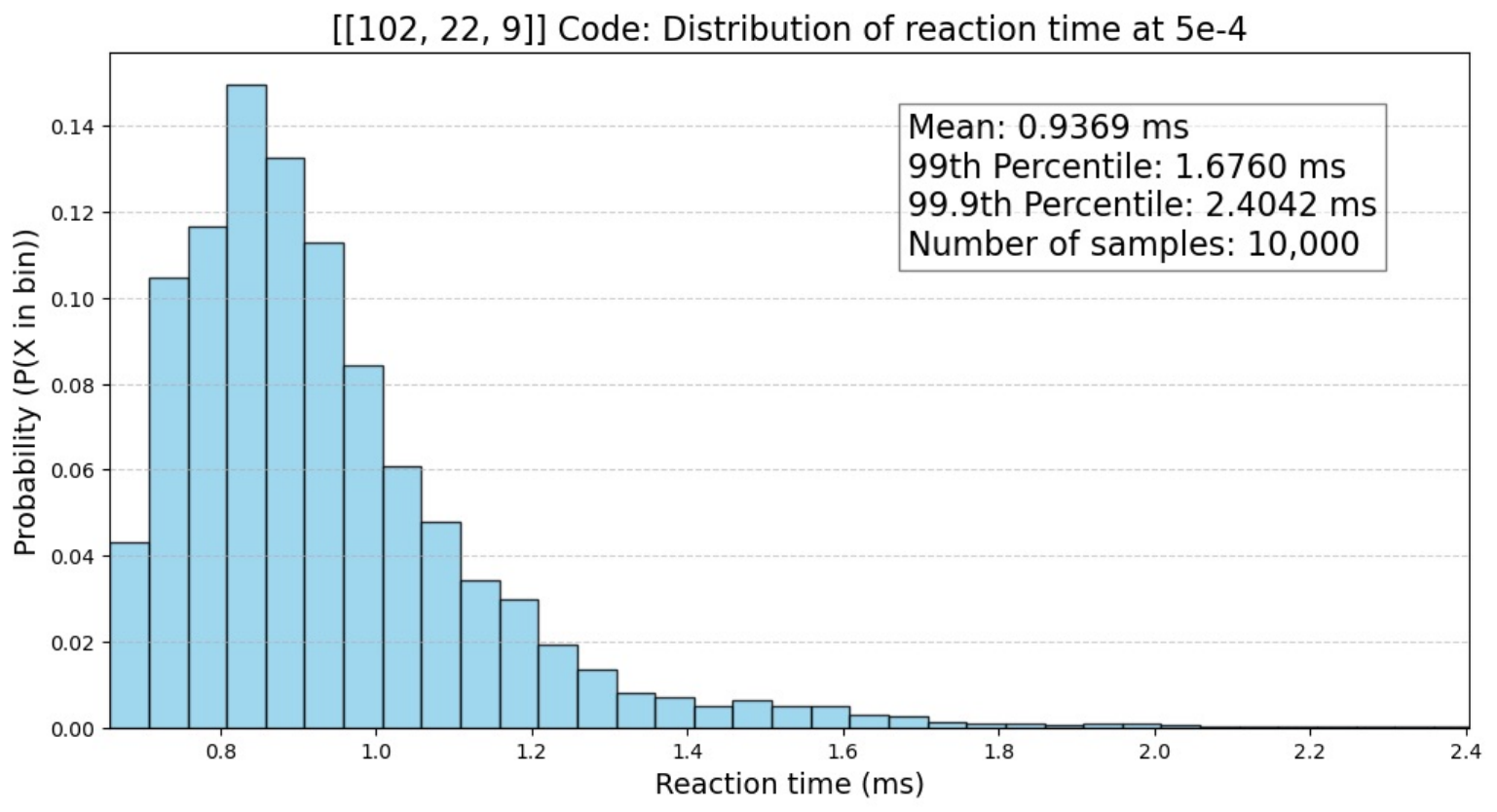}
    \vspace*{0.05in}

    \includegraphics[width=0.95\linewidth]{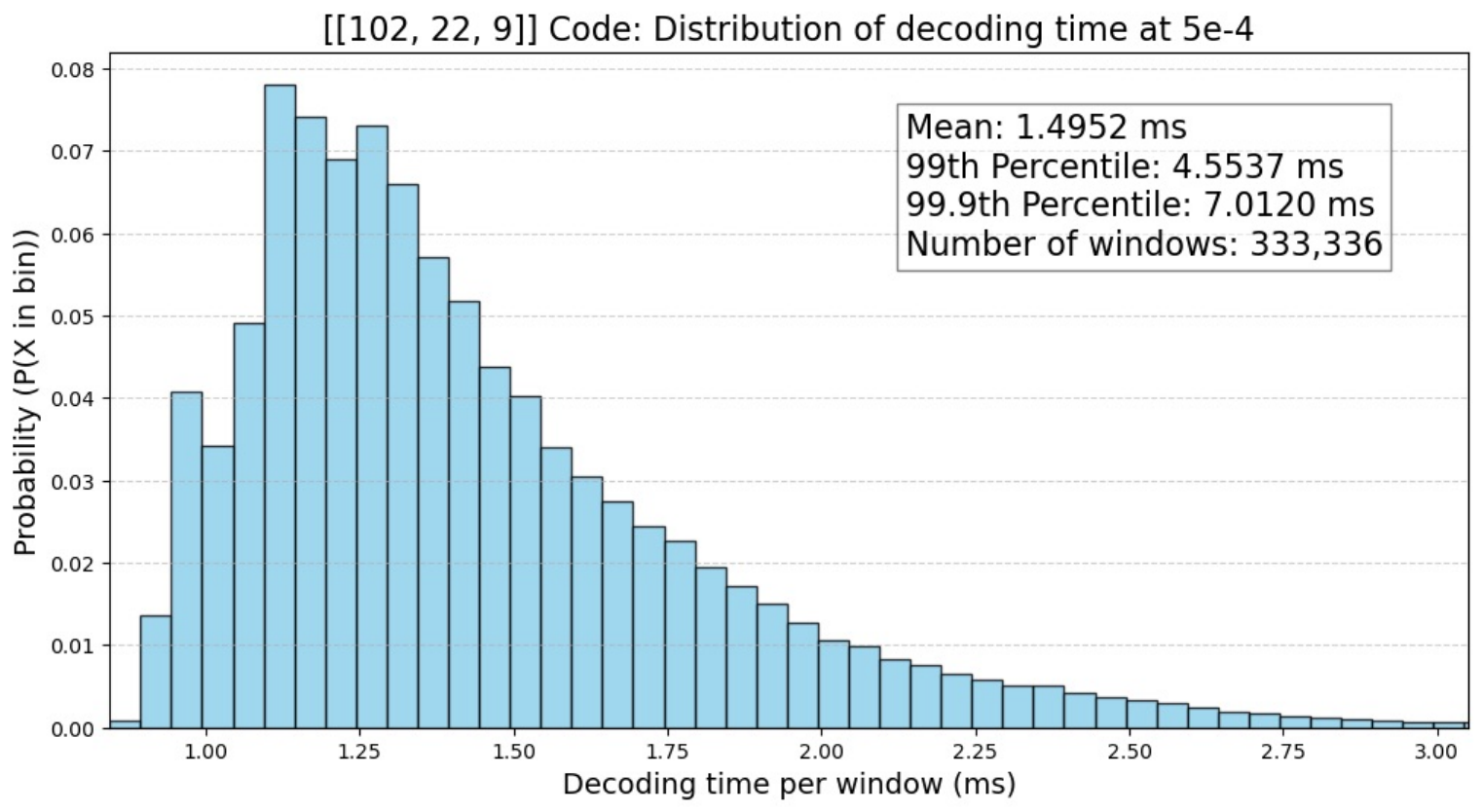}
    \caption{Probability distribution histograms for decoding time per window and reaction time at physical error rate $5\times 10^{-4}$ using a $(5,3)$ sliding window beam search decoder. Decoding time statistics are derived from a single run of $1,000,008$ SECs ($333,336$ windows). Reaction time statistics are generated from $10,000$ independent runs of $1,000,008$ SECs each.}
    \label{fig:decoding_and_reaction_time_5e-4}
\end{figure}

In \cref{sec:decoder}, we presented the probability distribution histograms for decoding time per window and reaction time at physical error rate $10^{-4}$ in \cref{fig:decoding_and_reaction_time_1e-4}. In this appendix,
we provide these decoding latency statistics at physical error rate $5 \times 10^{-4}$ in \cref{fig:decoding_and_reaction_time_5e-4}. These results, obtained using the $(5,3)$ sliding-window beam search decoder for both the \code{70} and \code{102} codes, show that the average decoding time per SEC and reaction time remain below 1ms. However, as expected at this higher noise level, the tail of the distribution is more pronounced compared to the $p = 10^{-4}$ case.

In \cref{tab:logical_error_rate_per_cycle}, we investigate the impact of the number of SECs on the logical error rate per SEC for the \code{102} code under the $(5,3)$ streaming configuration. We evaluated the decoder across 9, 18, 27, and 45 SECs, all of which are multiples of the code distance $d=9$. As shown in \cref{tab:logical_error_rate_per_cycle}, the logical error rate per SEC exhibits only a marginal increase as the number of SECs grows, demonstrating the stability of the streaming decoder over a high volume of SECs.

\begin{table*}[]
    \centering
    \begin{tabular}{|c|c|c|c|}
    \hline
       number of SECs & LER per SEC at $p=0.001$ & LER per SEC at $p=0.002$\\
        \hline
        45  & 6.84e-07 & 3.87e-05 \\
        \hline
        27  & 6.09e-07 & 3.75e-05 \\
        \hline
        18  & 6.49e-07 & 3.53e-05 \\
        \hline
        9  & 5.87e-07 & 3.49e-05 \\
        \hline
    \end{tabular}
    \caption{Logical Error Rate (LER) per SEC for the \code{102} code under the $(5,3)$ sliding window decoder. The number of SECs in the first column is chosen to be a multiple of the code distance. Physical error rates are set at $p = 0.001$ and $p = 0.002$.}
    \label{tab:logical_error_rate_per_cycle}
\end{table*}

\section{Tables of weight-reduced logical operators}
\label{appendix:Tables of weight-reduced logical operators}
\cref{tab:Symp_TransH} gives the weight-reduced logical operators for \code{54}.
\cref{tab:Symp_MEK} gives the weight-reduced logical operators for \code{70}.
\cref{tab:Symp_X_Totoro,tab:Symp_Z_Totoro} give the weight-reduced logical operators for \code{102}.

\begin{table*}
\begin{centering}
\begin{tabular}{|c|c|c|c|}
\hline 
${\cal L}_{x,1}$ & ${\cal L}_{x,2}$ & $i$ & $j$\tabularnewline
\hline 
\hline
$x^{3}+x^{5}+x^{13}+x^{14}+x^{20}+x^{22}$ & $x^{3}+x^{6}+x^{8}+x^{19}+x^{26}$ & 0 & 1\tabularnewline
\hline 
$x^{26}$ & $x^{4}+x^{7}+x^{9}+x^{11}+x^{15}+x^{17}+x^{18}+x^{19}+x^{20}+x^{26}$ & 1 & 0\tabularnewline
\hline 
\end{tabular}
\par\end{centering}
\caption{Table of generating polynomials for weight-reduced symplectic basis operators for the Q54 code.
Each $\bar{X}_i$ basis operator is taken to be the first row of the matrix $[\mathcal{L}_{x,1} | \mathcal{L}_{x,2}]$.
Similarly, each $\bar{Z}_j$ basis operator is taken to be $[\mathcal{L}_{x,2}^{-1} | \mathcal{L}_{x,1}^{-1}]$.
Note, that the use-case of this code is the existence of transversal logical $H$ gates, for which self-similarity of $\bar{X}_i$ and its symplectic partner $\bar{Z}_j$ is required.
Though it may not be obvious in this table, that self-similarity holds, up to multiplication by stabilizers.}
\label{tab:Symp_TransH}
\end{table*}

\begin{table*}
    \begin{centering}
    \begin{tabular}{|c|c|c|c|c|c|}
    \hline 
    ${\cal L}_{x,1}$ & ${\cal L}_{x,2}$ & ${\cal K}_{x,i}$ & $i$ & ${\cal K}_{z,j}$ & $j$\tabularnewline
    \hline 
    \hline
    \multirow{3}{*}{$1+xy^{4}+x^{4}y^{4}+x^{6}y^{4}+x^{2}+x^{6}y$} & \multirow{3}{*}{$x^{2}+x^{4}y^{2}+x^{5}y^{2}$} & $x^{2}$ & 3 & $\ensuremath{x}$ & 1\tabularnewline
    \cline{3-6} \cline{4-6} \cline{5-6} \cline{6-6} 
     &  & $x^{4}y$ & 5 & $\ensuremath{x^{3}}$ & 4\tabularnewline
    \cline{3-6} \cline{4-6} \cline{5-6} \cline{6-6} 
     &  & $y^{2}$ & 0 & $\ensuremath{x^{6}}$ & 2\tabularnewline
    \hline 
    $1+x^{2}+x^{3}+x^{2}y$ & $x^{2}+x^{3}y+x^{4}y+x^{5}y+xy^{3}$ & $xy$ & 1 & $x^{3}$ & 3\tabularnewline
    \hline 
    \multirow{2}{*}{$1+x^{2}+x^{6}y+xy^{2}+x^{3}y^{2}+y^{3}$} & \multirow{2}{*}{$x^{3}+x^{4}+y^{2}$} & $x^{3}$ & 2 & $x^{3}$ & 5\tabularnewline
    \cline{3-6} \cline{4-6} \cline{5-6} \cline{6-6} 
     &  & $xy^{4}$ & 4 & $x^{5}$ & 0\tabularnewline
    \hline 
    \end{tabular}
    \par\end{centering}
    \caption{Table of generating polynomials for a minimum-weight symplectic basis operators for the BB7-$[[70,6,9]]$ code.
    Each $\bar{X}_i$ basis operator is taken to be the first row of the matrix $\mathcal{K}_{x,i} [\mathcal{L}_{x,1} | \mathcal{L}_{x,2}]$.
    Similarly, each $\bar{Z}_j$ basis operator is taken to be $\mathcal{K}_{z,j} [\mathcal{L}_{x,2}^{-1} | \mathcal{L}_{x,1}^{-1}]$.
    Together the operators define a symplectic basis such that $\bar{X}_i \bar{Z}_j = (-1)^{\delta_{i,j}} \bar{Z}_j \bar{X}_i $.}
    \label{tab:Symp_MEK}
\end{table*}

\begin{table*}
\begin{centering}
\begin{tabular}{|c|c|c|c|}
\hline 
${\cal L}_{x,1}$ & ${\cal L}_{x,2}$ & $i$ & ${\cal K}_{x,i}$\tabularnewline
\hline 
\hline 
$x^{24}+x^{33}+x^{36}+x^{39}+x^{48}$ & $x^{12}+x^{21}+x^{27}+x^{33}+x^{39}+x^{48}$ & 20 & 1\tabularnewline
\hline 
\multirow{4}{*}{$\ensuremath{x^{2}+x^{5}+x^{29}+x^{36}+x^{39}+x^{46}}$} & \multirow{4}{*}{$\ensuremath{x^{11}+x^{18}+x^{35}+x^{45}}$} & 13 & 1\tabularnewline
\cline{3-4} \cline{4-4} 
 &  & 21 & $x^{9}$\tabularnewline
\cline{3-4} \cline{4-4} 
 &  & 14 & $x^{13}$\tabularnewline
\cline{3-4} \cline{4-4} 
 &  & 16 & $x^{44}$\tabularnewline
\hline 
$\ensuremath{x^{19}+x^{24}+x^{30}+x^{35}+x^{42}}$ & $\ensuremath{x^{21}+x^{26}+x^{27}+x^{39}}$ & 1 & 1\tabularnewline
\hline 
\multirow{2}{*}{$\ensuremath{x^{9}+x^{15}+x^{30}+x^{31}+x^{36}}$} & \multirow{2}{*}{$\ensuremath{x^{12}+x^{21}+x^{25}+x^{29}+x^{38}}$} & 12 & $1$\tabularnewline
\cline{3-4} \cline{4-4} 
 &  & 15 & $x^{47}$\tabularnewline
\hline 
$\ensuremath{x^{10}+x^{19}+x^{20}+x^{23}+x^{26}+x^{30}+x^{33}+x^{40}}$ & $\ensuremath{x^{25}+x^{43}}$ & 11 & 1\tabularnewline
\hline 
$\ensuremath{x^{9}+x^{19}+x^{20}+x^{24}+x^{25}+x^{30}+x^{36}}$ & $\ensuremath{x^{2}+x^{21}+x^{38}}$ & 10 & 1\tabularnewline
\hline 
\multirow{2}{*}{$\ensuremath{x^{11}+x^{24}+x^{27}+x^{48}}$} & \multirow{2}{*}{$\ensuremath{x^{3}+x^{30}+x^{33}+x^{40}+x^{44}}$} & 2 & $1$\tabularnewline
\cline{3-4} \cline{4-4} 
 &  & 0 & $x^{45}$\tabularnewline
\hline 
$x^{13}+x^{19}+x^{23}+x^{33}+x^{39}+x^{46}+x^{48}$ & $x^{4}+x^{25}+x^{39}$ & 9 & 1\tabularnewline
\hline 
$x^{2}+x^{18}+x^{27}+x^{34}+x^{35}+x^{36}+x^{37}$ & $x^{21}+x^{32}+x^{44}$ & 18 & 1\tabularnewline
\hline 
$x^{8}+x^{9}+x^{15}+x^{31}$ & $x+x^{10}+x^{29}+x^{33}+x^{48}$ & 7 & 1\tabularnewline
\hline 
$x^{5}+x^{14}+x^{21}+x^{37}+x^{46}$ & $x^{20}+x^{24}+x^{35}+x^{44}$ & 8 & 1\tabularnewline
\hline 
\multirow{2}{*}{$x^{5}+x^{25}+x^{28}+x^{35}+x^{36}+x^{46}$} & \multirow{2}{*}{$x^{18}+x^{31}+x^{34}+x^{48}$} & 5 & 1\tabularnewline
\cline{3-4} \cline{4-4} 
 &  & 3 & $\ensuremath{x^{37}}$\tabularnewline
\hline 
$x^{5}+x^{6}+x^{13}+x^{48}+x^{49}$ & $x^{5}+x^{18}+x^{45}+x^{49}$ & 19 & 1\tabularnewline
\hline 
$x^{5}+x^{6}+x^{14}+x^{21}+x^{22}+x^{23}+x^{28}$ & $x^{21}+x^{26}+x^{49}$ & 4 & 1\tabularnewline
\hline 
$x^{14}+x^{18}+x^{21}+x^{28}+x^{38}+x^{49}$ & $1+x^{20}+x^{31}+x^{41}+x^{44}$ & 17 & 1\tabularnewline
\hline 
$x^{15}+x^{31}+x^{39}+x^{40}+x^{41}+x^{47}$ & $x^{21}+x^{45}+x^{48}$ & 6 & 1\tabularnewline
\hline 
\end{tabular}
\par\end{centering}
\caption{Table of generating polynomials for weight-reduced $\bar{X}$ symplectic basis operators for the \code{102} code.
Each $\bar{X}_i$ basis operator is taken to be the first row of the matrix $\mathcal{K}_{x,i} [\mathcal{L}_{x,1} | \mathcal{L}_{x,2}]$.}
\label{tab:Symp_X_Totoro}
\end{table*}

\begin{table*}
\begin{centering}
\begin{tabular}{|c|c|c|c|}
\hline 
${\cal L}_{z,1}$ & ${\cal L}_{z,2}$ & $j$ & ${\cal K}_{z,j}$\tabularnewline
\hline 
\hline 
\multirow{2}{*}{$x^{3}+x^{21}$} & \multirow{2}{*}{$x^{6}+x^{13}+x^{16}+x^{20}+x^{23}+x^{26}+x^{27}+x^{36}$} & 20 & 1\tabularnewline
\cline{3-4} \cline{4-4} 
 &  & 8 & $x^{31}$\tabularnewline
\hline 
$x^{10}+x^{34}+x^{37}$ & $1+x^{16}+x^{35}+x^{41}+x^{42}+x^{43}$ & 13 & 1\tabularnewline
\hline 
\multirow{4}{*}{$x^{9}+x^{13}+x^{20}+x^{23}+x^{50}$} & \multirow{4}{*}{$x^{5}+x^{26}+x^{29}+x^{42}$} & 1 & 1\tabularnewline
\cline{3-4} \cline{4-4} 
 &  & 10 & $x$\tabularnewline
\cline{3-4} \cline{4-4} 
 &  & 12 & $x^{32}$\tabularnewline
\cline{3-4} \cline{4-4} 
 &  & 4 & $x^{45}$\tabularnewline
\hline 
$x+x^{8}+x^{12}+x^{16}+x^{44}$ & $x+x^{37}+x^{42}+x^{43}$ & 11 & 1\tabularnewline
\hline 
\multirow{2}{*}{$x^{6}+x^{7}+x^{12}+x^{45}$} & \multirow{2}{*}{$x^{3}+x^{9}+x^{14}+x^{42}+x^{49}$} & 2 & 1\tabularnewline
\cline{3-4} \cline{4-4} 
 &  & 3 & $x^{5}$\tabularnewline
\hline 
$x^{6}+x^{13}+x^{17}+x^{44}$ & $x^{5}+x^{6}+x^{13}+x^{14}+x^{49}$ & 9 & 1\tabularnewline
\hline 
\multirow{2}{*}{$x^{17}+x^{31}+x^{34}+x^{47}$} & \multirow{2}{*}{$x^{9}+x^{19}+x^{29}+x^{30}+x^{37}+x^{40}$} & 18 & 1\tabularnewline
\cline{3-4} \cline{4-4} 
 &  & 7 & $x^{21}$\tabularnewline
\hline 
\multirow{2}{*}{$x+x^{12}+x^{16}+x^{43}$} & \multirow{2}{*}{$x^{15}+x^{22}+x^{31}+x^{41}+x^{50}$} & 5 & 1\tabularnewline
\cline{3-4} \cline{4-4} 
 &  & 17 & $x^{42}$\tabularnewline
\hline 
$x^{3}+x^{6}+x^{7}+x^{10}+x^{44}$ & $x^{5}+x^{14}+x^{23}+x^{43}+x^{46}+x^{49}+x^{50}$ & 19 & 1\tabularnewline
\hline 
$x^{3}+x^{12}+x^{50}$ & $x+x^{7}+x^{10}+x^{13}+x^{14}+x^{19}+x^{23}$ & 14 & 1\tabularnewline
\hline 
$x^{3}+x^{14}+x^{22}+x^{29}+x^{43}$ & $1+x^{14}+x^{15}+x^{22}+x^{23}+x^{35}+x^{50}$ & 21 & 1\tabularnewline
\hline 
\multirow{2}{*}{$1+x^{10}+x^{27}+x^{34}$} & \multirow{2}{*}{$x^{6}+x^{9}+x^{16}+x^{40}+x^{43}+x^{50}$} & 15 & 1\tabularnewline
\cline{3-4} \cline{4-4} 
 &  & 6 & $x^{47}$\tabularnewline
\hline 
$x^{14}+x^{18}+x^{21}+x^{34}+x^{45}+x^{49}$ & $x^{5}+x^{15}+x^{16}+x^{50}$ & 16 & 1\tabularnewline
\hline 
$x+x^{12}+x^{27}+x^{37}+x^{50}$ & $x^{8}+x^{15}+x^{24}+x^{30}$ & 0 & 1\tabularnewline
\hline 
\end{tabular}
\par\end{centering}
\caption{Table of generating polynomials for weight-reduced $\bar{Z}$ symplectic basis operators for the \code{102} code.
Each $\bar{Z}_j$ basis operator is taken to be $\mathcal{K}_{z,j} [\mathcal{L}_{z,1} | \mathcal{L}_{z,2}]$.}
\label{tab:Symp_Z_Totoro}
\end{table*}

\begin{table}
    \begin{centering}
    \begin{tabular}{|c|c|c|c|c|}
    \hline 
    Weight & $X$ & $Z$ & $Y$ & Other\tabularnewline
    \hline 
    \hline 
    9 & 19 & 16 & - & -\tabularnewline
    \hline 
    10 & 37 & 34 & - & -\tabularnewline
    \hline 
    11 & 38 & 41 & - & -\tabularnewline
    \hline 
    12 & 74 & 85 & - & -\tabularnewline
    \hline 
    13 & 122 & 105 & - & -\tabularnewline
    \hline 
    14 & 208 & 184 & - & 4\tabularnewline
    \hline 
    15 & 327 & 325 & - & 19\tabularnewline
    \hline 
    16 & 446 & 460 & 1 & 65\tabularnewline
    \hline 
    17 & 399 & 394 & 4 & 249\tabularnewline
    \hline 
    18 & 116 & 138 & 12 & 528\tabularnewline
    \hline 
    19 & 7 & 11 & 11 & 1098\tabularnewline
    \hline 
    20 & - & - & 13 & 1701\tabularnewline
    \hline 
    21 & - & - & 23 & 2865\tabularnewline
    \hline 
    22 & - & - & 45 & 5127\tabularnewline
    \hline 
    23 & - & - & 100 & 6947\tabularnewline
    \hline 
    24 & - & - & 136 & 6375\tabularnewline
    \hline 
    25 & - & - & 253 & 4884\tabularnewline
    \hline 
    26 & - & - & 397 & 4516\tabularnewline
    \hline 
    27 & - & - & 458 & 3763\tabularnewline
    \hline 
    28 & - & - & 275 & 1672\tabularnewline
    \hline 
    29 & - & - & 63 & 313\tabularnewline
    \hline 
    30 & - & - & 2 & 13\tabularnewline
    \hline 
    \end{tabular}
    \par\end{centering}
    \caption{Table of all 43,725 Pauli operators of logical weight 1, 2, and 3 for the Q102 code, arranged by physical weight of their stabilizer-optimized representatives.
    This reflects data shown in \cref{fig:LW3_Totoro}.}
    \label{tab:LogicalWeight3Totoro}
\end{table}
\clearpage
\section{Limitations of architectures restricted to biplanar Tanner graphs}
\label{app:sec_biplanar_obstruction}

As shown in \cref{app:sec_code_database_table,sec:Correction of losses and leakages}, the best-performing codes we found for the walking cat architecture in the moving-qubit noise model, as judged by encoding rate and logical error rate, have check weight $8$ (\emph{e.g.}, \code{102}).
Moreover, the same conclusion holds under the additional design constraints used in our magic factories (\cref{subsec:ch2_factory}): among codes with strongly transversal Hadamard gate and block size small enough that the needed logical operators admit physical representatives of sufficiently low weight to justify a small cat factory,
every code we found, including \code{54}, had check weight $8$.

These codes, however, fall outside the design space of superconducting-circuit memory architectures, \emph{e.g.}, \cite{bravyi2024high}.
The original work on BB code memories focused on architectures whose syndrome-extraction Tanner graphs--barring experimental progress--are restricted to bipartite thickness $2$.
We call a bipartite graph \emph{biplanar} if its edge set can be partitioned into two planar bipartite graphs on the same vertex set,
equivalently if its bipartite thickness is at most $2$. This appendix explains why these codes fall outside that setting: any syndrome-extraction Tanner graph with check degree at least $8$ cannot be biplanar; and, maximally parallel
syndrome extraction circuits for such codes, to date, require a Tanner graph that is $8$-regular. We show that such tanner graphs are not biplanar.
This is a consequence of the following:

\begin{proposition}
\label{prop:biplanar_weight_eight_obstruction}
Let $n \in \N$ and let $G=(D \sqcup C,E)$ be a bipartite Tanner graph with $|D|=|C|=n$. If every check vertex in $C$ has degree at least $8$, then $G$ is not biplanar.
 Equivalently, the bipartite thickness of $G$ is at least $3$.
\end{proposition}

\begin{proof}
Assume for contradiction that $G$ is biplanar. Then we can write
\[
E = E_1 \sqcup E_2,
\]
where each $G_i=(D \sqcup C,E_i)$ is a planar bipartite graph on the same $2n$ vertices. Since $|D \sqcup C| = 2n$, Euler's formula for planar bipartite graphs gives
\[
|E_i| \leq 2|V(G_i)| - 4 = 4n - 4
\]
for $i=1,2$. Summing over the two planar layers yields
\[
|E| = |E_1| + |E_2| \leq 8n - 8.
\]
On the other hand, summing the check degrees gives
\[
|E| = \sum_{c \in C} \deg(c) \geq 8n > 8n - 8,
\]
which is a contradiction.
\end{proof}

This result means that one cannot implement the syndrome extraction circuit of a set of weight-$w$ stabilizer generators in a biplanar way for any $w \geq 8$.
With the exception of codes that admit a another set of stabilizer generators with lower weight (which may happen but we expect this to be rare), this eliminates BB codes and GB codes with weight $\geq 8$ from the biplanar framework.

\section{Proofs of circuit identities}
\label{app:Proofs of circuit identities}

\begin{figure}
    \centering
    \includegraphics[width=0.7\linewidth]{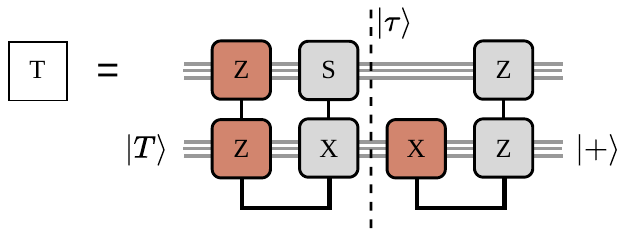}
    \caption{A $T$ gate using a $\ket{T}$ magic state, following Fig.~7 of~\cite{litinski2019game}.}
    \label{fig:T_with_T}
\end{figure}

\begin{figure}
    \centering
    \includegraphics[width=0.7\linewidth]{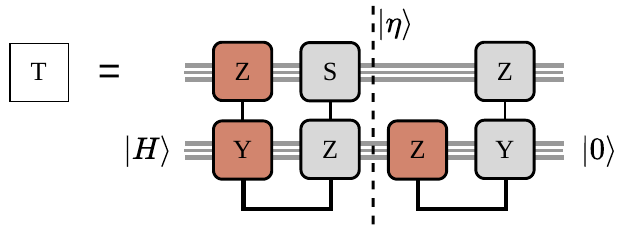}
    \caption{A $T$ gate using a $\ket{H}$ magic state.}
    \label{fig:T_with_H}
\end{figure}

\begin{proof}[Proof of \cref{prop:Tdagger_gate_by_measurements}]
Begin with \cref{fig:T_with_T}, from \cite{litinski2019game}.
Use the elementary identities $\ket{H} = SH\ket{T}$, $Z H S^{\dagger} = H S^{\dagger} Y$, $X H S^{\dagger} = H S^{\dagger} Z$, and $H S^{\dagger}\ket{+} = \ket{0}$; (Here and everywhere in this proof we ignore global phases).
Applying them to the lower register immediately yields the circuit of \cref{fig:T_with_H}.

Let's write the action of each step of \cref{fig:T_with_T} on an initial state $\ket{\psi}\ket{T} = \frac{1}{\sqrt{2}}\sum_{a,b\in\{0,1\}} c_a \ket{a} e^{i b\pi/4}\ket{b}$, up to the dashed line (we'll write only the summand, for brevity):
\begin{align}
e^{ib\pi/4}\left|a\right>\left|b\right>\,\underrightarrow{M_{Z\otimes Z}} & \,e^{i(m_{ZZ}\oplus a)\pi/4}\left|a\right>\left|m_{ZZ}\oplus a\right>\\
\underrightarrow{S^{m_{ZZ}}\otimes X^{m_{ZZ}}}\, & e^{ia\pi/4}\left|a\right>\left|a\right>
\end{align}
Here $M$ indicates a $Z\otimes Z$ measurement, and $m_{ZZ}$ its outcome.
We used the fact that $a\oplus b=m_{ZZ}$.
Thus:
\begin{align}
    \ket{\tau} = (I\otimes H S^{\dagger})\ket{\eta} = \sum_{a,b} e^{i a \pi/4} c_a \ket{a} \ket{a}
    \label{eq:T_injection_DashedLine}
\end{align}
In both figures, elements after the dashed line are {\em disentangling steps}, that follow the standard approach to disentangle the 2nd register of \cref{eq:T_injection_DashedLine}, leaving $T\ket{\psi}=\sum_{a} e^{i a \pi/4} c_a \ket{a}$.

In writing \cref{fig:architecture_overview_T_gates_by_meas}(a)-(b), we instead chose to measure the 2nd register of $\ket{\eta}=\sum_{a} e^{i a \pi/4} c_a \ket{a} \ket{Y_a}$ in a non-standard unbiased basis, $X$.
Using the elementary identity $\ket{Y_a} \propto (-i)^a\ket{X_{0}} + (-i)^{1\oplus a}\ket{X_{1}}$, we readily verify that the disentangling step of \cref{fig:architecture_overview_T_gates_by_meas}(a) instead leaves $\sum_{a} e^{-i a \pi/2} e^{i a \pi/4} c_a \ket{a}\ket{X_0} = T^{\dagger}\ket{\psi}\ket{X_0}$.
(Here $\ket{X_a}=H\ket{a}$ and $\ket{Y_a}=SH\ket{a}$ respectively.)
\end{proof}
\section{Ion loss rate}
\label{appendix:ion_loss_rate}

Since achieving low loss of ions is crucial for minimizing loss-related high-weight errors and ion reloading overheads, here we provide motivation for our estimate of $p_{\mathrm{loss}} = 10^{-7}$ per POC, which equates to one loss event per ion for approximately every 33 minutes for POC time of $200~\mathrm{\mu s}$, adapting the approach originally proposed in Ref.~\cite{wineland1998experimental} to infer the loss rate by estimating the collision rate between a trapped ion and background gas particles from its environment. 

In an isolated environment, a trapped ion in a Paul trap is lost when it escapes over the lowest saddle-point energy barrier, or equivalently in the pseudopotential approximation, when its secular kinetic energy is higher than the trap depth. In a well-designed cryogenic surface trap, electric field noise can be strongly suppressed, so that the dominant trap lifetime-limiting process is expected to be collisions between trapped ions and background neutral gas particles, as long as slow heating contributions from electric fields are balanced by scheduled cooling steps. We note that typical surface Paul traps have trap depths well over $10~\mathrm{meV}$, and $1~\mathrm{meV}$ equates to a range from tens of thousands to hundreds of thousands of phonons for typical mode frequencies in the order of one to several megahertz.

At the cryogenic temperature below 5 Kelvin, most of the background gas species are frozen or effectively adsorbed on cryogenic surfaces and residual background density is expected to be dominated by hydrogen molecules $\mathrm{H}_2$, which is the least effectively cryo-pumped species after helium; $\mathrm{H}_2$ typically enters the vacuum system from outgassing of vacuum system walls or components inside the vacuum while helium enters through vacuum leaks or permeation through glass viewports. Thus, a well-designed vacuum system is expected to have $\mathrm{H}_2$ as the leading contribution to background gas pressure; $\mathrm{H}_2$ outgassing rate inside the system can be substantially reduced by baking outgassing components at high temperature if allowed. Typical UHV systems operate with $10^{-11}~\,\mathrm{Torr}$ pressure at room temperature, and with good cryopumping, $\mathrm{H}_2$ vapor pressure can be as low $1 \times 10^{-18}\,\mathrm{Torr}$ near the target temperature of $5~\mathrm{K}$ ($6~\mathrm{K}$ in Ref.~\cite{borchert2019measurement}). We set a target $\mathrm{H}_2$ pressure level of $10^{-14}\,\mathrm{Torr}$ and show that it should be sufficient to reach the target ion lifetime and be feasible to realize. Using the ideal gas law, we obtain $\mathrm{H}_2$ gas density of approximately $1.9 \times 10^{4}~\mathrm{cm}^{-3}$ for partial pressure of $10^{-14}\,\mathrm{Torr}$ at 5 Kelvin.

Collision rate between a trapped ion and $\mathrm{H}_2$ is expressed by standard kinetic formula $n\langle\sigma v\rangle$, where $n$ is the background hydrogen gas density, and $\langle\sigma v\rangle$ is the average of the product of collisional cross-section $\sigma$ and relative velocity $v$ between the colliding $\mathrm{H}_2$ and the ion. The collisional cross-section will depend on whether the collision is inelastic or elastic. For inelastic collision, an upper limit is provided by the Langevin rate $2\pi nq\sqrt{\frac{\alpha}{\mu}}$, where $q$ is the electron charge and $\alpha$ is the polarizability of $\mathrm{H}_2$, both expressed in cgs units, and $\mu$ is the reduced mass of the ion-$\mathrm{H}_2$ system~\cite{wineland1998experimental}. Hydrogen gas has a mean polarizability of approximately $0.8~\mathring{\mathrm{A}}^3$, and thus for a heavy ion species such as ${}^{137}\mathrm{Ba}^{+}$, the aforementioned hydrogen gas density of $1.9\times10^{4}\,\mathrm{cm}^{-3}$ yields a Langevin rate of $2.8 \times 10^{-5}~\mathrm{s}^{-1}$, or one inelastic collision every 10 hours approximately. For comparison, a cryogenic ion trap that held a linear chain of 100 ${}^{171}\mathrm{Yb}^{+}$ ions for multiple hours at 4.5 Kelvin measured an inelastic collision rate of $2 \times 10^{-5} s^{-1}$ at 5 Kelvin (Fig.~10 in \cite{pagano2019cryogenic}), although the system employed a blade trap and not a surface trap.

On the other hand, elastic collision cross-section is larger, and Ref.~\cite{wineland1998experimental} reports a conservative estimate of $\langle \sigma v\rangle = \pi \Gamma(\frac{1}{3})\left(\frac{\pi\alpha q^2}{4\hbar}\right)^{2/3}\tilde{v}^{1/3}$ in cgs units where $\tilde{v} = \sqrt{2 k_B T / \mu}$ is the most probable relative speed between the ion and the colliding $\mathrm{H}_2$. Again for the aforementioned estimated $\mathrm{H}_2$ density at 5 Kelvin, the collision rate evaluates to $1.2 \times 10^{-4}\,\mathrm{s}^{-1}$, or one elastic collision every 2.3 hours. 

It should be pointed out that nominally a single elastic collision at 5 Kelvin does not provide sufficient kinetic energy for an ion to escape, as the large mass mismatch between $\mathrm{H}_{2}$ and heavy ion species such as ${}^{171}\mathrm{Yb}^{+}$ or ${}^{137}\mathrm{Ba}^{+}$ makes only a small fraction of the thermal energy transferred to the ion~\cite{pagano2019cryogenic}, with less than a few hundred milliKelvin in temperature units, or a few percent of $1~\mathrm{meV}$. So it would seem that inverting the elastic collision rate presents too conservative an estimate of the trapped ion lifetime. However, since an ion is never static in a Paul trap, it is possible to get enhanced heating from RF micromotion, where the colliding particle interrupts the ion's micromotion in such a way that the RF field ends up increasing the secular motional energy as it moves the ion in a new micromotion trajectory~\cite{major1968exchange}. Such a mechanism can greatly increase excess micromotion, especially if an ion is off the RF null as may be required during low-excitation transport through a junction~\cite{burton2023transport}. Also, given the Doppler laser cooling temperature of order one milliKelvin, an ion impacted by a single elastic collision with a $\mathrm{H}_2$ molecule might not be sufficiently cooled within the standard scheduled cooling time and might experience poor gate fidelity, such that the ion appears as a leakage or a loss within the circuit. In principle a hot data ion impacted by the collision can share its huge thermal energy with ancilla ions that it pairs with during two-qubit gates, but as explained in Sec.~\ref{sec:Correction of losses and leakages}, loss and leakage detection units should suppress the growth of such effective leakage/loss into high-weight errors. In this regard, it seems inverting the calculated elastic collision rate yields a reasonable and relevant estimate of the ion qubit lifetime within a programmed quantum circuit. Within this approach, our use of effective ion loss per every 33 minutes corresponds to roughly a factor of 5 higher elastic collision rate than what was calculated, which means we can still accommodate cryo-pumped vacuum system that yields vacuum pressure of $5 \times 10^{-14}\,\mathrm{Torr}$ at 5 Kelvin.

\end{document}